%% file: ustmain.tex
\documentclass{ustthesis}  
\pdfoutput=1
\usepackage{amssymb}
\usepackage{amsmath}
\usepackage{graphics}
\usepackage{physics}
\usepackage{hyperref}
\usepackage{mathrsfs}
\usepackage[b5paper]{geometry}
\usepackage{pdfpages}

\hypersetup{colorlinks=true,citecolor={blue},linkcolor={blue},urlcolor={blue}}
\newcommand{\mi}{\mathrm{i}}

\title{Exciton-Polaritons in Artificial Lattices and Electron Transport in Bose-Fermi Hybrid Systems}   

\author{Meng Sun}             
\college{Center for Theoretical Physics of Complex Systems, IBS School}  

\degree{Doctor of Philosophy}     
\degreedate{Mar. 26. 2020}         

\begin{document}

\setcounter{secnumdepth}{3}
\setcounter{tocdepth}{3}

\includepdf[pages=-]{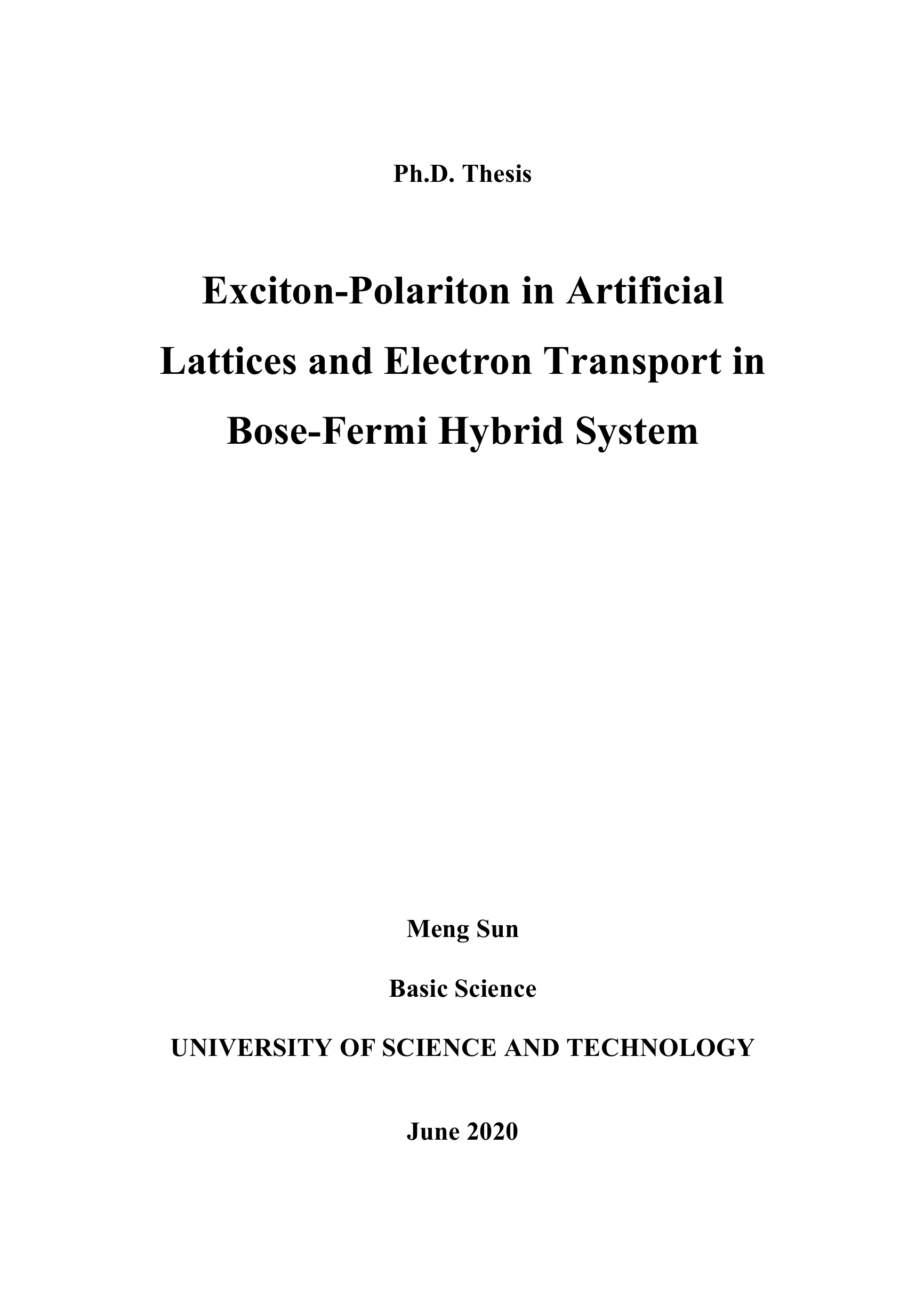}
\includepdf[pages=-]{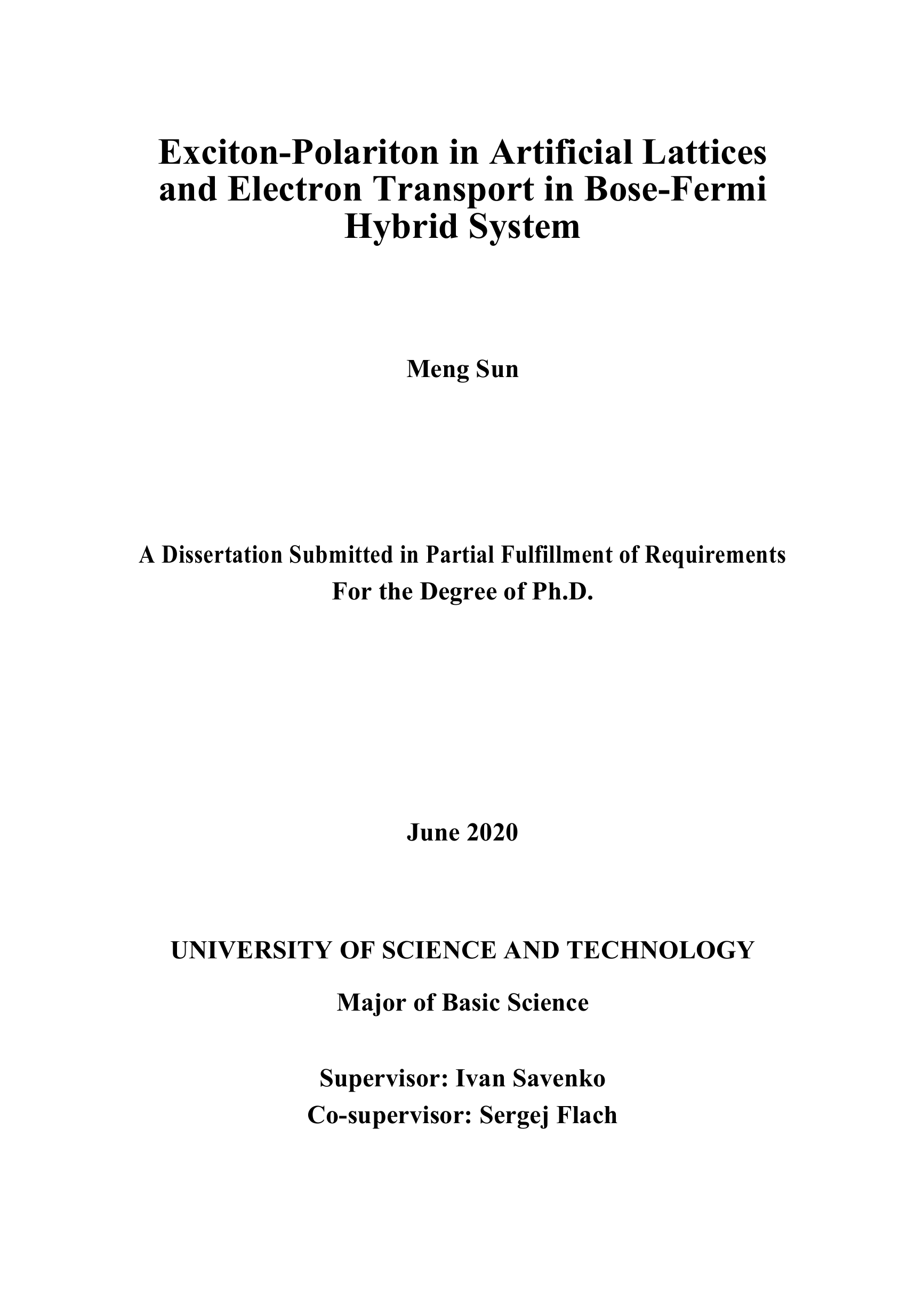}
\includepdf[pages=-]{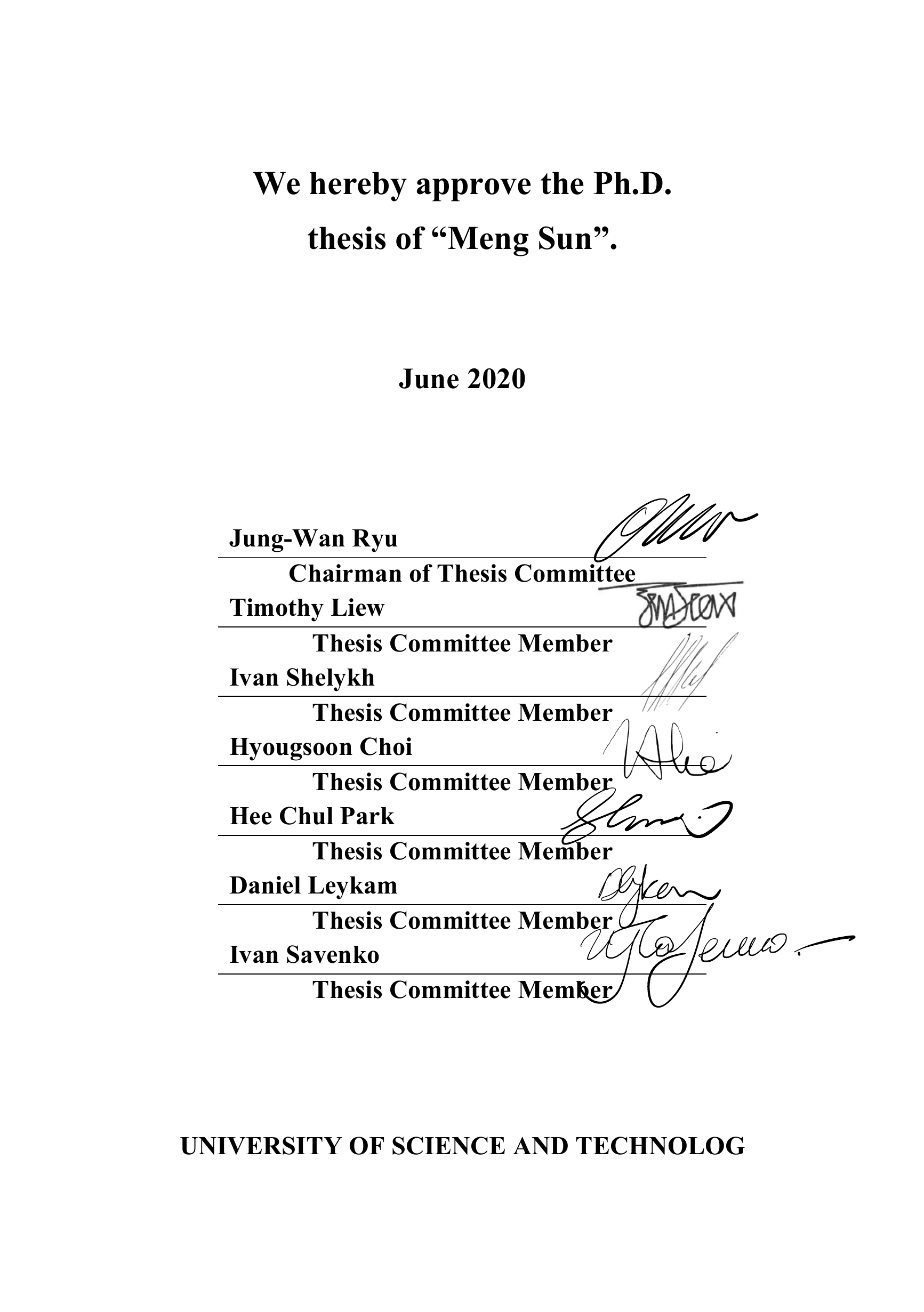}
\maketitle                  
\includepdf[pages=-]{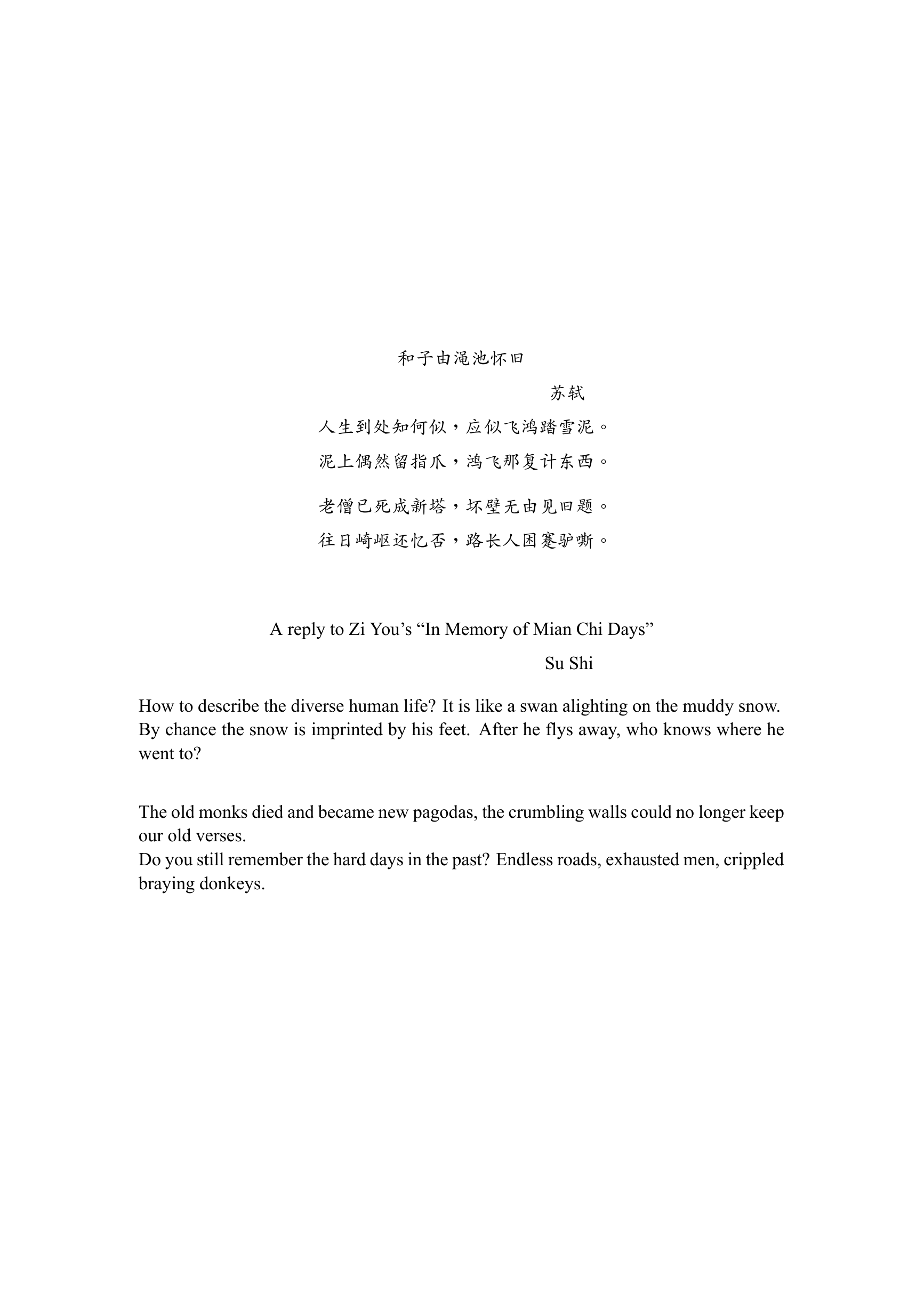}        
\include{acknowlegements_revised}   
\include{abstract_3-25_revised}          
\includepdf[pages=-]{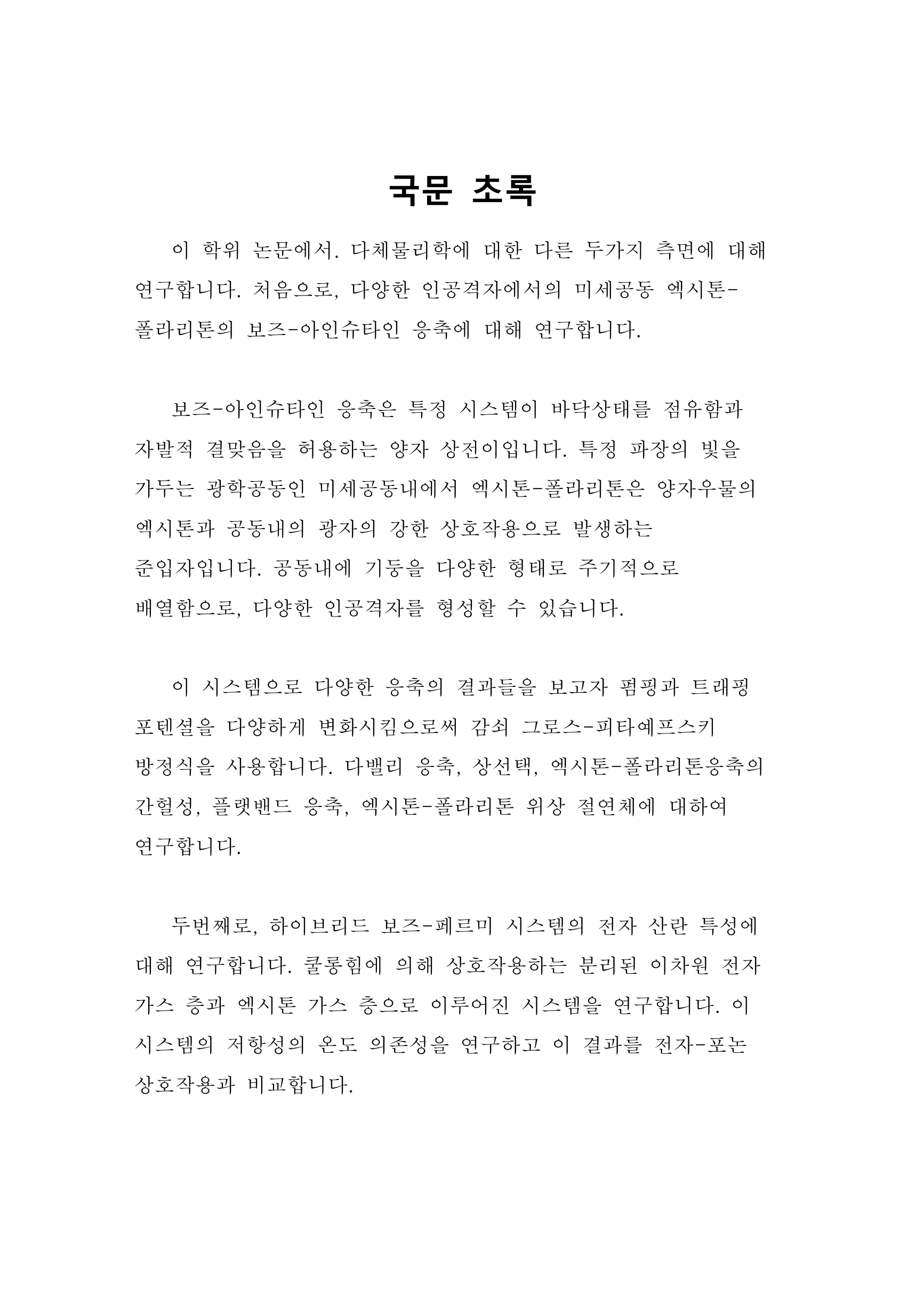}
\begin{romanpages}          
\tableofcontents            
\listoffigures              
\end{romanpages}            

\include{chapter1_2-26_revised} 
\include{chapter2_3-6_revised}
\include{chapter3_3-4_revised} 
\include{chapter4_2-28_revised}  
\include{chapter5_2-26_revised}  
\include{chapter6_3-18_revised}  
\include{conclusions_3-18_revised}

\appendix
\include{appendix2_revised} 
\include{appendix6_revised} 
\include{appendix7_revised2} 

\addcontentsline{toc}{chapter}{Bibliography}
\renewcommand{\bibname}{References}
\bibliographystyle{phaip}  
\bibliography{refs}        

\include{preface2}
\end{document}

%% file: acknowlegements_revised.tex
\begin{acknowledgements}
This work would not have been possible to complete without the support and encouragement of many people.

First and foremost, I would like to express my deepest gratitude to my supervisors Prof. Ivan Savenko and Prof. Sergej Flach.
When I first came to this institute, I was not very familiar with the topic of exciton-polariton physics, but they showed great patience in giving me plenty of complicated explanations that helped me to overcome some hard times in this period.
After I started the research, I was inspired by their broad vision about science and their powerful mathematical skills, both analytical and numerical.

I would also like to thank other collaborators in my institute who supported my Ph.D. program.
I had lots of discussions and exciting works with Prof. Daniel Leykam, Prof. Alexei Andreanov, Dr. Kristian Hauser Villegas, Dr. Anton Parafilo, and Dr. Sukjin Yoon.
I learned a lot from them from all our collaborations together.
Thanks to our institute, I had a lot of opportunities to meet and work with other internationl researchers.
I will never forget the edification given by Prof. Yuri Rubo, Prof. Timothy C. H. Liew, Prof. Vadim Kovalev, and Dr. Hugo Flayac, even though we were usually separated by a great distance.
Regarding to this thesis, I must thank Mr. Joel Rasmussen to help me with proofreading.

I started my Ph.D. in the middle of 2016, and during this program, of course, it was not always a pleasant time.
I must thank my companions, Mr. Ihor Vakulchyk and other Ph.D students in my institute.
We chatted together, played games together, and attended conferences and workshops together.
Without them, I think my memory about my Ph.D. program would be gray and dull.

I also want to thank the Center for Theoretical Physics of Complex Systems at the Institute for Basic Science.
Without their help and financial support, my life in Korea would have been much harder.

At last, I must thank my family for their endless encouragement and unconditional support.
It is their early education in my childhood that lead me to this scientific road.
\end{acknowledgements}

%% file: abstract_3-25_revised.tex
\begin{abstract}
In this thesis, we study two different aspects of many-particle physics.
In the first part, we study the Bose--Einstein condensation of microcavity exciton-polaritons in different artificial lattices.

Bose--Einstein condensation is a quantum phase transition, which allows the system to macroscopically occupy its ground state and develop coherence spontaneously.
Often studied in microcavities, which are optical cavities that trap light at specific wavelengths, exciton-polaritons are a kind of quasiparticle arising from the strong coupling between quantum well excitons and cavity photons.
By periodically aligning cavity pillars in different patterns, one can achieve different artificial lattice structures.

With this setup, we apply the driven-dissipative Gross--Pitaevskii equations to investigate the different consequences of the condensation by changing the pumping schemes and the design of the trapping potentials. Topics include multivalley condensation, phase selection and intermittency of exciton-polariton condensation, flat band condensation, and exciton-polariton topological insulators.

In the second part of this thesis, we focus on the electron-scattering properties of a hybrid Bose--Fermi system.
We consider a system consisting of a spatially separated two-dimensional electron gas layer and an exciton gas layer that interacts via Coulomb forces.
We study the temperature dependence of the system's resistivity with this interlayer electron--exciton interaction and compare the results with the electron--phonon interaction.
\end{abstract}

%% file: chapter1_2-26_revised.tex
\chapter{Exciton-polariton condensation in artificial lattices: An introduction}\label{CHONE}
In the last three decades, a two-dimensional system with strong light-matter interaction called a semiconductor microcavity has become a platform to observe Bose--Einstein condensation (BEC) in condensed matter physics.
The emergence of exciton-polaritons is the result of strong coupling between quantum well (QW) excitons and cavity photons in such semiconductor microcavities.
These quasi-particles were first predicted by Hopfield~\cite{Hopfield:1958aa} in the context of bulk semiconductors as the new eigenstates of a light-crystal Hamiltonian.
In 1992, the first experimental observation~\cite{Weisbuch:1992aa} of the strong coupling between excitons and photons in semiconductor microcavities was reported.
In 1996, it was proposed that in the ground state of the lower branch of exciton-polariton dispersion, quasi-BEC can form~\cite{Imamogu:1996aa}.
This prediction was later corroborated by several experimental works~\cite{Deng:2002aa,Kasprzak:2006aa,Balili:2007aa}.
Compared to conventional exciton BEC~\cite{Hanamura:1977aa} and cold atom BEC~\cite{Davis:1995aa,Pethick:2002tn}, polariton condensates have several advantages in different aspects.
\begin{itemize}
    \item \textit{Effective mass.}
In the vicinity of the ground state, the effective mass of polaritons is four orders of magnitude smaller than the mass of bare excitons.
This means that the critical temperature of the BEC for polaritons can be four orders of magnitude higher than the critical temperature of excitons~\cite{Pethick:2002tn}.
    \item \textit{Coherence.}
Due to the photonic component, polaritons can easily extend a coherent wave function in space despite the presence of crystal defects and disorder, which in the case of excitons can be easily localized.
    \item \textit{Lifetime.}
The lifetime of the polaritons ranges between $1$--$30$ ps~\cite{Wertz:2010aa} and up to $300$ ps~\cite{Sun:2017aa} with different Q factors of cavities and pumping in different materials.
This dynamical nature of the polariton condensates provides an experimental platform to study not only standard BEC physics but also non-equilibrium open systems consisting of highly degenerate interacting boson gases.
In contrast to equilibrium condensation, where only the lowest energy state can be macroscopically occupied, polaritons can form condensation in different states.
    \item \textit{Measurement.}
Experimentally, microcavity polaritons are one of the most accessible BEC systems.
This is because there is a one-to-one correspondence between the polaritons in mode $k_\parallel$ and the wavefunction of the emitted photon.
\end{itemize}

In the coming sections in this chapter, we will discuss some basic concepts about exciton-polaritons in BEC systems.
In Section~\ref{se:Ch1_microcavity_and_cp}, we will give a brief introduction to microcavities and cavity photons.
In Section~\ref{se:CH1_qw_ex}, we will discuss excitons, which are the matter part of polaritons, and then in Section~\ref{se:CH1_ex_pl} we will briefly review the basic properties of exciton-polaritons.
Last, we will discuss polariton condensates in Section~\ref{se:CH1_EP-BEC}.

\section{Microcavities and cavity photons}\label{se:Ch1_microcavity_and_cp}
\begin{figure}[ht]
    \centering
    \includegraphics[width=0.55\linewidth]{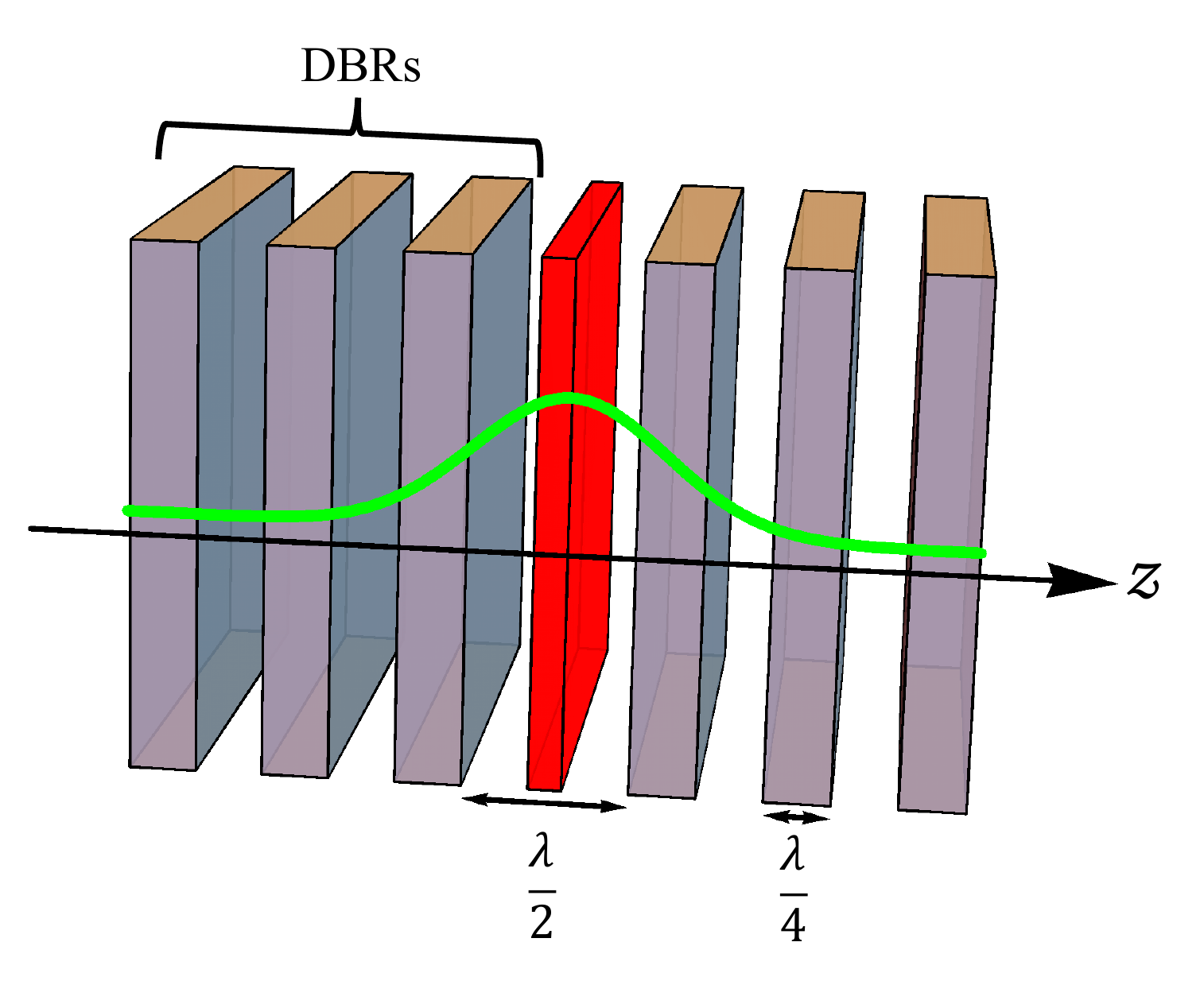}
    \caption[Microcavity schematic]{Schematic for a cavity along the $z$ direction. A cavity of thickness $\frac{\lambda}{2}$ is sandwiched by distributed Bragg reflectors (DBRs) of thickness $\frac{\lambda}{4}$. The QW is in the middle of the cavity (red). Light intensity (green curve) indicates that the cavity photons are localized inside the cavity.}
    \label{fig:Ch1_cavity}
\end{figure}
In Fig.~\ref{fig:Ch1_cavity}, we show a typical semiconductor microcavity structure.
The semiconductor QW is located in the middle of the cavity between two distributed Bragg reflectors (DBRs).
The DBRs are made of several layers with alternating high and low reflection indexes, and for each layer the optical thickness is $\frac{\lambda}{4}$, where $\lambda$ is the given wavelength of light.
The design of DBRs lets light of wavelength $\lambda$ have constructive interference when the light reflects in the interface, which creates a stop band to forbid transmission.
As a result, when the wavelength of the incident light is within the stop band, the DBRs can be regarded as a high-reflectance mirror.
With DBRs placed on either side of the cavity, which has an optical thickness of $n\times\frac{\lambda}{2}$ ($n\in \mathbf{N}$), the microcavity forms a resonance for the light of wavelength $\lambda$.
Such a resonance significantly enhances the amplitude of the light in the microcavity compared to free space, as shown in Fig.~\ref{fig:Ch1_cavity}.

The photon field for incident light in the planar cavity is confined only in the $z$ direction where the DBRs grow, while the $x$ and $y$ directions (also called in-plane) are free.
Considering light with incidence angle $\theta$ along the $z$ direction, the energy dispersion is
\begin{equation}
    E_{cav} = \frac{\hbar c}{n_c}\sqrt{k_\bot^2 + k_\parallel^2},
    \label{eq:Ch1_disp_cav}
\end{equation}
where $c$ is the speed of light, $n_c$ is the reflection index of the cavity, $k_\bot=n_c\frac{2\pi}{\lambda}$ is the wavevector in the direction where the photon field is confined ($z$ direction), and $k_\parallel$ is the in-plane wavevector.
With the refraction law
\begin{equation}
    \frac{\sin \theta_1}{\sin \theta_2} = \frac{n_2}{n_1},
    \label{eq:Ch1_refraction}
\end{equation}
we can get
\begin{equation}
    k_\parallel = k_\bot\tan\left[ \sin^{-1}\left( \frac{\sin \theta}{n_c} \right) \right].
    \label{eq:Ch1_k_p}
\end{equation}
When $k_\parallel \ll k_\bot$, we have $k_\parallel \approx \frac{2\pi \theta}{\lambda}$, and in this region, we can approximate the dispersion relationship by
\begin{equation}
    E_{cav} \approx \frac{\hbar c k_\bot}{n_c}\left( 1 + \frac{k_\parallel^2}{2k_\bot^2}\right),
    \label{eq:Ch1_E_k}
\end{equation}
with the definition: $E_{cav}\left( k_\parallel=0 \right) = \frac{\hbar c k_\bot}{n_c}$ and $m_{cav} = \frac{\hbar n_c k_\bot}{c} = E_{cav}\left(k_\parallel=0\right)\left(\frac{n_c}{c}\right)^{2}$.
Then Eq.~\eqref{eq:Ch1_E_k} can be simplified to
\begin{equation}
    E_{cav} = E_{cav}\left( k_\parallel =0\right) + \frac{\hbar^2 k_\parallel^2}{2 m_{cav}}.
    \label{eq:Ch1_E_k_2}
\end{equation}
As we can see in the calculations, due to the confinement, the energy of the photons has a parabolic dispersion and a finite effective mass in the in-plane direction.
Lastly, we need to mention that the typical effective mass of cavity photons is much smaller that the free electron mass, in most cases being $m_{cav}\approx 10^{-5} m_e$.

\section{Quantum well excitons}\label{se:CH1_qw_ex}
An exciton is a quasi-particle arising from the bound state of an electron in the conduction band and a hole in the valence band being attached to each other by Coulomb interaction.
It is electrically neutral and exists in different materials such as insulators and semiconductors.

Named after Gregory Wannier and Nevill Francis Mott, the Wannier--Mott exciton is one type of exciton typically found in semiconductors.
Due to the strong screening effect in solids and the small effective mass of the hole compared to the electron, the binding energy of Wannier--Mott excitons is around $10$--$100$ meV and the Bohr radius is around $1$--$10$ nm, which is larger than typical lattice spacing~\cite{Hanamura:1977aa}.

In exciton-polariton physics, excitons are usually confined in two-dimensional (2D) semiconductor QWs, in which the thickness of the QW is comparable to the Bohr radius of the exciton .
In most cases, the behaviour of QW excitons can be regarded as 2D quasi-particles.

\section{Exciton-polariton} \label{se:CH1_ex_pl}
The exciton-polariton is the consequence of strong coupling between light and matter, in which the light component is the cavity photon as discussed in Sec.~\ref{se:Ch1_microcavity_and_cp} and the matter component is the Wannier--Mott exciton from Sec.~\ref{se:CH1_qw_ex}.

\subsection{Basic Hamiltonian}
When considering exciton-polaritons, the QWs are often made from $\mathrm{InGaAlAs}$ alloys.
These materials generate $J=1$ heavy-hole excitons, where $J$ is the angular momentum on a given axis.
When the coupling strength between the exciton and cavity photon is much larger than the rate of decay and decoherence, it is claimed that excitons and cavity photons reach the strong coupling regime.
In this strong coupling regime, instead of treating excitons and cavity photons independently, we have to consider a new quasi-particle called an exciton-polariton (or polariton for short).

Neglecting the fast oscillating terms by the rotating wave approximation, one can write the system Hamiltonian with cavity photons and excitons in the style of second quantization, as
\begin{equation}
    \hat{H} = \sum_\mathbf{k} E_{cav}\left(\mathbf{k},k_\bot\right)\hat{a}^\dagger_\mathbf{k}\hat{a}_\mathbf{k}+E_{exc}\left(\mathbf{k}\right)\hat{b}_\mathbf{k}^\dagger\hat{b}_\mathbf{k}+\Omega\left( \hat{a}^\dagger_\mathbf{k}\hat{b}_\mathbf{k} + \hat{a}_\mathbf{k}\hat{b}^\dagger_\mathbf{k}\right),
    \label{eq:CH1_pl_ham}
\end{equation}
where $\hat{a}^\dagger_\mathbf{k}$ is the creation operator of cavity photons with the in-plane wave vector $\mathbf{k}$, for which we simplify our notation by $\mathbf{k}_\parallel \to \mathbf{k}$,
$\hat{b}_\mathbf{k}^\dagger$ is the creation operator of the QW excitons with the in-plane wave vector $\mathbf{k}$,
$\Omega$ is the exciton-photon dipole interaction strength usually called Rabi splitting,
and $E_{cav}\left(\mathbf{k},k_\bot\right)$ is the kinetic energy of the cavity photon.
By denoting the detuning parameter $\delta$ to show the energy difference between excitons and cavity photons, $\delta \equiv E_{exc}\left(\mathbf{k}=0\right) - E_{cav}\left( \mathbf{k}=0\right)$, we can write the Hamiltonian in matrix form as follows
\begin{equation}
    H =
    \begin{bmatrix}
        \frac{\hbar^2 k^2}{2m_{cp}} & \Omega \\
        \Omega &  \frac{\hbar^2 k^2}{2m_{ex}} - \delta
    \end{bmatrix}
    \label{eq:Ch1_pl_matrix},
\end{equation}
where $m_{cp}$ and $m_{ex}$ are the effective mass of cavity photon and exciton, respectively.

\begin{figure}[ht]
    \centering
    \includegraphics[width=0.65\linewidth]{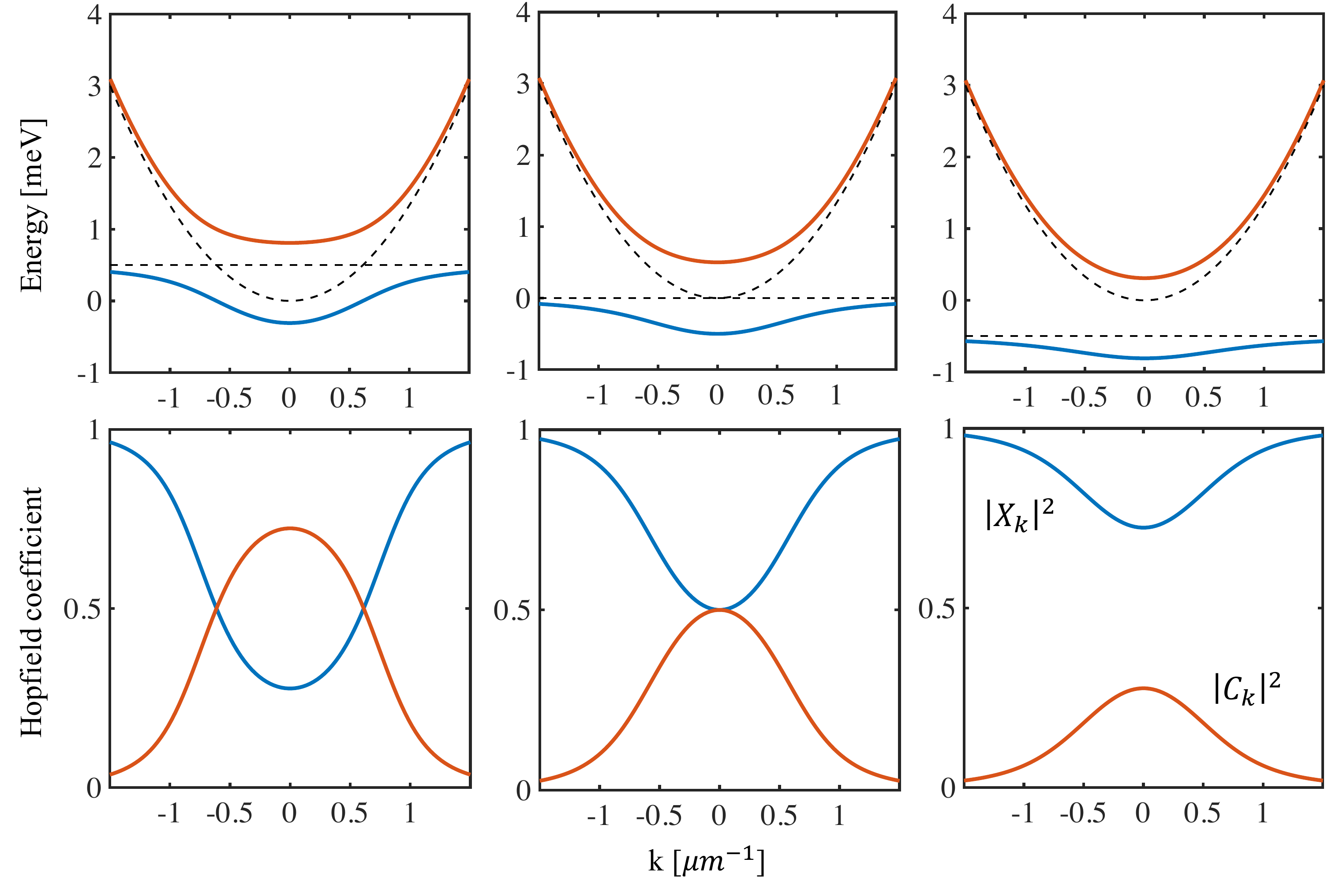}
    \caption[Polariton dispersion with different detuning]{Upper panel: Polariton dispersion from left to right with negative, zero, and positive detuning. Dashed lines correspond to the bare exciton and cavity photon mode. Lower panel: Corresponding Hopfield coefficients for the lower-branch polariton.}
    \label{fig:Ch1_disp}
\end{figure}

One can easily diagonalize the Hamiltonian in Eq.~\eqref{eq:CH1_pl_ham} with the two eigenvalues given by
\begin{equation}
    E_{lp,up} = \frac{1}{2}\left[ E_{cav} + E_{exc} \pm \sqrt{4\Omega^2+\left(E_{exc}-E_{cav}\right)^2}\right],
    \label{eq:Ch1_eigv}
\end{equation}
where $E_{lp}$ and $E_{up}$ are the lower-branch and upper-branch of the polariton having lower and higher eigenenergies, respectively.
The corresponding eigenvectors, which are usually called Hopfield coefficients, indicate the mixing of the excitonic and photonic components of the polariton and can be expressed by~\cite{Flayac:2012aa}
\begin{large}
\begin{eqnarray}
X_\mathbf{k}^L &= \frac{-\Omega \sqrt{2}}{\sqrt{4\Omega^2-\left(E_{cav}-E_{exc}\right)\left[ E_{exc}-E_{cav} + \sqrt{\left(E_{cav}-E_{exc} \right)^2+4\Omega^2}\right] }}, \\
X_\mathbf{k}^U &= \frac{\Omega \sqrt{2}}{\sqrt{4\Omega^2-\left(E_{exc}-E_{cav}\right)\left[ E_{exc}-E_{cav} + \sqrt{\left(E_{cav}-E_{exc} \right)^2+4\Omega^2}\right] }}, \\
C_\mathbf{k}^L &= \frac{\sqrt{4\Omega^2-\left( E_{cav}-E_{exc}\right) \left[E_{exc}-E_{cav} +\sqrt{\left( E_{cav}-E_{exc}\right)^2+4\Omega^2}\right]}}{\sqrt{2\left(  E_{cav}-E_{exc}\right)^2 +8\Omega^2}}, \\
C_\mathbf{k}^U &= \frac{\sqrt{4\Omega^2+\left( E_{cav}-E_{exc}\right) \left[E_{cav}-E_{exc} +\sqrt{\left( E_{cav}-E_{exc}\right)^2+4\Omega^2}\right]}}{\sqrt{2\left(  E_{cav}-E_{exc}\right)^2 +8\Omega^2}}.
\label{eq:Ch1_eigenvector}
\end{eqnarray}
\end{large}
For the polariton operators we have
\begin{eqnarray}
    \hat{c}_\mathbf{k} &= X_\mathbf{k}^L\hat{b}_\mathbf{k} + C_\mathbf{k}^L\hat{a}_\mathbf{k},\\
    \hat{c}_\mathbf{k}^{\dagger} &= X_\mathbf{k}^L\hat{b}^\dagger_\mathbf{k} + C_\mathbf{k}^L\hat{a}_\mathbf{k}^\dagger,\\
    \hat{d}_\mathbf{k} &= X_\mathbf{k}^U\hat{b}_\mathbf{k} + C_\mathbf{k}^U\hat{a}_\mathbf{k},\\
    \hat{d}_\mathbf{k}^\dagger &= X_\mathbf{k}^U\hat{b}_\mathbf{k}^\dagger + C_\mathbf{k}^U\hat{a}_\mathbf{k}^\dagger,
    \label{eq:Ch1_hope}
\end{eqnarray}
where $\hat{c}_\mathbf{k}$ ($\hat{d}_\mathbf{k}$) is the annihilation operator of the lower-branch (upper-branch) polariton,
and $X_\mathbf{k}^{L,U}$ and $C_\mathbf{k}^{L,U}$ represent the excitonic and photonic parts of the polariton, respectively, where $\abs{X_\mathbf{k}^{L,U}}^2 + \abs{C_\mathbf{k}^{L,U}}^2=1$.
In Fig.~\ref{fig:Ch1_disp}, we plot the dispersion of upper- and lower-branch polaritons with the corresponding lower-branch Hopfield coefficient with positive ($\delta=0.5$~meV), zero ($\delta = 0$~meV), and negative ($\delta=-0.5$~meV) detuning.
It should be noted that in exciton-polariton physics, and in particular exciton-polariton condensation, the lower branch is typically the main focus, so we will only discuss the lower-branch exciton-polaritons throughout this dissertation.

The effective mass of a polariton is the harmonic mean of the effective mass of its exciton and cavity photon.
For the lower-branch polaritons, we have
\begin{equation}
    \frac{1}{m_{lp}} = \frac{\abs{X_\mathbf{k}^L}^2}{m_{ex}} + \frac{\abs{C_\mathbf{k}^L}^2}{m_{cp}}.
    \label{eq:Ch1_mass}
\end{equation}
Due to the fact that the effective mass of excitons is much larger than that of cavity photons, the lower-branch polaritons at $\mathbf{k} \sim 0$ can be approximated by
\begin{equation}
    m_{lp} \approx \frac{m_{cav}}{\abs{C_\mathbf{k}^L}^2},~ ~ ~\textrm{when}~ \mathbf{k} \sim 0.
\end{equation}
\begin{equation}
    T_c = \left(\frac{n}{\zeta\left(3/2\right)}\right)^{2/3} \frac{2\pi\hbar^2}{mk_B} \approx 3.3125 \frac{\hbar^2 n^{2/3}}{mk_B},
    \label{eq:Ch1_t_c}
\end{equation}
where $n$ is the density of particle, $m$ is the mass of the particle and $\zeta$ is the Riemann zeta function.
Thus given the same particle density, the temperature required for polaritons to reach condensation is orders of magnitude higher than the temperature for excitons.

\subsection{Polariton decay}
Because the cavity photons can escape from the microcavity, which de-stabilizes the bound state between excitons and photons, polaritons have a finite lifetime.
Let us consider $\gamma_{cp}$ as the decay rate of the cavity photons due to leakage from imperfect DBRs and $\gamma_{ex}$ as the decay rate of the excitons.
Then the corresponding non-Hermitian Hamiltonian is
\begin{equation}
        H =
    \begin{pmatrix}
        \frac{\hbar^2 k^2}{2m_{cp}} - \mi\hbar\gamma_{cp} & \Omega \\
        \Omega &  \frac{\hbar^2 k^2}{2m_{ex}}-\mi\hbar\gamma_{ex} - \delta
    \end{pmatrix}.
    \label{eq:Ch1_lifetime}
\end{equation}
We consider two cases separately: the strong and weak coupling regimes.
The difference between strong and weak coupling depends on the coupling strength, $\Omega$, and the difference between the decay rates of the excitons and cavity photons, $\gamma_{cp} - \gamma_{ex}$.
For strong coupling regime, one has the condition such that $2\Omega \gg \hbar\left(\gamma_{cp}-\gamma_{ex} \right)$, which indicates that the excitation can coherently transfer between photon and exciton at least once.
\begin{figure}[ht]
    \centering
    \includegraphics[width=0.65\textwidth]{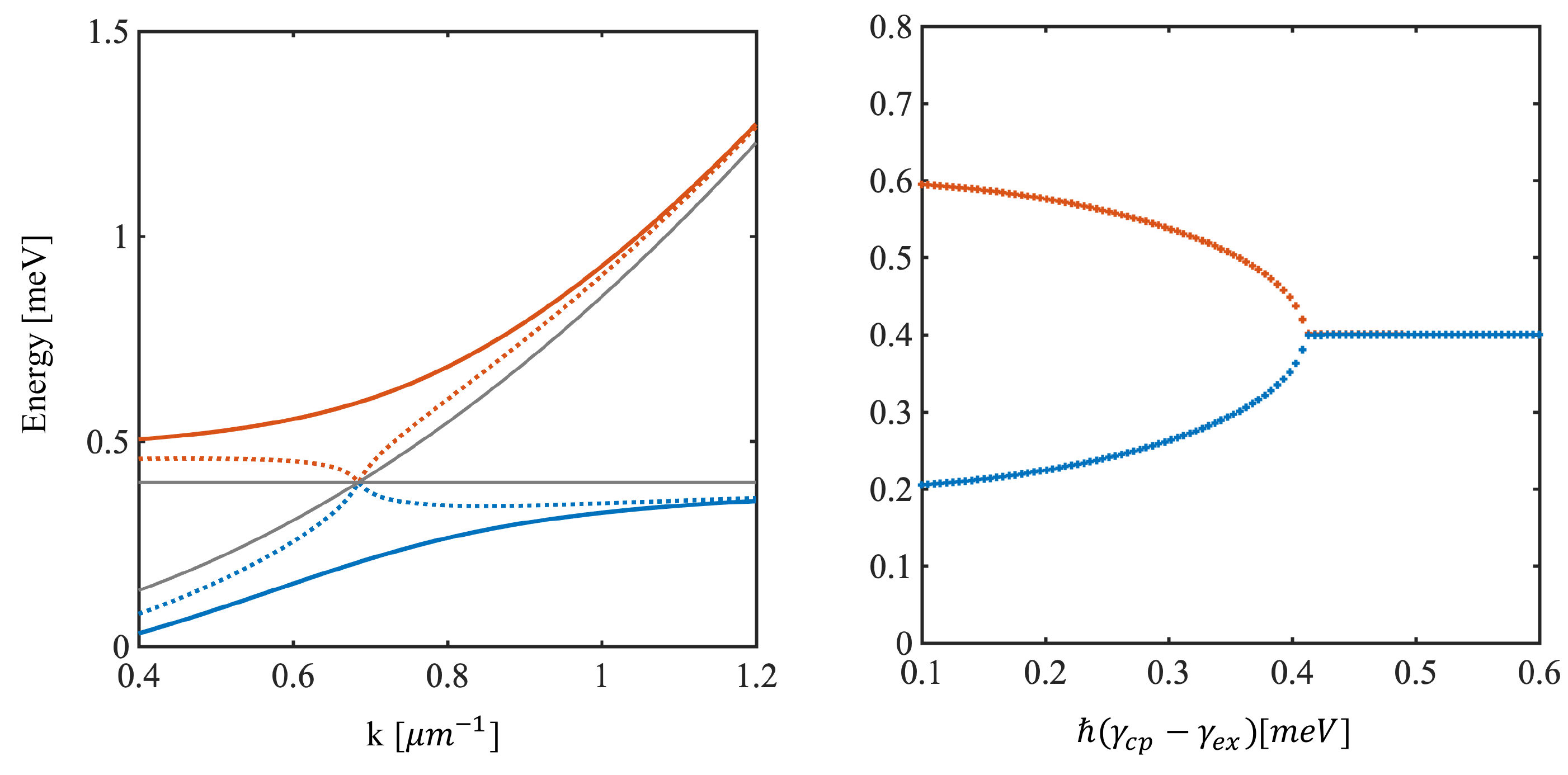}
    \caption[Polariton dispersion with decay]{Strong coupling regime (solid lines) and weak coupling regime (dotted lines) for the upper branch (red) and lower branch (blue). Left panel: Real parts of the eigenvalues. The grey lines show the uncoupled case for cavity photons and excitons. Right panel: The real part of eigenenergy for lower and upper branch of polariton at the point where the bare exciton and cavity photon cross as a function of decay rate difference between exciton and cavity photon.}
    \label{fig:Ch1_lifetime}
\end{figure}

In Fig.~\ref{fig:Ch1_lifetime} left panel, we plot the real part of the eigenvalues of Eq.~\eqref{eq:Ch1_lifetime} in different coupling regimes.
The parameters are chosen as follows: coupling strength $\Omega=0.2$~meV, detuning $\delta=0.4$~meV, exciton decay rate $\hbar\gamma_{ex} = 0.01$~meV, cavity photon decay rate $\hbar\gamma_{cp} = 0.1$~meV for the strong coupling case and $\hbar\gamma_{cp} = 0.41$~meV for thee weak coupling case.
In the right panel, we show the behaviour of the system gradually convert from the strong coupling to the weak coupling regime as increase the difference of the decay rate.
The energy gap is closed when $2\Omega=\hbar\left(\gamma_{cp}-\gamma_{ex}\right)$.

In the strong coupling limit, the two eigenvalues show an anticrossing behaviour and can be regarded as two polariton eigenmodes.
In polariton physics, we usually consider microcavities to have the following property: $\Omega \gg \hbar\gamma_{cp} \gg \hbar\gamma_{ex}$.
Then, Eq.~\eqref{eq:Ch1_eigv} can give us a nice approximation of the eigenenergies of the polariton.

\subsection{Exciton-polariton interaction}
One important difference between polaritons and cavity photons is the fact that polaritons can interact easily and directly with other particles due to their exciton component.
Polariton--polariton interaction can strongly affect the dynamics of the system, which leads to several fascinating nonlinear effects such as soliton behaviour and bistability.
Moreover, the excitonic part of polaritons can interact with excitations of surrounding lattices, which is known as interaction between polaritons and phonons.
By emitting and absorbing acoustic phonons, polaritons can transfer between different energy levels and reach BEC.

The interaction between exciton-polaritons is usually described by the following term
\begin{equation}
    \hat{H}_{pl-pl} = \frac{1}{2} \sum_{\mathbf{k,k^\prime ,q}} V_\mathbf{k,k^\prime,q} \hat{c}^\dagger_\mathbf{k+q}\hat{c}^\dagger_\mathbf{k^\prime -q}\hat{c}_\mathbf{k}\hat{c}_\mathbf{k^\prime},
    \label{eq:Ch1_pl-pl}
\end{equation}
where $V_\mathbf{k,k^\prime,q}$ accounts for the effective interaction strength between polaritons.
In the case when the momentum exchange is small, we can simplify the interaction as $V_\mathbf{k,k^\prime,0}\equiv V_{k}$.
One can further simplify the interaction term for a polariton system
\begin{equation}
    \label{eq:Ch1_pl-pl_detail}
    V_\mathbf{k} = \abs{X_k}^2\abs{X_k}^2 M_{ex},
\end{equation}
where $X_k$ is the Hopfield coefficient for polariton, and $M_{ex}$ is the exciton--exciton interaction which can be estimated by~\cite{Tassone:1999aa}
\begin{equation}
    \label{eq:Ch1_ex_ex}
    M_{ex} \approx 6 E_B \frac{a_B^2}{S}.
\end{equation}
Here $E_B$ is the exciton binding energy, $a_B$ is the exciton Bohr radius and $S$ is the sample surface.

The interaction between polaritons and phonons is given by~\cite{Deng:2010aa,Kavokin:2007aa}
\begin{equation}
    \hat{H}_{pl-ph} = \frac{1}{2} \sum_{\mathbf{k,q}}  V_{\mathbf{k,q}}^{ph} \hat{c}^\dagger_{\mathbf{k+q}}\hat{c}_\mathbf{k}\times \left( \hat{\mathscr{C}}_\mathbf{q,q_z} - \hat{\mathscr{C}}^\dagger_\mathbf{q,q_z} \right),
    \label{eq:Ch1_pl-ph}
\end{equation}
where $\hat{\mathscr{C}}_\mathbf{q,q_z}$ and $\hat{\mathscr{C}}_\mathbf{q,q_z}^\dagger$ are the phonon operators, $\mathbf{q_z}$ denotes the wavevector in the $z$ direction because the phonons are considered as three-dimensional particles unlike exciton-polaritons, and $V^{ph}_{\mathbf{k,q}}$ denotes the interaction strength between exciton-polaritons and phonons which is due to the interaction between excitons and phonons
\begin{equation}
    \label{eq:Ch1_ex_ph}
    V_{\mathbf{k,q}}^{ph}  = X_k^* X_\abs{\mathbf{k}+\mathbf{q}} \bra{k} \hat{H}_{ex-ph} \ket{k+q},
\end{equation}
where $X_k$ is the excitonic Hopfield coefficients of polariton.

\subsection{Exciton-polariton polarization}
Exciton-polaritons inherit pseudospin from the spin of their constituent excitons and cavity photons.
In the direction where the cavity grows, the total angular momentum of the electron in the conduction band is equal to $J_z^e=\pm \frac{1}{2}$, while that of the hole in the valance band is equal to $J_z^h = \pm \frac{1}{2}, \pm\frac{3}{2}$~\cite{Shelykh:2009aa}.
In the QW scenario, due to the confinement, the degeneracy in the different states is lifted.
Then the total angular momentum of an exciton in the ground state equals $J_z=\pm 1$ or $J_z =\pm 2$.
Moreover, because of the selection rules, the optical excitation on the excitons of state $J_z=\pm 2$ is strongly depressed, which means that they are not coupled with the photonic mode and do not form polaritons in the microcavity.
There are three main mechanisms of spin relaxation for excitons in semiconductors: (1) The Eliott--Yaffet mechanism~\cite{PhysRev.96.266} allows to transit between the light and dark exciton state, $J_z = \pm 1 \leftrightarrow J_z=\mp2$; (2) The D'yakonov--Perel mechanism~\cite{dyakonov1972spin} is caused by the spin-orbit interaction which also leads to the transition between $J_z = \pm 1 \leftrightarrow J_z=\mp2$; (3) The Bir--Aronov--Pikus mechanism~\cite{pikus1971exchange} involves the spin-flip exchange interaction of electrons and holes. Comparing to the two previous mechanisms, this mechanism is sufficiently enhanced in excitons~\cite{Maialle:1993aa}. This leads to the transition $J_z = +1 \leftrightarrow J_z=-1$.

\begin{figure}[ht]
    \centering
    \includegraphics[width=0.65\textwidth]{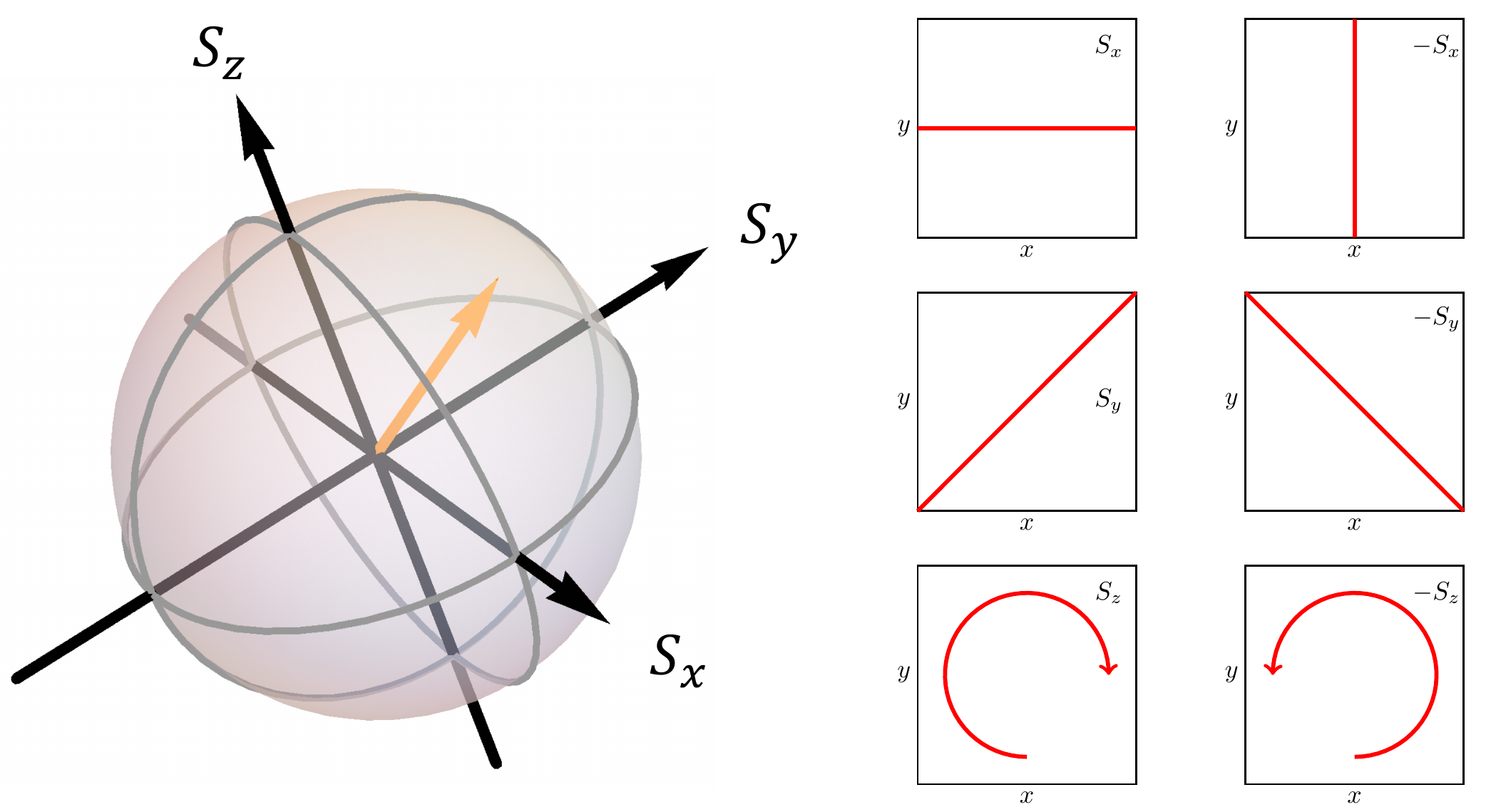}
    \caption[Pseudospin vector sphere]{Left: Pseudospin vector sphere. Right: Polarization of the $s_x$, $s_y$, and $s_z$ directions. The direction of the pseudospin vectors represent the polarization of the state. The north and south poles denote clockwise and anti-clockwise polarization, respectively, as shown in the bottom row of the right panel. The equator represents linear polarization, as shown in the upper two rows of the right panel.}
    \label{fig:Ch1_pseudospin}
\end{figure}

Let us introduce the pseudospin formalism that describes the polarization of the polariton mode.
Lower-branch polaritons at a given point $\mathbf{k}$ in reciprocal space can be described by a $2\times 2$ density matrix as
\begin{equation}
\rho_\mathbf{k} = N_\mathbf{k} \left[ \frac{\mathbf{I}}{2} + \mathbf{s}_\mathbf{k}\cdot \mathbf{\sigma}_\mathbf{k} \right],
\label{eq:Ch1_pseudospin}
\end{equation}
where $\mathbf{I}$ is the identity matrix, $\sigma_i$ is the Pauli matrix, $N_\mathbf{k}$ is the number of polaritons, and $\mathbf{s}_\mathbf{k}$ is the pseudospin of the polaritons.
In the strong coupling regime, these pseudospin components have a one-to-one correspondence to the Stokes parameters of the light emitted from the microcavity ~\cite{Kavokin:2004aa} with $\abs{\mathbf{s}}\leq \frac{1}{2}$ as shown in Fig.~\ref{fig:Ch1_pseudospin}.
In the general case, $s_z= \pm\frac{1}{2}$ denotes right- and left-circular polarizations, $s_x = \pm\frac{1}{2}$ denotes $X$- and $Y$-polarizations, and $s_y= \pm \frac{1}{2}$ denotes diagonal and anti-diagonal polarization.
Other points on the sphere represent the case of elliptical polarization.

Experimentally, polaritons are usually excited by coherent or incoherent optical pumping, and thus polariton polarization is inherited from the exciting light.
However, the initial states of the pseudospin of the polaritons can evolve with respect to time under the effect of an external magnetic field and internal effective magnetic field.
The temporal evolution of the density matrix $\rho_\mathbf{k}$ is described by
\begin{equation}
\mi\hbar \frac{d \rho_\mathbf{k}}{dt} = \left[ H_\mathbf{k}, \rho_\mathbf{k}\right],
\label{eq:Ch1_evolution_spin}
\end{equation}
with the Hamiltonian reading
\begin{equation}
    H_\mathbf{k} = E_k + \frac{\hbar}{2}\left( \Delta^{eff}_\mathbf{k} \cdot \sigma_\mathbf{k} \right),
    \label{eq:Ch1_evolution_spin2}
\end{equation}
where $E_k$ is the bare dispersion of the polariton and $\Delta_\mathbf{k}^{eff}$ is the effective magnetic field.
In microcavities with cylindrical symmetry in the linear regime, the local effective magnetic field acting on the polariton pseudospin field is called transverse-electric-transverse-magnetic (TE-TM) splitting.
As it points out in~\cite{Maialle:1993aa}, because of the long-range exchange interaction between electrons and holes~\cite{pikus1971exchange}, exciton has different energy when the non-zero in-plane wavevector is parallel and perpendicular to its dipole moment orientation. Even though the magnitude can hardly exceed a few $\mu$eV for bare exciton, in microcavities this contribution can be greatly amplified due to the coupling to the cavity photon mode, which is also split in TE- and TM-light polarization~\cite{Panzarini:1999aa}. Yet another contribution to the polariton TE-TM splitting is the $k$-dependence of the exciton oscillator strength. The oscillator strength for TE excitons varies as a function of $\cos\theta$ and for TM excitons varies as $\cos^{-1}\theta$, where $\theta$ is the angle of light propagating in the cavity.
If one neglects the oscillator contribution, the TE-TM polariton splitting magnitude (for the lower-branch polariton) can be estimated by
\begin{equation}
    \Delta_\mathbf{k}^{eff} = \abs{X_\mathbf{k}}^2 \Delta_\mathbf{k}^{ex} + \abs{C_\mathbf{k}}^2\Delta_\mathbf{k}^{cp},
    \label{eq:Ch1_TE-TM_effect}
\end{equation}
where $X_\mathbf{k}$ and $C_\mathbf{k}$ are the Hopfield coefficients for excitons and cavity photons, respectively, and $\Delta_\mathbf{k}^{ex}$ and $\Delta_\mathbf{k}^{cp}$ are the TE-TM splitting for the bare excitons~\cite{Maialle:1993aa} and cavity photons~\cite{Panzarini:1999aa}, respectively.
The magnitude of the effective TE-TM splitting is highly sensitive to the detuning of the two modes and the center frequency of the cavity photons.
In some cases, one may achieve $\Delta_\mathbf{k}^{eff} \sim 10^2 \Delta_\mathbf{k}^{ex}$~\cite{Shelykh:2009aa}.

\begin{figure}[ht]
    \centering
    \includegraphics[width=0.35\textwidth]{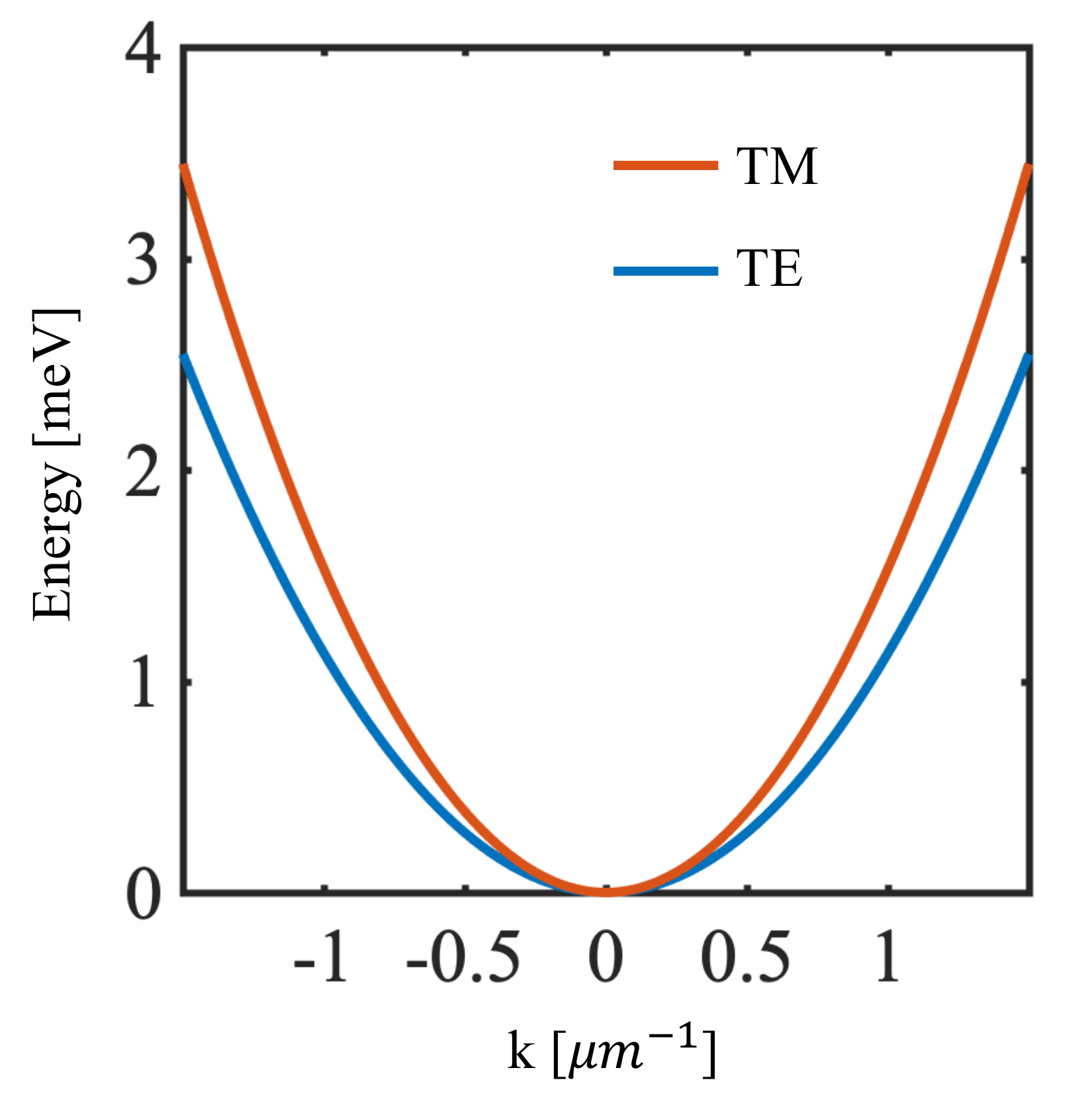}
    \caption[TE-TM Splitting]{Effective dispersion of TE and TM modes by solving Eq.~\eqref{eq:Ch1_TE_TM_Eig} with the following parameters:  $k_y = 0$ and $\gamma=0.2$~meV$\mu$m$^2$. }
    \label{fig:Ch1_TE_TM}
\end{figure}

The formula for TE-TM splitting can be derived in the following way.
When we consider the polarization of the condensate, the equation of motion can be generally written as
\begin{equation}
    \mi\hbar \partial_t \vec{\psi}\left( \mathbf{r},t\right) = \frac{\delta H}{\delta \vec{\psi}^* \left( \mathbf{r},t\right)},
    \label{eq:Ch1_TE-TM_EOM}
\end{equation}
where the order parameter of the condensate $\vec{\psi}\left(\mathbf{r},t \right)$ is a complex 2D vector and a function of position in the microcavity plane ($\mathbf{r}$) and time ($t$).
Without losing any generality, we consider the Hamiltonian with only the kinetic term, as~\cite{Rubo2013}
\begin{equation}
    H = \int d\mathbf{r} \frac{\hbar^2}{2}\left( \frac{1}{m_l} \abs{\nabla \cdot \vec{\psi}}^2 + \frac{1}{m_t} \abs{\nabla \times \vec{\psi}}^2\right),
    \label{eq:Ch1_TE-TM_effect_mass}
\end{equation}
where $m_l$ and $m_t$ are the longitudinal and transverse effective mass of the polaritons, respectively.
The 2D vector of the order parameter can be rewritten in the circular polarization basis $\psi_\pm$, which gives
\begin{equation}
    \vec{\psi} =  \left[ \frac{\left(\mathbf{x}+\mi\mathbf{y} \right)}{\sqrt{2}}\psi_+ +\frac{\left( \mathbf{x}-\mi\mathbf{y}\right)}{\sqrt{2}}\psi_-\right].
    \label{eq:Ch1_linear_circular}
\end{equation}
Transferring from the Cartesian coordinate system $\{\mathbf{x},\mathbf{y}\} $ to circular components $\{ \mathbf{z},\mathbf{z}^*\}$, we have
\begin{equation}
    \mathbf{z} = \frac{\mathbf{x}+\mi\mathbf{y}}{\sqrt{2}}, ~ ~ ~ \mathbf{z}^* = \frac{\mathbf{x}-\mi\mathbf{y}}{\sqrt{2 }}.
    \label{eq:Ch1_basis}
\end{equation}
Given Eqs.~\eqref{eq:Ch1_linear_circular} and~\eqref{eq:Ch1_basis}, we can rewrite the cross product and inner product in this new coordinate as
\begin{equation}
    \nabla \cdot \vec{\psi} = \frac{\partial \psi_+}{\partial \mathbf{z}^*}+ \frac{\partial \psi_-}{\partial \mathbf{z}}, ~ ~ ~ \nabla \times \vec{\psi} = \frac{\partial \psi_+}{\partial \mathbf{z}^*} - \frac{\partial \psi_-}{\partial \mathbf{z}}.
    \label{eq:Ch1_derivative}
\end{equation}
Considering Eqs.~\eqref{eq:Ch1_TE-TM_EOM} and~\eqref{eq:Ch1_derivative} and with the technique of integrating by parts, we get
\begin{eqnarray}
    \mi\hbar \frac{\partial \psi_+}{\partial t} &= -\frac{\hbar^2}{m^*} \left(  \frac{\partial^2}{\partial \mathbf{z} \partial \mathbf{z}^*} \psi_+ + \gamma \frac{\partial^2 \psi_-}{\partial \mathbf{z}^2} \right),\\
    \mi\hbar \frac{\partial \psi_-}{\partial t} &= -\frac{\hbar^2}{m^*} \left(  \frac{\partial^2}{\partial \mathbf{z} \partial \mathbf{z}^*} \psi_- + \gamma \frac{\partial^2 \psi_+}{\partial \mathbf{z}^{*2}} \right),
    \label{eq:Ch1_TE_TM_Z}
\end{eqnarray}
where $\frac{1}{m^*}= \frac{1}{2}\left(\frac{1}{m_l}+\frac{1}{m_t}\right)$ and $\gamma = \frac{m_t-m_l}{m_t+m_l}$.
Changing the variables back to $\{ \mathbf{x},\mathbf{y}\}$ coordinates, we have
\begin{eqnarray}
        \mi\hbar \frac{\partial \psi_\pm}{\partial t} &= -\frac{\hbar^2}{2m^*} \left[ \left(\frac{\partial^2}{\partial \mathbf{x}^2}+ \frac{\partial^2}{\partial \mathbf{y}^2} \right) \psi_\pm + \gamma \left( \frac{\partial}{\partial \mathbf{x}} \mp \mi\frac{\partial}{\partial \mathbf{y}}\right)^2 \psi_\mp \right].
    \label{eq:Ch1_TE_TM_XY}
\end{eqnarray}
We can represent this equation in matrix form on the basis of polarization $\{\psi_+,\psi_-\}$ as
\begin{equation}
    E
    \begin{bmatrix}
    \psi_+ \\
    \psi_-
    \end{bmatrix}
    =
    \frac{\hbar^2}{2m^*}
    \begin{bmatrix}
    k_x^2 + k_y^2 & \gamma\left( k_x -\mi k_y\right)^2 \\
    \gamma\left( k_x +\mi k_y\right)^2 & k_x^2 + k_y^2
    \end{bmatrix}
    \begin{bmatrix}
     \psi_+ \\
    \psi_-
    \end{bmatrix}.
    \label{eq:Ch1_TE_TM_Eig}
\end{equation}
The result of the eigenvalue problem is shown in Fig.~\ref{fig:Ch1_TE_TM}.

Finally, we remark that in Eq.~\eqref{eq:Ch1_TE_TM_Eig}, we only consider the free dispersion case.
To describe more sophisticated problems, we need to consider extra terms like
\begin{equation}
    \mathcal{H} = H - \mu n + \mathcal{H}_{int} + \mathcal{H}^\prime,
    \label{eq:Ch1_TE_TM_extra}
\end{equation}
where $H$ is from Eq.~\eqref{eq:Ch1_TE-TM_effect_mass}, $\mu$ is the chemical potential, $n=\vec{\psi}^* \cdot \vec{\psi}$ is the exciton-polariton density, $\mathcal{H}_{int}$ is the interaction between particles, and $\mathcal{H}^\prime$ stands for the other possible perturbations.

\section{Exciton-polariton condensation}\label{se:CH1_EP-BEC}
In this section, we will first discuss the basic concepts of Bose--Einstein condensation and the experimental evidence of exciton-polariton condensation.
Then we will introduce the methods applied in the study of exciton-polariton condensation.


\subsection{Bose--Einstein condensation}
In quantum mechanics, bosons are particles that follow the Bose–Einstein statistics.
One important feature of bosons is that they are allowed to accumulate in a single degenerate quantum state.
According to Bose--Einstein statistics, at absolute zero, all particles should remain in their ground state.
Historically, in 1925, S. Bose~\cite{Bose:1924aa} and A. Einstein~\cite{Einstein:2006aa} proposed that a new phase transition should occur for non-interacting bosons at low temperature.
For many years, Bose--Einstein condensates were unreachable due to technological limits in cooling down the particles to their critical temperature.
Finally, in 1995, the first experimental observation of BEC was made and was later awarded the Nobel prize~\cite{Anderson:1995aa}.

Theoretically, BEC is a phase transition characterized by the macroscopic occupation of particles in their ground states.
Such a phase transition happens when the order parameter, i.e. chemical potential, becomes zero.

Let us consider $N$ non-interacting bosons at temperature $T$ in volume $L^d$, where $L$ is the system size and $d$ is the dimension of the system.
Then the distribution of the particles is given by
\begin{equation}
    f_B\left( \mathbf{k} ,T, \mu \right) = \frac{1}{\exp\left( \frac{E\left( \mathbf{k}\right) -\mu}{k_B T}\right)-1},
    \label{eq:Ch1_BEC_statistics}
\end{equation}
where $\mathbf{k}$ is the wavevector, $E\left( \mathbf{k}\right)$ is the particle dispersion (with $E\left(0\right)=0$), $k_B$ is the Boltzmann constant, and $\mu$ is the chemical potential ($\mu <0$).

For a fixed number of particles, in the normalization condition, we have
\begin{equation}
    N\left( T,\mu \right) = \frac{1}{\exp\left( -\frac{\mu}{k_B T}\right)-1} + \sum_{\mathbf{k}\neq 0} f_B \left( \mathbf{k},T,\mu\right),
    \label{eq:Ch1_BEC_distribtion}
\end{equation}
where we separate the particles in ground ($\mathbf{k}=0$) and excited ($\mathbf{k}\neq 0$) states.
In the thermodynamic limit, we can replace the summation with an integral and get the total particle density by
\begin{equation}
    n\left( T, \mu \right) = \lim _{L\to +\infty} \frac{N\left(T,\mu \right)}{L^d} = n_0 + \frac{1}{\left( 2\pi\right)^d} \int_0^{+\infty} f_B \left( \mathbf{k},T,\mu \right) d^d k,
    \label{eq:Ch1_BEC_density}
\end{equation}
with the ground state density,
\begin{equation}
    n_0 \left( T,\mu \right) = \lim_{L\to +\infty} \frac{1}{L^d} \frac{1}{\exp\left( -\frac{\mu}{k_B T} \right)-1}.
    \label{eq:Ch1_BEC_density0}
\end{equation}
When $\mu \neq 0$, the ground state density goes to zero.
The integral part of Eq.~\eqref{eq:Ch1_BEC_density} increases as the chemical potential $\mu$ approaches zero from negative infinity.
This means that if one increases the particle density ($n$) in the system, the chemical potential ($\mu$) will also increase.
The system reaches its critical density as follows,
\begin{equation}
    n_c\left(T\right) = \lim_{\mu \to 0} \frac{1}{\left( 2\pi \right)^d} \int_{0}^{+\infty} f_B \left( \mathbf{k},T \right)d^d k.
    \label{eq:Ch_1_critical_density}
\end{equation}
This integral can be calculated analytically in the parabolic dispersion case, which is $E\left(\mathbf{k}\right) = \frac{\hbar^2 k^2}{2m}$.
The result converges for $d>2$ and diverges for $d \leq 2$;
this means that if the system dimension is less than or equal to 2, the system can hold an infinite number of bosons while the chemical potential is non-zero.
Thus, BEC cannot happen in the $d \leq 2$ case.
When $d>2$, extra particles will collapse to the ground state when the system reaches its critical density.
This gives the density of the ground state as
\begin{equation}
    n_0\left(T\right) = n\left(T\right) -n_c\left(T\right).
    \label{eq:Ch1_groundstate_density}
\end{equation}
The appearance of the macroscopic occupation of the ground state indicates that BEC takes place.

In exciton-polariton physics, the particles are confined in the cavity, as shown in Fig.~\ref{fig:Ch1_cavity}.
This system can usually be regarded as a one-dimensional (1D) or 2D system.
From the previous discussion, we know that in 1D or 2D infinite homogeneous systems, Bose--Einstein condensate cannot exist in principle.
However, if the size of the system is finite, a quasi-condensation state is possible because of the cut off of the integral as discussed in~\cite{Bagnato:1991aa,Hohenberg:1967aa,PhysRevLett.17.1133}.

\subsection{BEC of exciton-polaritons}
The first experimental demonstration of Bose--Einstein condensation utilized dilute atomic gases; the first success was obtained in rubidium vapours~\cite{Anderson:1995aa} in 1995.
Following this early achievement, BEC based on exciton-polaritons was claimed~\cite{Kasprzak:2006aa} in 2006.
\begin{figure}[ht]
    \centering
    \includegraphics[width=0.65\textwidth]{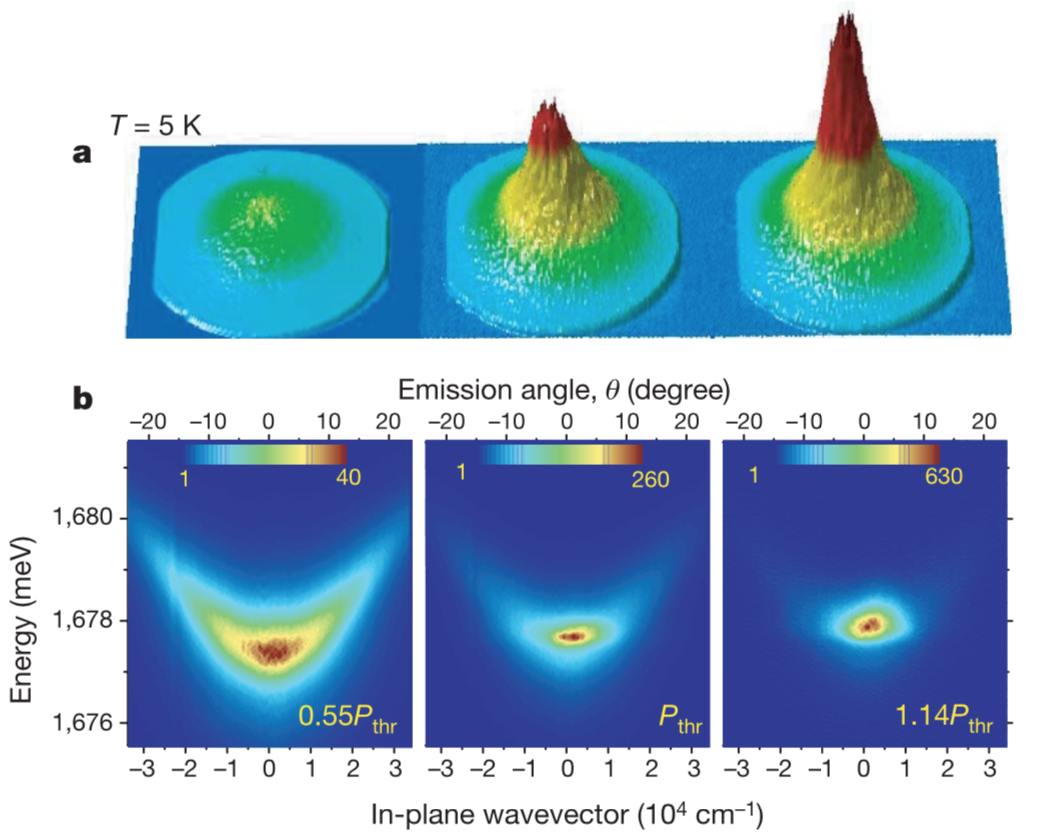}
    \caption[Polaritons in BEC]{Bose--Einstein condensation of exciton-polaritons at $5$ K. (a) Pseudo-3D images of far-field emission with increasing pumping intensity. (b) Energy-resolved spectra results corresponding to (a). A narrowing of the particle distribution is shown at $k=0$ when the pumping is above the threshold. The figure is taken from~\cite{Kasprzak:2006aa}.}
    \label{fig:Ch1_exciton-polariton_BEC}
\end{figure}

In the work performed by Kasprzak et al.~\cite{Kasprzak:2006aa}, they applied a $\mathrm{CdTe}$/$\mathrm{CdMgTe}$-based microcavity at a temperature of $5$ K.
In Fig.~\ref{fig:Ch1_exciton-polariton_BEC}, they present particle density results based on the angular distribution of the spectrally integrated emissions.
From left to right, the pumping intensity increases gradually and crosses the pumping threshold.
As one can see in Fig.~\ref{fig:Ch1_exciton-polariton_BEC}(a), when the pumping is below the threshold (left panel), the far-field emission shows a smooth distribution centered around the ground state, i.e. $k=0$.
With increasing pumping intensity, from around the threshold (middle panel) to above the threshold (right panel), one can see that the emission from the ground state ($k=0$) increases sharply and becomes dominant.
This evidence reflects that the polaritons begin to macroscopically occupy the ground state when the pumping is above the threshold.
In Fig.~\ref{fig:Ch1_exciton-polariton_BEC}(b), the authors show the spectrum vs. angle-resolved results.
By increasing the pumping intensity, particle density at the ground state rises significantly.
One can also observe that the ground state energy also increases slightly due to particle interaction.

The previous experiment we discussed is based on a planar microcavity with no internal potential to confine the polaritons.
Instead of a potential profile of the microcavity, confinement in this case derives from the pumping because the size of the condensate strongly depends on the size of the pumping spot. However, with several modern techniques~\cite{Schneider:2017dq,Lai:2007aa,Cerda-Mendez:2010aa,Balili:2006aa} to modify the energy dispersion of cavity photons and QW excitons, one can achieve a strong trapping potential that can lead to single-mode or multi-mode exciton-polariton BEC.
Compared to the planar cavity case, the character of the BEC will be more obvious with the help of internal potential.
Through a periodic engineering of the microcavity, one can get a system with a particular potential as an analogue of different lattices to study the coherence properties and the interaction between different condensation modes.

\begin{figure}[ht]
    \centering
    \includegraphics[width=0.65\textwidth]{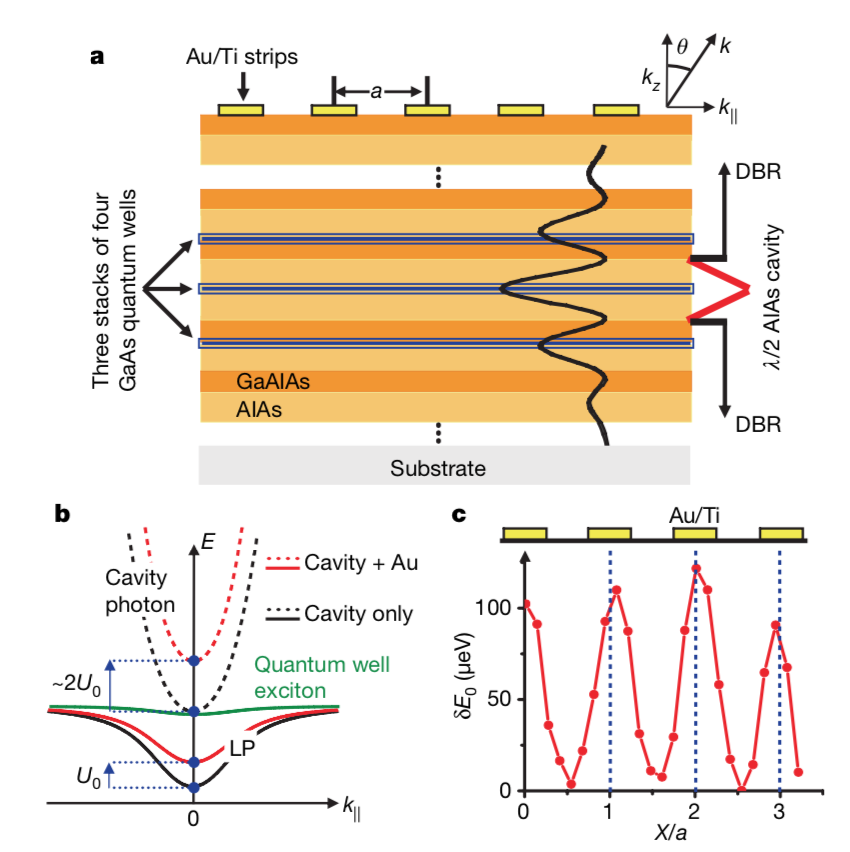}
    \caption[Polariton condensation in a periodic structure]{(a) Schematic of a cavity polariton array formed by placing periodic thin metallic strips ($\mathrm{Au}$/$\mathrm{Ti}$) on top of the cavity. (b) Dispersion for the cavity photon mode and lower-branch polariton mode. (c) Spatial energy modulation for the lower-branch polariton mode ($\delta E_0$) is detected from the position-dependent central energy of emission from lower-branch polaritons. The figure is taken from~\cite{Lai:2007aa}.}
    \label{fig:Ch1_periodic}
\end{figure}

One noticeable example was reported by Lai et al.~\cite{Lai:2007aa}.
In this work, they generated an array of trap potential around $U_0\approx200~\mu$eV by periodically applying thin metallic strips ($\mathrm{Au}$/$\mathrm{Ti}$) on the top of a microcavity, as shown in Fig.~\ref{fig:Ch1_periodic}(a).

Under the metallic layer mask, the resonance energy of the cavity photons is increased by $2U_0$ (at $k_\parallel=0$) compared to the case with a bare cavity, according to transfer-matrix calculation~\cite{Yeh:1988aa}.
As shown in Fig.~\ref{fig:Ch1_periodic}(b), when the cavity photons covered by the metallic layer couple with the excitons, the resulting lower-branch polaritons blueshift by about $U_0$ (at $k_\parallel=0$), compared to the case in which excitons are coupled with photons in the bare cavity region.
In Fig.~\ref{fig:Ch1_periodic}(c), the measured spatial modulation of the lower-branch polariton energy is shown, where the modulation is about $U_0/2\approx 100~\mu$eV. 
The energy difference between this measurement and the prediction in Fig.~\ref{fig:Ch1_periodic}(b) is due the limited spatial resolution of the optical detection system.

\begin{figure}[ht]
    \centering
    \includegraphics[width=0.65\textwidth]{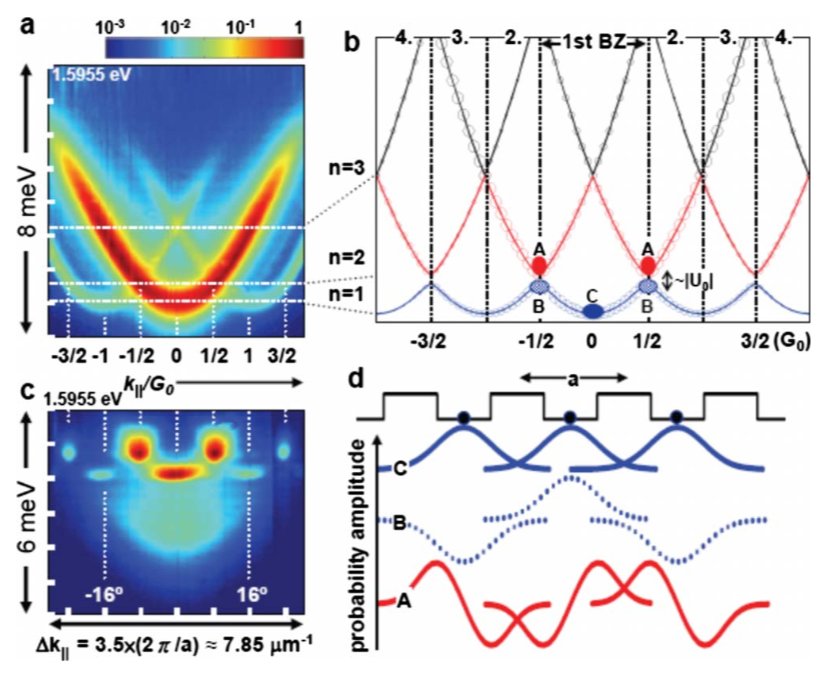}
    \caption[Polariton band structure]{Energy-momentum dispersion and real space wavefunction of polaritons in a 1D array. (a) Time-integrated energy vs. in-plane momentum for the polariton array with detuning $E_{cav}\approx E_{exc}$ when the pumping is below the threshold. (b) Scheme of the band structure for the polariton array with lattice constant $a$. The gap between the first and second bands is about $\abs{U_0}\approx 200~\mu$ eV. (c) Energy vs. momentum of the polariton condensate array with detuning $E_{cav}-E_{exc}\approx 6$ meV when the pumping is above the threshold. (d) Scheme of the Bloch wavefunction for states A, B, and C labeled in (b). The figure is taken from~\cite{Lai:2007aa}.}
    \label{fig:Ch1_bandstructure}
\end{figure}

Then, the team excited the microcavity periodically masked by the metallic layers by a laser pulse near the QW exciton resonance.
With a large in-plane wave number of the laser pumping, one can make sure that the polariton coherence introduced by the laser is lost by the polariton-phonon scattering process before the polaritons reach the ground state at $k_\parallel=0$. 
This result is shown in Fig.~\ref{fig:Ch1_bandstructure}.

In Fig.~\ref{fig:Ch1_bandstructure}(a), the energy vs. in-plane momentum ($E$ vs. $k_\parallel$) dispersion relation for the 1D polariton array is shown.
In this situation, the detuning is close to zero, i.e. $E_{cav}\approx E_{exc}$, and the pumping is below the threshold.
To introduce the band structure for the polaritons in this array system, they applied a ``nearly free polariton" approximation in the presence of a periodic square-well potential.
Given a 1D periodic square potential $U\left(x\right)$ with a lattice constant, one can use Bloch theory to obtain the extended band structure, as presented in Fig.~\ref{fig:Ch1_bandstructure}(b).
The band gaps at the edge of the Brillouin zone due to the barrier potential are around $\abs{U_0}\approx 200~\mu$eV.
This result well reproduces the observed polariton energy dispersion in momentum space below the pumping threshold in Fig.~\ref{fig:Ch1_bandstructure}(a).
The absence of band gaps in the observed dispersion picture is due to the finite lifetime of the polaritons, which provides a large broadening of the emission lines (around $500~\mu$eV).

In Fig.~\ref{fig:Ch1_bandstructure}(c), the energy spectrum in momentum space is shown in the case with detuning $E_{cav}-E_{exc}\approx 6$ meV and pumping that is above the threshold.
In this situation, polariton emissions occur in two states with an energy difference of about $1$ meV.
Peaks of the emissions can be observed at $k_\parallel=0$ and $k_\parallel= \pm \frac{G_0}{2}$, and other weaker emissions can also be found at $k_\parallel = \pm G_0$ and $k_\parallel= \pm \frac{3G_0}{2}$, where $G_0=\frac{2\pi}{a}$ is the primitive reciprocal lattice vector.
The two states of emissions stand for the zero and $\pi$ states corresponding to $k_\parallel = 0,\pm G_0$ and $k_\parallel = \pm \frac{G_0}{2},\pm \frac{3G_0}{2}$, respectively.

A schematic of the spatial distribution based on the Bloch approach is drawn in Fig.~\ref{fig:Ch1_bandstructure}(d) for the states A, B, and C labeled in Fig.~\ref{fig:Ch1_bandstructure}(b).
The zero state labeled as C carries an s-like wave with a maximum amplitude in the potential wells and shares the identical phase between adjacent wells.
Meanwhile the states A and B at $k_\parallel=\pm \frac{G_0}{2}$ correspond to the $\pi$ state which has $\pi$ phase difference between adjacent wells.
However, compared to state A, state B is an unstable state, which can be examined by its different density distribution in real space.

The experimental realization in~\cite{Lai:2007aa} triggered a great scientific interest in investigating exciton-polaritons in different artificial lattices.
This is mainly because exciton-polariton gases under periodic lattice potential are a promising system to simulate many-body physics having relatively strong interaction, relatively small effective mass, and a variety of techniques to fabricate the lattice system.
To understand what happens in these many-body systems, we need to introduce some theoretical tools to describe system evolution in the next section.

\subsection{The driven-dissipative Gross--Pitaevskii equation}
In a previous subsection, we introduced an interaction term to describe the polariton-polariton collision in Eq.~\eqref{eq:Ch1_pl-pl}.
This interaction term makes the dynamics of the exciton-polaritons nontrivial and is responsible for a number of nonlinear and quantum effects.
To investigate the many-body problem, one possible way is to apply the Hartee or mean-field approach and assume that the wavefunction is a symmetric product of a single-particle wavefunction.
Following the analysis in a text~\cite{Pethick:2002tn}, in a fully condensed situation, all the bosons are in the same single-particle state, and therefore we may write the $N$-particle wavefunction, $\Psi\left(\mathbf{r}_1,\mathbf{r}_2,...,\mathbf{r}_N\right)$, as
\begin{equation}
    \Psi\left(\mathbf{r}_1,\mathbf{r}_2,...,\mathbf{r}_N\right) = \prod_{i=1}^N \phi\left( \mathbf{r}_i\right).
    \label{eq:Ch1_many_wf}
\end{equation}
For each wavefunction in the single-particle state, we have the usual normalized condition
\begin{equation}
    \int d\mathbf{r}\abs{\phi\left(\mathbf{r}\right)}^2 = 1.
\end{equation}
To take interaction between the particles into account, we introduce an effective interaction term $U_0\delta\left( \mathbf{r}-\mathbf{r}^\prime\right)$.
Then the effective Hamiltonian may be written as
\begin{equation}
    H = \sum_{i=1}^N \left(\frac{\mathbf{p}_i^2}{2m} +V\left(\mathbf{r}_i\right) \right) + U_0 \sum_{i<j}\delta\left( \mathbf{r}_i - \mathbf{r}_j \right),
    \label{eq:Ch1_Ham_MP}
\end{equation}
where $V\left( \mathbf{r}\right)$ is the external potential.
The expectation value of the Hamiltonian in the state from Eq.~\eqref{eq:Ch1_many_wf} is given by
\begin{eqnarray}
        E &=& \bra{\Psi\left( \mathbf{r}_1,\mathbf{r}_2,...,\mathbf{r}_N \right)} H \ket{\Psi\left( \mathbf{r}_1,\mathbf{r}_2,...,\mathbf{r}_N \right)} \\ 
        &=& N \int d\mathbf{r} \left[ \frac{\hbar^2}{2m}\abs{\nabla \phi\left(\mathbf{r} \right)}^2 + V \left(\mathbf{r}\right) \abs{\phi\left(\mathbf{r}\right)}^2 + \frac{N-1}{2} U_0 \abs{\phi\left(\mathbf{r}\right)}^4\right]. \nonumber
        \label{eq:Ch1_Exp_Energy}
\end{eqnarray}
The factor before the interaction term, $C_N^2 =\frac{N\left(N-1\right)}{2}$, indicates all possible combinations of the interaction between two bosons in an $N$-particle system.

We introduce the concept of the wavefunction $\psi\left(\mathbf{r}\right)$ for the condensed state,
\begin{equation}
    \psi\left( \mathbf{r} \right) = \sqrt{N} \phi\left(\mathbf{r}\right),
    \label{eq:Ch1_order_parameter}
\end{equation}
as the order parameter.
With this new parameter, we can rewrite the energy of the system as
\begin{equation}
    E\left( \psi \right) = \int d\mathbf{r} \left[ \frac{\hbar^2}{2m}\abs{\nabla \psi\left(\mathbf{r} \right)}^2 + V \left(\mathbf{r}\right) \abs{\psi\left(\mathbf{r}\right)}^2 + \frac{1}{2} U_0 \abs{\psi\left(\mathbf{r}\right)}^4\right],
    \label{eq:Ch1_Exp_Energy_order}
\end{equation}
where we neglect the $N^{-1}$ term by assuming $N$ is large. 

As a next step, we minimize the energy in Eq.~\eqref{eq:Ch1_Exp_Energy_order} with respect to two independent variables $\psi\left(\mathbf{r}\right)$ and $\psi\left( \mathbf{r}\right)^*$ under the condition that the total number of particles is constant.
We first define the Lagrange multiplier as
\begin{equation}
    \mathcal{L}\left(\psi\left(\mathbf{r}\right),\psi^*\left(\mathbf{r}\right),\mu\right) = E\left(\psi\left(\mathbf{r}\right),\psi^*\left(\mathbf{r}\right)\right)-\mu \left(\int d\mathbf{r} \psi\left(\mathbf{r}\right) \psi^*\left(\mathbf{r}\right) -N \right),
    \label{eq:Ch1_Mulipler}
\end{equation}
and find the variation with respect to $\psi\left(\mathbf{r}\right)^*$ as
\begin{eqnarray}
        \delta_{\psi^*\left(\mathbf{r}\right)} \mathcal{L} &=& \delta_{\psi^*\left(\mathbf{r}\right)}\int d\mathbf{r} \left[ \frac{\hbar^2}{2m}\abs{\nabla \psi\left(\mathbf{r} \right)}^2 + V \left(\mathbf{r}\right) \abs{\psi\left(\mathbf{r}\right)}^2 + \frac{1}{2} U_0 \abs{\psi\left(\mathbf{r}\right)}^4 - \mu \psi\left(\mathbf{r}\right) \psi^*\left(\mathbf{r}\right)\right] \nonumber \\
        &=&  \int d\mathbf{r} \left[ \frac{\hbar^2}{2m}\nabla^2 \psi\left(\mathbf{r} \right) + V \left(\mathbf{r}\right) \psi\left(\mathbf{r}\right) +  U_0 \abs{\psi\left(\mathbf{r}\right)}^2 \psi\left(\mathbf{r}\right)- \mu \psi\left(\mathbf{r}\right) \right].
        \label{eq:Ch1_Mulipler2}
\end{eqnarray}
In order to get this result, we use the integrating by parts method and neglect the surface term by assuming that the size of the system is finite.
To get the extreme value for the energy with the constant number of particles constraint, we require that
\begin{equation}
     \delta_{\psi^*\left(\mathbf{r}\right)}\mathcal{L}\left(\psi\left(\mathbf{r}\right),\psi^*\left(\mathbf{r}\right),\mu\right) = 0.
\end{equation}
This gives us a time-independent Gross--Pitaevskii equation,
\begin{equation}
    -\frac{\hbar^2}{2m}\nabla^2 \psi\left(\mathbf{r}\right) + V\left(\mathbf{r}\right) \psi\left(\mathbf{r}\right) + U_0 \abs{\psi\left(\mathbf{r}\right)}^2 \psi\left(\mathbf{r}\right) = \mu \psi\left(\mathbf{r}\right),
    \label{eq:Ch1_Gross_Pitaevskii_t_independent}
\end{equation}
where $\mu$ is the chemical potential.
To get a time-dependent Gross--Pitaevskii equation, the stationary conditions $\psi\left(\mathbf{r},t\right)$ must develop in time as $e^{-\mi\mu t/\hbar}$, which gives
\begin{equation}
    \mi\hbar \frac{\partial \psi \left(\mathbf{r},t\right)}{\partial t} = -\frac{\hbar^2}{2m} \nabla^2 \psi\left(\mathbf{r},t \right) + V \left(\mathbf{r}\right) \psi\left(\mathbf{r},t\right) + U_0 \abs{\psi\left(\mathbf{r},t\right)}^2 \psi\left(\mathbf{r},t\right).
    \label{eq:Ch1_GP_time_dep}
\end{equation}

In exciton-polariton BEC, the number of particles in the condensate is not constant due to the finite lifetime of polaritons.
To maintain the condensation, one needs to introduce external pumping to the system in experiment.
Accordingly, we need to extend Eq.~\eqref{eq:Ch1_GP_time_dep} to a more generalized form to better describe the lower-branch polariton field.

If the Rabi frequency $\Omega$ is much larger than the other energy scales in the system, the generalized Gross--Pitaevskii equation for polaritons with a coherent pumping is~\cite{Carusotto:2013aa},
\begin{equation}
    \mi\hbar \frac{\partial}{\partial t} \psi\left(\mathbf{r},t\right) = \left[ -\frac{\hbar^2}{2m}\nabla^2 + V\left(\mathbf{r}\right) + \alpha \abs{\psi\left(\mathbf{r},t\right)}^2 -\frac{\mi \gamma}{2} \right] \psi\left(\mathbf{r},t\right) +  \mi P\left(\mathbf{r},t\right),
    \label{eq:Ch1_coherent_pump}
\end{equation}
where $\psi\left(\mathbf{r},t\right)$ represents the lower-branch polaritons, $\alpha$ is the interaction strength, $\gamma$ is the polariton decay, and $P\left(\mathbf{r},t\right)$ is the coherent pumping.

Equation~\eqref{eq:Ch1_coherent_pump} describes the case where the microcavity is driven by a coherent, quasi-resonant pump.
In this case, the microscopic details of the system are under control and one can develop an \textit{ab inito} description of the system~\cite{Carusotto:2013aa}.
However, another pumping scheme is widely applied in polariton physics, called incoherent pumping.
In incoherent pumping, the lower-branch polaritons do not inherit any information from the pumping source, such as phase or frequency, because of the relaxation process toward the bottom of the lower polariton branch.
Incoherent pumping usually pumps the system far above the bottom of the lower polariton branch by optical~\cite{PhysRevB.76.201305,Deng15318,PhysRevB.72.201301} or electrical means~\cite{Schneider:2013aa,PhysRevB.77.113303}.
The incoherent polaritons start to accumlate significantly in the bottleneck region in momentum space where the dispersion of polariton changes drastically.
The energy relaxation process toward the ground state is provided by the scattering of hot polaritons and phonons.
Due to the reducing of the density of state for lower polariton in the bottom region, the relaxation by phonon--polariton scattering is quite slow compared to polariton--polariton scattering in a strong enough pumping intensity.
For polariton--polariton collisions, two polaritons collide in the bottleneck region with each other.
One of the polaritons is scattered to the bottom of the lower polariton branch and the other one is scattered to the region where polariton is more excitonic.
Given the bosonic statistics of polaritons, this relaxation process turns out to be stimulated as soon as the density of polaritons at the bottom becomes unit.
This indicates that when the stimulation overcomes the losses, a macroscopic coherent population of polaritons starts to accumulate in the ground state i.e., the condensate appears.

In practice, to describe incoherent pumping, one should introduce an amplification term in the equation of motion Eq.~\eqref{eq:Ch1_GP_time_dep} to include the stimulated scattering into the condensate and an external rate equation to describe the polariton reservoir density

\begin{eqnarray}
        \mi\hbar \frac{\partial}{\partial t} \psi\left(\mathbf{r},t\right) &=& \left[ -\frac{\hbar^2}{2m}\nabla^2 + V\left(\mathbf{r}\right) + \alpha \abs{\psi\left(\mathbf{r},t\right)}^2 -\frac{\mi \gamma}{2} \right] \psi\left(\mathbf{r},t\right)\nonumber \\ &+& \frac{\mi}{2}R\eta_R\left(\mathbf{r},t\right) \psi\left(\mathbf{r},t\right),
        \label{eq:Ch1_inchoerent_pumping1}\\
        \frac{\partial}{\partial t} \eta_R\left(\mathbf{r},t\right) &=& I\left(\mathbf{r},t\right) - R \eta_R\left(\mathbf{r},t\right) \abs{\psi\left(\mathbf{r},t\right)}^2 - \gamma_R \eta_R \left(\mathbf{r},t\right).
        \label{eq:Ch1_inchoerent_pumping2}
\end{eqnarray}
Equation~\eqref{eq:Ch1_inchoerent_pumping2} describes the dynamics of the polariton reservoir density in the bottleneck region, $\eta_R\left(\mathbf{r},t\right)$, classically, where $I\left(\mathbf{r},t\right)$ is the intensity of the incoherent pumping, and $\gamma_R$ is the decay of the reservoir particles.
The reservoir is coupled to the condensed polaritons via the term $R\eta_R \left(\mathbf{r},t\right)\abs{\psi\left(\mathbf{r},t\right)}^2$, where $R$ is the phenomenological coupling strength between reservoir particles and condensed particles.

Equation~\eqref{eq:Ch1_coherent_pump} and Eqs.~\eqref{eq:Ch1_inchoerent_pumping1} \& \eqref{eq:Ch1_inchoerent_pumping2} are two important sets of equations to simulate the dynamics of exciton-polariton condensates.
By modifying the pumping terms and different shapes of the potential, we can realize various features of exciton-polariton systems, which we will discuss in the next several chapters.

\section{Outline of the thesis}
This thesis is organised as follows.
We discuss polariton condensation in simple artificial lattices in Chapter~\ref{CHTWO}.
Specifically, in Section~\ref{Ch2}, by engineering the potential of cavity photons and excitons separately, we get a non-trivial dispersion for polaritons in the ground state and study the dynamics of the system with an incoherent pumping model.
In Section~\ref{Ch3}, we consider an extended Gross--Pitaevskii equation with complex-valued potential and complex non-linearity to describe the gain and loss of polaritons in a 1D chain.
In this model, we find that condensation takes place in either the $0$- or $\pi$-state of the ground state, which is different from~\cite{Lai:2007aa}.
Next, in Chapter~\ref{CHTHREE}, we focus on the polariton in complex artificial lattices.
In Section~\ref{Ch4}, we provide a novel way to excite the compact localized condensation of exciton-polaritons in a Lieb lattice by Laguerre--Gaussian pumping and check the dynamics of the system.
In Section~\ref{Ch5}, we develop a method to generate topological edge states using a local magnetic field in a graphene lattice.
Following these treatments of exciton-polaritons in artificial lattices, in Chapter~\ref{Ch6}, we study the transport properties of 2D electron systems interacting with exciton or exciton-polariton condensates.
We show with this new type of interaction that the resistivity of the 2D material can be orders of magnitude higher than that from traditional phonon-electron interaction.
Finally, we conclude this thesis with a summary and further prospects related to the work in Chapter~\ref{Ch7}.

%% file: chapter2_3-6_revised.tex
\chapter{Exciton-Polariton in simple lattice}\label{CHTWO}
In this chapter, we devote our focus on the exciton-polariton in the simple lattice which has one lattice site in per unit cell.
In the tight-binding model with the nearest neighbour hopping approach, this indicates one has one simple band in the spectrum.
In the following sections, instead of the discrete lattice approach, we consider the exciton-polariton in the continuous limit to include long-range interaction.

At the beginning of this chapter, we investigate the dispersion of exciton-polaritons in the case that the coupling between exciton part and cavity photon part is in two separate periodic potentials.
We find a nontrivial ground state in the middle of the Brillouin zone. We further investigate the corresponding dynamics of exciton-polaritons in this system.

In the rest part of this chapter, we extend the Gross--Pitaevskii equation with complex-valued potential and complex-valued nonlinearity.
The imaginary part of the potential describes the gain and loss of the exciton-polaritons, and the imaginary term of the nonlinear interaction defines the saturation of the gain from the reservoir.
We find that in such a configuration, the condensate phase may change if we tune the real or imaginary part of the nonlinearity.

\section{Multivalley engineering in semiconductor microcavities}\label{Ch2}
%
%
As we know, photonic and electronic systems support many universal phenomena.
To mention a few, topological photonics has recently risen from ideas in the study of topological insulators~\cite{Lu:2014aa}, and the field of spintronics has been an inspiration for optical analogues for the optical spin Hall effect~\cite{Leyder:2007aa} and the development of photonic spin switches~\cite{Lagoudakis:2002aa,Amo:2010aa}.
While the advantages of spintronics for information processing remain promising, the flourishing field of valleytronics proposes to encode information in the valley degree of freedom of multivalley semiconductors~\cite{Behnia:2012aa,Nebel:2013aa}, including transition metal dichalcogenides (TMDs)~\cite{Xu:2014aa,Jones:2014aa,He:2014aa,Chernikov:2014aa}.
This raises the question of whether valleytronics is itself a universal concept that can also appear in suitably engineered photonic systems.

Recently, several works have attempted to hybridize light confined in planar microcavities with TMDs~\cite{Vasilevskiy:2015aa,Lundt:2016aa} resulting in exciton-polaritons with large binding energies.
Indeed, such a system is highly promising as a nonlinear photonic system operating at room temperature; however, the valleytronic features of multivalley semiconductors that occur at wave vectors given by the inverse crystal lattice constant are uncoupled from optical modes that are restricted to lower in-plane wave vectors inside the light cone.
Thus, one needs a different approach to engineer multiple valleys that are suitable for nonlinear optical valleytronics.

For this purpose, exciton-polaritons remain a good candidate as their relatively large micron-scale de Broglie wavelength gives them the advantage to be strongly manipulated by micron-scale potentials in microcavities.
Such potentials can be manipulated either by spatial modulation of the photon energy~\cite{Kaitouni:2006aa,Lai:2007aa} or the exciton energy~\cite{Balili:2007aa,Amo:2010aa,Assmann:2012aa,Cristofolini:2013aa,Askitopoulos:2013aa}.
Periodic potential arrays have been introduced~\cite{Lai:2007aa,Cerda-Mendez:2010aa} with various lattice geometries~\cite{Kim:2013aa,Cerda-Mendez:2013aa,Winkler:2016aa}, leading to several different phenomena, for example gap solitons~\cite{Ostrovskaya:2013aa,Tanese:2013aa}, flatbands~\cite{Jacqmin:2014aa}, and Bloch oscillations \cite{Flayac:2011aa,Flayac:2013aa}.
Researchers have also suggested several new devices \cite{Flayac:2013aa,Marsault:2015aa} and (theoretically) non-trivial topological properties~\cite{Karzig:2015aa,Nalitov:2015aa,Bardyn:2015aa,Bardyn:2016aa}.

In this chapter, we consider the behavior of exciton-polaritons in a microcavity where both the optical and excitonic components are separately manipulated by two periodic potentials.
These potentials can be achieved by ``proton implantation'' \cite{Schneider:2016aa}, in which the properties of QWs and semiconductor microcavities are spatially patterned after growth.
The different localization of photons and excitons by their respective potentials theoretically allows for a nontrivial overlap of their wave functions that depends on the in-plane momentum.
Remarkably, we can realize momentum-dependent coupling between excitons and cavity photons, which gives rise to the formation of nontrivial dispersion with degenerate ground states at non-zero momenta at the bottom of different valleys in the reciprocal space.

We further show that when considering TE-TM splitting, different valleys have different polarizations, analogous to the spin-valley coupling that forms the basis of valleytronics in 2D semiconductor systems.
For additional effects that arise from the unusual exciton-polariton dispersion in our system, we consider the behaviour of the system under incoherent excitation conditions.
Considering the inherent excitation case, it is known that exciton-polaritons may undergo BEC~\cite{Kasprzak:2006aa}, characterized by the breaking of U(1) phase symmetry and the appearance of a macroscopic coherent low-energy state.
Other symmetries may also be broken under BEC, both in exciton-polariton systems and other systems including spin symmetry breaking~\cite{Sadler:2006aa,Ohadi:2012aa}, translational symmetry breaking~\cite{Kanamoto:2003aa}, and angular momentum symmetry breaking~\cite{Butts:1999aa,Liu:2015aa}.
In our system, we find that there is also a spontaneous breaking of linear momentum symmetry, where the condensates can spontaneously choose between different valleys in the dispersion.

%
%
\subsection{Dispersion of lattice exciton-polaritons}
\begin{figure}[ht]
\centering
	\includegraphics[width=0.79\linewidth]{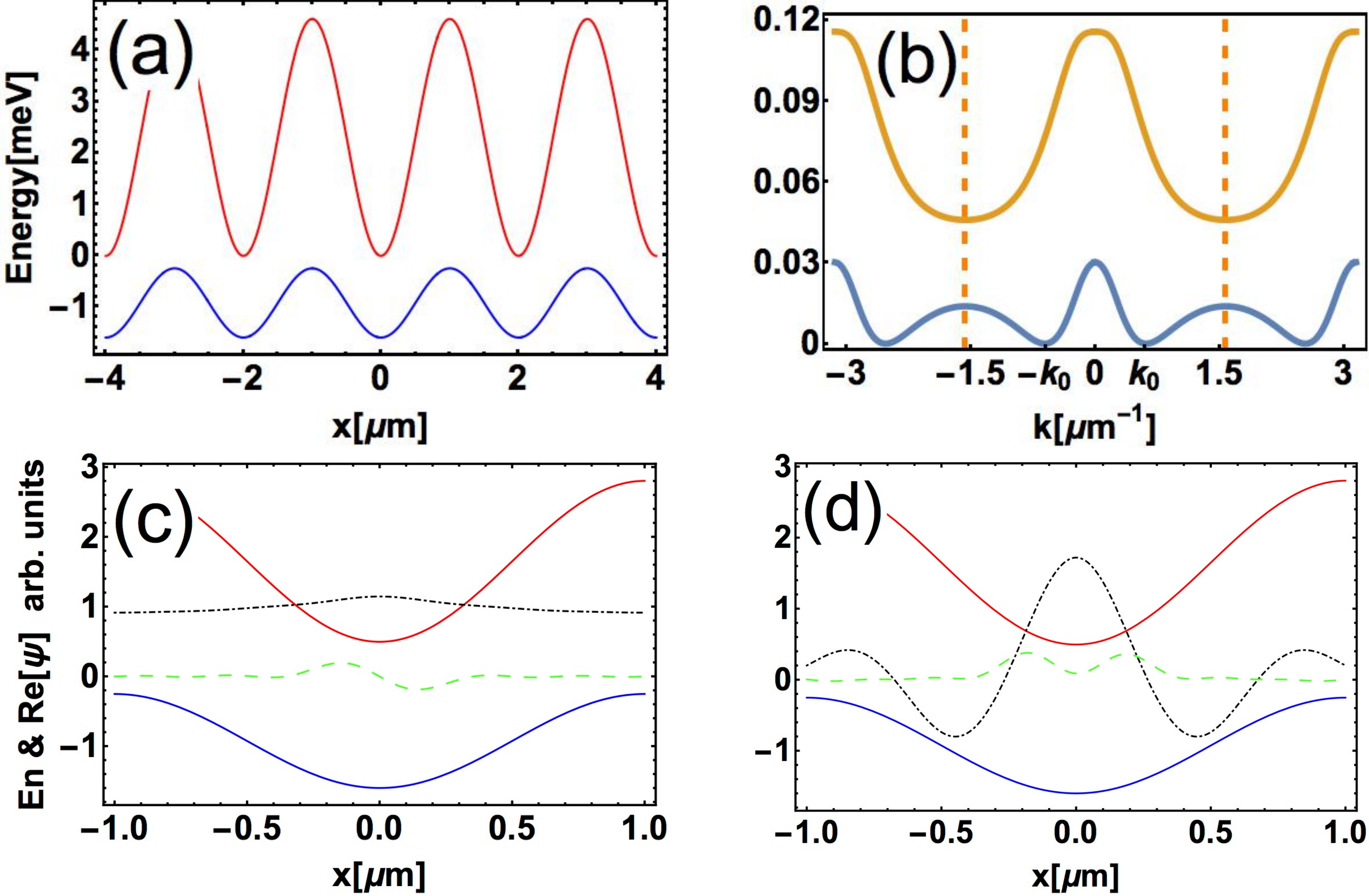}
	\caption[Potential setting, eigenenergy, and eigenstate of the system]{(a) Potential profiles for QW excitons (red) and cavity photons (blue). (b) Dispersion of the lowest-energy exciton-polaritons (blue), and the imaginary part of the polariton energy as a function of $k$ (yellow). The dashed vertical lines label the edges of the first Brillouin zone. (c, d) Real part of the wave function of the cavity photons (black dot-dashed line) and excitons (green dashed line) in a single lattice period with (c) $k=0$ and (d) $k$ at the edge of the Brillouin zone. Red and blue curves are the potentials of the excitons and cavity photons. The figure is taken from~\cite{Sun:2017ab}.}
	\label{fig:CH2_Fig1}
\end{figure}

Let us begin by considering a 1D system of cavity photons and QW excitons~\cite{Tanese:2013aa}, which have potentials with the same periodicity but different alignment in energy, as shown in Fig.~\ref{fig:CH2_Fig1}(a).
Using the Bloch theory and the model of coupled harmonic oscillators, we can apply the central equation in~\cite{kittel:1966aa} and solve the eigenvalue problem of the system.
In a brief form this equation reads:
\begin{eqnarray}
	\begin{pmatrix}
		\lambda_{C}-\mi\hbar/\tau_C-E & \Omega \\
		\Omega & \lambda_{X}-\mi\hbar/ \tau_X -E
	\end{pmatrix}
	C_k
	 + \sum_{G}
	\begin{pmatrix}
		\tilde{V}_{C}\left( G \right) & 0 \\
		0 & \tilde{V}_X\left( G \right)
	\end{pmatrix}
	C_{k-G}=0,
	\label{eq:CH2_SP1}
\end{eqnarray}
where $\lambda_{C}=\frac{\hbar^{2}k^{2}}{2m_{C}}$ and $\lambda_{X}=\frac{\hbar^{2}k^{2}}{2m_{X}}$ are the kinetic energy terms of the photonic and excitonic counterparts, respectively.
Parameters $\tau_{C,X}$ are the lifetimes of the cavity photons and excitons, $\Omega$ is the exciton-photon (Rabi) coupling constant, and $\tilde{V}_C\left( G \right)$ and $\tilde{V}_{X}\left( G \right)$ are the Fourier components of the potentials for the cavity photons and excitons in real space.
The summation is over $G$ , which is the reciprocal lattice vector of different order.
The reciprocal lattice vector is the same because we apply the same periodicity for the exciton and photon potential. The vector $C_{k}$ denotes the exciton and photon components in the polariton mode with different $k$, and $E$ is the eigenenergy of the exciton-polariton mode.

After solving the eigenvalue problem with a truncation of a finite number of $G$ (see details in Appendix~\ref{AP:A1}), we find the dispersion of the system of exciton-polaritons in $k$-space, as shown in Fig.~\ref{fig:CH2_Fig1}(b), and the wave functions of the photons and excitons, as shown in Fig.~\ref{fig:CH2_Fig1}(c, d).
One should note that the excitonic dispersion is nearly flat on the $\mu m^{-1}$ scale, such that excitons are well localized on the minima of their potential.

In the meantime, cavity photons can also be localized depending on their momentum, which makes the coupling of photons and excitons momentum-dependent.
As Fig.~\ref{fig:CH2_Fig1}(b) shows, the dispersion is characterized by two minima at non-zero wave vectors, $k=k_0$.
In the following sections, we will show that this peculiar dispersion leads to non-trivial effects such as spontaneous momentum symmetry breaking upon exciton-polariton condensation.

%
%
\subsection{Polariton BEC in the thermal equilibrium limit}
Before considering the structure of the dispersion in 2D lattices, it is helpful for us to understand the consequences of the dispersion shown in Fig.~\ref{fig:CH2_Fig1}(b) for the 1D scenario.
Here, we begin by considering the behavior of the system under non-resonant pumping, with which polariton condensation can be expected in the lattice~\cite{Lai:2007aa}.
Because of their finite lifetime, exciton-polaritons are non-equilibrium systems and so would not necessarily form in the ground state~\cite{Maragkou:2010aa}; however, at high densities, energy relaxation is typically enhanced to the ground state~\cite{Winkler:2016aa}.
In this section, we give a qualitative argument in the limit of thermal equilibrium~\cite{Sun:2017aa}.

From Eq.~\eqref{eq:CH2_SP1} we get the dispersion relationship, $E_k$, in the linear regime [shown in Fig.~\ref{fig:CH2_Fig1}(b)]. Then the Hamiltonian of the system can be written as
\begin{equation}
	\label{eq:CH2_H-2}
	\hat{\mathcal{H}}=\sum_k  E_k \hat{a}_k^\dag \hat{a}_k +\alpha \hat{a}_k^\dag \hat{a}_k^\dag \hat{a}_k \hat{a}_k + 2\alpha \sum_{k' \neq k} \hat{a}_k^\dag \hat{a}_{k'}^\dag \hat{a}_k \hat{a}_{k'},
\end{equation}
where we introduce polariton--polariton interaction with the strength $\alpha$. The factor $2$ in Eq.~\eqref{eq:CH2_H-2} is characteristic of the momentum space scattering processes~\cite{Whittaker:2005kg} and can be considered as a permutation of $\hat{a}_k^\dag \hat{a}_{k'}^\dag \hat{a}_k \hat{a}_{k'}$.
\begin{figure}[ht]
	\centering
	\includegraphics[width=0.65\linewidth]{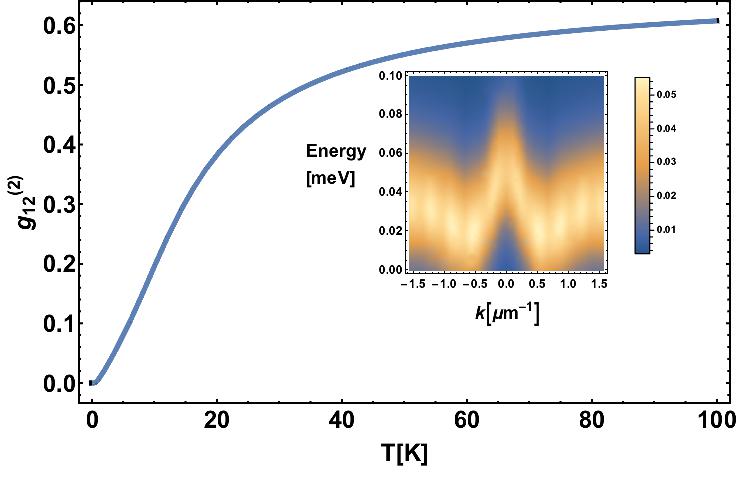}
	\caption[Temperature-dependent $g_2$ function]{Second-order correlation function as a function of temperature. The total number of exciton-polaritons is fixed at $n=100$. The inset plots the low-density result with the total number of polaritons restricted to $n=3$, compared with the dispersion in Fig.~\ref{fig:CH2_Fig1}(b). The figure is taken from~\cite{Sun:2017ab}.}
	\label{fig:CH2_TMFig}
\end{figure}

At zero temperature, one can expect that only the two lowest energy momentum states at $k_1= -k_0$ and $k_2=k_0$ are populated, where $k_0$ is the right minimum valley of the blue curve in Fig.~\ref{fig:CH2_Fig1}(b). 
Then the energy of the system can be written as
\begin{align}
	E\left( n_1 , n_2 \right) &= nE_{k_1} + \alpha \left( n_1^2 +n_2^2 - n + 4n_1 n_2 \right),
	\label{eq:CH2_E-2}
\end{align}
where we define $n_{k_1} = n_1 $, $n_{k_2} = n_2$, and the total population $n =n_1 +n_2$.
It is easy to see that when $\rho=\left( n_1-n_2 \right)/n = \pm 1$, the system achieves its minimum energy. 
In other words, at zero temperature, one can expect that the system would spontaneously choose the state with all the polaritons at either $k_1$ or $k_2$. 
This result can be further confirmed by calculating the second-order correlation function and spectrum corresponding to the Hamiltonian in Eq.~\eqref{eq:CH2_H-2}, as shown in Fig.~\ref{fig:CH2_TMFig} (see Appendix~\ref{AP:A2} for details of the calculation).

%
%

\subsection{Non-equilibrium model of polariton BEC}

In models with weak energy relaxation, it is not necessary to reach the actual ground state of the system~\cite{Krizhanovskii:2009aa,Maragkou:2010aa} due to the finite lifetime of exciton-polaritons.
In this section, we further investigate the behaviour of the system using a stochastic quantum treatment and accounting for various scattering processes (see Appendix~\ref{AP:A4}).
We consider an $\rm{InGaAlAs}$ alloy-based microcavity and use the following parameters during calculation: sound velocity $c_s=5370$ m/s~\cite{Hartwell:2010aa}, and $\gamma=i\hbar/\tau=\hbar/18$ ps$^{-1}$~\cite{Gao:2012aa}.

\begin{figure}[ht]
\centering
	\includegraphics[width=0.99\linewidth]{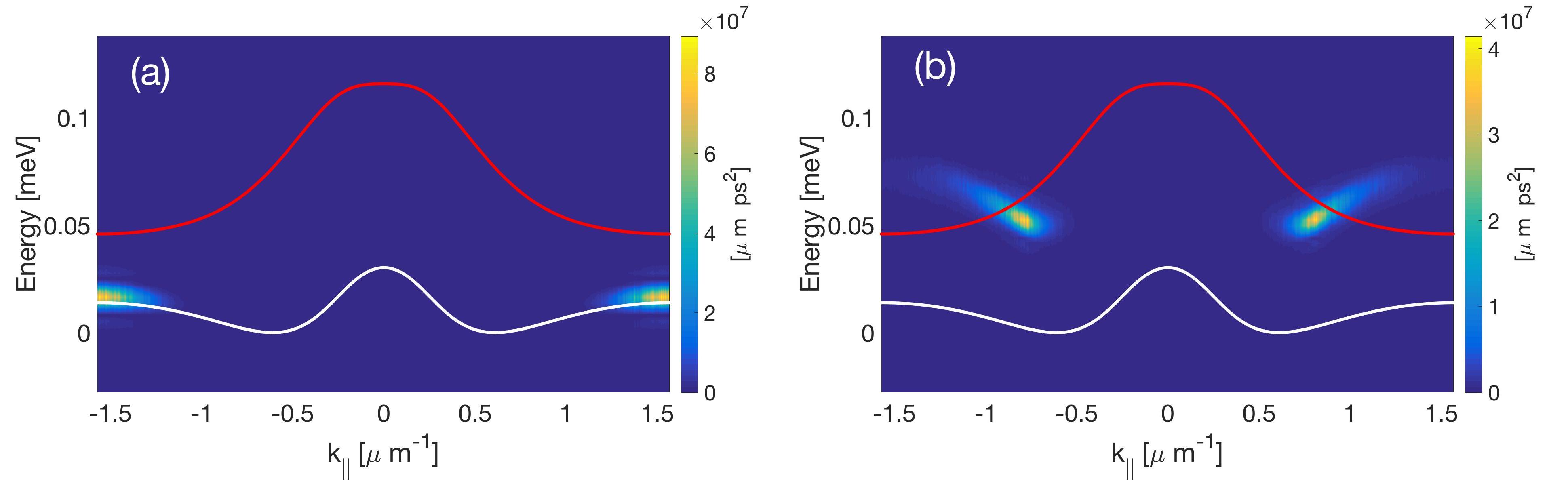}
	\caption[Exciton-polariton spectra]{Distribution of exciton-polaritons localized in the potential profile shown in Fig.~\ref{fig:CH2_Fig1}(b) in the regime of homogeneous incoherent excitation of the system at steady state (1500 ps) including acoustic phonon-assisted scattering.
	Polariton--polariton interaction is switched (a) off and (b) on by setting $\alpha$ to zero and non-zero, respectively.
Red curves indicate the $k$-dependence of the decay rates, and white curves show the exciton-polariton dispersion in the linear regime.
	Exciton-polariton condensates form in the reciprocal space at $k_\parallel\approx \pm 0.74$ $\mu$m$^{-1}$. The figure is taken from~\cite{Sun:2017ab}.}
\label{fig:CH2_Fig4}
\end{figure}

In Fig.~\ref{fig:CH2_Fig4}(a), we turn off the polariton--polariton interaction and see that, in this case, there is no blueshift and the particles occupy mostly the edge of the Brillouin zone. 
This happens because particle lifetime increases with an increase of $|k_{\parallel}|$ with a corresponding decrease of the decay rate, see red curves in Fig.~\ref{fig:CH2_Fig4}.
However, if we account for interaction, we achieve degenerate condensation at $k=\pm k_0$ points due to the interplay of particle lifetime and interactions, as shown in Fig.~\ref{fig:CH2_Fig4}(b). 
Exciton-polaritons also blueshift in energy [compared with Fig.~\ref{fig:CH2_Fig4}(a)] due to this interaction.

One interesting point is that if we change the potential profiles for the excitons and photons, we can achieve different points of condensation; in particular, we can make particles condense at $k=0$ and $k=k_{BZ}$ (see Appendix~\ref{AP:A5}).

Although the condensation of polaritons to non-zero momentum states has been observed previously~\cite{Liu:2015aa,Maragkou:2010aa,Kusudo:2013aa}, in those observations it was a purely non-equilibrium effect.
In our work, on the other hand, condensation to non-zero momentum takes place even in the limit of thermal equilibrium, that is, with strong energy relaxation. 
Furthermore, since the non-zero momentum states are the true ground state of the system, they are likely to be highly stable after they have formed, particularly in polariton systems close to thermal equilibrium. 
This may include the previously developed long-lifetime inorganic microcavities~\cite{Nelsen:2013aa} as well as organic systems~\cite{Kena-Cohen:2010aa} with faster energy relaxation processes~\cite{Lanty:2008aa}.

\subsection{Polariton polarization and dispersion in a 2D lattice}
\begin{figure}[ht]
	\includegraphics[width=0.99\linewidth]{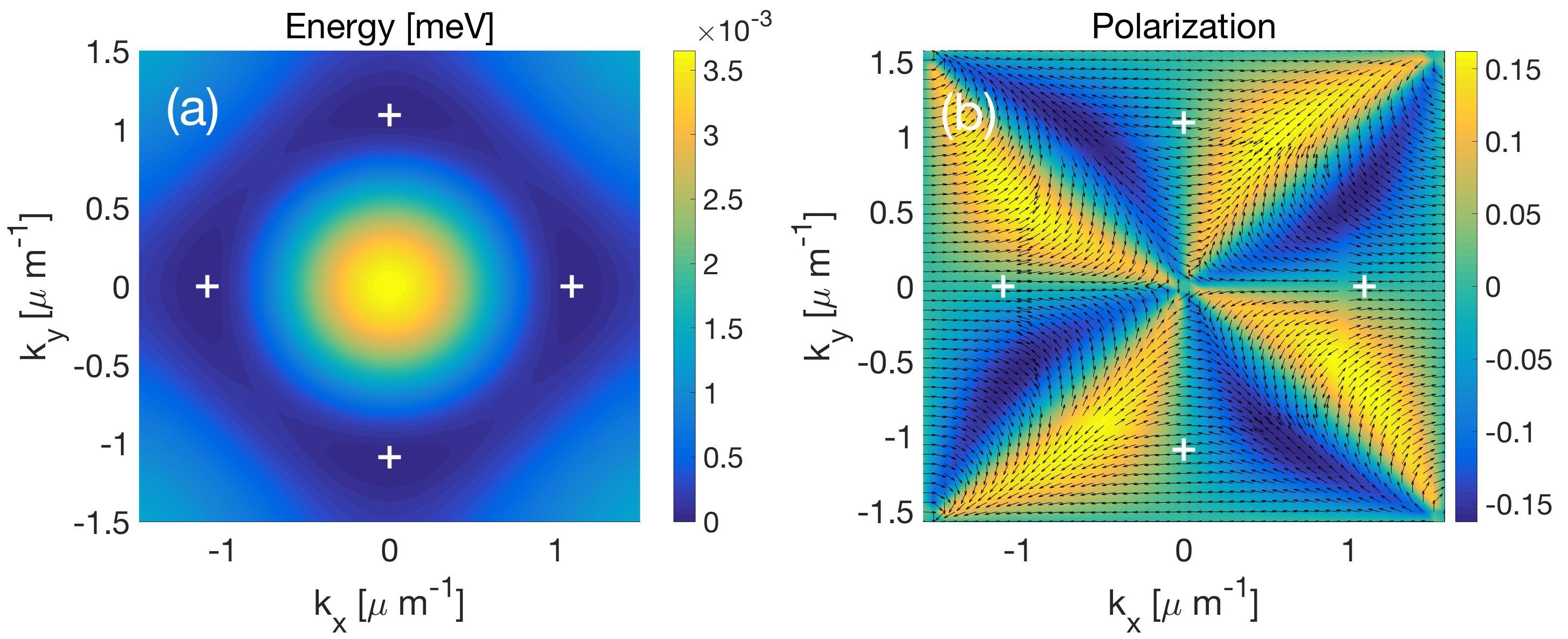}
\caption[Illustration of multivalley coupling]{Illustration of multivalley coupling. (a) Energy dispersion in 2D in the first Brillouin zone, and (b) corresponding polarization chart. The arrows in (b) show the polarization in $x$- and $y$-directions, and colors represent the polarization in the $z$-direction. The white crosses mark the minima of the energy dispersion. The figure is taken from~\cite{Sun:2017ab}.}
\label{fig:CH2_Fig2D}
\end{figure}

In this section, we extend the dispersion calculation into a 2D square lattice structure.
Figure~\ref{fig:CH2_Fig2D}(a) shows the energy of the system ground state in the first Brillouin zone (see also Appendix~\ref{AP:A5}).
Here we can identify four energy minima at the bottom of different valleys in the reciprocal space.
A fundamental feature of 2D semiconductors for valleytronics is the spin-valley coupling that allows different valleys to be excited with light in different polarizations.
We introduce the TE-TM splitting in our model to investigate its consequences. The corresponding Hamiltonian is
\begin{equation}
\label{eq:CH2_TE_TM}
\mathcal{H}_{TE-TM}=\left(\begin{array}{cc}0&\Delta\left(\frac{\partial}{\partial x}-\mi\frac{\partial}{\partial y}\right)^2\\\Delta\left(\frac{\partial}{\partial x}+\mi\frac{\partial}{\partial y}\right)^2&0\end{array}\right).
\end{equation}

After accounting for this splitting, we obtain the polarization structure of the lowest energy band, as shown in Fig.~\ref{fig:CH2_Fig2D}(b).
Here we note that the different valleys have different polarizations, which implies that they can be selectively excited by a resonant excitation with specific polarization.
The geometry of the lowest energy pseudospin should allow to form such patterns as skyrmions~\cite{Flayac:2013aa,Cilibrizzi:2016aa} or spin whirls~\cite{Cilibrizzi:2015aa} under pulsed-resonant excitation.

%
%
\subsection{Summary}
In this chapter, we have considered the formation of exciton-polaritons in a semiconductor microcavity with separate spatially patterned potentials for cavity photons and excitons.
This separated confinement of cavity photons and excitons allows us a momentum-dependent coupling that gives rise to a unique shape of the dispersion, in which degenerate ground states appear at non-zero momenta.
We studied two different limits corresponding to strong and weak energy relaxation.
In the limit of strong energy relaxation, a simple equilibrium theoretical model predicts spontaneous symmetry-breaking in momentum space.
In the limit of weak energy relaxation, a non-equilibrium model accounting for phonon scattering processes shows that the system gives non-equilibrium condensation at a non-zero wave vector.
At last, considering exciton-polaritons in a 2D square lattice, we predicted the formation of a multivalley-dispersion.
Here, different valleys exhibit different polarizations, which, in principle, allows us to selectively excite the system with a polarized laser, and forms a foundation for exciton-polariton valleytronics.

%% file: chapter3_3-4_revised.tex
\section{Phase selection and intermittency of exciton-polariton condensates in 1D periodic structures}\label{Ch3}
Since the discovery of the superfluid--Mott insulator transition with cold atoms in optical lattices~\cite{Jaksch:1998aa,Greiner:2002aa}, systems with bosons in periodic potentials have drawn much attention for both fundamental and applied interests.
Compared to the cold atoms, exciton-polaritons in semiconductor microcavities~\cite{Weisbuch:1992aa,Deng:2002aa,Kavokin:2005kj} possess substantially smaller effective masses and can condense not only at liquid helium~\cite{Kasprzak:2006aa,Balili:2007aa,Lai:2007aa} but also up to room temperature~\cite{Baumberg:2008aa,Lerario:2017aa}.
This advantage makes polaritons in artificial periodic potentials an excellent alternative platform for studying many-body physics, gap solitons~\cite{Tanese:2013aa,Buller:2016aa}, topological polariton states~\cite{Karzig:2015aa,Gulevich:2016aa}, and classical~\cite{Ohadi:2017aa} and quantum~\cite{Liew:2018aa} simulators.

There is a significant difference between polariton BEC and traditional BEC.
In the former, external pumping (coherent or incoherent) is required to create and maintain polaritons due to their finite lifetime in the microcavity, and this pumping usually prohibits the particles from reaching thermal equilibrium, so that steady-state condensates can be formed in excited states with many-body correlations.
Particularly, for polariton BEC in periodic potentials, several non-trivial condensations have been reported, for example $\pi$-condensation in 1D lattices~\cite{Lai:2007aa} and $p$- and $d$-condensation in 2D lattices~\cite{Kim:2011aa}, as well as mixed condensates~\cite{Zhang:2015aa}.

Moreover, polariton condensation in the presence of distributed gain and loss for the single-particle states is expected to be accompanied by the formation of spontaneous currents~\cite{Nalitov:2017aa}.
Another recent finding is polariton condensation in flat bands of 1D~\cite{Baboux:2016aa} and 2D~\cite{Klembt:2017aa,Whittaker:2018ab,Sun:2018aa} periodic systems.
Such condensates provide a strong enhancement of the effects of polariton--polariton interaction due to the reduced kinetic energy of the particles.

In this chapter, we consider a 1D polariton system in a complex periodic potential and complex nonlinearity to account for polariton--polariton interaction and gain saturation.
We show that for a detailed description of the system, it is necessary to consider the imaginary part of the periodic potential, which describes the distributed gain and losses~\cite{Winkler:2016aa} of the single-particle states in the microcavity.
By carefully tuning the parameters of the complex potential (such as the height and width of its imaginary part), one can control the state of the system and demonstrate that several conceptually different situations are possible.

In the case of a relatively large bandwidth (i.e., a large energy difference between the $0$- and $\pi$-states for the single-particle spectrum), we find that the condensation changes from $0$-state to $\pi$-state (or vice versa) with increasing interaction between the particles.
This counter-intuitive effect takes place since the state with maximal gain becomes unstable due to the strongly interacting particles, while the state with minimal gain starts to accumulate bosons.
Consequently, the total number of particles in the condensate is not maximized.

Another interesting result is the formation of propagating dark solitons when the crossover between $0$- and $\pi$-state condensates happens.
At a certain magnitude of polariton--polariton interaction, comparable with the bandwidth, and as a result of soliton propagation, the polaritons are distributed quasi-homogeneously along the dispersion curve instead of accumulating in a single quantum state as in typical condensation.
In this case, the polaritons occupy the band more or less uniformly, and short correlations in space and time manifest as intermittency of the condensate state.


%
%
\subsection{Theoretical model}
We study the solutions to the 1D generalized Ginzburg--Landau equation (GLE)~\cite{Keeling:2008js},
\begin{equation}\label{eq:CH4_GPeq}
    \mi\hbar\partial_t\psi =
    -\frac{\hbar^2}{2 m^*}\partial^2_x\psi + V\left(x\right)\psi
    +\left(\alpha-\mi\beta\right)|\psi|^2\psi,
\end{equation}
where $\psi\left(x,t\right)$ is the wave function of polariton condensate, $V(x)=V(x+a)$ is the complex periodic potential with the lattice constant $a$, $m^*$ is the polariton effective mass, $\alpha$ is the polariton--polariton interaction constant~\cite{Ciuti:1998aa,Tassone:1999aa,Magnusson:2011aa}, and $\beta>0$ accounts for the gain-saturated nonlinearity of the system~\cite{Keeling:2008js, Karpov:2015aa}.
By scaling the wave function $\psi\rightarrow\psi/\sqrt{\beta}$, one can set $\beta=1$ to obtain the dimensionless interaction constant $\alpha/\beta$.
To get Eq.~\eqref{eq:CH4_GPeq}, one can consider the steady-state condition of Eq.~\eqref{eq:Ch1_inchoerent_pumping2} and substitute the result into Eq.~\eqref{eq:Ch1_inchoerent_pumping1}
\begin{eqnarray}
    \label{eq:CH4_GPEQ1}
\mi\hbar \frac{\partial}{\partial t} \psi\left(\mathbf{r},t\right) &=& \left[ -\frac{\hbar^2}{2m}\nabla^2 + V\left(\mathbf{r}\right) + \alpha \abs{\psi\left(\mathbf{r},t\right)}^2 -\frac{\mi \gamma}{2} \right] \psi\left(\mathbf{r},t\right)\nonumber
    \\ &+& \frac{\mi}{2}R\frac{I\left(\mathbf{r},t\right)}{R \abs{\psi\left(\mathbf{r},t\right)}^2 + \gamma_R} \psi\left(\mathbf{r},t\right).
\end{eqnarray}
Then by expanding the last term to the first order in the limit $\frac{\gamma_R}{R\abs{\psi}^2}\gg 1$, one can get
\begin{eqnarray}
    \label{eq:CH4_GPEQ2}
\mi\hbar \frac{\partial}{\partial t} \psi\left(\mathbf{r},t\right) &=& \left[ -\frac{\hbar^2}{2m}\nabla^2 + V\left(\mathbf{r}\right) + \alpha \abs{\psi\left(\mathbf{r},t\right)}^2 -\frac{\mi \gamma}{2} \right] \psi\left(\mathbf{r},t\right)\nonumber
    \\ &+& \frac{\mi}{2}R I\left(\mathbf{r},t\right)\gamma_R\psi\left(\mathbf{r},t\right) -\frac{\mi R^2}{2\gamma_R}\abs{\psi\left(\mathbf{r},t\right)}^2\psi\left(\mathbf{r},t\right).
\end{eqnarray}
In Eq.~\eqref{eq:CH4_GPeq}, the potential is complexed value, by comparing with Eq.~\eqref{eq:CH4_GPEQ2}, we have
\begin{equation}
    \textrm{Re}\left[ V\left(x\right)\right] = V\left(\mathbf{r}\right),~ ~ ~ \textrm{Im}\left[V\left(x\right)\right]= -\frac{1}{2}\left( \gamma - RI\left(\mathbf{r},t\right)\gamma_R \right)
\end{equation}
and similarly
\begin{equation}
    \beta = \frac{R^2}{\gamma_R}.
\end{equation}

By applying the generalized GLE~\eqref{eq:CH4_GPeq}, one has (i) a simple form, as compared to more detailed descriptions involving the incoherent exciton reservoir, and (ii) a minimal set of parameters, which allows us to obtain good qualitative insight into the physics of exciton-polariton condensation in periodic potential.
We want to note that when the system reaches a steady-state on long time scales, the interaction and dissipative parameters get rescaled by taking the reservoir steady-state into account.
However, in the general form of Eq.~\eqref{eq:CH4_GPeq}, the periodicity of the potential and the presence of the two types of non-linearities (interaction and dissipation) remain unchanged, which is a great benefit of the Ginzburg--Landau approach.
Another advantage of Eq.~\eqref{eq:CH4_GPeq} is that it can be easily compared with the complex GLE without a periodic potential, which is a well-established model for the study of nonlinear phenomena in different areas, see for example~\cite{Aranson:2002aa} for a review.

\begin{figure}[ht]
\centering
\includegraphics[width=0.75\textwidth]{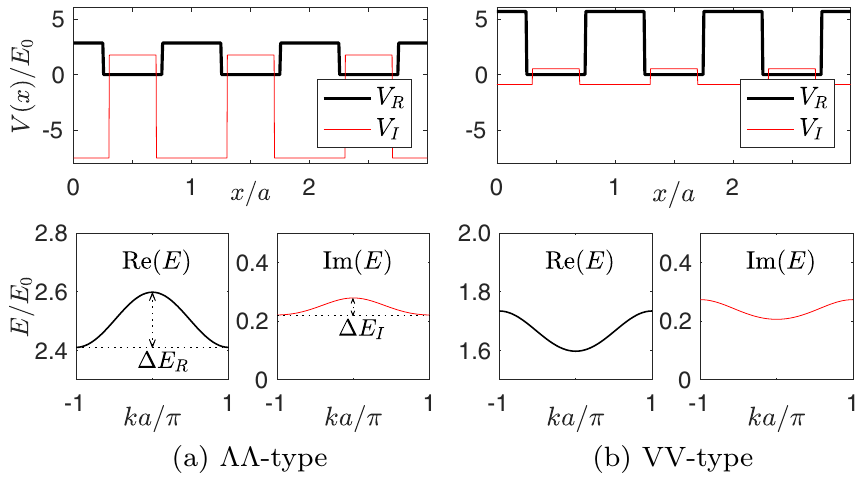}
\caption[Setup of 1D complex potentials and corresponding eigenvalues]{Complex potentials and dispersion of the first band. The real and imaginary parts are plotted with thick black lines and thin red lines, respectively. (a) The $\Lambda\Lambda$ case with the following parameters: $U = 14(\hbar^2/m^* a^2)$, $W = 3.26~U$, $\Gamma=2.64~U$. (b) The VV case with the following parameters: $U=28(\hbar^2/m^* a^2)$, $W=0.25~U$, $\Gamma= 0.16~U$. The energy bandwidth ratio ${\Delta}E_R/{\Delta}E_I \approx 3$ (a) and $2$ (b). In both cases, $a_R = 0.5a$  and $a_I = 0.4a$. The energies are measured in units of $E_0=\pi^2\hbar^2/2m^*a^2$. The figure is taken from~\cite{Yoon:2019aa}.}
\label{CH4_fig:1-1}
\end{figure}

To describe the complex potential $V(x)=V_R(x)+{\mi}V_I(x)$ in one unit cell ($0\le x <a$), we introduce
%
\begin{subequations}\label{eq:CH4_CoxV}
\begin{align}
  V_R(x) &= U\,\Theta\!\left(\left | x -\frac{a}{2} \right |-\frac{a_R}{2} \right), \\
  V_I(x) &= W\,\Theta\!\left(\frac{a_I}{2}-\left | x -\frac{a}{2} \right | \right)-\Gamma,
\end{align}
\end{subequations}
where $\Theta(x)$ is the Heaviside step function, $U$ is the height of the potential barriers, $W$ describes the local gain from the pumping, and $\Gamma$ defines the uniform losses of the system (due to finite polariton lifetime in the microcavity).
Parameters $a_R$ and $a_I$ are the widths of the real potential wells and the imaginary potential barriers, respectively.
By varying the parameters, one can get the potential in Fig.~\ref{CH4_fig:1-1}.

Complex potential [Eq.~\eqref{eq:CH4_CoxV}] reflects the experimental situation in which the system is pumped from the excitonic reservoirs created in the barriers.
Due to exciton-polariton interaction, the particles move into the wells (similar to the case of 2D lattices of trapped polariton condensates~\cite{Ohadi:2017aa}), so that the gain part of the potential (parameter $W$) is located at the wells.
The excitonic reservoirs also increase the barrier heights.
It should be noted that a uniform microcavity subject to a periodic pumping with period $a$ is described by the same model [Eq.~\eqref{eq:CH4_CoxV}].
This pumping not only produces periodic gain (periodic imaginary part), but also periodic repulsive potential (periodic real part).
This setup has a clear benefit in that period $a$ can be tuned.
Since the gain is maximized in the potential wells, the $0$-state, which mainly resides in the wells, has bigger gain than the $\pi$-state, which mainly resides in the barriers~\cite{Lai:2007aa}.

It is important that, depending on the parameters of the potential in Eq.~\eqref{eq:CH4_CoxV}, the single-polariton spectrum can be categorized into four qualitatively different types.
We classify them as $\Lambda\Lambda$ [see Fig.~\ref{CH4_fig:1-1}(a), lower panel], VV [Fig.~\ref{CH4_fig:1-1}(b), lower panel], V$\Lambda$, and $\Lambda$V, depending on the position of the energy minimum (the real part of the eigenvalue) and the position of the gain minimum (the imaginary part of the eigenvalue) for the first band.
For example, the $\Lambda\Lambda$-type corresponds to the case when the minimum energy and minimum gain are both at the edge of the first Brillouin zone at $k=\pm \pi/a$ [see Fig.~\ref{CH4_fig:1-1}(a)].
We also define the effective widths of the bands $\Delta E_R$ and $\Delta E_I$.

\begin{figure}[!htb]
\centering
\includegraphics[width=0.85\textwidth]{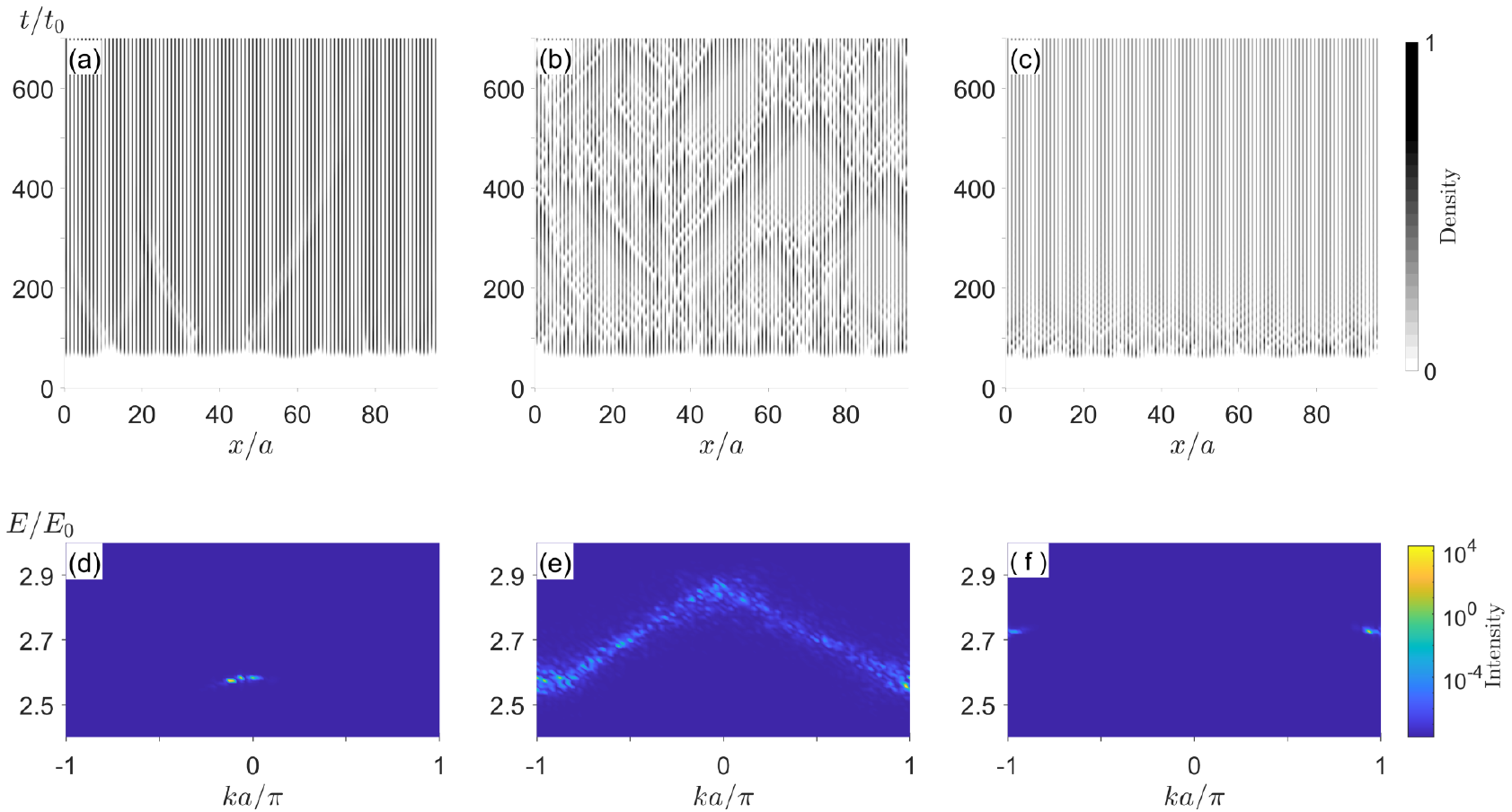}
\caption[Spatiotemporal patterns of the condensate density for the $\Lambda\Lambda$ case]{(a--c) Spatiotemporal patterns of condensate density $|\psi(x,t)|^2$, and (d--f) corresponding intensity $|\psi(k,E)|^2$ for the $\Lambda\Lambda$ case. The dimensionless interaction constant ($\alpha/\beta$) is equal to $0$, $2$, and $6$ in left, middle, and right panels, respectively. Time is measured in units of $t_0 \equiv \hbar/E_0$. The figure is taken from~\cite{Yoon:2019aa}.}
\label{CH4_fig:2-1}
\end{figure}

In what follows, we will consider the formation of the polariton condensate to be slightly above the threshold, when the losses in the system, as governed by the parameter $\Gamma$, are sufficient to make the first band the only one that possesses a positive imaginary part of the eigenvalue.
Thus, the particles are expected to condense into this band.
The complex potential in Fig.~\ref{CH4_fig:1-1} is then set by these two approaches:
(i) detuning the width and depth of the well to get ground state dispersion in the $\Lambda\Lambda$ (or VV) case, and (ii) changing the magnitude of the overall shift of the imaginary part to make the first band the only one with a positive imaginary part of the eigenvalue.
We also consider the first band to be well separated from the other bands, thus disregarding transitions between bands.

%
%
\subsection{Spatiotemporal dynamics of the condensate}

\begin{figure}[ht]
\centering
\includegraphics[width=0.85\textwidth]{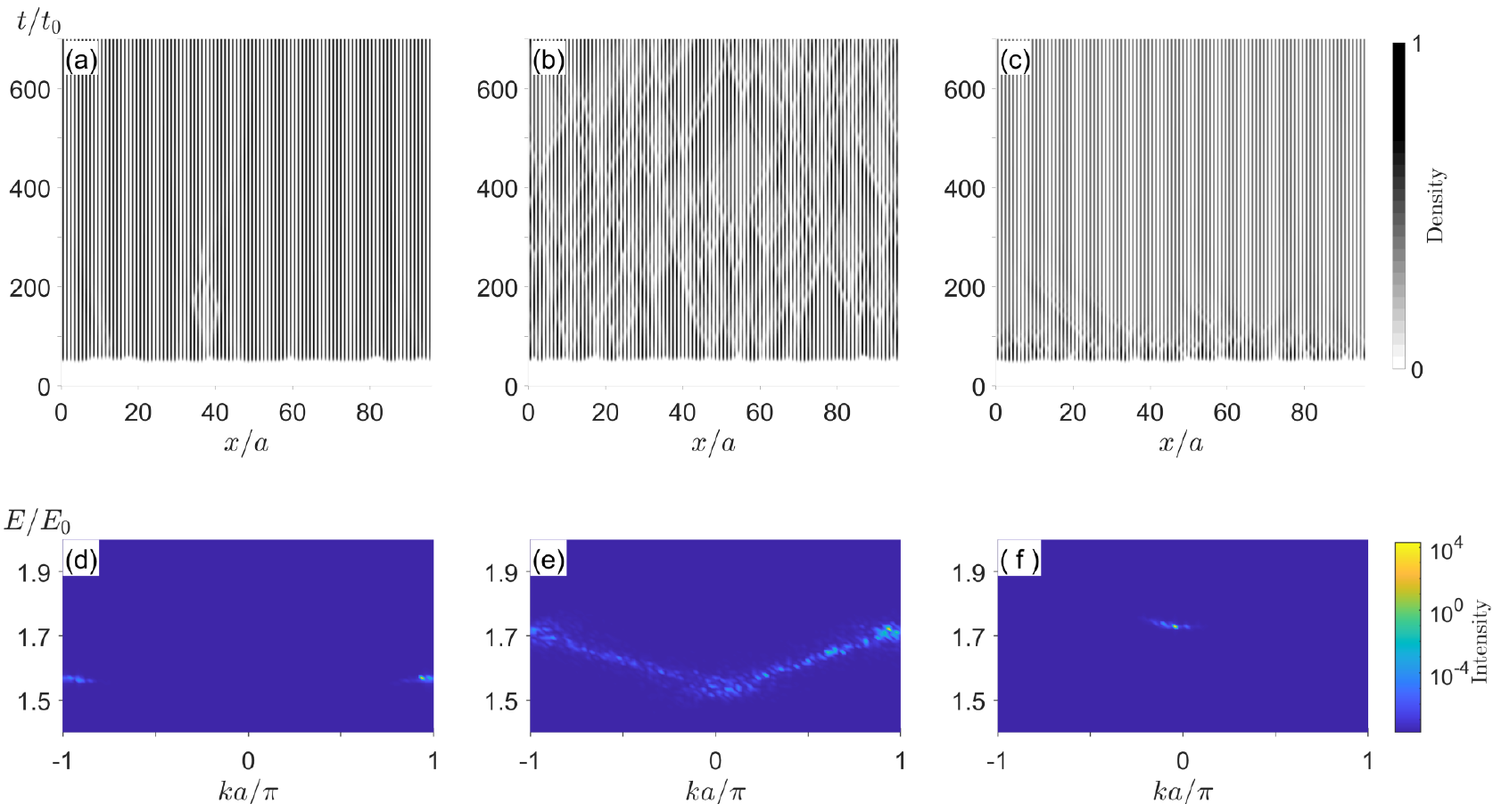}
\caption[Spatiotemporal patterns of the condensate density for the VV case]{(a--c) Spatiotemporal patterns of condensate density $|\psi(x,t)|^2$, and (d--f) corresponding intensity $|\psi(k,E)|^2$ (d, e, f) for the VV case. The dimensionless interaction constant ($\alpha/\beta$) is equal to $0$, $0.65$, and $2$ in left, middle, and right panels, respectively. The figure is taken from~\cite{Yoon:2019aa}.
}
\label{CH4_fig:2-2}
\end{figure}

For the non-interacting case, i.e. $\alpha = 0$ in Eq.~\eqref{eq:CH4_GPeq}, one expects to obtain the condensate in the state with the maximum gain.
Then the number of particles reaches its maximum in this state and can be stabilized by finite gain-dissipation parameter $\beta$.
Therefore, we can expect $0$-state condensation in the $\Lambda\Lambda$ and V$\Lambda$ cases, and $\pi$-state condensation in the $\Lambda$V and VV cases.
By checking the numerical simulations of Eq.~\eqref{eq:CH4_GPeq}, we can show that in the V$\Lambda$ and $\Lambda$V cases, this scenario of condensation persists even in the presence of strong repulsive polariton--polariton interaction.
However, the polariton--polariton interaction has a dramatic effect on the condensate loaded in the $\Lambda\Lambda$ and VV cases, leading to a fundamental reconstruction of the condensate state by increasing interactions.
Therefore, we will mostly concentrate on the VV and $\Lambda\Lambda$ configurations.

In Figs.~\ref{CH4_fig:2-1} and~\ref{CH4_fig:2-2}, we show the spatiotemporal dynamics of the condensate for the $\Lambda\Lambda$ and VV cases, respectively.
The upper rows of these figures show the spatiotemporal patterns of the condensate density, $|\psi(x,t)|^2$, and the lower ones show the polariton emission intensities, $|\psi(k,E)|^2$, for a small initial random seed of noise.
We calculated 50 trajectories and checked that different small initial random seeds give qualitatively similar pictures.
In the density plots, shown in the panels (a), (b), and (c) in both figures, the local maxima of the condensate density (dark black vertical lines along the time axis) are at the centers of the wells of the real potential $V_R$, where the maximal gain is attained [see also Fig.~\ref{CH4_fig:4}(a,b)].

From left to right in Figs.~\ref{CH4_fig:2-1} and \ref{CH4_fig:2-2}, the dimensionless parameter $\alpha/\beta$ increases and we observe considerable changes in the spatiotemporal density patterns and condensation states.
When $\alpha/\beta=0$, the condensate is formed in the state with the maximum gain: the $0$-state in the $\Lambda \Lambda$ case and the $\pi$-state in the VV case.
Defects appearing at the early stage of evolution dissipate away at later times.
When $\alpha/\beta$ takes an intermediate value, polaritons no longer accumulate in the state with a well-defined wave vector, but rather they distribute along the whole band.
As one can see from the corresponding panels [Figs.~\ref{CH4_fig:2-1}(b) and~\ref{CH4_fig:2-2}(b)], strong spatiotemporal chaos is present in this case.
Surprisingly, with further increase of the $\alpha/\beta$ ratio, well-defined condensation takes place again, but now the condensate is formed at the minimum of the dispersion, i.e. in the state with the smallest gain; this corresponds to the $\pi$-state in the $\Lambda \Lambda$ case and the $0$-state in the VV case.
This result is in contrast with the one obtained in the zero-interaction regime.
It should be noted that the $\pi$-state in our system is different from the one discussed in~\cite{Lai:2007aa}, where it corresponded to the minimum of the second band.

The density patterns presented in Figs.~\ref{CH4_fig:2-1}(c) and~\ref{CH4_fig:2-2}(c) resemble those of the spatiotemporal intermittency in the 1D complex Ginzburg--Landau equation (CGLE)~\cite{Chate:1994aa,Melo:1993aa,Hecke:1998aa},
which is
\begin{equation}\label{CH4_eq:NLSE}
\mi \partial_t A = \mi A + (c_1+ \mi )\partial_x^2 A + (c_3 - \mi)|A|^2 A \ .
\end{equation}
The nonlinear term here takes the same form as in Eq.~\eqref{eq:CH4_GPeq} and the linear terms for $c_1>0$ correspond to the complex-valued energy dispersion of the $\Lambda \Lambda$ case, even though the dispersion in Fig.~\ref{CH4_fig:1-1} is not a quadratic but a periodic function.
The parameters $c_1$ and $c_3$ play similar roles as ${\Delta}E_R/{\Delta}E_I$ and $\alpha/\beta$ in our system, respectively.
However, the shapes of the real and imaginary parts of the dispersion in Fig.~\ref{CH4_fig:1-1} are not exactly proportional to each other, and the \textit{continuous} translation symmetry of the CGLE is reduced to a \textit{discrete lattice} translation symmetry due to the periodic potential.

Another important difference is that even though the offset term $\mi A$ in Eq.~\eqref{CH4_eq:NLSE} admits only a restricted range of wave numbers ($-1<k<1$) to have a positive gain with a maximum at $k=0$, the edge points $k=\pm 1$ have zero gain, and therefore the condensate cannot be formed at the edge.
Instead, the polariton system is characterized by the edge points of the first Brillouin zone $k = \pm \pi/a$ with finite gain and the singularity in the density of states.
It is this feature that allows the formation of the polariton condensate at the edge.

In the system described by the CGLE, the spatiotemporal intermittent phase appears in the transition from a plain-wave state to a turbulent (chaotic) state~\cite{Chate:1994aa}.
In our case, the phase landscape is different: the spatiotemporal intermittency appears in the transistion from a $0$-state Bloch wave to a $\pi$-state Bloch-wave.
Nonetheless, the spatiotemporal intermittency pattern is quite similar to the conventional one.
We note that the intermittency phase exhibits some features of deterministic chaos within a single trajectory.
Similar intermittent polariton states have recently been reported in spontaneously-formed periodic structures under resonant driving~\cite{Gavrilov:2018aa}.

\begin{figure}[ht]
\centering
\includegraphics[width=0.85\textwidth]{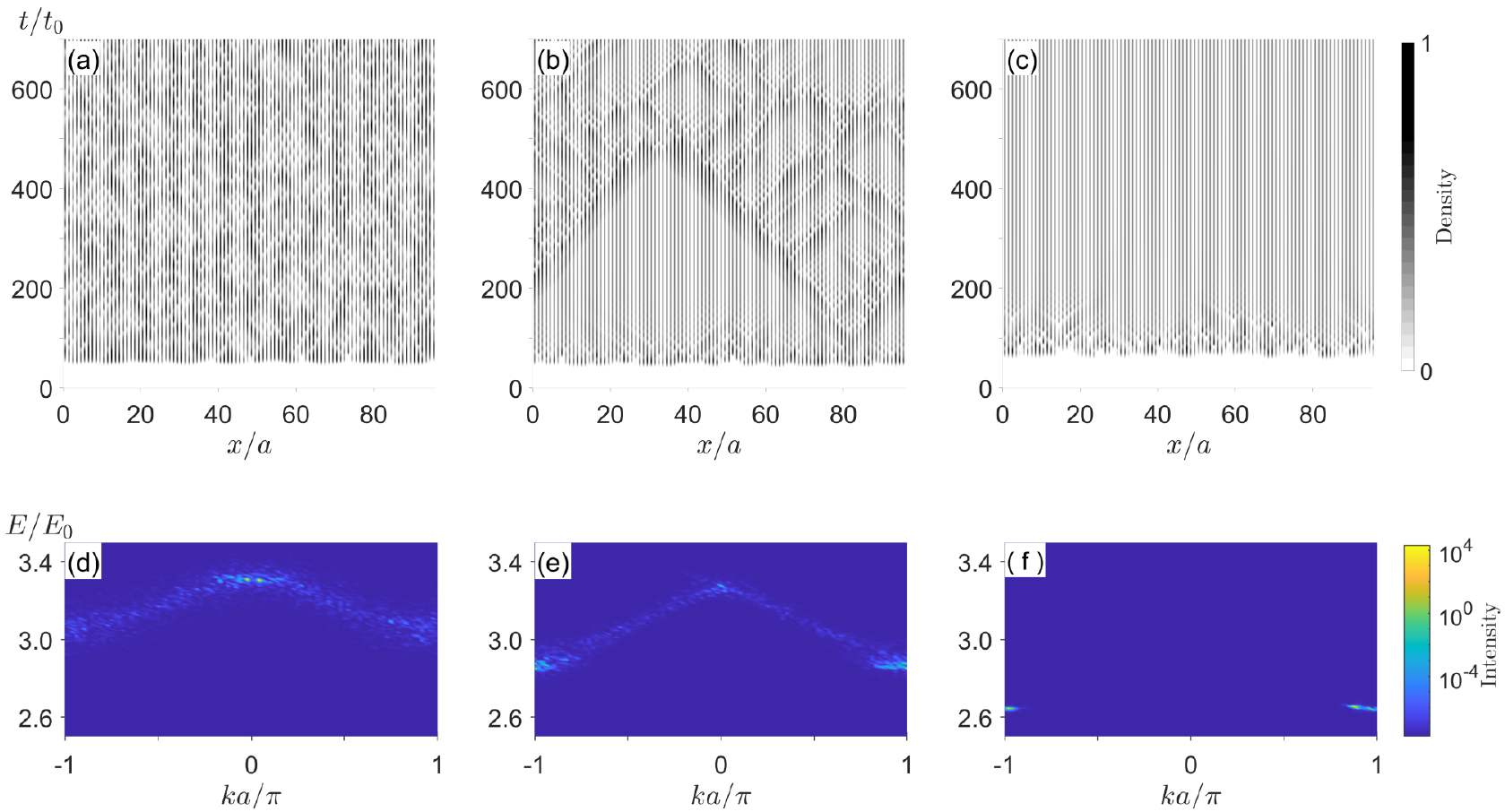}
\caption[Comparing spatiotemporal patterns with fixed nonlinearity]{(a--c) Spatiotemporal patterns of condensate density $|\psi(x,t)|^2$, and (d--f) corresponding intensity $|\psi(k,E)|^2$ for the $\Lambda\Lambda$ case with $\alpha/\beta = 4$ fixed. From left to right: the ratios of the real and imaginary parts of the energy bandwidths are ${\Delta}E_R/{\Delta}E_I \approx 1$, $2$, and $4$; $W = 3.82~U$, $3.43~U$, and $3.18~U$ with $U=14(\hbar^2/m^* a^2)$; and $\Delta E_R \approx  0.107~E_0$, $0.164~E_0$, and $0.201~E_0$ with $E_0=\pi^2\hbar^2/2m^*a^2$, respectively. The figure is taken from~\cite{Yoon:2019aa}.}
\label{CH4_fig:3-1}
\end{figure}

The crossover from spatiotemporal intermittent dynamics to the $\pi$-condensate in the $\Lambda\Lambda$ case (or to the $0$-condensate in the VV case) is not always possible.
The formation of a new condensate phase depends not only on the polariton interaction strength $\alpha/\beta$, as has been discussed above, but also on the ratio ${\Delta}E_R/{\Delta}E_I$.
Figure~\ref{CH4_fig:3-1} demonstrates (for the $\Lambda\Lambda$ case) that a sufficiently large value of the ratio ${\Delta}E_R/{\Delta}E_I$ is required to reach a $\pi$-state from the mixture of spatiotemporal intermittency states.

%
%
\subsection{Spatiotemporal intermittency}
The transition from a regular/laminar state to an irregular/turbulent state via spatiotemporal intermittency has been suggested to be related to the directed-percolation process~\cite{Pomeau:1986aa}.
Following directed percolation studies, spatial intermittency can be quantified by measuring the distribution of the lengths of laminar domains, and this is expected to reveal a power-law behavior at the threshold of spatiotemporal intermittency~\cite{Chate:1987aa, Chate:1994aa}.

The spatiotemporal pattern we observe is not exactly the same as in conventional spatiotemporal intermittency~\cite{Pomeau:1986aa}, because, in our system, the intermittent pattern appears halfway between two ordered phases ($0$- and $\pi$-phases).
On both sides of the intermittency domain, there is only one stable condensate state.
Inside the intermittency domain, both $0$- and $\pi$-phases of the condensate are stable, but they possess different basins of attraction.
The basin of attraction of the new phase grows with increasing polariton--polariton interaction, while the basin of attraction of the old phase shrinks down.
Near the boundary of the intermittency domain, a phase with a smaller basin of attraction is created by the propagating solitons, which have the form of small domains of this less-probable phase.

%
%
%
\begin{figure}[ht]
\centering
\includegraphics[width=0.75\textwidth]{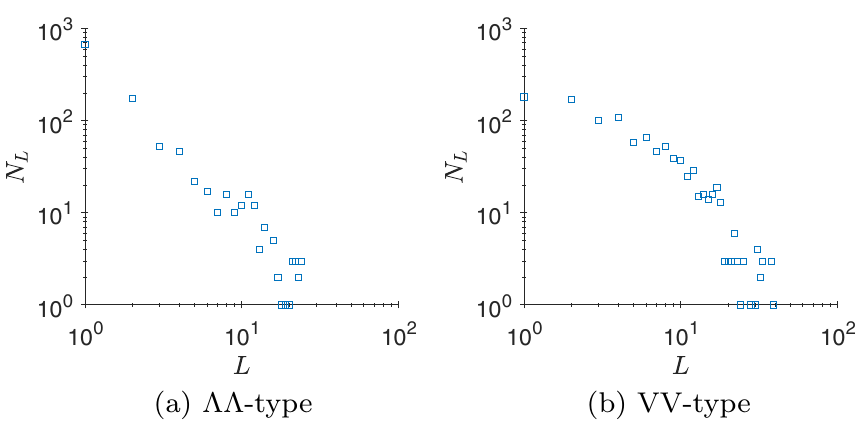}
\caption[Distributions of the number of defect domains]{Distributions of the number of defect domains ($N_L$) for the spatiotemporal patterns presented in (left) Fig.~\ref{CH4_fig:2-1}(b), and (right) Fig.~\ref{CH4_fig:2-2}(b).
$N_L$ is sampled from $t/t_0=200$ to $t/t_0=700$ with a time interval of $t/t_0=5$.
Both horizontal and vertical axes are in log scale. The figure is taken from~\cite{Yoon:2019aa}.}
\label{CH4_fig:1}
\end{figure}

However, the spatiotemporal pattern in our case is quite similar to the one appearing in the transition to turbulence.
We also consider the distribution of lengths of the defect domains, which represent domains of dark solitons in the $\Lambda \Lambda$ case and domains of the absence of dark solitons in the VV case.
Figure~\ref{CH4_fig:1} shows the distributions of $N_L$ (number of defect domains of length $L$).
Here $N_L$ is sampled at $t/t_0 = 200\to 700$ with a time interval of $5$. We have checked that the linearly declining behavior (in log-log scale) is insensitive to the choice of smaller time intervals.
We indeed observe a behavior close to power-law, which is the fingerprint of spatiotemporal intermittency.

%
%
\subsection{Soliton dynamics}
\begin{figure}[ht]
\centering
\includegraphics[width=0.85\textwidth]{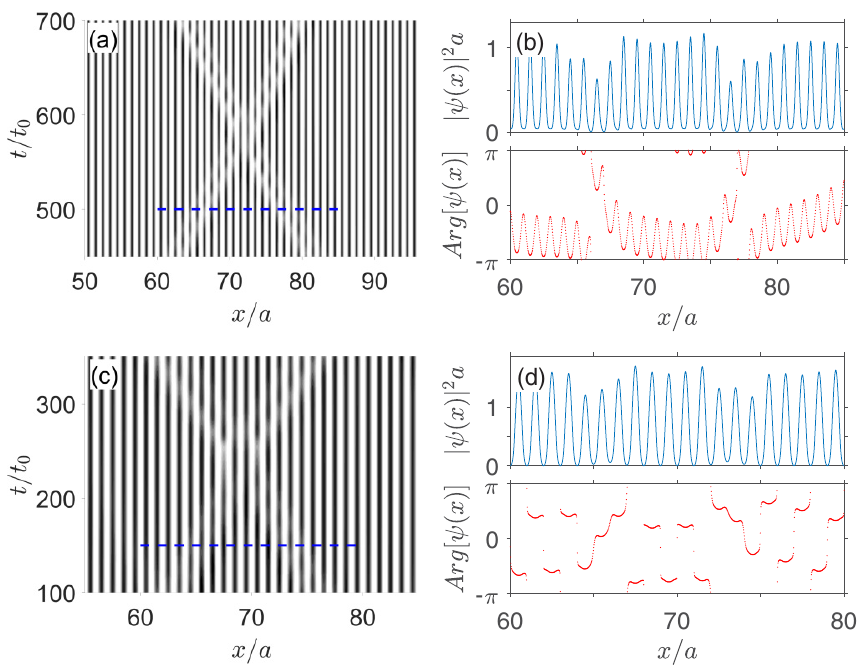}
\caption[Propagation and collision of solitons]{Propagation and collision of a pair of solitons in (a,b) the $\Lambda\Lambda$ case and (c,d) the VV case, for $\alpha/\beta=0.4$ when the number of solitons is small. (b,d) Upscaled particle density profiles and the phases of the wave functions just before the collisions marked by the thick dashed blue lines in (a,c), respectively. The figure is taken from~\cite{Yoon:2019aa}.}
\label{CH4_fig:4}
\end{figure}

At weak polariton--polariton interaction, the long-range order of the $0$-condensate in the $\Lambda\Lambda$ case and the $\pi$-condensate in the VV case is destroyed by the formation of propagating defects.
Each defect extends only over a few lattice constants, and they are characterized by the suppression of condensate occupation and by the phase slips.

As an example, Fig.~\ref{CH4_fig:4} shows the collision events of a pair of such defects propagating towards each other.
Individual collisions are seen only for weakly interacting polaritons when the concentration of defects is small.
Each defect is characterized by a depletion of particle density together with an abrupt change in the phase of the wave function at two edge points of the defect, where the phase change is about $\pi$ in the $\Lambda\Lambda$ case and smooth in the VV case.
The fact that the defects maintain their properties after the collision indicates that they can also be considered as dark solitons.
We note, however, that these solitons differ from the Bekki--Nozaki hole solutions of the CGLE~\cite{Bekki:1985aa} and the dissipative Gross--Pitaevskii equation for polariton mean field~\cite{Xue:2014aa}.
In our case, solitons represent the building blocks of a new condensate phase.

Soliton density increases with $\alpha/\beta$, and some of them attach to each other to form wide soliton domains, which one can see in Figs.~\ref{CH4_fig:2-1}(b) and~\ref{CH4_fig:2-2}(b).
In the $\Lambda\Lambda$ case, with a further increase of polariton--polariton interaction and for a sufficiently large ${\Delta}E_R/{\Delta}E_I$ ratio, a complete array of dark solitons is formed as in Fig.~\ref{CH4_fig:2-1}(c).
The wave function phase changes by $\pi$ per every lattice constant and the quasi-long-range order appears again, manifesting the formation of a $\pi$-condensate phase.

%
%
\subsection{Summary}
We have shown that interacting exciton-polaritons loaded into a one-dimensional microcavity with a periodic potential and periodic distribution of losses can condense into nontrivial states, where losses are not minimized but rather maximized.
Under certain conditions, polaritons can form a space-time intermittency phase, which separates two condensate phases with minimal and maximal losses.
The reconstruction of the condensate wave function takes place by a proliferation of dark solitons along the periodic structure.
The nuclei of the new condensate phase, which are characterized by the maximization of losses, are formed with increasing polariton--polariton interaction, and they can be seen as a result of dark solitons gluing  together.

%% file: chapter4_2-28_revised.tex
\chapter{Exciton polariton in complex lattices}\label{CHTHREE}
In this chapter, we consider the lattice with more than one site per unit cell which supports more than one band in the tight-binding model.

In the first part of the chapter, we study the formation of the compact localized state in exciton-polaritons condensation. Considering the Lieb lattice, we appliy a resonant Laguerre--Gaussian pulse to excite the condensate in the compact localized state and checked the evolution of the state with and without a background homogeneous incoherent pumping.

In the second part of the chapter, we considered the topological properties in exciton-polariton lattice system. We showed that by arranging local magnetic quantum dots (QDs) into a graphene pattern and considering TE-TM splitting, a non-trivial topological edge state can be found.
Further, by changing the magnitude of the local magnetic field, we found that the Chern number can change from $\pm 2$ to $\mp 1$.

\section{Excitation of localized condensates in the flat band of exciton-polariton Lieb lattice}\label{Ch4}
The full quench of single-particle kinetic energy is the main feature of dispersionless or flat bands (FBs)~\cite{Derzhko:2015aa,Leykam:2018aa,Leykam:2018ab}.
In many-body physics, even weak interations between particles become important in FB systems.
One example of interesting fermionic correlations is the fractional quantum Hall effect that appears in flat Landau levels.
Particles with bosonic statistics are also expected to dramatically change their properties under FB settings.
Due to the high degeneracy of the FB energy level, one can construct compact localized states (CLSs) that extend over a few lattice sites only for a specific tight-binding model.
The first such observation in a 2D dice lattice was found by Sutherland~\cite{Sutherland:1986aa}.
At a low concentration of bosonic particles, they can be distributed over several CLSs in such a way that their wave functions do not overlap, so that the total energy is minimized in the case of repulsive interaction between particles.
Thus, depending on the number of occupied sites, the bosons can develop a supersolid phase that features periodic density modulation~\cite{Huber:2010aa}.

This chapter applies an exciton-polariton approach to address the loading of bosons of finite lifetime into a FB and the related effects.
The exciton-polaritons represent strongly coupled states of microcavity photons and semiconductor QW excitons~\cite{Kavokin:2017aa}.
Driven-dissipative condensates of exciton-polaritons have been reliably observed in semiconductor microcavities~\cite{Kasprzak:2006aa,Balili:2007aa}; following this,
the potential of polariton condensates in artificial lattices for both applied and fundamental research has been well studied.
Examples include $\pi$-condensates at the edges of bands in 1D periodic potentials~\cite{Lai:2007aa}, and $d$-condensates in 2D square lattices~\cite{Kim:2011aa}.
Various methods of polariton trapping have been employed.
In particular, polariton condensates subject to spatially periodic acoustic phonon fields have been successfully created and studied~\cite{Cerda-Mendez:2010aa,Cerda-Mendez:2013aa}.
It was also shown that the periodic long-range order in a polariton condensate under resonant excitation can appear spontaneously~\cite{Gavrilov:2018aa}.

Interest has blossomed in exciton-polariton condensation in more complicated artificial periodic potentials, which target topologically protected~\cite{Karzig:2015aa,Nalitov:2015aa,Bardyn:2015aa,St-Jean:2017aa,Li:2018aa,Solnyshkov:2018aa} and single-particle FBs.
Such condensation has been studied in honeycomb~\cite{Jacqmin:2014aa}, kagome~\cite{Masumoto:2012aa,Gulevich:2016aa}, 1D Lieb~\cite{Baboux:2016aa} and 2D Lieb~\cite{Klembt:2017aa,Whittaker:2018ab} lattices.
In FB systems, the coherence length of polariton condensates only extends to a small number of lattice sites; two possible explanations for this are that the potential disorder limits the range or more simply that the fragmentation is a generic feature of non-equilibrium condensation in FBs.

The lattice considered here is a 2D Lieb lattice with a geometry similar to the one in~\cite{Klembt:2017aa}.
We investigate the combined effect of distributed dissipation and exciton-photon coupling on the band structure.
In the following sections, by examining both the energy and lifetime of the particles, we identify possible candidate states for condensation in each band. We show that while no perfect FB exists in this continuous, ``non-tight-binding'' system, the concept of long-lived strongly localized states, as maintained by the destructive interference of propagating waves, is still valid to some degree.

%
%
\subsection{2D Lieb lattice and band structure}
\begin{figure}[ht]
\centering
\includegraphics[width=0.6\linewidth]{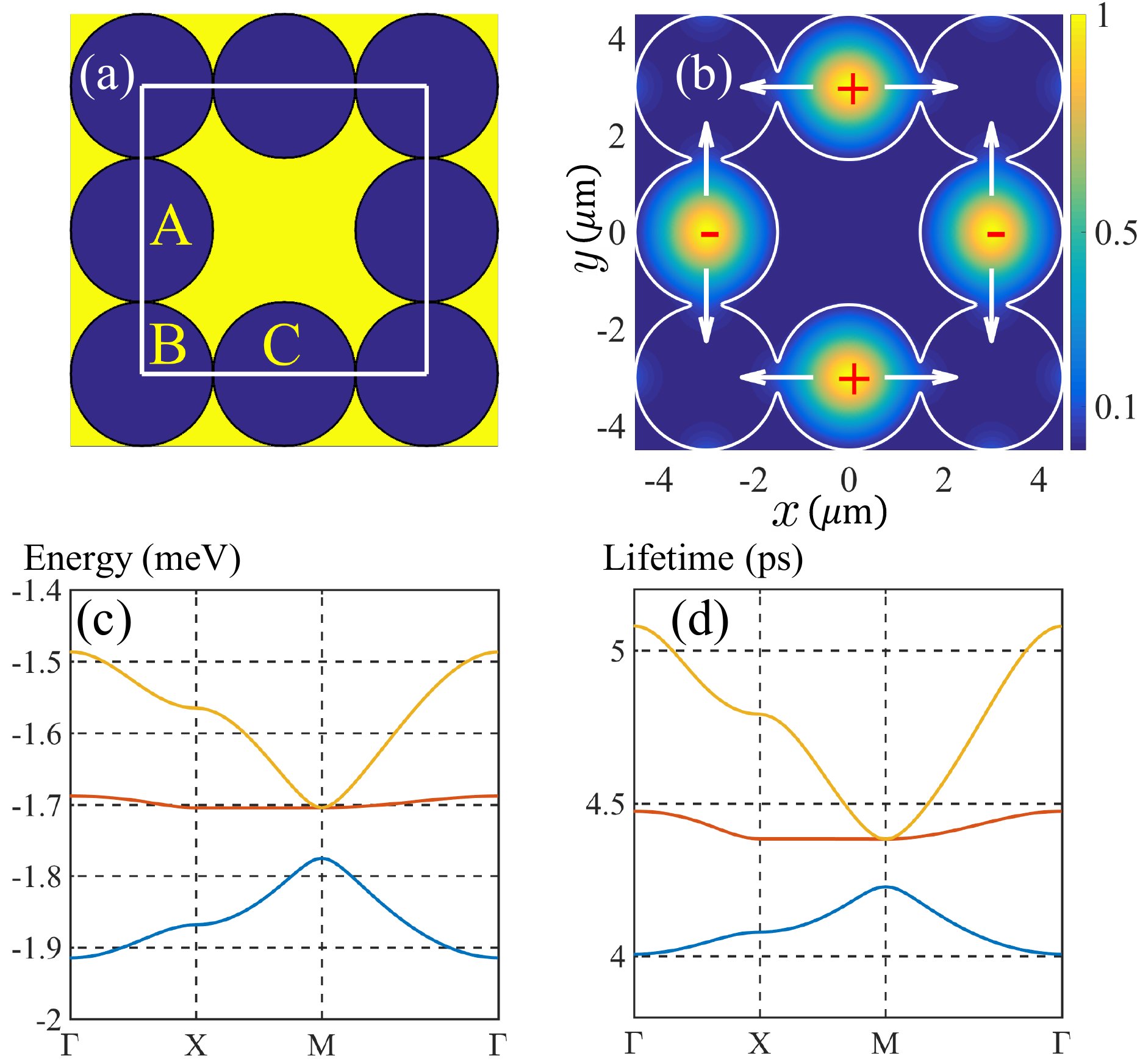}
\caption[Lieb lattice and corresponding spectrum]{(a,b) System schematics and (c,d) single-particle spectra.
(a) The Lieb lattice plaquette including three pillars (QWs) per unit cell: A, B, and C.
(b) Probability density of the photonic component of the single-polariton (Bloch) state of the nearly flat (second) band at the $\Gamma$-point.
Signs indicate the wave function phase. The weak population of the B sites is not visible.
A CLS possesses a similar structure and will not propagate along the arrow directions due to destructive interference caused by the $\pi$-phase difference at sites A and C.
(c) The real part of the energy of the single-particle Bloch bands.
(d) The lifetimes of the Bloch states (the inverse imaginary part of the eigenvalues).
 The figure is taken from~\cite{Sun:2018aa}.}
\label{fig:CH3_1}
\end{figure}
The exciton-polariton condensate wave function can be written as $\Psi=(\varphi,\chi)^\mathrm{T}$, where $\varphi$ and $\chi$ are the photonic and excitonic components, respectively.
The mean-field Hamiltonian of the system reads (we set $\hbar=1$)
\begin{equation}
  \hat{H} =
  \begin{pmatrix}
  -\frac{\nabla^2}{2m_c}+V(\mathbf{r}) & \Omega \\
  \Omega & \delta -\frac{\mi}{2\tau_x} -\frac{\nabla^2}{2m_x}+\alpha_x|\chi|^2
  \end{pmatrix},
  \label{CH3_Ham}
\end{equation}
where $m_c$ and $m_x$ are the microcavity photon and exciton effective masses, respectively, $\Omega$ is the Rabi frequency, $\alpha_x$ is the exciton--exciton interaction strength, $\tau_x$ is the exciton lifetime, and $V(\mathbf{r})=V_r(\mathbf{r})-{\mi}V_i(\mathbf{r})$ is the complex-valued potential experienced by the photonic component separately from the excitonic component~\cite{Sun:2017ab}.
The real part of the potential $V_r$ is defined by the QWs that form the Lieb lattice [see Fig.~\ref{fig:CH3_1}(a)]. The imaginary part $V_i$ describes the distributed losses in the system.
We set $V_r=0$ and $V_i=0.1\,\mathrm{meV}$ inside the wells, while  $V_r=30\,\mathrm{meV}$ and $V_i=2.1\,\mathrm{meV}$ in the barriers.
It should be noted that photon lifetime is expected to be nonuniform; indeed, the barriers are usually produced by a partial etching of the DBRs, which introduces additional photon leakage from the barrier area.
The diameter of each QW is $3\,\mu\mathrm{m}$ and the lattice constant is $a=6\,\mu\mathrm{m}$.
The other parameters are $\hbar\Omega=4.25\,\mathrm{meV}$, $m_c=3.2\times10^{-5}\,m_e$, $m_x=10^5\,m_c$, $\tau_x=100$ ps, and the detuning is $\delta=-4.0\,\mathrm{meV}$.
Here, $\tau_x^{-1}\ll-2\mathrm{Im}\{V(\mathbf{r})\}$, so that the losses in the polariton system are controlled by the photonic component.

The unit cell of the Lieb lattice is composed of three QWs labeled A, B, and C, as shown in Fig.~\ref{fig:CH3_1}(a).
It is well known that in the framework of a tight-binding model the system spectrum possesses a FB.
The CLS in the tight-binding FB is located on the A and C sites of a single .
The phases on A and C are shifted by $\pi$, and the CLS is maintained due to the destructive interference of waves propagating from A and C sites to B~\cite{Vicencio:2015aa}.
The Bloch state of the second (nearly) FB at the $\Gamma$-point, as shown in Fig.\ref{fig:CH3_1}(b), has a similar structure except that it extends over the whole lattice.
This state also shows a $\pi$ phase shift between A and C sites, in addition to a very weak excitation of the B sites.

Figure~\ref{fig:CH3_1}(c) shows the three lowest bands representing the spectrum of noninteracting polaritons ($\alpha_x=0$).
It is clear that the continuous model [Eq.~\eqref{CH3_Ham}] does not lead to a perfect FB.
The middle band, which is flat within the tight-binding model with nearest-neighbor hopping, shows a small but finite dispersion.

Another key feature of this system concerns the dispersion of losses in the bands, as shown in Fig.~\ref{fig:CH3_1}(d).
For the lowest band, the state with the smallest losses occurs at the corner of the Brillouin zone (M point) with the wave vector $k_x={\pm}k_y={\pm}\pi/a$.
For the middle (nearly flat) band, minimum dissipation takes place at $k=0$ ($\Gamma$ point).
The wave function of this state corresponds to highly occupied A and C sites and nearly empty B sites, as shown in Fig.~\ref{fig:CH3_1}(b).

%
%
\subsection{Laguerre--Gaussian resonant pumping}
We excite the compact localized condensate (CLC) of the middle band in Fig.~\ref{fig:CH3_1}(c), i.e. the FB, by exposing the Lieb lattice structure to a short resonant (ring-shaped) Laguerre--Gaussian pulse centered at one.
The polariton wave function in this case evolves according to
\begin{equation}\label{CH3_Evol}
  \mi\dot{\Psi} = \hat{H}\Psi+\begin{pmatrix} \mi P(\mathbf{r},t)\\{0}\end{pmatrix},
\end{equation}
where the pulse profile is given by~\cite{Kim:1999aa}
\begin{equation}\label{CH3_Pulse}
  P(\mathbf{r},t)=P_0 \frac{(x\pm{\mi}y)^2}{R^2}
  \exp\!\left[-\frac{r^2}{R^2}-\mi\omega_0t\right]\!\theta(t)\theta(t_p-t).
\end{equation}
Here, $P_0$ is the pulse amplitude, $R$ is the  radius of the pulse ring, $\omega_0$ is the frequency of the pulse coinciding with the frequency of the FB at the $\Gamma$ point, $\theta(t)$ is the Heaviside step function, and $t_p$ is the pulse duration.
Transport of polaritons to the B sites should be blocked due to the $\pi$-phase difference of the wave functions on the A and C sites.
The phase and intensity plot in Fig.~\ref{fig:CH3_2}(a) shows that we can achieve this $\pi$ phase difference by centering the pump beam at the center of the unit cell.

%
%
\subsection{Dynamics of the CLC}
\begin{figure}[ht]
\centering
\includegraphics[width=0.79\linewidth]{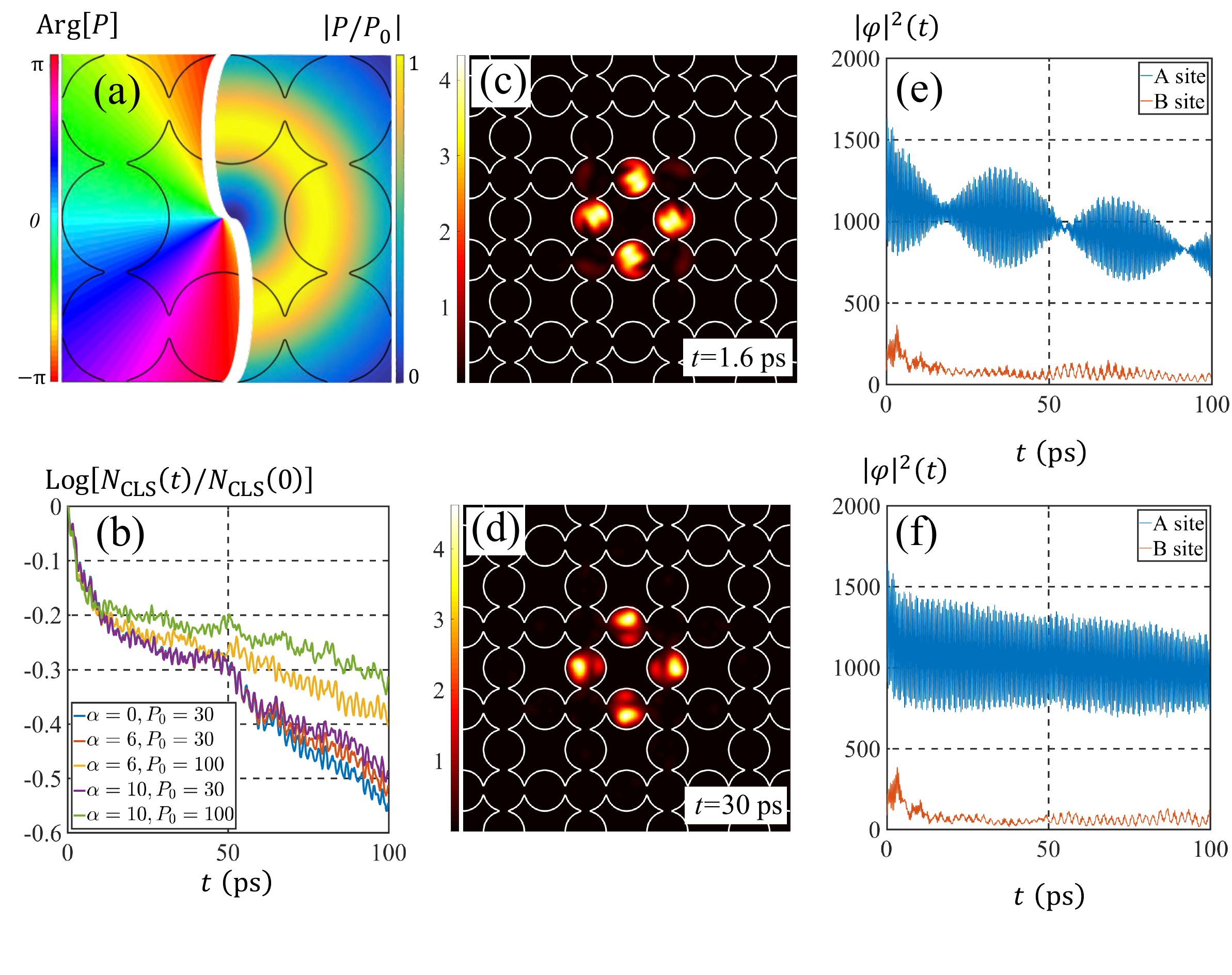}
    \caption[Laguerre--Gaussian pulse and corresponding CLC dynamics]{(a) Phase and intensity of the Laguerre--Gaussian pulse with a radius of $R=1.5\,\mu$m centered at the Lieb lattice plaquette. (b) Decay of the CLC for different magnitudes of the interaction strengths $\alpha_x$ in units of $\mu\mathrm{eV}\cdot\mu\mathrm{m}^2$ and the coherent strength $P_0$ in units of $\mathrm{meV}\cdot\mu\mathrm{m}^{-1}$.
(c--f) Dynamics of the CLC in the absence of dissipation.
(c,d) Snapshots of CLC particle density distribution $\left(|\varphi(\mathbf{r},t)|^2+|\chi(\mathbf{r},t)|^2\right)d^2r$
at two different times for $\alpha_x=10\,\mu\mathrm{eV}\cdot\mu\mathrm{m}^2$.
(e,f) Rabi oscillations of the photonic component from the A and B sites for $\alpha_x=0$ (e) and $\alpha_x=10\,\mu\mathrm{eV}\cdot\mu\mathrm{m}^2$ (f) with $P_0=100\,\mathrm{meV}\cdot\mu\mathrm{m}^{-1}$. The figure is taken from~\cite{Sun:2018aa}.
}
\label{fig:CH3_2}
\end{figure}

To characterize the CLC dynamics, it is convenient to use the function
\begin{equation}\label{CH3_NCLS}
N_\textrm{CLS}\left(t\right)=N_c(t)+N_x(t)=\int_{\textrm{A,C}}\left(|\varphi|^2+|\chi|^2\right) d^2r
\end{equation}
that measures the total number of particles residing at the A and C sites of the excited plaquette.
We trace the evolution of the system just after the pulse is switched off at $t=0$.
Figure~\ref{fig:CH3_2}(b) shows the particle decay rate in the CLC for different intensities of interaction strength and coherent pumping.

A counterintuitive result in Fig.~\ref{fig:CH3_2}(b) is a decrease of particle loss from CLC with increasing polariton--polariton interaction strength $\alpha_x$, or alternatively, with increasing the coherent pumping amplitude $P_0$, which increases the number of particles in the condensate and elevates the role of interaction.
Despite the repulsive nature of exciton--exciton interaction, as the figure shows it has a focusing effect on the CLC in the Lieb lattice.

Other than a gradual decay of the excited CLC, we observe fast Rabi oscillations of particle number and more complex short and long-time dynamics.
To highlight these effects arising from the two-component (exciton and photon) nature of polaritons and their continuous, ``non-tight-binding'' propagation, we also consider CLC dynamics in the absence of dissipation.
Figure~\ref{fig:CH3_2}(c) and (d) show snapshots of the particle density $\left(|\varphi(\mathbf{r},t)|^2+|\chi(\mathbf{r},t)|^2\right)d^2r$ at $t=1.6\,\mathrm{ps}$ and $t=30\,\mathrm{ps}$, respectively.
Due to the shape of the Laguerre--Gaussian pulse, the condensates excited in the A and C wells are smaller than the well size.
%
%
%

We can also see a slow modulation of the amplitude of the Rabi oscillations of the photonic component, which can be measured experimentally~\cite{Dominici:2014aa}.
Figure~\ref{fig:CH3_2}(e) and (f) show the time dependence of the total number of photons in an A site (the same as in a C site), as well as in a B site.
For the interaction-free case [Fig.~\ref{fig:CH3_2}(e)], one can see that the condensate dynamics at the A and C sites is characterized by fast Rabi oscillations with a slow beating of their amplitude.
The beating half-period $t_b$ is about $30\,\mathrm{ps}$; this matches the width of the FB ${\Delta}E_f\simeq0.02\mathrm{meV}\simeq\hbar/t_b$, and thus the effect likely results from the finite band width.
In the presence of polariton--polariton interaction, the beatings of the Rabi oscillations on the A(C) sites are smoothed out [Fig.~\ref{fig:CH3_2}(f)]. Note that occupation in the B sites is very low in both cases.

%
%
\subsection{Prolonging the CLC}
Polariton lifetime in etched microcavities is typically short, which makes it hard to maintain and manipulate condensates for times longer than several ps.
It follows from Fig.~\ref{fig:CH3_2}(d) that the lifetime of particles in the CLS (second band at the $\Gamma$ point) is $\tau_\textrm{CLS}\approx4.5$ ps.
One way to increase the lifetime would be to use microcavities with higher Q factors; alternatively, losses can also be compensated for by incoherent background pumping to maintain the CLC
When incoherent background pumping is present, the evolution of the system is described by the equations
\begin{subequations}\label{CH3_INC}
\begin{align}
 \mi\begin{pmatrix}\dot{\varphi}\\\dot{\chi}\end{pmatrix} &=
 \hat{H}\begin{pmatrix}{\varphi}\\{\chi}\end{pmatrix}
 +\frac{{\mi}cn_r}{2}\begin{pmatrix}{0}\\{\chi}\end{pmatrix}
 +\begin{pmatrix}{\mi P(\mathbf{r},t)}\\{0}\end{pmatrix}, \\[5pt]
 \dot{n}_r &= I - \tau_r^{-1} n_r-c|\chi|^2n_r,
\end{align}
\end{subequations}
where $n_r$ is the density of the reservoir particles, $\tau_r=10\,\mathrm{ps}$ is their lifetime, $c=0.005$ ps$^{-1}\mu$m$^2$ is a phenomenological reservoir-system coupling rate, and $I$ is the intensity of the homogeneous incoherent pumping.
The reservoir dissipation rate $\tau_r^{-1}$ is considered to be of the same order of magnitude as the polariton dissipation rate $-2\mathrm{Im}\{V(\mathbf{r})\}$, which is usually assumed for polariton systems under non-resonant pumping~\cite{Ostrovskaya:2012aa,Ma:2016aa} including polariton lattices~\cite{Ma:2015aa,Yulin:2016aa}.
To avoid polariton excitation in the first and third (and higher) bands, we consider intensity $I$ to be below the polariton condensation threshold.
Below, we use the following threshold intensity $I_{th}=(c\tau_r \tau_x)^{-1}$ as a reference.
\begin{figure}[ht]
\centering
\includegraphics[width=0.55\linewidth]{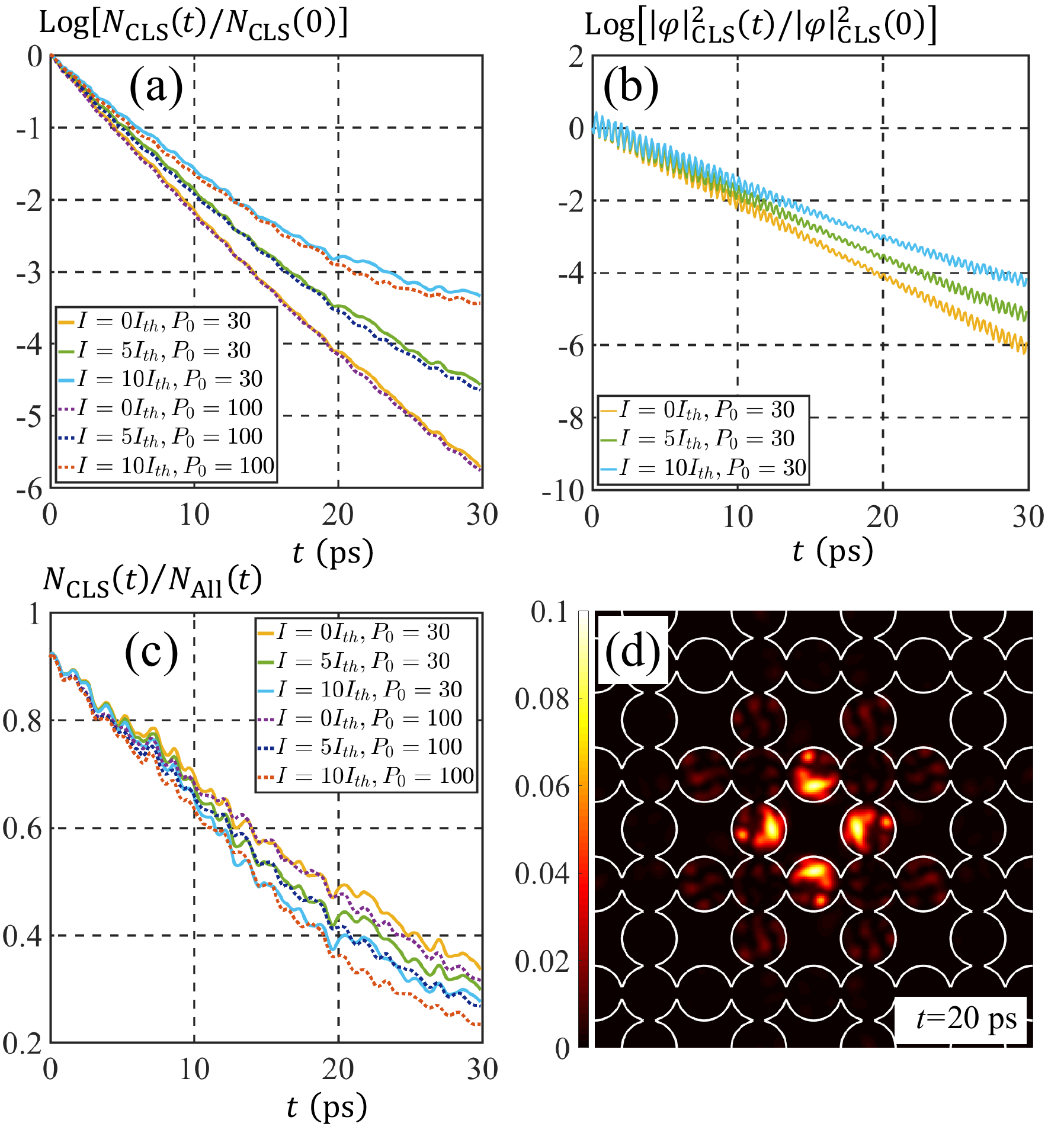}
\caption[CLC with incoherent pumping]{(a) Decay of the CLC for different incoherent pumping intensities $I$ (and various coherent pumpings $P_0$). The Laguerre--Gaussian resonant pulse radius is $R=1.5~\mu$m. (b) Photonic decay of the CLC for different incoherent pumping intensities. (c) Evolution of the ratio of CLC particles and the total number of particles for different incoherent pumping intensities. (d) Snapshot of the particle density in the CLC at 20~ps. The figure is taken from~\cite{Sun:2018aa}.}
\label{fig:CH3_3}
\end{figure}

Figure~\ref{fig:CH3_3}(a) shows the decay of particles residing in the CLC for different pumping intensities (both incoherent and coherent) together with a reference curve of the decay at $I=0$.
The increase of $I$ compensates for the decay of particles from the CLC.
The corresponding photonic decay also shows a similar behavior, as seen in Fig.~\ref{fig:CH3_3}(b).
One can see from both panels that the Rabi oscillations persist in the presence of incoherent background pumping, indicating that the CLC maintains its coherence.

Several drawbacks accompany the use of incoherent pumping.
First, it leads to the excitation of particles in other (non-flat) bands and thus increases the occupation of the B sites. Second, although background pumping preserves the CLC for longer times, it also generates noise.
Figure~\ref{fig:CH3_3}(c) shows the ratio of particles in the CLC to the total number of particles in the system: the larger the $I$, the worse the signal-to-noise ratio.
At $I=10I_\mathrm{th}$ and after 20~ps, about $60\%$ of the polaritons already escaped from the CLC.
Despite this, the four CLC QWs containing only $40\%$ of the polaritons remain the most populated wells, as shown in Fig.~\ref{fig:CH3_3}(d).
We can conclude then that the background pumping prolongs the CLC for one order of magnitude longer times than single-polariton lifetimes.

%
%
\subsection{Discussion}
Using an example of a realistic 2D exciton-polariton Lieb lattice with distributed losses, we have shown that the (nearly) flat band in this system possesses small but finite dispersion, both in the energy and the lifetime of the states.
We have demonstrated the possibility to excite compact localized condensates in this nearly FB using resonant Laguerre--Gaussian pulses.
In spite of the small dispersion of the band, the localization and coherence of the compact localized condensates remain well defined.
They exhibit unusual dynamics, following from modulated fast Rabi oscillations.
This coherent excitation of CLCs opens new possibilities to use polariton Lieb lattices as platforms for network computations; in particular, it permits to construct CLC graphs, similar to recent proposals for classical~\cite{Berloff:2017aa,Ohadi:2017aa} and quantum~\cite{Liew:2018aa} simulators.
In the presence of an incoherent homogeneous background pumping, the coherent compact localized condensates can be maintained for times much longer than regular polariton lifetimes.
Thus, both the phase and polarization of localized condensates may be able to be used to encode information in the future.

%% file: chapter5_2-26_revised.tex
\section{Exciton-Polariton Topological Insulator with an Array of Magnetic Dots}\label{Ch5}

Robust transport of particles in topologically nontrivial systems follows from the existence of protected edge states at the interfaces between media having different topological properties~\cite{Hasan:2010aa,Qi:2011hb,Chiu:2016aa}.
Such edge modes are robust against disorder, which makes them appealing to both theoretical and experimental research. Phase transitions to topological insulating phases were originally discovered in the context of the integer quantum Hall effect~\cite{Klitzing:1980aa}, where a nonzero quantized Chern number is associated with disorder-robust quantization of the particle current in strong external magnetic fields.
Later this idea was extended to the anomalous quantum Hall and quantum spin Hall effects~\cite{Haldane:1988aa, Kane:2005aa, Bernevig:2006aa, Konig:2007aa}, where an external net magnetic field is not required. In these cases, the field is replaced by other effects such as time-reversal symmetry breaking from an internal magnetization of the material~\cite{Karplus:1954aa, Chang:2013aa} or strong spin-orbit coupling. Topological phases have now been extended to various physical systems, including photonics, cold atoms, and acoustics~\cite{Wang2009,Jotzu:2014aa, Yang:2015aa}.

Recently, an exciton-polariton topological Chern insulator has been demonstrated ~\cite{Klembt:2018aa}. In contrast to other systems, the nontrivial topological phase in this work arose from the following main features: an artificial lattice made of micrometer-sized pillars, excitonic spin-orbit coupling or photonic TE-TM splitting, and an external magnetic field in the Faraday configuration~\cite{Bardyn:2015aa, Nalitov:2015aa}. The effects are as follows. The lattice creates a finite Brillouin zone with energy bands in reciprocal space, the spin-orbit coupling creates a momentum-dependent energy splitting between spin states without opening a bandgap, and the external magnetic field induces a Zeeman splitting between spin-up and spin-down exciton-polaritons. All three effects are required to create a topological bandgap that contains the protected edge states, in contrast to other platforms, where a lattice combined with a uniform or staggered magnetic field is sufficient.


To date, theoretical and experimental works regarding exciton-polariton topological insulators have considered a uniform external magnetic field~\cite{Gulevich:2016aa,Nalitov:2015aa,Bardyn:2015aa,Yi:2016aa,PhysRevB.93.085438}.
In order to observe a measurable bandgap, a large external magnetic field has been employed, requiring superconducting coils and cryogenic temperatures~\cite{Klembt:2018aa}.
Such a strong external magnetic field is a significant drawback that limits potential applications in real devices.
In particular, the essential advantages of microcavity samples (e.g. their compactness and potential for room-temperature operation) are neutralized.
It is therefore crucial to explore application-friendly mechanisms and techniques by which topological polaritons can be created without an external magnetic field.

Existing proposals have been based on such effects as the nonlinear dynamics under resonant pumping~\cite{Mandal:2019aa}, time or spin-dependent pumping~\cite{Sigurdsson:2019aa}, and spontaneous symmetry breaking~\cite{Karzig:2015aa,Janot:2016aa}.
All these proposals, though, have technological limitations; for example, using the nonlinear terms due to particle interaction requires strong resonant pumping of the sample in order to excite the edge as a perturbation~\cite{Bardyn:2015aa,PhysRevB.93.085438}.
This spin-dependent pumping demands each lattice site to be treated independently, and this makes experimental design tricky.

The following sections introduce an easily realizable alternate design for an polariton topological insulator in which the external uniform magnetic field is replaced by an internal inhomogeneous magnetization, produced by a magnetic material (MM) embedded within the microcavity.
Such magnetic quantum dots are not only interesting by itself~\cite{Vavassori:2000aa}, the applications can found in superconductors~\cite{Erdin:2002aa,Otani:1993aa,Lyuskyutov:2005aa} and electron system~\cite{Li:2007aa}.
We show that the resulting staggered magnetic field is capable of opening a topological bandgap, and can yield topological transitions between different Chern numbers ($\pm 2$ to $\mp 1$) as either the magnetic field or spin-orbit coupling strength are varied, similar to the uniform external field case~\cite{Bleu:2017aa}.
Our approach can eliminate the need for superconducting coils and other equipment, thereby bringing practical device applications with exciton-polariton topological insulators closer.



\subsection{System schematic}
\begin{figure}[ht]
    \centering
	\includegraphics[width=0.55\textwidth]{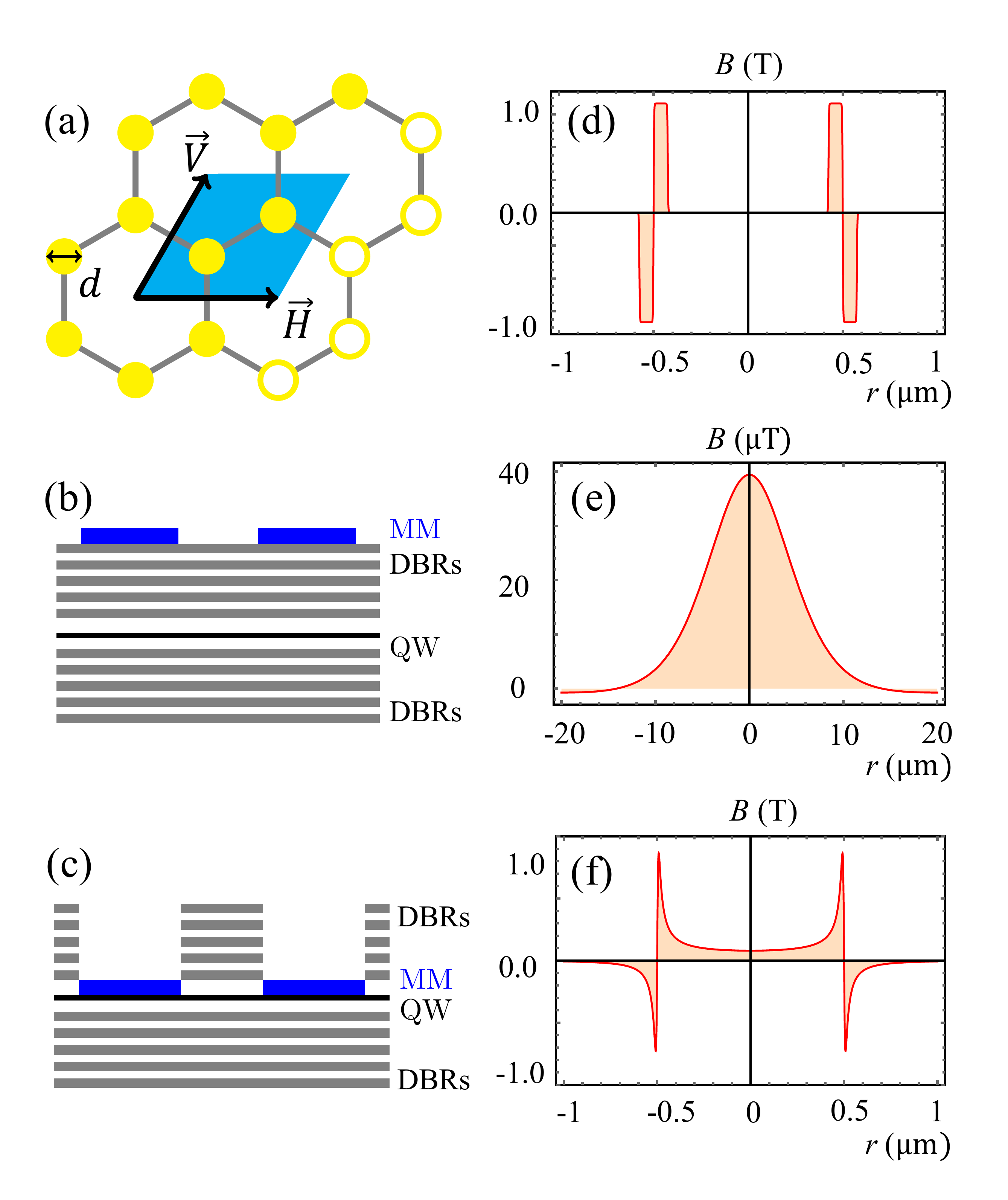}
	\caption[TI system and magnetic field]{Left panels: System schematic: (a) A honeycomb lattice made of pillars of the etched cavity; yellow empty rings denote the region of the edge. The direction of the edge is along the $\vec{V}$ direction;
	(b) Setup using magnetic material (MM) placed on the top DBR;
	(c) Setup using MM placed at the top of the QW (close to the exciton layers).
	Right panels: Magnetic field profile in the case of a single MM disk with radius $R = 0.5$~$\mu$m  (d) Magnetic field used in numerical calculations, rectangle functions with width $dw$; (e) Magnetic field for case (b) and 10 $\mu$m away from the QW; (f) Magnetic field for case (c) and 8~nm away from the QW. The figure is taken from~\cite{Sun:2019ab}.}
\label{fig:Ch5_Fig1}
\end{figure}

Let us consider polaritons loaded in the honeycomb lattice shown in Fig.~\ref{fig:Ch5_Fig1}(a).
In recent experiments~\cite{Klembt:2018aa}, an external homogeneous magnetic field perpendicular to the lattice was applied to break time reversal symmetry.
To get rid of the setup required to generate a sufficiently strong magnetic field, we use an MM layer to represent an array of sub-micron magnetic QDs. When magnetized, they can replace an external field.
%
The first case to consider locates the MM on the pillars from the top, thus covering the upper DBR.
Then the MM represents an array of magnetic QDs [\ref{fig:Ch5_Fig1}(b)].
%
%
%

The magnetic field from a single disc-shaped QD of radius $R$
reads~\cite{Erdin:2002aa,Lyuksyutov:2005aa}
\begin{eqnarray}\label{eq:Ch5_EqDotMF}
B_z(r,z)=2\pi \mu_0 M R\int_0^\infty J_0(rq)J_1(Rq)\mathrm{e}^{-|z|q}q dq,
\end{eqnarray}
where $\mu_0 = 4\pi\times10^{-7} H/m$ is the vacuum permeability, $M=8\times 10^4Oe\cdot m$ is the 2D magnetization perpendicular to the disc-shaped QD, $J_i$ are the Bessel functions, $r$ is the distance from the center of the QD, $q$ is the in-plane momentum, and $z$ is the vertical distance from the MM to the QW.

The integral in Eq.~\eqref{eq:Ch5_EqDotMF} can be calculated semi-analytically, using special functions~\cite{A.-P.-Prudnikov:2003aa}, giving%
\begin{eqnarray}
B_z=\frac{\left[
f^2(R^2-r^2-z^2)E(f)
+(1-f^2)4rRK(f)
\right]}{4\mu_0^{-1}M^{-1} (Rr)^{3/2}(1-f^2)f^{-1}},~~ \label{eq:Ch5_Bfield}
\end{eqnarray}
where $f=\sqrt{\frac{4rR}{z^2+(r+R)^2}}$, and $K(f)$ and $E(f)$ are the elliptic integrals of the first and second kind, respectively.

Figure~\ref{fig:Ch5_Fig1}(e) shows the calculated profile of the $z$-projection of the magnetic field.
Since the distance between the layers of the excitons and the MM amounts to $z\sim$10 $\mu$m in this case (according to a typical size of a microcavity), the resulting effective magnetic field acting on the excitons is only of the order of $\mu$T, which is too small to create a useful topological bandgap. This is because $B_z$ in Eq.~\eqref{eq:Ch5_Bfield} decays exponentially with $z$.

To achieve a sufficient magnetic field, the separation between the MM and QW must be reduced.
For this, the scheme in Fig.~\ref{fig:Ch5_Fig1}(c) is considered, where the microcavity is etched to form a honeycomb lattice and an MM layer is deposited in the etched regions.
This allows the MM to be placed very close to the QWs, where the excitons reside, leading to interlayer separations of only $z \sim $1--20~nm and strong magnetic fields ($\sim 1$T).
The corresponding magnetic field profile [Fig.~\ref{fig:Ch5_Fig1}(f)] shows a rather different shape, as it is strongly localized at the micropillar/MM boundary, where it also changes sign.
Consequently, the magnetic field is said to be \emph{staggered} with zero net flux.
The profile in Fig.~\ref{fig:Ch5_Fig1}(f) is from a MM of radius $R=0.5\mu$m that produces a $\sim 1$~T magnetic field at a distance of 8~nm.

The process to fabricate this setup is as follows. First, grow the cavity with a multilayered structure. Second, etch micropillars to form the desired lattice potential. Third, deposit the MM onto the structure from the top. This way, part of the MM will be on the edges of the pillars.
However, as our calculations explicitly demonstrate, such remote sources of magnetic field will not substantially contribute and can therefore be neglected.
Lastly, by heating the sample above the Curie temperature and exposing it to a strong magnetic field, we can magnetize the sample and make it a permanent magnet.


\subsection{Transport of exciton-polaritons }
To determine the ability of this kind of staggered magnetic field configuration to create topological edge states, we numerically compute the band structure and Chern numbers of the polariton honeycomb lattice.
Due to computational limitations, we neglect the weak magnetic field inside and far away from the micropillars and approximate $B_z$ with a step function profile of width $dw$, as shown in Fig.~\ref{fig:Ch5_Fig1}(d).
%
We describe the time evolution of the exciton-polaritons in the cavity by the Gross--Pitaevskii equation
\begin{eqnarray}
    \label{eq:Ch5_EOM}
    \mi\hbar \frac{\partial \psi_{\pm}}{\partial t} &=& -\frac{\hbar^2}{2m_{eff}}\nabla^2 \psi_{\pm} + V \psi_{\pm} + \Delta_{\pm}^{eff} \psi_{\pm} \nonumber \\
    &+& \beta^{eff} \left( \partial_x \mp \mi \partial_y \right)^2 \psi_\mp, \label{eq:Ch5_GPE}
\end{eqnarray}
where $\psi_\pm$ are the wave functions of polaritons with up- and down-polarization, $m_{eff}$ is the effective mass, the first term in the r.h.s. stands for the free particle propagation, $V$ is the potential of the honeycomb lattice with the lattice constant $3$~$\mu$m and site diameter $d = 0.75$~$\mu$m, and $\Delta^{eff}_\pm$ is the effective Zeeman splitting due to the presence of the MM with the shape presented in Fig.~\ref{fig:Ch5_Fig1}(d).
In the calculations, we first assume the lateral size of the MM to be the same as the well of the honeycomb lattice, and second, that the width of the rectangle functions is [Fig.~\ref{fig:Ch5_Fig1}(d)] $dw = 0.15$~$ \mu$m. Also, $\beta^{eff}$ is an effective TE-TM splitting.
We choose the Hopfield coefficients to be $\abs{X_H}^2 = 1 - \abs{C_H}^2 = 0.3$.
This gives us an effective polariton mass of $m_{eff} \approx m_C/\abs{C_H}^2$, with the effective mass of cavity photons as $m_C =3.23\cdot10^{-5}m_e$, where $m_e$ is free electron mass.
For polaritons, we have a similar (to excitons) definition of the effective Zeemann splitting and the TE-TM splitting: $\Delta_\pm^{eff} = \abs{X_H}^2 \Delta_\pm$ and $\beta^{eff} = \beta \abs{C_H}^2$.
In order to quantitatively estimate the magnetic field, we use the Zeeman splitting determined by the relation $\Delta_+-\Delta_- = g_x \mu_B |\mathbf{B}|$, where $g_x \mu_B\approx 180$~$\mu$eV$\cdot$T$^{-1}$ for excitons~\cite{Schneider:2013aa}.
As a result, for a Zeeman splitting term equaling $\Delta = \abs{\Delta_\pm}=1$~meV, the peak magnetic strength approximately equals to $5.5$~T, which means that the MM is $1$~nm away from the QW.


\subsection{Phase diagram and edge modes}
First, we compute the bulk bands of Eq.~\eqref{eq:Ch5_GPE} by assuming a periodic structure and the Bloch wave Ansatz $\psi_{\pm} = u_{\pm}(\mathbf{k},\mathbf{r}) e^{\mi \mathbf{k}\cdot \mathbf{r} + \mi E t/\hbar}$.
Second, we find the bulk Bloch wave eigenstates and the Chern number~\cite{Fukui:2005aa} with
\begin{equation}
C =  \sum_{E_n < E_g} C_n = \frac{1}{2\pi \mi}\sum_{E_n < E_g} \oint_{\mathrm{BZ}}F_{\mu\nu}^n d^2k,
\label{eq:Ch5_Chern}
\end{equation}
where the Berry connection $A_n\left(k\right)$ ($n = 1,2$) in the $n$th band below the energy gap ($E_n<E_g$) and the associated field strength $F_{\mu\nu}\left(k\right)$ are defined as
\begin{eqnarray}
    A_\mu &=& \bra{n\left(k\right)}\partial_\mu\ket{n \left(k\right)},\nonumber \\
    F_{\mu\nu}\left(k\right) &=& \partial_\mu A_\nu\left(k\right) - \partial_\nu A_\mu\left(k\right),
\end{eqnarray}
where $\ket{n\left(k\right)} $ is the eigenvector of the $n$th Bloch band, and the inner product denotes integration over the unit cell.
Unit vector $\mu$ and $\nu$ denote the directions of the two reciprocal lattice vectors.

\begin{figure}[ht]
    \centering
    \includegraphics[width=0.55\textwidth]{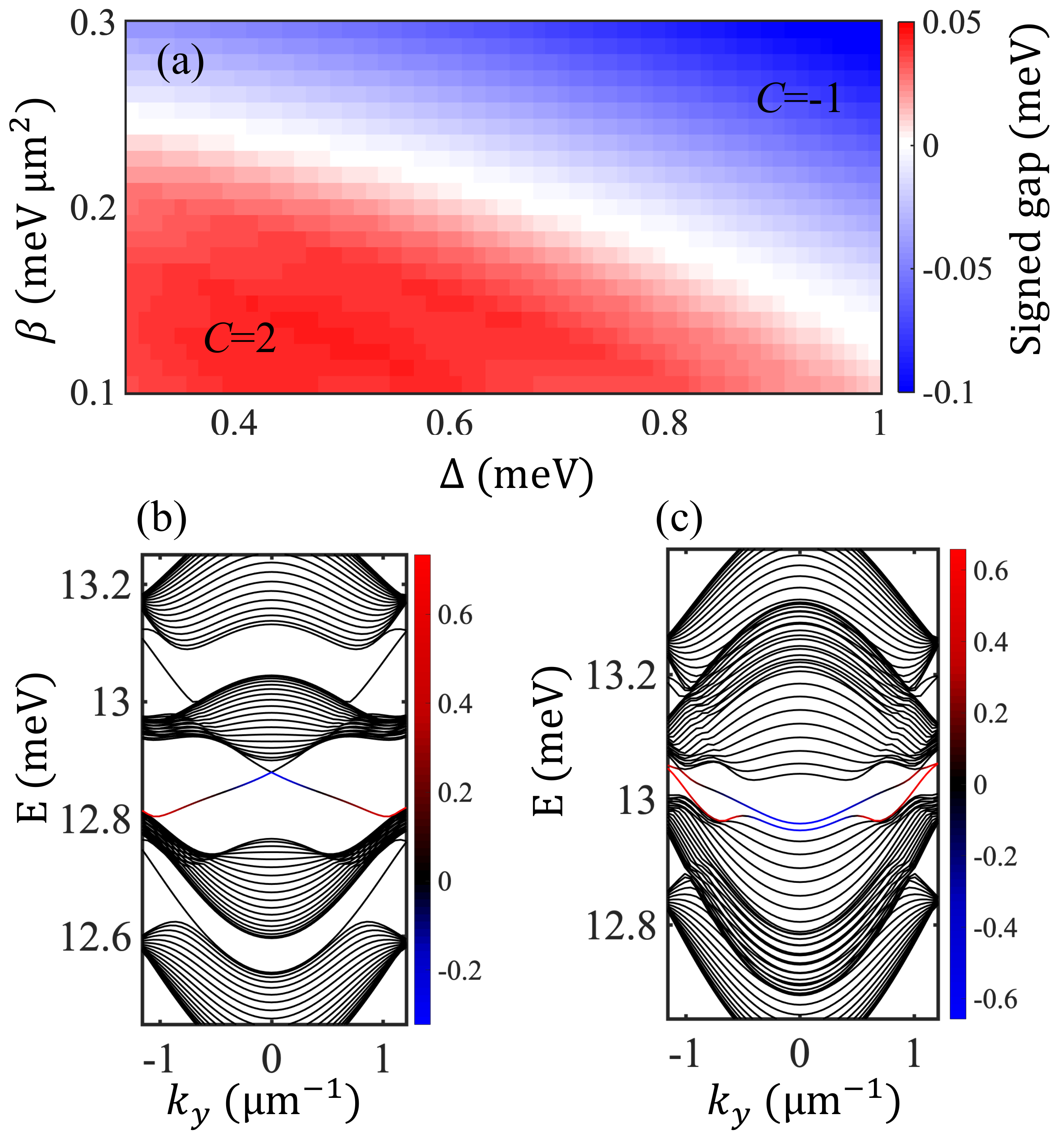}
    \caption[Chern number phase diagram and edge states]{(a) Phase diagram for the Chern numbers with a TE-TM splitting of $0.1 \backsim 0.3$~meV$\mu$m$^2$ and a Zeeman splitting of $0.3 \backsim 1.0$~meV. The color bar indicates the size of the direct gap multiplied by the sign of the Chern number in units of meV. (b) and (c) Spectra of semi-infinite systems with Chern numbers $C = -1$ and $C = 2$, respectively. The color bar indicates the polarization imbalance $I = \frac{\abs{\psi_+}^2-\abs{\psi_-}^2}{\abs{\psi_+}^2+\abs{\psi_-}^2}$. For the $C=-1$ case, the edge mode is calculated at $\Delta=0.95$~meV and $\beta=0.27$~meV$\cdot\mu$m$^2$. For the $C=2$ case, the edge mode is calculated at $\Delta=0.46$~meV and $\beta=0.13$~meV$\mu$m$^2$. The figure is taken from~\cite{Sun:2019ab}.}
    \label{fig:Ch5_Fig2}
\end{figure}

Solving Eq.~\eqref{eq:Ch5_Chern} for the system described in Eq.~\eqref{eq:Ch5_EOM}, we plot a Chern number diagram in Fig.~\ref{fig:Ch5_Fig2}(a), using the following parameters: $\Delta_\pm=\left[0.3\sim1.0\right]$~meV and a TE-TM splitting of $\beta = \left[ 0.1\sim0.3\right] $~meV$\mu$m$^{2}$.
As shown in the diagram, the system can possess different Chern numbers depending on the parameters. The phase transition between Chern numbers $C=2$ and $C=-1$ is due to a closing and opening of the gap at the $M$ point.
A similar behavior has been reported in~\cite{Bleu:2017aa} for the case of a homogeneous external magnetic field.

To confirm the existence of topological edge states, we also compute the energy spectrum of a semi-infinite structure (with 20 unit cells in the $\vec{H}$ direction and twisted boundary conditions in the $\vec{V}$ direction).
Typical spectra for the two topological phases are plotted in Fig.~\ref{fig:Ch5_Fig2}(b,c), which shows that the bulk bands and bandgaps host one and two edge states, respectively, in the gap between the second and third bands.
%
%
%
%
%
%

\subsection{Comparison with the homogeneous case}
\begin{figure}[ht]
    \centering
    \includegraphics[width=0.45\textwidth]{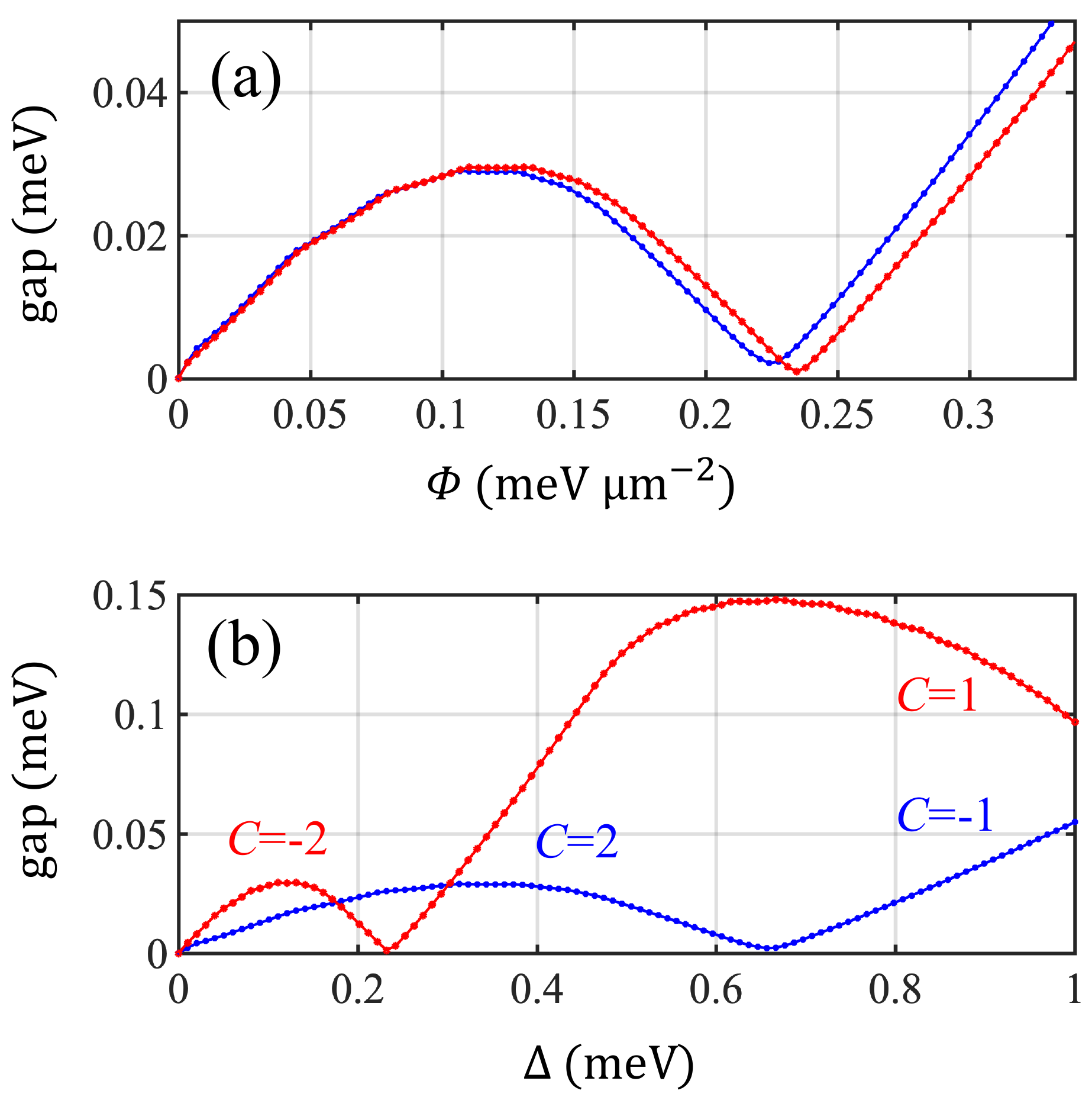}
    \caption[Bandgap between MM and external magnetic field]{(a) Bandgap as a function of the absolute flux. (b) Bandgap as a function of the peak value of the Zeeman splitting. Blue and red lines show the MM and homogeneous magnetic field cases, respectively. The TE-TM splitting is fixed to $\beta = 0.2$~meV$\cdot\mu$m$^2$. The figure is taken from~\cite{Sun:2019ab}.}
    \label{fig:Ch5_gap}
\end{figure}

One key difference between systems with MM and with an external magnetic field is the magnitude of the magnetic flux.
In the MM case, on account of the strong localization of the magnetic field, the total net flux within one unit cell is almost zero.
To compare the MM and the external magnetic field cases, we define the absolute value of the flux on the plane by the integral over a single unit cell region as
\begin{equation}
    \Phi = \frac{\oint_S \abs{{\Delta_{\pm}}}ds}{S},
\end{equation}
where $S$ is the area of the unit cell.
Since the Zeeman splitting is proportional to the magnetic field, we can compare the bandgaps between the 2nd and 3rd levels as functions of the absolute flux, $\Phi$, in the two cases.
We assume that the direction of the magnetic field is the same as the direction of the outer edge of the MM.
Figure~\ref{fig:Ch5_gap}(a) shows that, given the same value of the absolute magnetic flux, the gap closes and reopens in a similar manner.

Alternatively, we can compare the magnitude of the gap as a function of the peak value of the magnetic field, i.e., the peak intensity of the Zeeman splitting.
Figure~\ref{fig:Ch5_gap}(b) demonstrates the sizes of the gaps in the MM and homogeneous cases as functions of the peak value of the Zeeman splitting.
We also label the Chern numbers in both cases before and after the gap is closed.
In the homogeneous case, the gap closes much earlier than in the MM case.
%
%
%


\subsection{Cavity with a MM exposed to an external magnetic field}
\begin{figure}[ht]
\centering
\includegraphics[width=0.45\textwidth]{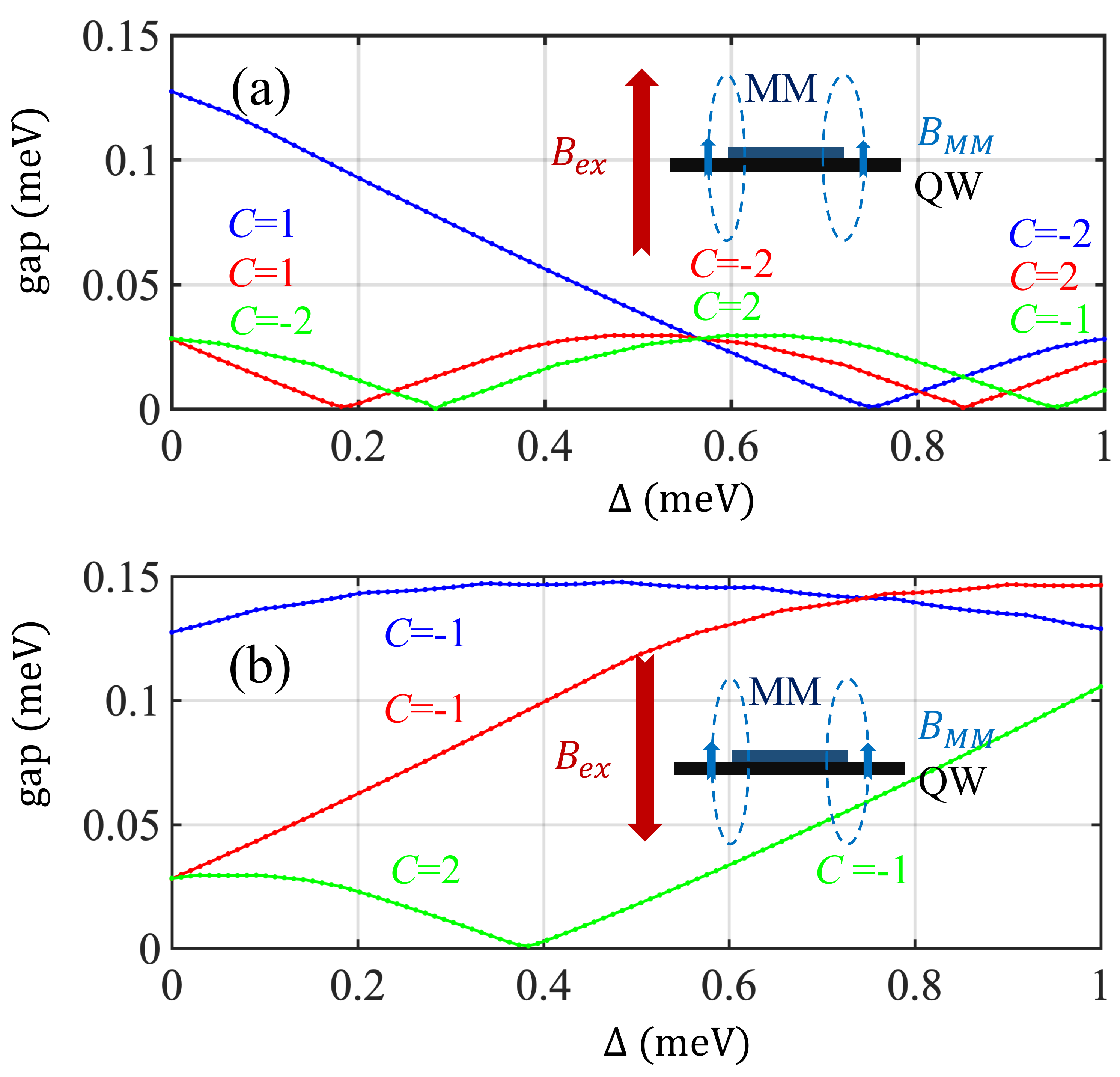}
\caption[Bandgap of MM under external magnetic field]{The mutual effect of MM and an external magnetic field with (a) parallel and (b) antiparallel magnetic field directions, as drawn in the insets.
The x-axis is the intensity of the local magnetic field of MM.
Different colors reflect different magnitudes of external magnetic field,
$B_{ex}\approx 2.8$, $1.7$, and $0.5$~T, corresponding to the Zeeman splittings $\Delta_{ex}=0.5$ (blue), $0.3$ (red), and $0.1$~meV (green).
The Chern numbers are labeled before and after each gap.
The TE-TM splitting is fixed at $\beta = 0.2$~meV$\cdot\mu$m$^2$. The figure is taken from~\cite{Sun:2019ab}.}
    \label{fig:Ch5_fig4}
\end{figure}

With MM embedded in a cavity, a homogeneous magnetic field can still be applied to the sample. In this case, the resulting field represents a superposition of two magnetic field profiles (Fig.~\ref{fig:Ch5_fig4}).

The first case to consider is one in which the external magnetic field $\mathbf{B}_{ex}$ is parallel to the outside edge of the MM, $\mathbf{B}_{MM}$, as shown in the inset of Fig.~\ref{fig:Ch5_fig4}(a).
Here, we see an additional phase transition in red and green curves.
Specifically, we can classify two distinct regimes.
The first one is when the MM is weak ($\Delta < 0.2$~meV). With an increasing homogeneous magnetic field ($\Delta_{ex}$, from green to red curve), the system undergoes a phase transition from Chern number $C=-2$ to $C=1$.
The second one is when the MM is moderately strong ($0.3$~meV$<\Delta<0.8$~meV), in which the Chern number changes from $C=2$ to $C=-2$ as $\Delta_{ex}$ increases.

The second case is the antiparallel case as shown in Figure~\ref{fig:Ch5_fig4}(b).
Red and blue curves show that the gap size in the $C=-1$ case can be efficiently enlarged.
Gap sizes are vital in topological polaritonic systems, since the finite lifetime of the exciton-polaritons can broaden the bandwidth of the spectrum and diminish and destroy nontrivial topological properties.
We conclude that the combination of the two magnetic fields reduces the required strength of the external magnetic field to generate a topological insulator state with a given bandgap.
%
%
%



\subsection{Discussion}
This Chapter showed that a local magnetic field from the presence of a magnetic material can be sufficiently strong to open a gap at the Dirac point and allow for the observation of nontrivial topological states in an exciton-polariton system loaded in a honeycomb lattice.
With intensity changes of the embedded magnetic field or TE-TM splitting, the system undergoes a phase transition between two nontrivial states with the Chern numbers $\pm2$ and $\mp1$.

The key advantage of this setup is the size of the system, which can be much smaller than those requiring a homogeneous external magnetic field.
This can be highly beneficial for future experiments and device applications.
Furthermore, we have studied the Chern numbers and gap sizes as functions of the magnetic flux strength and the peak value of the magnetic field, with results showing that designs utilizing MM and a regular homogeneous field both demonstrate similar behavior.

Next, the joint effect of an internal MM field with an external magnetic field was explored.
Depending on the relative direction of the two fields, one can switch between different Chern numbers.
This switching can be performed ``on the fly", since it depends only on a small change of the external magnetic field, thereby enabling control over the number and/or the direction of the topological edge states.
By reversing the direction of the external magnetic field, one can also keep the Chern number the same but enlarge the size of the gap significantly, thus increasing the speed of the edge state.
This allows us to propagate polaritons over longer distances before they decay due to their finite lifetime.

%% file: chapter6_3-18_revised.tex
\chapter{Bogolon-mediated electron scattering in hybrid Bose--Fermi systems}\label{Ch6}
%
%
In this chapter, we discuss the second topic of this thesis.
We show that when a 2D electron gas is coupled with a condensate of, e.g. indirect excitons, the contribution to the electron resistivity from the interaction with the excitations above the condensate can be orders of magnitude higher than the typical phonon contribution.
\section{Background}
Electron scattering in solid-state nanostructures plays a crucial role in their 2D transport~\cite{Kawamura:1992aa,Hwang:2008aa}, dramatically modifying electric conductivity.
Conventionally, there are two principal electron scattering mechanisms: disorder- or impurity-mediated~\cite{Jena:2007aa,Gibbons:2009aa}, and lattice phonon-mediated~\cite{Kawamura:1992aa}. The former processes are more pronounced at low environmental temperatures.
In the case of an attracting impurity, electrons can be captured, and thus the number of electrons decreases. Otherwise, repulsive centers decrease the electron mean free path and scattering time~\cite{Shi:2012aa,Bourgoin:1992aa,Eshchenko:2002aa,Palma:1995aa,Boev:2018ab}.
With an increase of temperature, electron scattering accompanied by acoustic and optical phonons of the crystal lattice becomes more efficient~\cite{Gummel:1955aa, Lax:1960aa, Abakumov:1976aa} and at some point dominant.

Conventional scattering mechanisms are also present in various new hybrid structures, which are the focus of current research~\cite{Cotlet:2016aa,Laussy:2010aa,Sau:2010aa,Alicea:2010aa,Mourik:2012aa}.
Hybrid Bose--Fermi systems represent a layer of fermions, usually a 2D electron gas (2DEG), coupled to another layer or layers of bosons such as excitons, exciton-polaritons, and Cooper pairs in superconductors.
In these systems, research interests are two-fold. One is devoted to high-temperature boson-mediated  superconductivity~\cite{Skopelitis:2018aa} and other condensation phenomena in interacting structures, including the Mott phase transition from an ordered state to electron-hole plasma~\cite{Kochereshko:2016aa}.
The other is devoted to finding additional new mechanisms of fermion scattering in the 2DEG, thus modifying the temperature dependence of the kinetic coefficients. Such possibilities explain the motivation to study electron transport in hybrid systems.

In this chapter, we will consider two different systems of fermion layers, specifically a layer of graphene (electrons with linear dispersion) and a layer of normal metal (electrons with parabolic dispersion).
Boson layers in both of the systems are expected as a condensed exciton gas, and the interactions between boson and fermion layers are described by Coulomb forces~\cite{Boev:2018ab,Kochereshko:2016aa, Matuszewski:2012aa}.
When the boson gas is in a condensed state, the corresponding interaction can be regarded as a counterpart to phonon-mediated scattering~\cite{Kovalev:2011aa, Kovalev:2013aa, Batyev:2014aa}.
Such 2D condensation has been reported in various solid-state systems~\cite{Butov:2003aa,Kasprzak:2006aa,Schneider:2013aa},
in which the lattice vibrations turn out to be not the only \textit{sound} available.
In the presence of BEC~\cite{Butov:2003aa}, other excitations come into play, commonly referred to as Bogoliubov quasi-particles or bogolons, which have linear dispersion at small momenta.

In the following sections, we show that an additional principal mechanism of electron scattering appears, stemming from the inter-layer electron--exciton interaction, or bogolon-mediated scattering.
Also, we demonstrate that the difference between acoustic phonon-related and bogolon-assisted scattering is more than just the magnitude of the sound velocity.
%


%
%
\section{System schematic}
\begin{figure}[ht]
\centering
\includegraphics[width=0.69\textwidth]{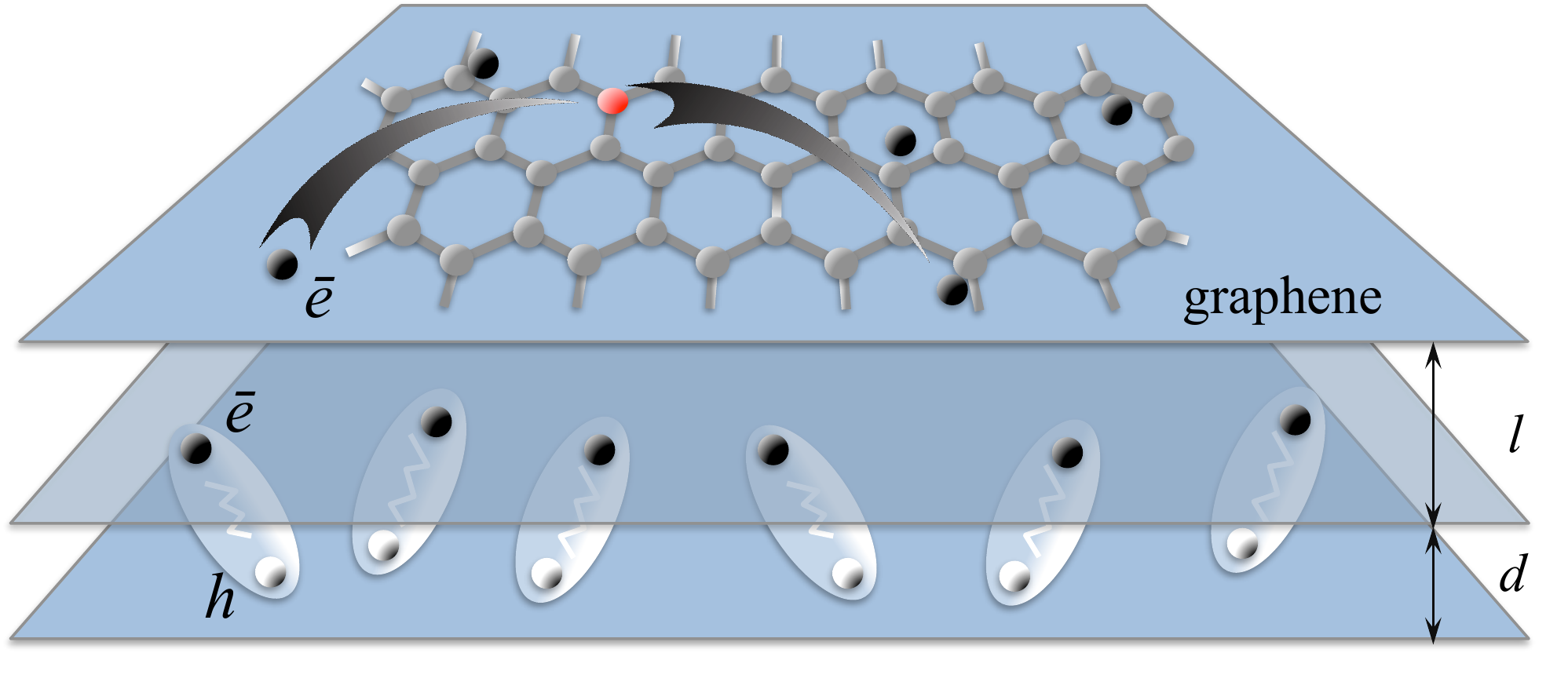}
\caption[Hybrid system schematic]{System schematic. A fermion layer (here, graphene) is located at distance $l$ from a 2D dipolar exciton gas residing in two parallel layers at distance $d$ from each other. Particles couple via Coulomb interaction. The figure is taken from~\cite{Sun:2019aa}.}
\label{fig:Ch6_1}
\end{figure}

Let us first consider a hybrid system consisting of a fermion layer separated by distance $l$ from a double QW that contains a dipolar exciton gas, where the distance between the layers of electrons and holes is $d$ (see Fig.~\ref{fig:Ch6_1}).
Electron--exciton interaction in this system can be described by the Hamiltonian
\begin{equation}\label{CH6_eq.1}
V=\int d\mathbf{r}\int d\mathbf{R}\Psi^\dag_\mathbf{r}\Psi_\mathbf{r}g\left(\mathbf{r}-\mathbf{R}\right)\Phi^\dag_\mathbf{R}\Phi_\mathbf{R},
\end{equation}
where $\Psi_\mathbf{r}$ and $\Phi_\mathbf{R}$ are the quantum field operators of the electrons and excitons, respectively, $g\left(\mathbf{r}-\mathbf{R}\right)$ is the Coulomb interaction between an electron and an exciton, $\mathbf{r}$ is the electron coordinate within the fermion layer, and $\mathbf{R}$ is the exciton center-of-mass coordinate.
We disregard the internal structure of the excitons to focus on their collective motion.

\begin{figure}[ht]
    \centering
    \includegraphics[width=0.60\textwidth]{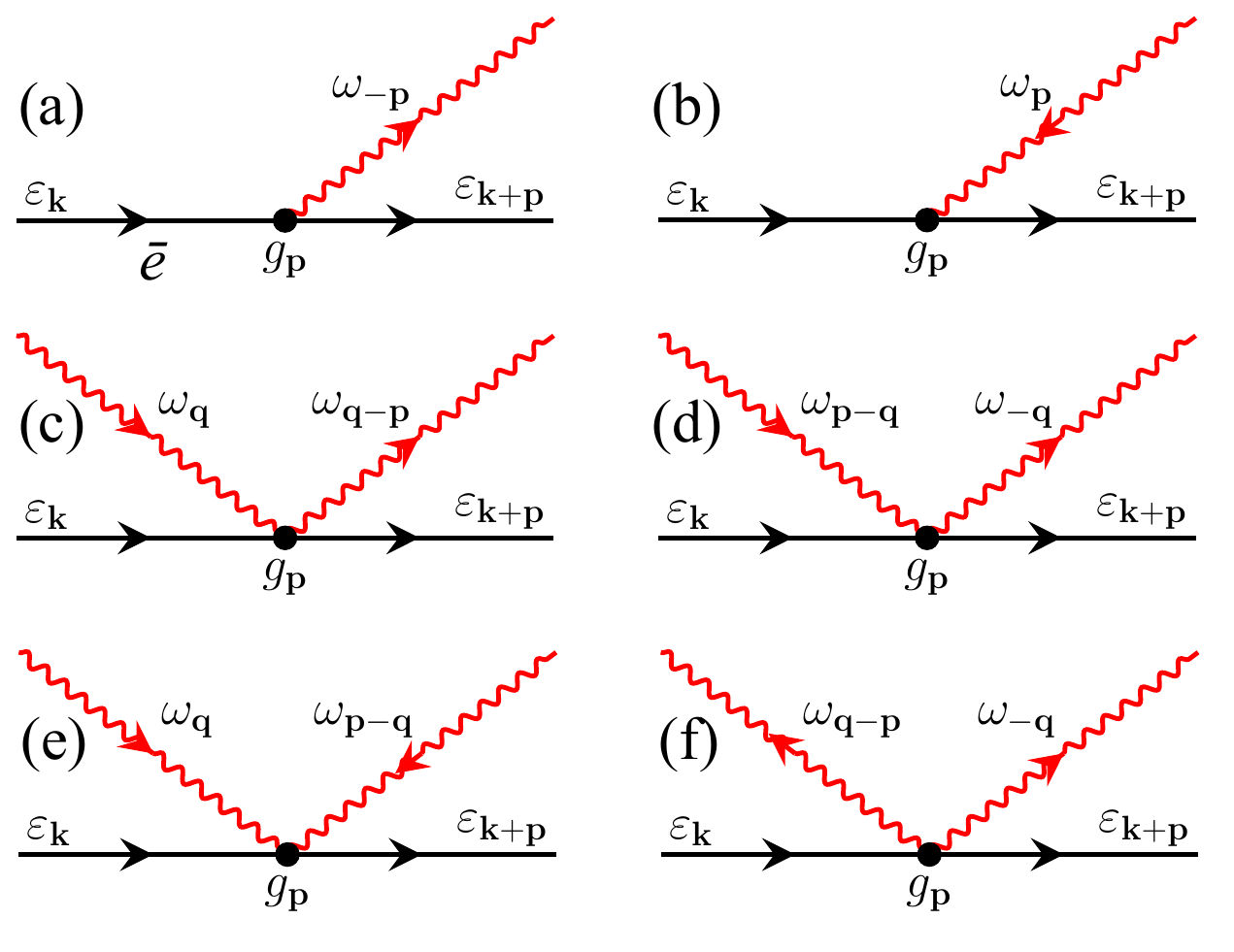}
    \caption[Electron scattering diagrams]{Schematic of electron scattering as mediated by bogolon emission and absorption processes. The black lines represent the electrons and the red wavy lines are the bogolons. (a,b) Single-bogolon scattering events from Eq.~\eqref{CH6_eq.5-a}. (c--f) Double-bogolon scattering events from Eq.~\eqref{CH6_eq.5-b}. The figure is taken from~\cite{Villegas:2019aa}.}
    \label{fig:CH6_2}
\end{figure}

We assume that the temperature of the system is below the critical temperature at which the excitons become a degenerate Bose gas~\cite{Fogler:2014aa}. This temperature is given by $kT_c =\frac{2\pi\hbar^2}{m_x}n_x$, where $n_x$ and $m_x$ are the exciton density and effective mass, respectively.
We can then use the model of a weakly interacting non-ideal Bose gas and write the exciton field operators as $\Phi_\mathbf{R}=\sqrt{n_c}+\phi_\mathbf{R}$, where $n_c$ is the condensate density.
In other words, we separate the condensed and non-condensed particles.
Substituting this into Eq.~\eqref{CH6_eq.1} and taking into account the selection rules, we find the electron--bogolon interaction potential by
\begin{eqnarray}
V_1 &=& \sqrt{n_c}\int d\mathbf{r}\Psi^\dag_\mathbf{r}\Psi_\mathbf{r} \int d\mathbf{R}g\left(\mathbf{r}-\mathbf{R}\right)\left[\varphi^\dag_\mathbf{R}+\varphi_\mathbf{R}\right],\label{CH6_eq.2-a} \\
V_2 &=& \int d\mathbf{r} \Psi^\dagger_\mathbf{r} \Psi_\mathbf{r} \int d\mathbf{R} g\left(\mathbf{r}-\mathbf{R}\right) \phi^\dagger_\mathbf{R} \phi_\mathbf{R}. \label{CH6_eq.2-b}
\end{eqnarray}
Furthermore, we take the Fourier transform of the operators in Eq.~\eqref{CH6_eq.2-a} and Eq.~\eqref{CH6_eq.2-b}, using
\begin{equation}\label{CH6_eq.3}
    \varphi^\dag_\mathbf{R}+\varphi_\mathbf{R}= \frac{1}{L}\sum_{\mathbf{p}} e^{\mi\mathbf{pR}} \left[(u_\mathbf{p}+v_{-\mathbf{p}})b_\mathbf{p}+(v_\mathbf{p}+u_{-\mathbf{p}})b^\dag_{-\mathbf{p}}\right],
\end{equation}
where $b^\dag_{\mathbf{p}}$ and $b_{\mathbf{p}}$ are the creation and annihilation operators of the bogolons, respectively, with the coefficients reading~\cite{Giorgini:1998aa}
\begin{eqnarray}\label{CH6_eq.4}
    u^2_{\mathbf{p}}=1+v^2_{\mathbf{p}}&=&\frac{1}{2}\left(1+\left[1+\frac{(Ms^2)^2}{\omega^2_{\mathbf{p}}}\right]^{1/2}\right),\\\nonumber
    u_{\mathbf{p}}v_{\mathbf{p}}&=&-\frac{Ms^2}{2\omega_{\mathbf{p}}}.
\end{eqnarray}
Here $M$ is the exciton mass, $s=\sqrt{\kappa n_c/M}$ is the sound velocity of the bogolons,
$\omega_k=sk(1+k^2\xi^2)^{1/2}$ is their spectrum, $\kappa=e_0^2d/\epsilon$ is the Fourier image of the exciton--exciton interaction strength, $e_0$ is electron charge, $\epsilon$ is the dielectric function, and $\xi=\hbar/(2Ms)$ is the healing length of the condensation.
Combining Eqs.~\eqref{CH6_eq.2-a}--\eqref{CH6_eq.3} with the Fourier transformation of operator for electron $\Psi_\mathbf{r}=\frac{1}{L}\sum_\mathbf{k} c_\mathbf{k} e^{\mi\mathbf{k}\mathbf{r}}$ yields
\begin{eqnarray}
V_1 &=& \frac{\sqrt{n_c}}{L} \sum_{\mathbf{k,p}} g_\mathbf{p} \left[ \left( v_\mathbf{p} + u_\mathbf{-p} \right)b^\dagger_\mathbf{-p} + \left( u_\mathbf{p} + v_\mathbf{-p}\right)b_\mathbf{p} \right] c^\dagger_\mathbf{k+p} c_\mathbf{k}, \label{CH6_eq.5-a} \\
V_2 &=& \frac{1}{L^2}\sum_{\mathbf{k},\mathbf{p},\mathbf{q}} g_\mathbf{p}\bigg[ u_{\mathbf{q}-\mathbf{p}} u_\mathbf{q} b^\dagger_{\mathbf{q}-\mathbf{p}} b_\mathbf{q} + u_{\mathbf{q}-\mathbf{p}} v_\mathbf{q} b^\dagger_{\mathbf{q}-\mathbf{p}} b^\dagger_{-\mathbf{q}} \nonumber \\
&+& v_{\mathbf{q}-\mathbf{p}} u_\mathbf{q} b_{-\mathbf{q}+\mathbf{p}} b_\mathbf{q} + v_{\mathbf{q}-\mathbf{p}} v_\mathbf{q} b_{-\mathbf{q}+\mathbf{p}} b^\dagger_{-\mathbf{q}}\bigg] c^\dagger_{\mathbf{k}+\mathbf{p}} c_\mathbf{k},\label{CH6_eq.5-b}
\end{eqnarray}
where $g_\mathbf{p}$ is the Fourier image of the electron--exciton interaction, which we will discuss in a later section, and $L$ is the size of the system.
Illustrations of the processes in Eqs.~\eqref{CH6_eq.5-a} and ~\eqref{CH6_eq.5-b} are presented in Fig.~\ref{fig:CH6_2}, which depicts the scattering of an electron mediated by the absorption or emission of bogolons.

\section{Electron--exciton interaction}
\begin{figure}
    \centering
    \includegraphics[width=0.55\textwidth]{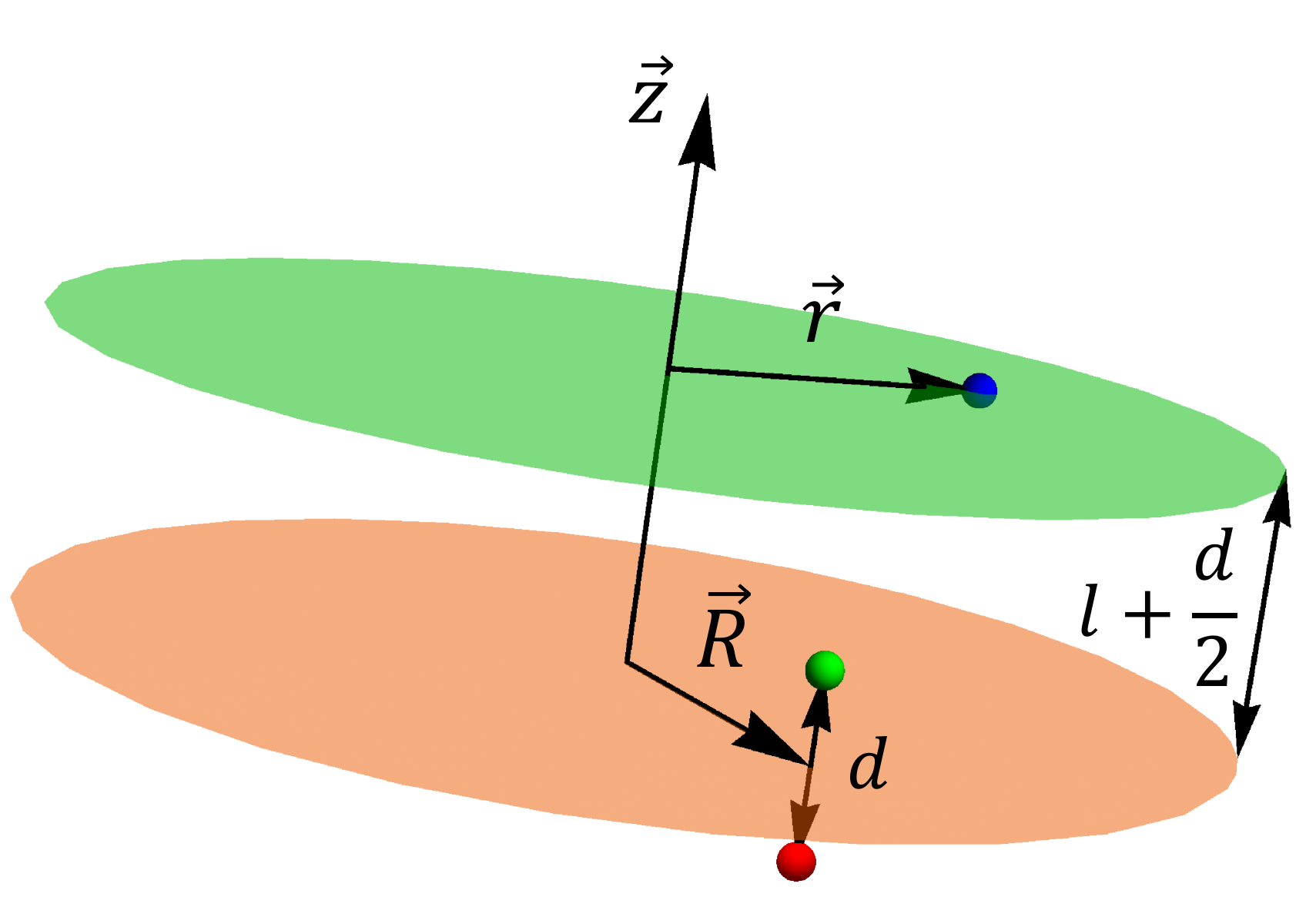}
    \caption[Electron and exciton interaction in the hybrid system]{Schematic of the interaction between an electron and exciton (electron--hole pair) in the hybrid system. The blue dot in the green disk is an electron with coordinate $\mathbf{r}$ in the fermion layer. For the exciton, we consider the center of mass for the electron--hole pair, labeled by $\mathbf{R}$ in the orange plane. The green (red) dot represents the electron (hole) of the exciton. As shown in Fig.~\ref{fig:Ch6_1}, parameters $l$ and $d$ are the distances between layers in the $z$-direction. Assuming the effective mass of the electron and hole is the same, then the distance between the green and orange planes is $l+\frac{d}{2}$.  }
    \label{fig:CH6_gk}
\end{figure}

To derive the formula of the interaction term in momentum space, we consider the system as shown in Fig.~\ref{fig:CH6_gk}.
The electron located in the fermion layer is represented by the blue dot in the green disk.
Vector $\mathbf{r}$ represents the location of the electron in the fermion layer, where the origin is the point where the $z$-axis intersects with the green plane.
For the exciton, we consider the center of mass represented by $\mathbf{R}$ in the orange disk.

By the notations $r_{e-e}$ ($r_{e-h}$) representing the distance between electron--electron (electron--hole), $e_0$ representing the unit of electron charge and $\epsilon$ representing the dielectric constant for the given material, considering the formula of Coulomb potential we have
\begin{eqnarray}
g\left(r\right) &=& \frac{e_0^2}{4\pi\epsilon}\left( \frac{1}{r_{e-e}}-\frac{1}{r_{e-h}}\right) \\ \nonumber
&=& \frac{e_0^2}{4\pi \epsilon} \left( \frac{1}{\sqrt{l^2+\left( \mathbf{r} - \mathbf{R} \right)^2}} - \frac{1}{\sqrt{ \left(l+d\right)^2+\left( \mathbf{r} -\mathbf{R} \right)^2 }}\right), \label{eq:CH6_gk_real_space}
\end{eqnarray}
where, as shown in Fig.~\ref{fig:CH6_gk}, the 2D vectors $\mathbf{r}$ and $\mathbf{R}$ are the positions of the electron and exciton in their respective layers, and $l$ and $d$ are the distances in the $z$-direction between layers.

Renewing the variable $\mathbf{r}$ by the definition $\mathbf{r} \equiv \mathbf{r} - \mathbf{R}$ and then applying the Fourier transform, we have
\begin{eqnarray}
 g\left( \mathbf{k} \right) &=& \frac{e_0^2}{4\pi \epsilon} \int d\mathbf{r} \left[ \left( l^2 + \mathbf{r}^2 \right)^{-\frac{1}{2}} -\left( \left( l+d \right)^2 + \mathbf{r}^2 \right)^{-\frac{1}{2}} \right] e^{-\mi \mathbf{k} \cdot \mathbf{r}} \\ \nonumber
 &=& \frac{e_0^2}{4 \pi \epsilon} \int_0^\infty dr \int_0^{2\pi} d\theta \left[ \frac{re^{-\mi k r \cos\theta}}{\sqrt{l^2 + r^2}} - \frac{re^{-\mi k r \cos\theta}}{\sqrt{\left( l + d \right)^2 +r^2 }} \right] \\ \nonumber
 &=& \frac{e_0^2}{4 \pi \epsilon} \int_0^\infty dr \left[ \frac{r}{\sqrt{l^2 + r^2}} \frac{r}{\sqrt{\left( l+d \right)^2 + r^2}}\right] 2\pi \mathcal{J}_0\left( k r \right) \\ \nonumber
 &=& \frac{e_0^2}{2 k \epsilon } e^{-kl} \left( 1 - e^{-kd} \right),
\end{eqnarray}
where $\mathcal{J}_0$ is the Bessel function of first kind, and $k$ is the magnitude of $\mathbf{k}$.
We can further approximate this interaction by,
\begin{equation}
    g\left(\mathbf{k}\right) = \frac{e_0^2}{2k \epsilon}e^{-kl}.
    \label{eq:CH6_gk_k}
\end{equation}
We will use this formula in later calculations.
%
%
%
\section{Graphene case}
In the graphene case, we will discuss the single-bogolon scattering process as described by Eq.~\eqref{CH6_eq.5-a}.
\subsection{Particle transport}
We use the Boltzmann transport theory \cite{Kawamura:1990aa} to calculate the resistivity of electrons in graphene, which is given by
\begin{equation}\label{CH6_eq.6}
\rho^{-1} = e_0^2 D\left(E_F\right)\frac{v_F^2}{2}\langle \tau \rangle,
\end{equation}
where $v_F$ is the Fermi velocity, $E_F$ is the Fermi energy, and the density of states of graphene at the Fermi level reads $D\left(E_F\right) =\left(g_s g_v/2\pi \hbar^2\right)E_Fv_F^{-2}$, where $g_{s,v}=2$ are the spin and valley $g$-factors, respectively.
We can write the energy-averaged relaxation time as~\cite{Kawamura:1990aa,PhysRevB.45.3612,Hwang:2008aa}
\begin{equation}\label{CH6_eq.7}
\langle \tau \rangle = \frac{\int d \varepsilon D\left(\varepsilon \right) \tau\left( \varepsilon \right) \left[ -\frac{df^0\left(\varepsilon\right)}{d \varepsilon}\right]}{\int d \varepsilon D\left(\varepsilon\right) \left[ -\frac{df^0\left(\varepsilon\right)}{d \varepsilon} \right]},
\end{equation}
where $f^0\left(\varepsilon\right)=\{\exp[(\varepsilon-\mu)/(k_BT)]+1\}^{-1}$ is the Fermi distribution function, $\mu$ is the chemical potential, $k_B$ is the Boltzmann constant, and the energy-dependent inverse relaxation time reads
\begin{equation}\label{CH6_eq.8}
  \frac{1}{\tau\left(\varepsilon\right)} = \sum_\mathbf{k'}\left(1 -\cos\theta_\mathbf{kk'}\right) W_\mathbf{kk'} \frac{1-f^0\left(\varepsilon'\right)}{1-f^0\left(\varepsilon\right)}.
\end{equation}
Here, $\theta_\mathbf{kk'}$ is the scattering angle between $\mathbf{k}$ and $\mathbf{k'}$, $\varepsilon=\hbar v_F\abs{\mathbf{k}}$ is the dispersion of graphene, and $W_\mathbf{kk'}$ is the probability of transition from an initial electron state $\mathbf{k}$ to the final state $\mathbf{k}'$, as given by
\begin{equation}\label{CH6_eq.9}
    W_\mathbf{kk'}=\frac{2\pi}{\hbar}\sum_\mathbf{q} \abs{C_\mathbf{q}}^2 \Delta\left(\varepsilon, \varepsilon'\right),
\end{equation}
where $C_\mathbf{q}$ is the scattering matrix element, and
\begin{equation}\label{CH6_eq.10}
  \Delta\left(\varepsilon,\varepsilon'\right)=N_q\delta\left(\varepsilon-\varepsilon'+\hbar\omega_q\right)+\left(N_q+1\right)\delta\left(\varepsilon-\varepsilon'-\hbar\omega_q\right),
\end{equation}
where $N_q=\{\exp[\hbar\omega_q/(k_BT)]-1\}^{-1}$ is the Bose distribution function.
Summing up, the energy-dependent relaxation time reads
\begin{eqnarray}\label{CH6_eq.11}
\frac{1}{\tau\left(\varepsilon\right)} &=& \frac{e_0^4 d^2 n_c}{8 \pi\epsilon^2 \hbar} \int d \mathbf{k'}\left(1-\cos\theta_{\mathbf{kk'}}\right) \int d \mathbf{q} e^{-2\abs{q}l}\abs{u_\mathbf{q}+v_\mathbf{q}}^2 \nonumber \\
&\times& \frac{1-f^0\left(\varepsilon'\right)}{1-f^0\left(\varepsilon\right)}\Delta\left(\varepsilon,\varepsilon'\right)\delta \left(\mathbf{q}-\mathbf{k}+\mathbf{k'}\right).
\end{eqnarray}
Using Eq.~\eqref{CH6_eq.4} and assuming a linear dispersion of bogolons $\omega_\mathbf{q}=s\abs{\mathbf{q}}$ (which is legitimate at $q\ll \xi^{-1}$ ), we find:
\begin{eqnarray} \label{CH6_eq.12}
\frac{1}{\tau\left(\varepsilon\right)} &=& \sum_{n=1,2}\frac{e_0^4 d^2 n_c }{8\pi \epsilon^2 \hbar^3 v_F^2} \int_0^{2\pi} d\theta \varepsilon_n \left( 1-\cos\theta\right)  \frac{1-f^0\left(\varepsilon_n \right) } {1-f^0\left(\varepsilon\right)} \nonumber \\ 
&\times& e^{-2l\lambda}\left( \sqrt{1+\frac{M^2s^2}{\hbar^2\lambda^2}} - \frac{Ms}{\hbar \lambda}\right) \frac{N_\lambda + \delta_{n,2}}{|F'_n \left( \varepsilon_n \right) |} ,
\end{eqnarray}
where $\lambda\equiv\abs{\mathbf{k}-\mathbf{k'}} = \sqrt{k^2+k'^2-2kk'\cos\theta}$ and thus is a function of $k$, $k'$, and $\theta$, $\varepsilon_n$ are two roots of the equation $F_{1,2} \left(\varepsilon'\right)=\varepsilon-\varepsilon'\pm \hbar \omega_\lambda=0$, $F'_{n}\left(\varepsilon'\right)$ is its first derivative, and $\delta_{n,2}$ is the Kronecker delta. Specifically, $n=2$ corresponds to the bogolon emission process.
Substituting Eq.~\eqref{CH6_eq.12} into the average lifetime from Eq.~\eqref{CH6_eq.7}, we can numerically calculate the conductivity as in Eq.~\eqref{CH6_eq.6}. Before doing so though, let us first analytically consider the limiting cases of high and low temperatures.

%
%
\subsection{High-temperature limit}
Let us analyze Eq.~\eqref{CH6_eq.12} and find the principal dependence of conductivity on $T$ at high temperatures, $T_{BG} \ll T \ll E_F/k_B$, where we denote the Bloch--Gr\"uneisen temperature as $T_{BG}=2\hbar sk_F/k_B$.
Since $T \gg T_{BG}$, we have $\hbar \omega_{\mathbf{q}} \ll k_B T$.
In this case, the Bose--Einstein distribution can be approximated as $N_q \sim k_B T / \hbar \omega_\mathbf{q}$, and $\Delta\left(\varepsilon,\varepsilon'\right) = \left( 2k_B T / k_B T \right) \delta\left( \varepsilon -\varepsilon' \right)$.
Then we find the energy-dependent relaxation time through
\begin{eqnarray} \label{CH6_eq.13}
\frac{1}{\tau\left(\varepsilon\right)} = \frac{e_0^4d^2k_BT}{8\pi^2\epsilon^2\hbar^2v_F^2} \int^{2\pi}_0 d\theta \left( 1- \cos\theta\right) \varepsilon e^{-\lambda l}\left( \sqrt{ \frac{1}{s^2 \lambda^2} + \frac{M^2}{\hbar^2 \lambda^4}} -\frac{M}{\hbar \lambda} \right).
\end{eqnarray}
One should notice that the integral in Eq.~\eqref{CH6_eq.13} is temperature-independent.
Under the limit $T\ll E_F/k_B$, the contribution from the Fermi energy in Eq.~\eqref{CH6_eq.7} is dominant.
This gives us $\langle \tau \rangle \approx \tau\left(E_F\right)\sim T^{-1}$. Substituting this expression in Eq.~\eqref{CH6_eq.6}, we find that the resistivity linearly depends on the temperature, as in the case of phonon-assisted relaxation~\cite{Hwang:2008aa}.
Indeed, the temperature should still be smaller than the exciton condensation temperature. Otherwise, bogolon-mediated relaxation cannot exist.

%
%
\subsection{Low-temperature limit}\label{sec:low}
To investigate the principal $T$-dependence of resistivity at low temperatures, we will use the Bloch--Gr\"uneisen formalism as described in~\cite{Ziman:2001aa, Zaitsev:2014aa}.
We start from the Boltzmann equation
\begin{equation}\label{CH6_1}
e_0\textbf{E}\cdot\frac{\partial f}{\hbar\partial \textbf{p}}=I\{f\},
\end{equation}
where $f$ is the electron distribution, $\mathbf{p}$ is the wave vector ($p\equiv\abs{\mathbf{p}}$), $\mathbf{E}$ is the perturbing electric field, and $I\{f\}$ is the collision integral (see Appendix \ref{AP:CH6} for the explicit form of $I$ and other details of derivation).
For relatively weak electric fields, $f$ can be expanded into
\begin{eqnarray}\label{CH6_f}
f=f^0(\varepsilon_p)-\left(-\frac{\partial f^0}{\partial\varepsilon_p}\right)f^{(1)}_\textbf{p},
\end{eqnarray}
where the correction $f^{(1)}_\textbf{p}$ has the dimensionality of energy.
Without a loss of generality, we set the electric field along the $x$-axis and use the ansatz
\begin{eqnarray}\label{CH6_fp}
f^{(1)}_\textbf{p}= v_F\frac{e_0E_xp_x}{k_F}\tau(\varepsilon_p).
\end{eqnarray}
After some algebra, we find the resistivity in the form
\begin{eqnarray}
\label{CH6_rhoV1}
\rho\propto\frac{1}{\tau_0}&=&\frac{\hbar\xi_I^2}{8\pi^2k_FM}\frac{1}{k_BT}\int_0^\infty dq q^4e^{-2ql}q(\Gamma_--\Gamma_+)_{k_F}\nonumber\\
&{}&{}\times N_q(1+N_q),
\end{eqnarray}
where $\tau_0$ is the effective scattering time, $\xi_I= e_0^2d\sqrt{n_c}/2\epsilon$, and
\begin{eqnarray}
\label{CH6_gamma2}
\Gamma_\pm=\frac{2|v_Fk_F\pm sq|(2v_Fsk_F-v_F^2q)}{\hbar v_F^3k_F q\sqrt{\pm 4k_Fsv_Fq+4k_F^2v_F^2-v_F^2q^2}}.
\end{eqnarray}
The subscript $k_F$ in the expression $(\Gamma_--\Gamma_+)_{k_F}$ in Eq.~\eqref{CH6_rhoV1} means that all the electron wave vectors $p$ are to be substituted by $k_F$.

For temperatures much lower than the Bloch--Gr\"{u}neisen temperature, we find the following expression:
\begin{eqnarray}\label{CH6_f_tau}
\frac{1}{\tau_0}=\frac{I_0\xi_I^2k_F^2}{4\pi^2\hbar \alpha^4v_F^2M}\left(\frac{k_BT}{E_F}\right)^4,
\end{eqnarray}
where $I_0\approx 26.2$ is a dimensionless factor.
In terms of the resistivity,
\begin{eqnarray}
\label{CH6_EqResist}
\rho=\frac{\pi\hbar^2}{e_0^2E_F}\frac{1}{\tau_0}=(1.0\times 10^6\;\Omega)\left(\frac{k_BT}{E_F}\right)^4.
\end{eqnarray}
In this estimation, we use a dimensionless parameter $\Tilde{l}=lk_BT/(\hbar s)$, which is determined by the interlayer distance $l$, the sound velocity $s$ (which is in turn determined by the condensate density), and temperature. We also use the condition $\Tilde{l}\ll 1$ to get an analytical dependence at low $T$.

For temperatures far lower than room temperature ($k_BT_R\approx 26$ meV), we have $\Tilde{l}\ll 1$.
If $T\ll T_{BG}$, where $T_{BG}\ll E_F/k_B$ since $s\ll v_F$, we find the precise form of what we mean by \textit{low temperatures}: $k_BT/E_F<10^{-2}$.
For typical $E_F\sim 10^{-1}$ eV, this gives $T_{BG}=183$ K and  $T<18$ K (for the particular range of distances between the layers $l$ up to $50$ nm).

Moving forward, it is interesting to compare the formula in Eq.~\eqref{CH6_EqResist} rewritten in a different form as
\begin{eqnarray}\label{CH6_f_tau_app}
\frac{1}{\tau_0}=\frac{5I_0e_0^6}{8\pi^2\epsilon^2v_F^2}\frac{n_cd}{M}\frac{1}{E_Fk_F}\left(\frac{k_BT}{\hbar s}\right)^4,
\end{eqnarray}
with the phonon-mediated scattering case~\cite{Hwang:2008aa} as
\begin{eqnarray}
\label{CH6_phonon}
\frac{1}{\tilde{\tau}_0}=\frac{D^24!\zeta(4)}{2\pi\rho_mv_{ph}}\frac{1}{E_Fk_F}\left(\frac{k_BT}{\hbar v_{ph}}\right)^4,
\end{eqnarray}
where $\rho_m$ is the density of graphene, and $\zeta$ is the Riemann zeta-function. We see that both the inverse times have the same $T$-dependence at low temperatures with the phonon velocity $v_{ph}$ replaced by the sound speed $s$ in the bogolon-mediated scattering case.

Moreover, the result presented in~\cite{Hwang:2008aa} [and Eq.~\eqref{CH6_phonon}] assumes that the dominant contribution to the scattering comes from the longitudinal acoustic phonons.
A more recent study~\cite{Kaasbjerg:2012aa} shows that the transverse acoustic phonons dominate at low temperatures.
As a result, the resistivity obeys the power law $\rho_{ph}\propto T^\alpha$ with $\alpha\sim 6$, even in the absence of screening~\cite{Hwang:2007aa}, and this can additionally impair the impact of the phonon-related scattering.
Screening in our hybrid system case is a nontrivial issue requiring separate consideration.
Here we may simply note that the screening can likely be disregarded for certain $l$.
As a result, we conclude that at low temperatures, the $T$-dependence of the resistivity due to the bogolon-mediated scattering events is fundamentally different from the phonon case.
Since the bogolons have smaller temperature exponent $T^4$ than phonons, we imagine the former to dominate (at $T\ll T_{BG}$).

It should be emphasized that we do not have to put $\tilde{l}\ll 1$. However, the general case does not allow for the analytical extraction of the temperature dependence of resistivity out of the integration, and therefore a numerical approach is required.

%
%
\subsection{Numerical treatment}
To build the plots, we use Eqs.~\eqref{CH6_eq.6},~\eqref{CH6_eq.7}, and~\eqref{CH6_eq.12} with parameters typical for GaAs-based structures: $\epsilon=12.5\epsilon_0$ where $\epsilon_0$ is a vacuum permittivity, $M=0.52m_0$ where $m_0$ is the free electron mass, $d=10$ nm, $l=10$ nm, and $v_F=10^8$~cm/s~\cite{Castro-Neto:2009aa,Das-Sarma:2011aa}.

\begin{figure}[ht]
    \centering
    \includegraphics[width=0.79\textwidth]{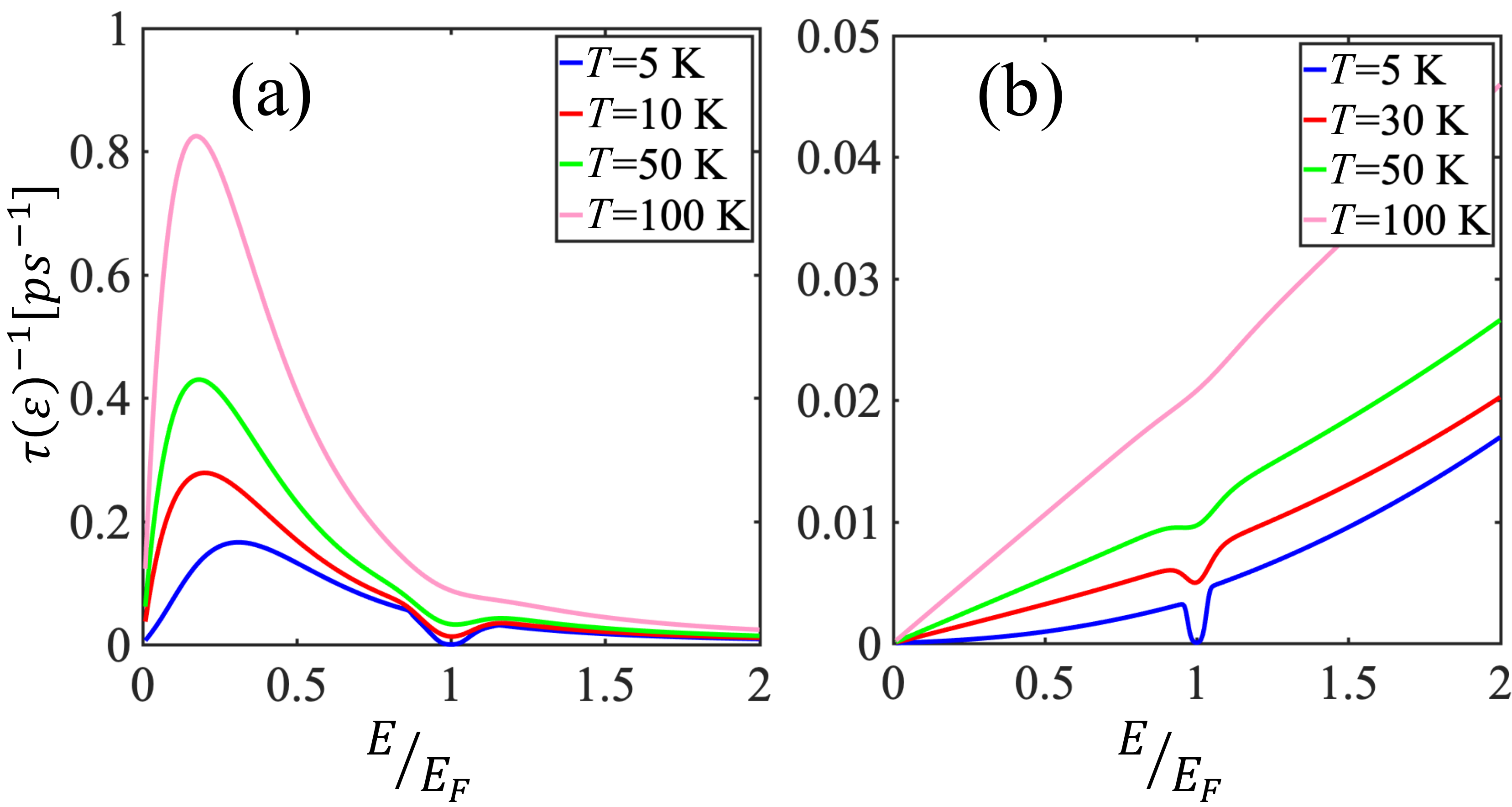}
    \caption[Energy-dependent inverse relaxation time]{Energy-dependent inverse relaxation time of electrons in the (a) single-bogolon (emission and absorption) processes, and (b) phonon-assisted processes at different temperatures. In panel (a), $n_c=10^{11}$ cm$^{-2}$ and thus $s\approx 7\times 10^6$~cm/s. The figure is taken from~\cite{Sun:2019aa}.}
    \label{fig:CH6_3}
\end{figure}
Figure~\ref{fig:CH6_3} shows the inverse energy-dependent relaxation time as a function of energy for different temperatures.
We compare here the bogolon-mediated scattering with the acoustic phonon-assisted relaxation under certain conditions~\footnote{The formulas for phonons are taken from \cite{Hwang:2008aa} with the following parameters: graphene mass density $\rho=7.6\times 10^{-8}$~g$/$cm$^2$, phonon velocity $v_{ph}=2\times 10^6$~cm/s, deformation potential $D=6.8$~eV from \cite{Kaasbjerg:2012aa}, and electron density $n=10^{12}$~cm$^{-2}$.}.
We find some similarities between the bogolon- and phonon-mediated processes. In both cases, the inverse lifetime grows with increasing temperature due to the increase of the number of fermions and bosons (bogolons or phonons) in the system.
We also observe low-temperature dips at the Fermi energy, which are due to a sharpening of the Fermi surface.

Despite these similarities, there is a fundamental difference between the two principal channels of scattering, originating from the mechanisms of electron--phonon and electron--bogolon interaction.
The former derives from the crystal lattice deformation potential theory, while the latter has an electric nature and the matrix element contains the Coulomb interaction term.

\begin{figure}[ht]
\centering
    \includegraphics[width=0.79\textwidth]{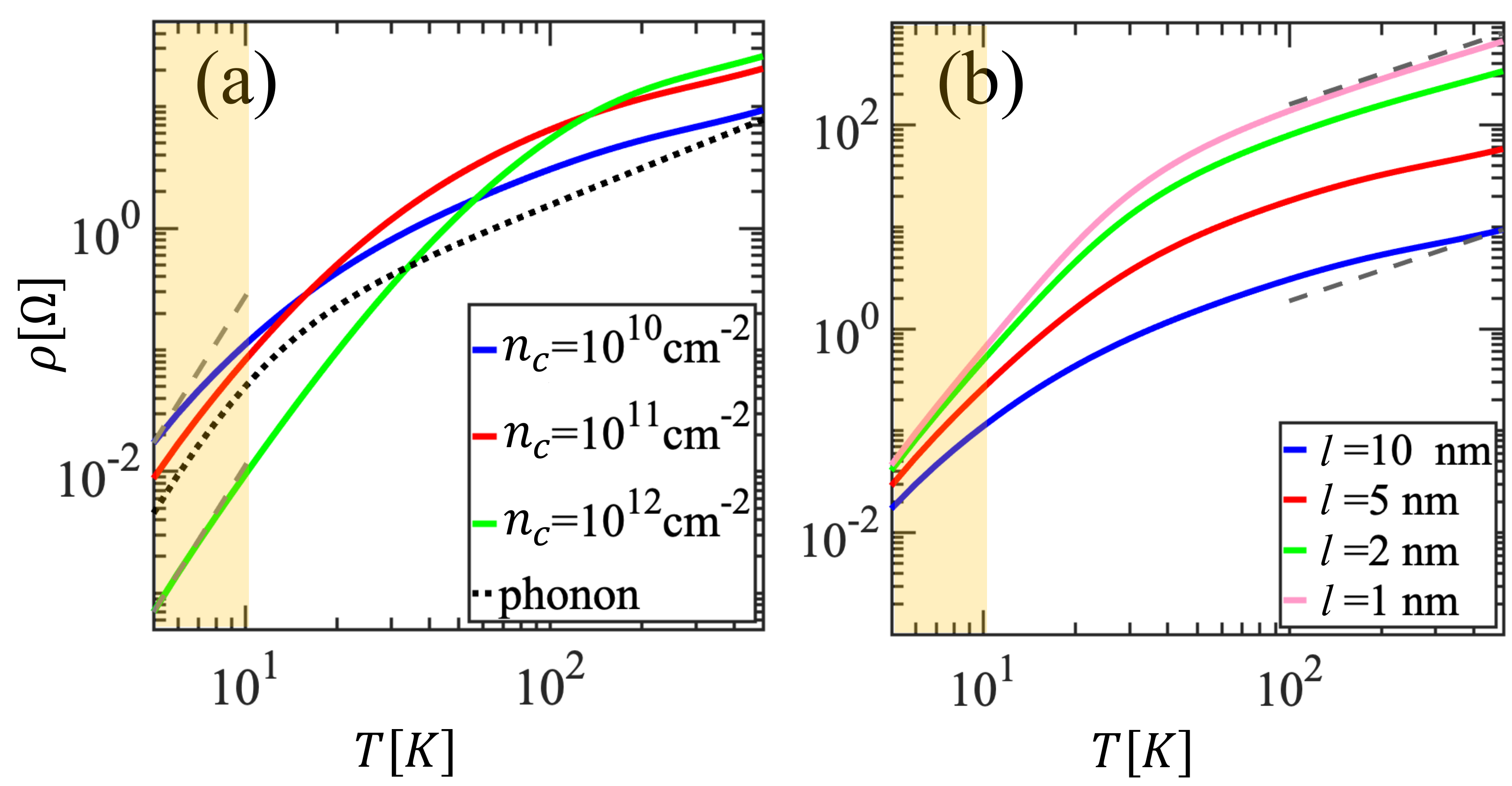}
    \caption[Temperature-dependent bogolon-mediated resistivity]{Bogolon-mediated resistivity of graphene as a function of temperature for (a) different particle densities in condensate $n_c$ at $l=10$ nm, and for (b) different interlayer distances $l$ at $n_c=10^{10}$~cm$^{-2}$.
    The dashed grey lines stand for the low- and high-temperature analytics, indicating $\sim T^4$ and $\sim T$  behavior in (a) and (b) respectively.
    The black dotted line in (a) shows the phonon-mediated resistivity for comparison.
    The yellow-shaded regions highlight the temperature regime in which the condensation of indirect excitons in GaAs structures was experimentally reported. The figure is taken from~\cite{Sun:2019aa}.}
    \label{fig:CH6_4}
\end{figure}

Figure~\ref{fig:CH6_4} demonstrates the behavior of graphene resistivity as a function of temperature for different condensate densities and interlayer spacings.
We also compare this with the phonon-mediated resistivity.
All the curves show $\sim T^4$ dependence at low temperatures and $\sim T$ dependence at high temperatures.
Consequently, the primary behavior of resistivity is deceptively similar to the phonon-assisted case, as reported in~\cite{Hwang:2008aa}.
In the case of bogolons, different $n_c$ affect the sound velocity, and the Bloch--Gr\"{u}neisen temperature changes accordingly: $T_{BG} \approx 54$, $190$, and $540$~K for densities $n_c=10^{10}$, $10^{11}$, and $10^{12}$~cm$^{-2}$, respectively.
This is the reason why we obtain a better agreement between the numerical results and the $T^4$-analytics in the high-density regime.
Further, Fig.~\ref{fig:CH6_4}(b) shows that by decreasing $l$, we can increase the Coulomb interaction strength, thereby increasing the resistivity of graphene.

It should be noted that the parameters of a GaAs-based material were considered here.
Indirect excitons in such materials only condense at temperatures less than $10$ K (yellow regions in Fig.~\ref{fig:CH6_4}); however, a higher $T_c$, for which we predict a linear temperature dependence of resistivity, might be achieved in other materials or systems.
For example, the critical temperature for degenerate exciton Bose gas can possibly reach $\sim 100$~K in MoS$_2$~\cite{Fogler:2014aa}.
Another potential candidate is exciton-polaritons, where quasi-condensation has been reported even at room temperature~\cite{Lerario:2017aa}, even though in this situation one needs to consider the non-equilibrium physics which need to be investigated in the future.

%
%

%
\section{Metal Case}
In this section, we consider a system consisting of a 2DEG with a parabolic dispersion of electrons and a layer of Bose-condensed exciton gas~\cite{Butov:2003aa,Fogler:2014aa,Butov:2017aa}.
\subsection{Single-bogolon scattering}
To investigate the principal $T$-dependence of single bogolon mediated resistivity at low temperatures, we follow the routine discussed in Section~\ref{sec:low} by writing the Boltzmann equation
\begin{eqnarray}
\label{CH7_Eq1}
e_0\textbf{E}\cdot\frac{\partial f}{\hbar\partial \textbf{p}}=I\{f\},
\end{eqnarray}
where $f$ is the electron distribution function, $\mathbf{p}$ is the wave vector, $\mathbf{E}$ is an external electric field, and $I\{f\}$ is the collision integral involving single-bogolon scattering processes, as shown in Fig.~\ref{fig:CH6_2}(a) and (b) (see Appendix~\ref{AP:CH7} for the explicit form of $I$ and other details of derivation).
For relatively weak electric fields, we apply Eq.~\ref{CH6_f} to express the function $f$,
%
%
%
where for the 2DEG case we express the term $f^{\left(1\right)}_\mathbf{p}$ as
\begin{equation}
\label{CH7_EqAnsatz}
\phi_\textbf{p}= (e_0E_x)(\hbar m^{-1}p_x)\tau(\varepsilon_p),
\end{equation}
where $m$ is the effective electron mass in the 2DEG and $\tau(\varepsilon_p)$ is the relaxation time.
The factor $e_0E_x$ is the force acting on the electron while $\hbar m^{-1}p_x$ is the electron velocity.
The function $\phi_\mathbf{p}$ therefore gives the work done by the electric field on the electron during time $\tau(\varepsilon_p)$.

Using Eqs.~\eqref{CH7_Eq1} and~\eqref{CH7_EqAnsatz}, we find the average value of the scattering time (see Appendix~\ref{AP:CH7}) and then the single bogolon mediated resistivity, which is
\begin{eqnarray}
\label{CH7_rho1bgen}
\rho^{(1)}=\frac{\pi\hbar^3\xi_I^2}{e_0^2ME_F}\sum_{n=0}^\infty\frac{(-2)^nl^n\gamma_n}{n!(\hbar s)^{n+4}}(k_BT)^{n+4},
\end{eqnarray}
where $\xi_I= e_0^2d\sqrt{n_c}/2\epsilon$, $E_F$ is the Fermi energy, $\gamma_n=(n+3)!\zeta(n+3)/[(2\pi)^2k_BT_\textrm{BG}]$, $T_\textrm{BG}=2\hbar sk_F/k_B$ is the Bloch--Gr\"{u}neisen temperature with $k_F$ the Fermi wave vector, $k_B$ the Boltzmann constant, and $\zeta(x)$ the Riemann zeta function.
The leading term in Eq.~\eqref{CH7_rho1bgen}, if $T\ll T_\textrm{BG}$, reads
\begin{eqnarray}
\label{CH7_rho1b}
\rho^{(1)}\approx\frac{\pi\hbar^3\xi_I^2}{e_0^2ME_F}\frac{3!\zeta(3)}{(2\pi)^2k_BT_\textrm{BG}}\left(\frac{k_BT}{\hbar s}\right)^4,
\end{eqnarray}
and hence the single bogolon resistivity behaves as $\rho^{(1)}\propto T^4$ at low temperatures.

%
%
\subsection{Double bogolon scattering}
Double bogolon resistivity can also be derived from Eq.~\eqref{CH7_Eq1}. The collision integral now expresses the net scattering into a state with momentum $\hbar\mathbf{p}$, involving a pair of bogolons, as shown in Fig.~\ref{fig:CH6_2}(c--f) (see also Appendix~\ref{AP:CH7}).
After some delicate derivations we find
\begin{eqnarray}
\label{CH7_rho2bmain}
\rho^{(2)}=\frac{M^2s^2}{8\pi^2e_0^2m^3v_F^5}\int\limits_{L^{-1}}^\infty\frac{k^2g_k^2dk}{\sinh^2\left[\frac{\hbar sk}{2k_BT}\right]}\ln(kL),
\end{eqnarray}
where $v_F$ is the Fermi velocity.
To derive Eq.~\eqref{CH7_rho2bmain}, we used the approximation $v_F\gg s$ and introduced the infrared cut-off $L^{-1}$ for the wave vector integrals, which is necessary for convergence.
The physical meaning of this cut-off is the absence of fluctuations with wavelengths larger than $L$.
The cut-off can also be related to the critical temperature of the BEC in a finite trap of length $L$~\cite{Bagnato:1991aa}.
Indeed, BEC cannot form in infinite homogeneous 2D systems at finite temperatures~\cite{Hohenberg:1967aa}, and thus a trapping with a characteristic size of $L$ is required~\cite{Butov:2017aa}.

We can further extract the temperature dependencies for the two limits (see Appendix~\ref{AP:CH7}).
At low temperatures $T\ll T_\textrm{BG}$,
\begin{eqnarray}
\label{CH7_rho2b}
\rho^{(2)}\approx\frac{s^2e_0^2d^2}{2v_F^5\epsilon^2}\left(\frac{T}{T_\textrm{BG}}\right)^3\frac{\pi}{6(2l)^3}\ln \left(\frac{L}{2l}\right),
\end{eqnarray}
while at high temperatures $T\gg T_\textrm{BG}$,
\begin{eqnarray}
\label{CH7_highT}
\rho^{(2)}\approx\frac{s^2e_0^2d^2}{2v_F^5\epsilon^2}\left(\frac{T}{T_\textrm{BG}}\right)^2\frac{1}{(2l)^3}\ln \left(\frac{L}{2l}\right).
\end{eqnarray}
\begin{figure}[ht]
\centering
\includegraphics[width=0.65\textwidth]{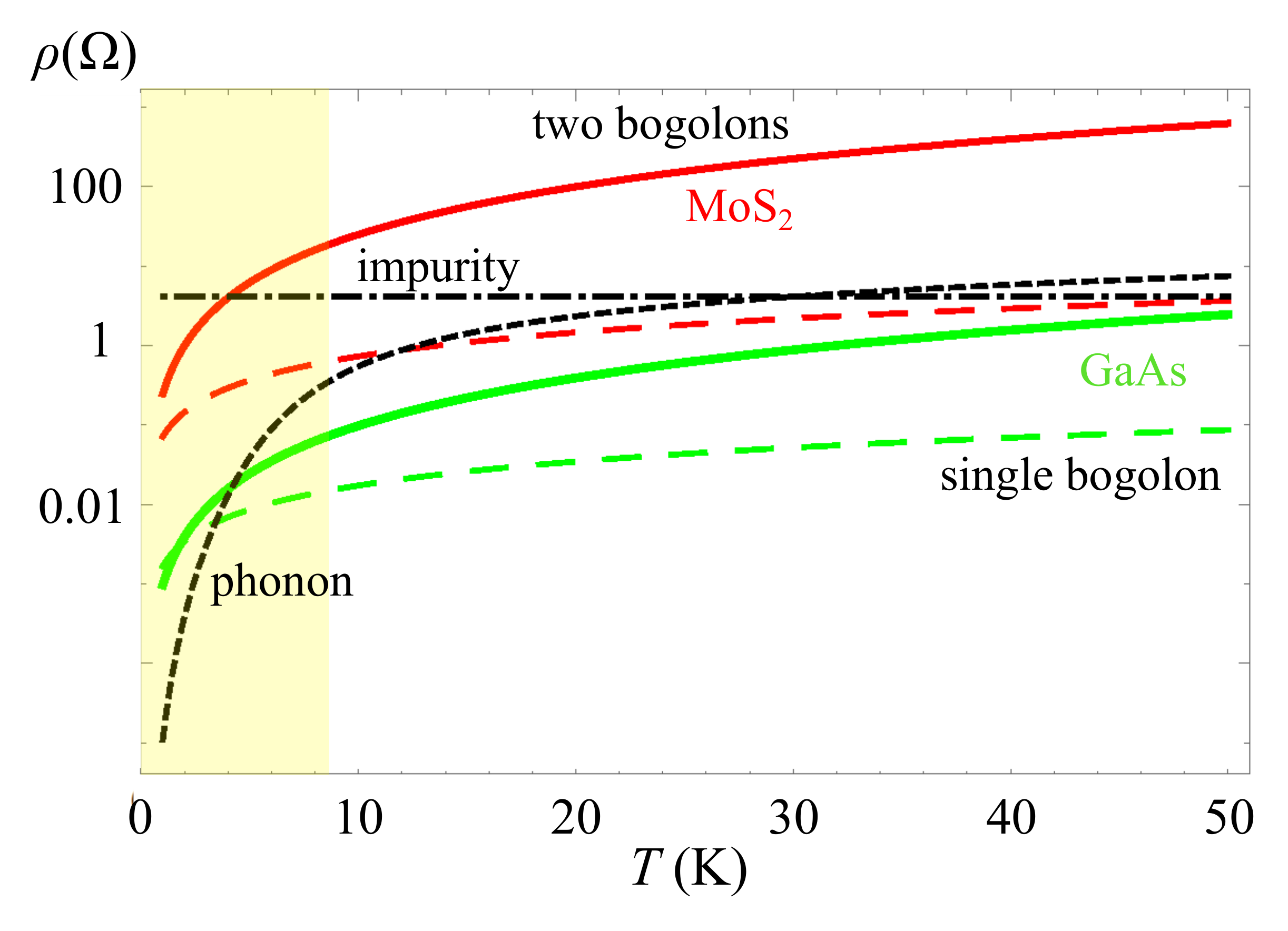}
\caption[Temperature-dependent resistivity of $\textrm{MoS}_2$ and GaAs]{Resistivity as a function of temperature for MoS$_2$ (red) and GaAs (green) exciton condensates.
Colored solid and dashed curves stand for the double-bogolon and single-bogolon contributions, respectively.
Black dash-dotted and dashed curves show the impurity and phonon contributions, respectively. We set $n_e=10^{13}$~cm$^{-2}$ and $n_c=10^9$~cm$^{-2}$. The figure is taken from~\cite{Villegas:2019aa}.}
\label{CH7_Fig3}
\end{figure}
%
%
%

%
%
\subsection{Discussion}
Compared to Eq.~\eqref{CH7_rho1b}, we conclude that the double-bogolon process should dominate over the single-bogolon process at very low temperatures.
However, in this range, scattering due to impurities is usually the strongest, which hinders the possibility to observe low-temperature asymptotics.
Figure~\ref{CH7_Fig3} shows the temperature behavior of different principal contributions to resistivity. In order to compare bogolon-mediated scattering with the scattering from phonons and impurities, we employed the theoretical and experimental results reported elsewhere~\cite{Kawamura:1992aa, Macleod:2009aa, Min:2012aa,Mendez:1984aa,Hirakawa:1986aa,Gold:1990aa}
and the parameters characteristic of a 2DEG in GaAs and excitons in both GaAs and MoS$_2$ materials\footnote{The values of the parameters were taken from~\cite{Kaasbjerg:2013aa,Basu:1980aa,Mair:1997pt} as follows.
Dielectric constants: $\epsilon_\textrm{GaAs}=12.5\epsilon_0$, $\epsilon_{\textrm{MoS}_2}=4.89\epsilon_0$; effective electron masses ($m_0$ is the bare electron mass): $m_\textrm{GaAs}=0.067m_0$, $m_{\textrm{MoS}_2}=0.47m_0$; exciton masses: $M_\textrm{GaAs}=0.517m_0$, $M_{\textrm{MoS}_2}=0.499m_0$; exciton sizes: $d_\textrm{GaAs}=10.0$ nm, $d_{\textrm{MoS}_2}=3.5$ nm; deformation potential: $D_\textrm{GaAs}=12$ eV.
}.

Figure~\ref{CH7_Fig3} shows the dependence of resistivity on temperature in the range$1-50$~K.
We set the concentration of impurities as $n_i=10^9$ cm$^{-2}$, which is attainable in high-quality $\textrm{GaAs}$ 2DEG~\cite{Hwang:2008ab,Manfra:2014aa}.
The yellow-shaded region highlights the low-temperature regime $T\ll T_\textrm{BG}$, where for both $\textrm{GaAs}$ and $\textrm{MoS}_2$ materials, $T_\textrm{BG}\approx 80$~K.
In this regime, the impurity scattering dominates even in high-quality $\textrm{GaAs}$ 2DEG;
however, when $T>14$~K we see that the double-bogolon scattering contribution to the resistivity can become an order of magnitude larger than all other contributions, if the double QW is made of $\textrm{MoS}_2$ material.
Scattering processes from impurities, phonons, and single bogolons have comparable contributions in the temperature range $20-50$~K.
The critical temperature of exciton quasi-condensation (or the formation of a degenerate Bose gas) in $\textrm{GaAs}$ has been reported to be around $T_c\sim 1-7$~K~\cite{Butov:2003aa}, and it is predicted to reach $T_c\sim 100$~K in $\textrm{MoS}_2$~\cite{Fogler:2014aa}.
We can alternatively put the structure from Fig.~\ref{fig:Ch6_1} in a microcavity, and instead of indirect excitons consider exciton-polaritons---for this, the same treatment is possible except for a different effective mass and the appearance of the Hopfield coefficients. In such a scenario, degenerate Bose gas (quasi-condensation or superfluidity) has been reported even at room temperature~\cite{Lerario:2017aa}.
One can also consider 2DEG in graphene instead of GaAs, where the scattering from impurities is suppressed and mobility is high.

In the conventional approach to hybrid 2DEG--BEC systems, the double-bogolon interaction, Eq.~\eqref{CH6_eq.5-b}, has been disregarded as it relates to second-order perturbation theory in fluctuations above the macroscopically-occupied ground state.
Figure~\ref{CH7_Fig3} demonstrates that this widespread approximation is not valid in the context of the  exciton condensates in MoS$_2$ material.
For GaAs exciton layers, the impurity is dominant for temperatures from $T\sim0-3$~K, at which the condensates can exist.
Hence, we will focus on MoS$_2$ in what follows.

\begin{figure}[ht]
\centering
\includegraphics[width=0.65\textwidth]{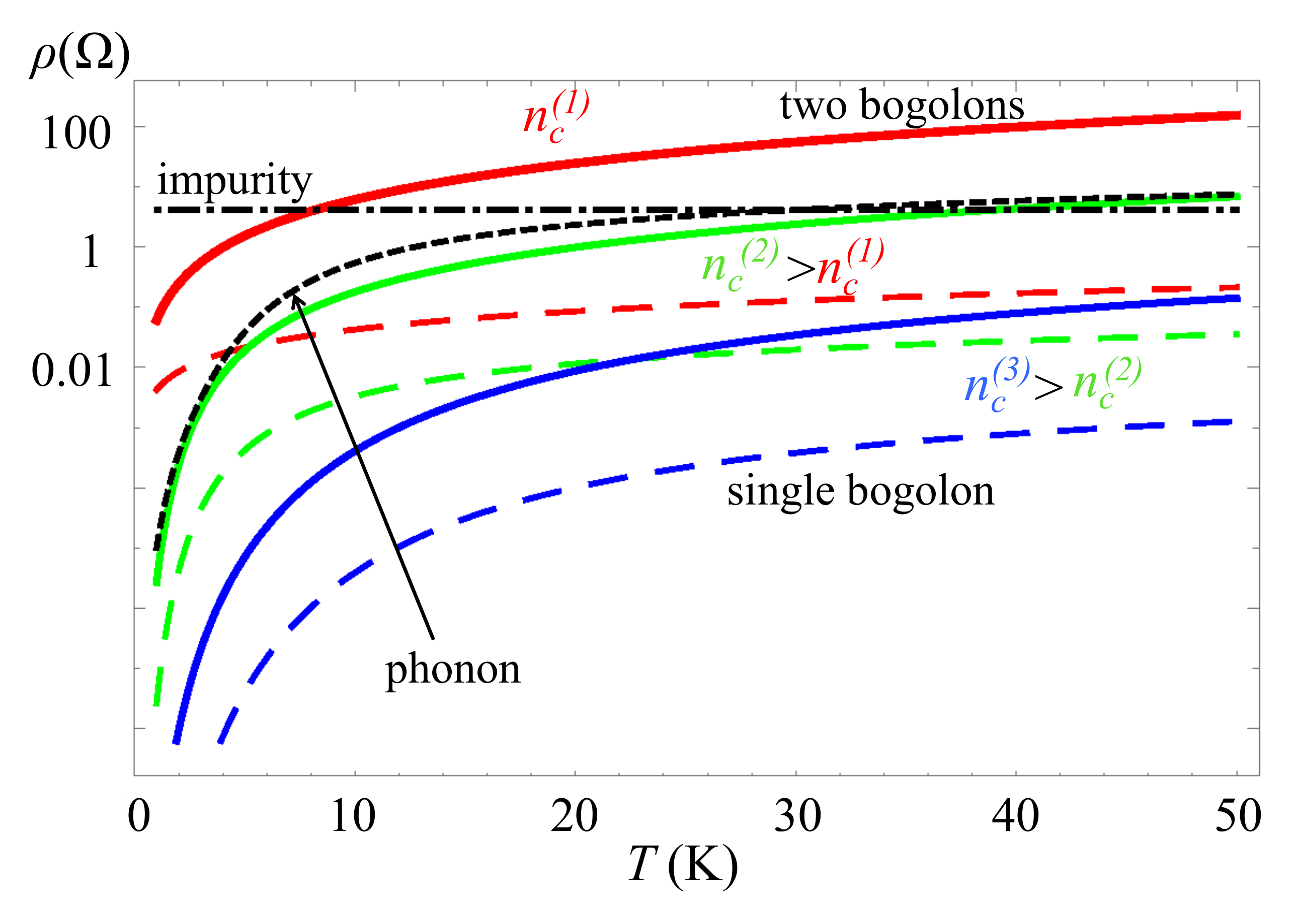}
\caption[Temperature-dependent resistivity with different condensate densities]{Temperature dependence of resistivity for various MoS$_2$ condensate densities: $n_c=10^8$~cm$^{-2}$ (red), $n_c=10^{10}$~cm$^{-2}$ (green), and $n_c=10^{11}$~
cm$^{-2}$ (blue).
The corresponding Bloch--Gr\"{u}neisen temperatures are $T\sim17$, $174$, and $549$~K, respectively. Black dash-dotted and dashed curves show the impurity and phonon contributions, respectively.
The electron density is fixed at $n_e=5\times10^{12}$~cm$^{-2}$. The figure is taken from~\cite{Villegas:2019aa}.}
\label{CH7_Fig4}
\end{figure}

Figure~\ref{CH7_Fig4} demonstrates the dependence of resistivity on condensate densities in the MoS$_2$-based exciton layer.
One can see that both single- and double-bogolon contributions increase as the condensate density decreases at relatively low temperatures (blue to green, green to red).
Such a tendency can be understood from Eqs.~\eqref{CH7_rho1b} and~\eqref{CH7_rho2b}, giving $\rho^{(1)}\sim \xi_I^2/(T_\textrm{BG}s^4)\sim n_c^{-1.5}$ and $\rho^{(2)}\sim s^2/T^3_\textrm{BG}\sim n_c^{-0.5}$.
Note that a similar behavior in the single-bogolon case happens in exciton BEC--graphene structures~\cite{Sun:2019aa}.
At $T\gg T_\textrm{BG}$ though, $\rho^{(2)}$ becomes $n_c$-independent, as follows from Eq.~\eqref{CH7_highT}; this could be seen by the red, green, and blue solid curves in Fig.~\ref{CH7_Fig4} starting to approach each other at higher temperatures.
We further note that the BEC-related screening from impurity and phonon processes has no significant effect, and so we plotted only two related curves.

Here we emphasize that these observations are valid as long as $n_c$ is macroscopically large, for the following two reasons.
First, in our calculations, we assume that the bogolon dispersion is linear (see Appendix~\ref{AP:CH7}), which only remains valid for $n_c\gtrsim 10^8$ cm$^{-2}$. %
Second, even if this assumption could be relaxed, we note that replacing the exciton field operator through $\Phi_\mathbf{R}=\sqrt{n_c}+\varphi_\mathbf{R}$, where $n_c$ is treated as an ordinary complex number instead of an operator, represents a mean field approach which is an essential ingredient of the Bogoliubov theory. 
Thus, we cannot expand our conclusions to the $n_c\rightarrow 0$ limit.
Despite this, we expect that in this limit, the bogolon contribution should vanish and be replaced by the bare exciton contribution, as dictated by $u_{\mathbf{p}}=1$ and $v_{\mathbf{p}}=0$ in Eqs. \eqref{CH6_eq.5-a} and~\eqref{CH6_eq.5-b}.

The dependence of resistivity on the separation between layers $l$ turns out rather strong, as expected, while the dependence on sample size $L$ (for double-bogolon scattering, where we introduced the infrared cut-off) is weak~(see Appendix~\ref{AP:CH7}).
This allows us to optimize sample design by varying $l$.

\begin{figure}[ht]
\centering
\includegraphics[width=0.65\textwidth]{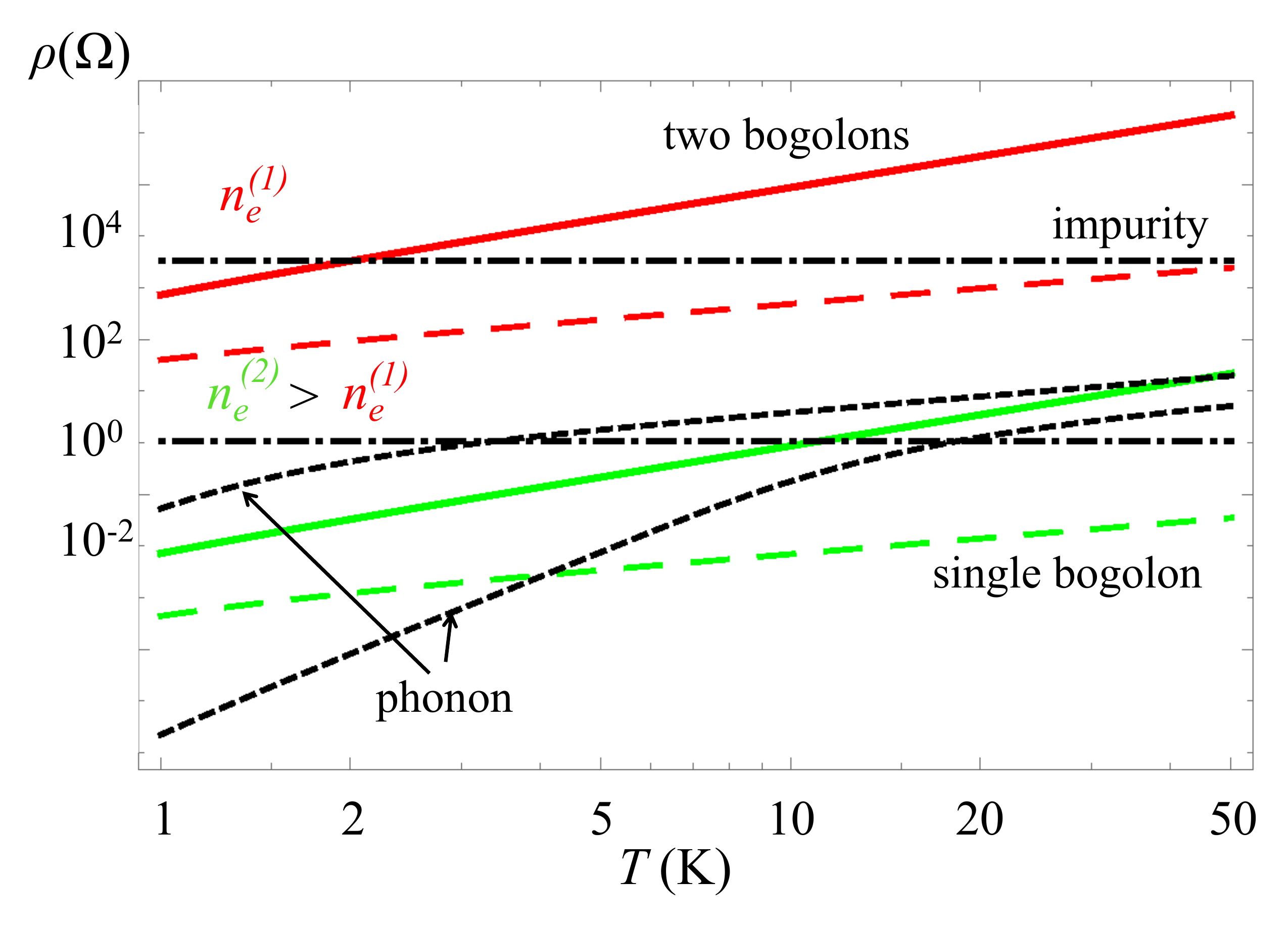}
\caption[Resistivities of single- and double-bogolon processes]{Temperature dependence of single-bogolon (dashed colored) and double-bogolon (solid colored) resistivities at different electron densities: $n_e=10^{11}$ (red) and $n_e=10^{13}$~cm$^{-2}$ (green). The corresponding Bloch--Gr\"{u}neisen temperatures are $T\sim2$ and $25$~K, respectively.
Black dash-dotted and dashed curves show the impurity and phonon contributions, respectively.
The density of the MoS$_2$ condensate is $n_c=10^8$~cm$^{-2}$. The figure is taken from~\cite{Villegas:2019aa}.}
\label{CH7_Fig5}
\end{figure}

The bogolon, impurity, and phonon-mediated resistivities all depend on the density of charge carriers in the 2DEG (see Fig.~\ref{CH7_Fig5}), and therefore generally decrease with an increase of $n_e$.
However, they do so at different rates.
For example, at $n_e=10^{13}$~cm$^{-2}$, impurity-induced scattering is dominant in the temperature range $T\sim 0-20$~K, while both the double-bogolon- and phonon-induced scattering become dominant (with comparable contributions) at $T\sim 20-50$~K.
At the lower density $n_e=10^{11}$~cm$^{-2}$, double-bogolon scattering starts to give the largest contribution at $T\gtrsim 5$~K, even reaching two orders of magnitude greater than the impurity contribution at $50$~K.

The dominance of the double-bogolon channel over the single-bogolon scattering in the MoS$_2$ exciton layer can be understood from an analysis of the matrix elements in the Fermi golden rule.
In the single-bogolon case, there appears a small factor $(u_\mathbf{p}+v_{-\mathbf{p}})\sim\sqrt{1+A^2}-A$, where $A=(Ms)/(\hbar \lambda)$ (see Appendix~\ref{AP:CH7}).
In other words, $(u_\mathbf{p}+v_{-\mathbf{p}})\sim(p\xi)^2\ll 1$.
In particular, in MoS$_2$, this factor is sufficiently small to compensate the large value of $\sqrt{n_c}$.
In contrast, there is no such cancellation effect in the double-bogolon terms, where there appears the product $u_\textbf{p}v_\textbf{p}\sim (p\xi)^{-1}\gg 1$ instead of $u_\mathbf{p}+v_{-\mathbf{p}}$.
Here we can compare with phonon scattering, where this cancellation effect does not take place, so that single-phonon scattering has larger contribution than double-phonon scattering.
This argument illustrates the difference between bogolon and phonon-assisted scattering, which is due to the difference in the origin of Coulomb interaction between the particles.

Experimentally, it might be difficult to resolve all the different contributions to the total resistivity, especially at low temperatures.
However, using the analytical formula Eq.~(\ref{CH7_highT}), we predict that the high-temperature resistivity should behave as $\rho\sim T^2$, if the double-bogolon scattering gives the dominant contribution.
This is in contrast to phonons giving $\rho\sim T$, and impurities with nearly absent $T$-dependence.


\section{Summary}
In this chapter, we have studied the transport of electrons in graphene and 2D metals coupled with a 2D Bose-condensed dipolar exciton gas via Coulomb interaction.

In the graphene case, we calculated the energy-dependent relaxation time of electrons, accompanied by the emission and absorption of a Bogoliubov excitation.
We further showed that bogolon-mediated scattering not only gives a significant correction to the phonon-assisted relaxation, but it also prevails given specific system geometry and temperatures.

In the 2DEG case, we calculated the resistivity by extending the Bloch--Gr\"{u}neisen approach and provided analytical formulas for the single- and double-bogolon scattering channels. We discovered that double-bogolon scattering becomes the dominant mechanism in hybrid systems in a certain range of temperatures.
%

%% file: conclusions_3-18_revised.tex
\chapter{Summary \& Outlook}\label{Ch7}
We finish this thesis with a general summary about the works we have done and try to give a short discussion about the future possible topics in related fields.

In chapter~\ref{CHONE}, we introduced the basic concepts behind exciton-polaritons~\cite{Deng:2010aa,Kavokin:2007aa,Shelykh:2009aa,Bramati:2013aa}.
We started from a microcavity system in which exciton-polaritons can be found, and then discussed the system's basic Hamiltonian.
We gave a short review of experimental exciton-polariton Bose--Einstein condensation and then discussed the related numerical model to treat such condensation~\cite{Carusotto:2013aa,Keeling:2008js,Wouters:2007aa}.

In Chapter~\ref{CHTWO}, we first considered polaritons in a system in which the cavity photons and excitons were spatially separated in the microcavity.
This special confinement gave us a momentum-dependent coupling between the excitons and cavity photons, which gave rise to unique shape of the dispersion for the ground state.
Further investigations revealed that in the two-dimensional case with TE-TM splitting, a multivalley-dispersion of the ground state was formed~\cite{Sun:2017ab}.
This gives a new technique to achieve valley polarization of exciton-polaritons compared to other methods by applying transition metal dichalcogenide (TMD) monolayers~\cite{Krol:2019aa,Dufferwiel:2017aa,Sun:2017ct}.
Secondly, we theoretically examined the formation of polariton condensation in a one-dimensional microcavity wire with a periodic, complex-valued potential~\cite{Yoon:2019aa}.
With a generalized Gross--Pitaevskii equation, we showed that condensation can occur in a nontrivial state compared to the known result~\cite{Lai:2007aa}, and under certain conditions, a space-time intermittency phase may appear between two distinct condensate phases~\cite{Hecke:1998aa,Chate:1994aa,PhysRevLett.120.033901}.

In Chapter~\ref{CHTHREE}, we considered the exciton-polaritons in complex artificial lattices~\cite{Sun:2018aa,Ko:2020aa,Sun:2019ab}.
We studied polariton condensation in a two-dimensional system with a Lieb lattice-shaped potential and investigated the dynamics of the resulting compact localized condensation.
We demonstrated that, unlike the previous works~\cite{Baboux:2016aa,Klembt:2017aa,PhysRevLett.120.097401}, we can use a Laguerre--Gaussian pulse to pump the system near the flat band frequency to excite the compact localized condensates.
This gives us a possible application for Laguerre-Gaussian beam in exciton-polariton physics other than generating the vortex~\cite{Kwon:2019aa}.
We further showed that with incoherent homogeneous background pumping, the coherent compact localized condensates can be maintained longer than the lifetime of the polaritons.
This proposal permits one to construct graphs of compact localized states, similar to the proposals for classical~\cite{Berloff:2017aa} and quantum simulators~\cite{Liew:2018aa,Buluta:2009aa,Georgescu:2014aa}.
Later in this chapter, we considered a polariton system with a honeycomb lattice-shaped potential, compared to the early works~\cite{Bardyn:2015aa,Karzig:2015aa,Klembt:2018aa} we exhibited that it is possible to use a local magnetic field generated by magnetic materials embedded in the microcavity to open a gap in the vicinity of the Dirac point.
Between this gap, we observed one or two nontrivial topological states in the polariton system.
We further showed that the Chern number can be changed by tuning the magnitude of the magnetic field or the intensity of the TE-TM splitting similar to~\cite{Nalitov:2015aa,Bleu:2016aa}.

Following these concrete works regarding polariton condensation, there are still some fundamental, intriguing questions in this field.
One interesting topic is the formation and annihilation of vortices in polariton systems~\cite{Keeling:2008js,Lagoudakis:2008aa,Tosi:2011kv,Kwon:2019aa}.
It is known that the decay of the first-order spatial coherence, the recombination of thermally excited vortices, and the temperature dependence of a fractional condensate will be decisive experimental signatures indicating the crossover from BEC to the Berezinskii--Kosterlitz--Thouless phase transition~\cite{Kosterlitz_1972,Kosterlitz_1973,berezinsky1970destruction,berezinsky1972destruction}.
Because of the finite lifetime of polaritons, this crossover in the open driven-dissipative system still needs to be understood~\cite{Roumpos2012}.

Another interesting topic is the compact localized states in polariton systems~\cite{Huber:2010aa,Jacqmin:2014aa,Whittaker:2018ab,Klembt:2017aa}.
With the tight-binding model, the existence of compact localized states gives flat bands, which are a signal of macroscopic degeneracy and diverging density of states for the corresponding Hamiltonian~\cite{Leykam:2018aa,Leykam:2018ab}.
In polariton compact localized condensation, we have an opportunity to study the consequence of interaction in a perturbation-sensitive system~\cite{Kuno_2020,danieli2020manybody}.
Besides, due to the flatness of the dispersion, energy conservation is automatically protected in polariton--polariton interaction, which may be a good environment to enhance the interaction in polariton systems.

As we know, the polariton is a quasiparticle in the strong coupling regime.
Recently, there has been a growing interest to increase the coupling strength between matter and light, which is known as ultrastrong (or deep strong) coupling~\cite{Anappara:2009aa,Gunter:2009aa,Frisk-Kockum:2019aa,Forn-Diaz:2019aa}.
This novel regime reveals more quantum properties in polariton systems and opens a new topic in this field.

As our final work in this thesis, we looked at a different field than in previous chapters. Chapter~\ref{Ch6}  considered the transport of electrons coupled with two-dimensional Bose-condensed dipolar exciton gas by Coulomb interaction~\cite{Kovalev:2013aa,Batyev:2014aa,Boev:2018ab,Sun:2019aa,Villegas:2019aa}.
This Bose-condensed exciton gas can be easily transformed into a polariton BEC system due to the similarity of their interaction terms~\cite{Deng:2010aa,Kavokin:2007aa}.
We calculated resistivity by extending the Bloch--Gr\"{u}neisen approach and provided analytical formulas for both single-bogolon and double-bogolon scattering channels.
Compared to the normal phonon interaction channel, we found that two-bogolon scattering becomes the dominant mechanism in the system at a certain temperature range~\cite{Hwang:2008aa}.
This mechanism provides an opportunity to use bogolon pairs to generate Cooper pairs in the Bardeen--Cooper--Schrieffer phase~\cite{Laussy:2010aa}, which is a subject of future research.

%% file: appendix2_revised.tex
\chapter{Appendix: Multivalley engineering in semiconductor microcavities}

\section{Bloch theory for exciton-photon lattices}\label{AP:A1}

Here, we present details of the bare exciton-polariton dispersion calculation in the framework of the Bloch theory.
In the main text of the manuscript, we solve the eigenvalue problem of the system with Eq.~\eqref{eq:CH2_SP1}. This equation corresponds to the eigenvalue problem of the following matrix:
\begin{tiny}
	\begin{equation} 	\label{AP2_EqMatSM}
		\hspace*{-10mm}
		\begin{pmatrix}
			\tilde E_C( k+G)+\tilde{V}_{C}\left( 0 \right) & \tilde{V}_{C}\left( G \right) & \tilde{V}_{C}\left( 2G \right) & 0 & 0 & \Omega \\
			\tilde{V}_{C}\left( -G \right) & \tilde E_C(k) +\tilde{V}_{C}\left( 0 \right) & \tilde{V}_{C}\left( G \right) & 0 & \Omega & 0 \\
			\tilde{V}_{C}\left( -2G \right) & \tilde{V}_{C}\left( -G \right) & \tilde E_C(k-G) + \tilde{V}_{C}\left( 0 \right) & \Omega & 0 & 0 \\
			0 & 0 & \Omega & \tilde E_X(k-G) +\tilde{V}_{X}\left( 0 \right) & \tilde{V}_{X}\left( G \right) & \tilde{V}_{X}\left( 2G \right) \\
			0 & \Omega & 0 & \tilde{V}_{X}\left( -G \right) & \tilde E_X(k)+\tilde{V}_{X}\left( 0 \right) & \tilde{V}_{X}\left( G \right)\\
			\Omega & 0 & 0 & \tilde{V}_{X}\left( -2G \right) & \tilde{V}_{X}\left( -G \right) & \tilde E_X(k+G)+\tilde{V}_{X}\left( 0 \right)
		\end{pmatrix},
	\end{equation}
\end{tiny}
where ${\tilde{E}_C(q)}=\frac{\hbar^2q^2}{2m_C}-\frac{i\hbar}{\tau}$ and ${\tilde{E}_X(q)}=\frac{\hbar^2q^2}{2m_X}-\frac{i\hbar}{\lambda}$. Then, $\tilde{V}_C\left( 0,-G,G \right)$ are the 0th, --1st, and 1st order terms of the Fourier series for the periodic potential of the cavity photon, and $\tilde{V}_X\left( 0,-G,G \right)$ are the 0th, --1st, and 1st order terms of the Fourier series for the excitonic potential. 
The matrix in Eq.~\eqref{AP2_EqMatSM} is reduced to the summation of terms in Eq.~\eqref{eq:CH2_SP1} from $-G$ to $G$. In our real calculations, we did the summation over the terms from $-150\cdot G$ to $150\cdot G$.


\section{Simplified model of equilibrium polariton condensation}\label{AP:A2}

Following Eqs.~\eqref{eq:CH2_H-2} and~\eqref{eq:CH2_E-2}, for the case of a fixed number of particles and low temperature, we can write the probability of occupation of different modes, referred to as the probability distribution function (PDF), as
\begin{equation}
	p\left( n_1, n_2 \right)=\frac{1}{\mathcal{Z}}e^{-E\left( n_1,n_2 \right)/k_B T},
	\label{AP2_ProD}
\end{equation}
where $\mathcal{Z}=\sum_{n_1,n_2}e^{-E\left( n_1,n_2 \right)/k_B T}$ is the partition function, $T$ is the temperature, and $k_B$ is Boltzmann's constant. The second-order correlation function then reads
\begin{equation}
	g_{12}^{(2)} =\frac{\langle \hat{a}_1^\dag \hat{a}_2^\dag \hat{a}_1 \hat{a}_2 \rangle}{\langle \hat{a}_1^\dag \hat{a}_1 \rangle \langle \hat{a}_2^\dag \hat{a}_2 \rangle} = \frac{\langle n_1 n_2 \rangle}{\langle n_1 \rangle \langle n_2 \rangle},
	\label{AP2_G_2}
\end{equation}

where $\langle n_1\rangle$, $\langle n_2\rangle$, and $\langle n_1n_2\rangle$ can be calculated from the PDF $p(n_1,n_2)$, see Fig.~\ref{fig:CH2_TMFig}. 
At zero temperature, $g_{12}^{(2)}=0$, confirming our earlier arguments on the choice of the state required for energy minimisation. 
With increasing temperature, $g_{12}^{(2)}$ rises as the system may be excited out of the ground state.

Allowing for the population of many modes in reciprocal space (instead of the two that we considered), the probability of occupation of any quantum state can be found by a straightforward generalisation of Eq.~\eqref{AP2_ProD}; 
in principle, the full exciton-polariton intensity distribution can then be obtained by summing over the PDF. However, in practice, the size of the Hilbert space grows exponentially with the number of particles in the system, and therefore it becomes possible to use our simple treatment to evaluate the equilibrium photoluminescence spectrum in the low-density regime with only a few particles in the system (see inset in Fig.~\ref{fig:CH2_TMFig}). 
The spectrum here was phenomenologically broadened in both energy and wave vector, accounting for the finite lifetime of the polaritons and the finite size of a typical condensate, respectively.

\section{Nonequilibrium model of polariton condensation}\label{AP:A4}
Now we present details on the nonequilibrium model.
Interaction with the reservoir of acoustic phonons in a semiconductor crystal lattice is described by the microscopic Fr\"ohlich Hamiltonian~\cite{Tassone:1997aa}:
\begin{eqnarray}
	\hat{\mathcal{H}}_\textrm{int}=\sum_{\mathbf{q},k}G_{\mathbf{q}}\hat{b}_{\mathbf{q}}\hat{a}^\dagger_{k+q_x}
	\hat{a}_k+G_{\mathbf{q}}^*\hat{b}^\dagger_{\mathbf{q}}\hat{a}_{k+q_x}\hat{a}^\dagger_k,
	\label{AP2_neq}
\end{eqnarray}

where parameters $G_{\mathbf{q}}$ are the exciton--phonon interaction strengths evaluated elsewhere~\cite{Hartwell:2010aa}.
The phonon wavevector here is
$\mathbf{q}=\mathbf{e}_xq_x+\mathbf{e}_yq_x+\mathbf{e}_zq_z$, where $\mathbf{e}_x$,
$\mathbf{e}_y$, and $\mathbf{e}_z$ are unit vectors: $\mathbf{e}_x$ is in the direction of the 1D polariton system, $\mathbf{e}_z$ is in the structure growth direction, and $\mathbf{e}_y$ is perpendicular to both.
The phonon dispersion relation, $\hbar\omega_{\mathbf{q}}=\hbar
c_s\sqrt{q_x^2+q_y^2+q_z^2}$, is determined by the sound velocity, $c_s$.

The equations of motion for the polariton macroscopic wavefunction, $\psi$, and the exciton reservoir occupation number, $n_\textrm{R}$, read~\cite{Savenko:2013aa,Wouters:2007aa}:
\begin{eqnarray}
	\label{AP2_EqGPEnR}
	&&i\hbar\frac{\partial\psi(x,t)}{\partial t}={\cal F}^{-1}\left[E_k\psi_k+{\cal S}_k(t)\right] +\frac{i\hbar}{2}\left[Rn_\textrm{R}(x,t)-\gamma_0-\frac{2i}{\hbar}\alpha|\psi({x},t)|^2\right]\psi(x,t)\nonumber \\
	&&~~~~~~~~~~~~~~~~~~  +\sum_k\left[{\cal T}_{-k}(t)+{\cal T}^*_k(t)\right]e^{-ikx}\psi(x,t);\\
	\label{AP2_EqGPEnR2}
	&&\frac{\partial n_\textrm{R}(x,t)}{\partial t}=-(\gamma_\textrm{R}+R|\psi(x,t)|^2)n_\textrm{R}+P,
\end{eqnarray}
where ${\cal F}^{-1}$ stands for the inverse Fourier transform, $E_k$ is the free polariton dispersion, $\psi_k$ is the Fourier image of the macroscopic wavefunction, $P$ and $\gamma_\textrm{R}$ are the incoherent reservoir homogeneous pumping intensity and inverse lifetime of the reservoir, respectively, and $R$ is the system-reservoir excitation exchange rate.
The term ${\cal S}_k(t)$ corresponds to the emission of phonons by a condensate stimulated by polariton concentration.

The stochastic term ${\cal T}_{q_x}$ in the last line of Eq.~\eqref{AP2_EqGPEnR} is defined by the correlations:
\begin{align}
	\left<{\cal T}_{q_x}^*(t){\cal T}_{q_x^\prime}(t^\prime)\right>&=\sum_{q_y,q_y}\left|G_{{q_x,q_y,q_z}}\right|^2n_{q_x,q_y,q_z}\delta_{q_x,q_x^\prime}\delta(t-t^\prime);\notag\\
	\left<{\cal T}_{q_x}(t){\cal T}_{q_x^\prime}(t^\prime)\right>&=
	\left<{\cal T}_{q_x}^*(t){\cal T}_{q_x^\prime}^*(t^\prime)\right>=0,
	\label{AP2_EqThermal}
\end{align}
where $n_{\mathbf{q}}$ is the temperature-dependent density of phonons in the state with wavevector $\mathbf{q}$.

Solving Eq.~\eqref{AP2_EqGPEnR} numerically and averaging over different stochastic realizations of the phonon field, we obtained the result shown in Fig.~\ref{fig:CH2_Fig4} of the main text.


\section{Energy Band Structure}\label{AP:A5}
Here we discuss the energy band structure of the exciton-polariton lattice [Fig.~\ref{fig:CH2_Fig1}(a)] as well as an alternative configuration.
\begin{figure}[ht]
	\centering
	\includegraphics[width=0.45\linewidth]{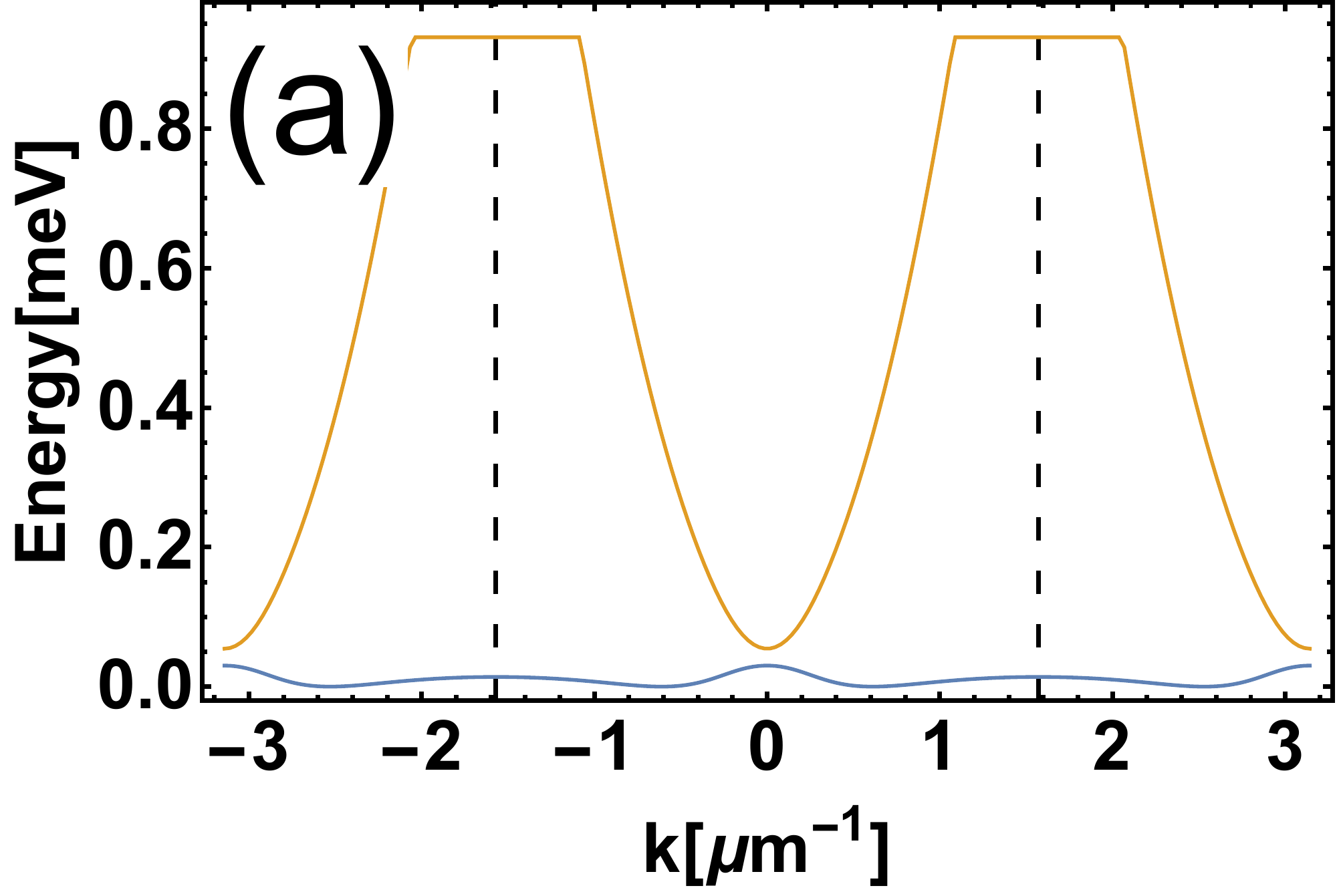}
	\includegraphics[width=0.45\linewidth]{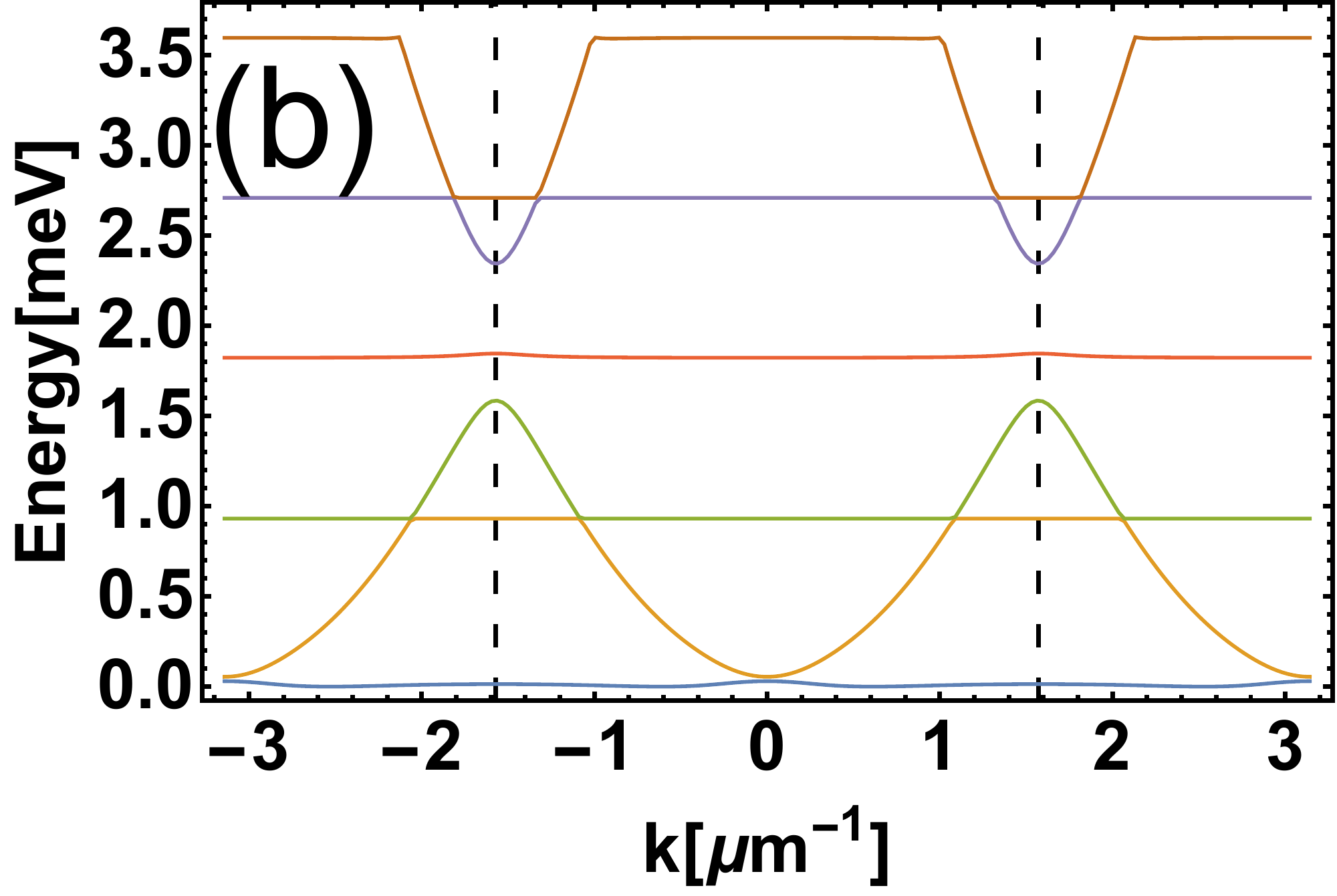}
	\caption[Band spectrum]{Energy bands in (a) level 1 and level 2, and from (b) level 1 to level 6.  Picture from \cite{Sun:2017ab}.}
	\label{fig:AP2_s-1}
\end{figure}
\begin{figure}[ht]
	\centering
	\includegraphics[width=0.45\linewidth]{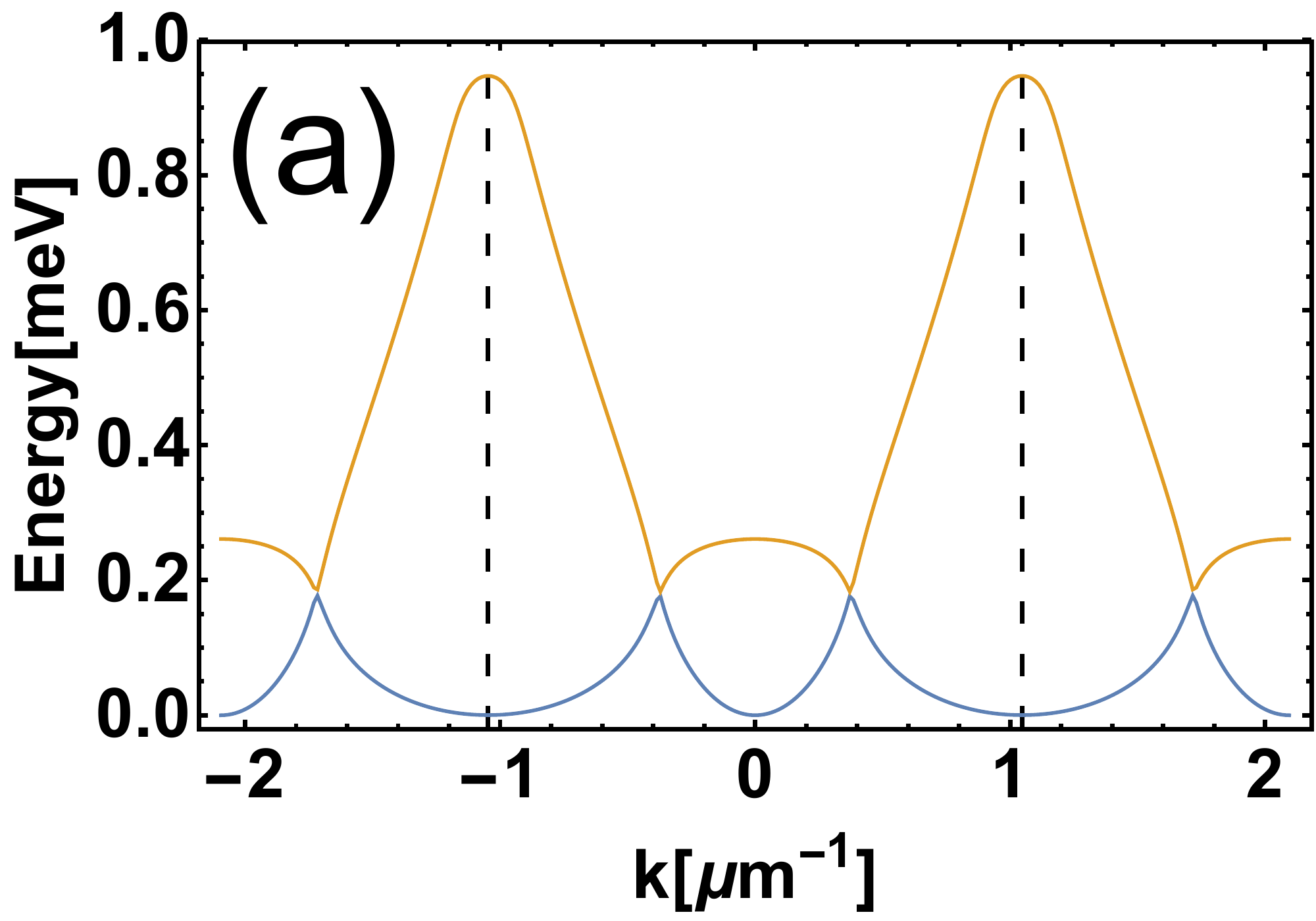}
	\includegraphics[width=0.45\linewidth]{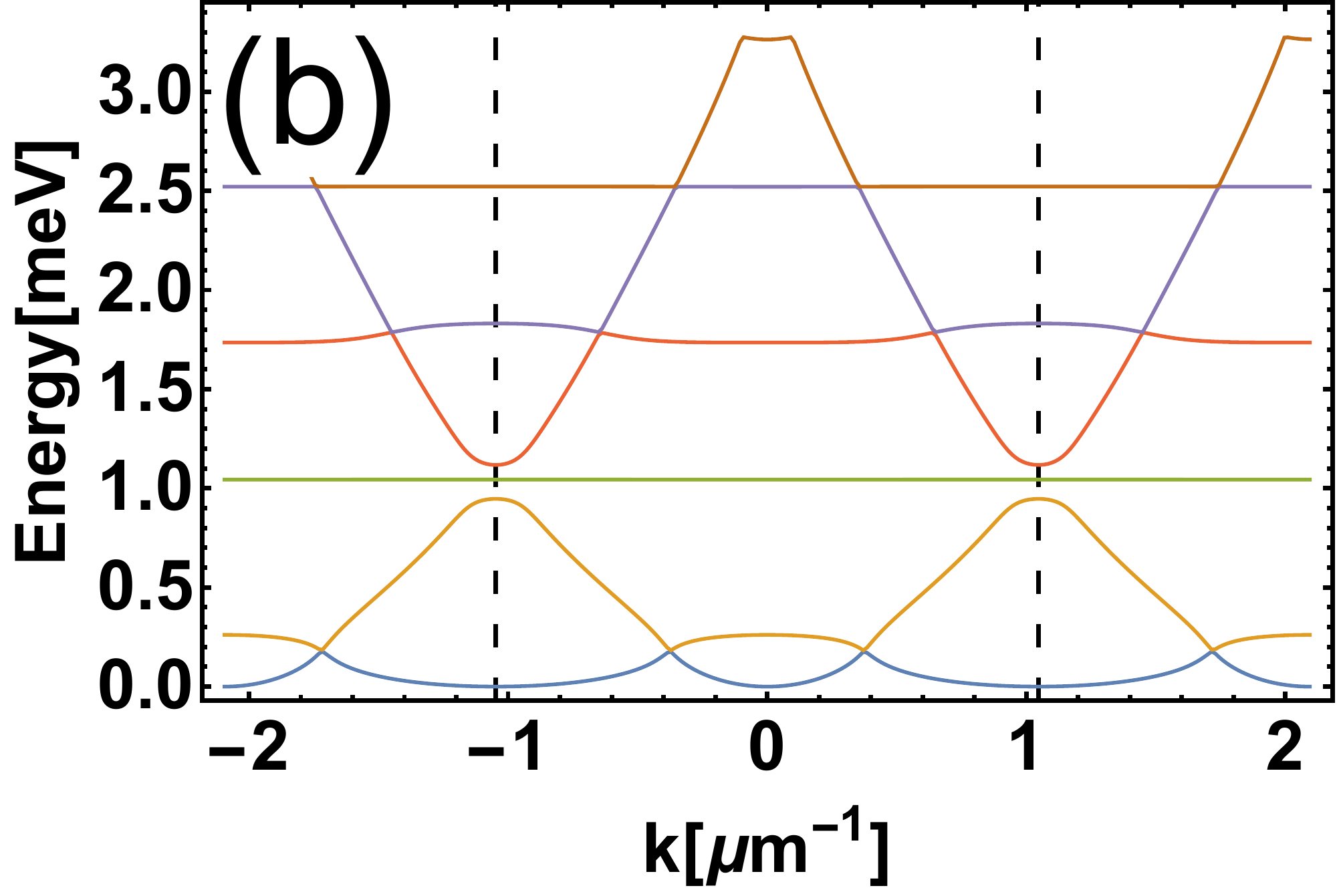}
	\caption[Band spectrum in the alternative configuration ]{Energy bands in the case where the minima in $k$-space are at the $\Gamma$ point and the edge of the Brillouin zone, for (a) level 1 and level 2 bands, and (b) level 1 to level 6 bands. The figure is taken from~\cite{Sun:2017ab}.}
	\label{fig:AP2_s-2}
\end{figure}
The parameters for these plots are as follows. Lattice period: $T=2.0$ $\mu$m; potential profile for the cavity photons: sine function $-1.1\sim0.25$ meV; potential profile for the excitons: sine function $-0.95 \sim 460.29$ meV; exciton-photon coupling constant: $\Omega=0.7$ meV; decay rate of the cavity photons: $\gamma =0.42$ meV; decay rate of the excitons: $0.04 $ meV; effective mass of a cavity photon: $5*10^{-5} m_e$; effective mass of an exciton: $0.22 m_e$, where $m_e$ is the free electron mass.

Let us also change the parameters of the potential energies of the excitons and photons to see which dispersion we can achieve in $k$-space.
\begin{figure}[ht]
	\centering
	\includegraphics[width=0.555555\linewidth]{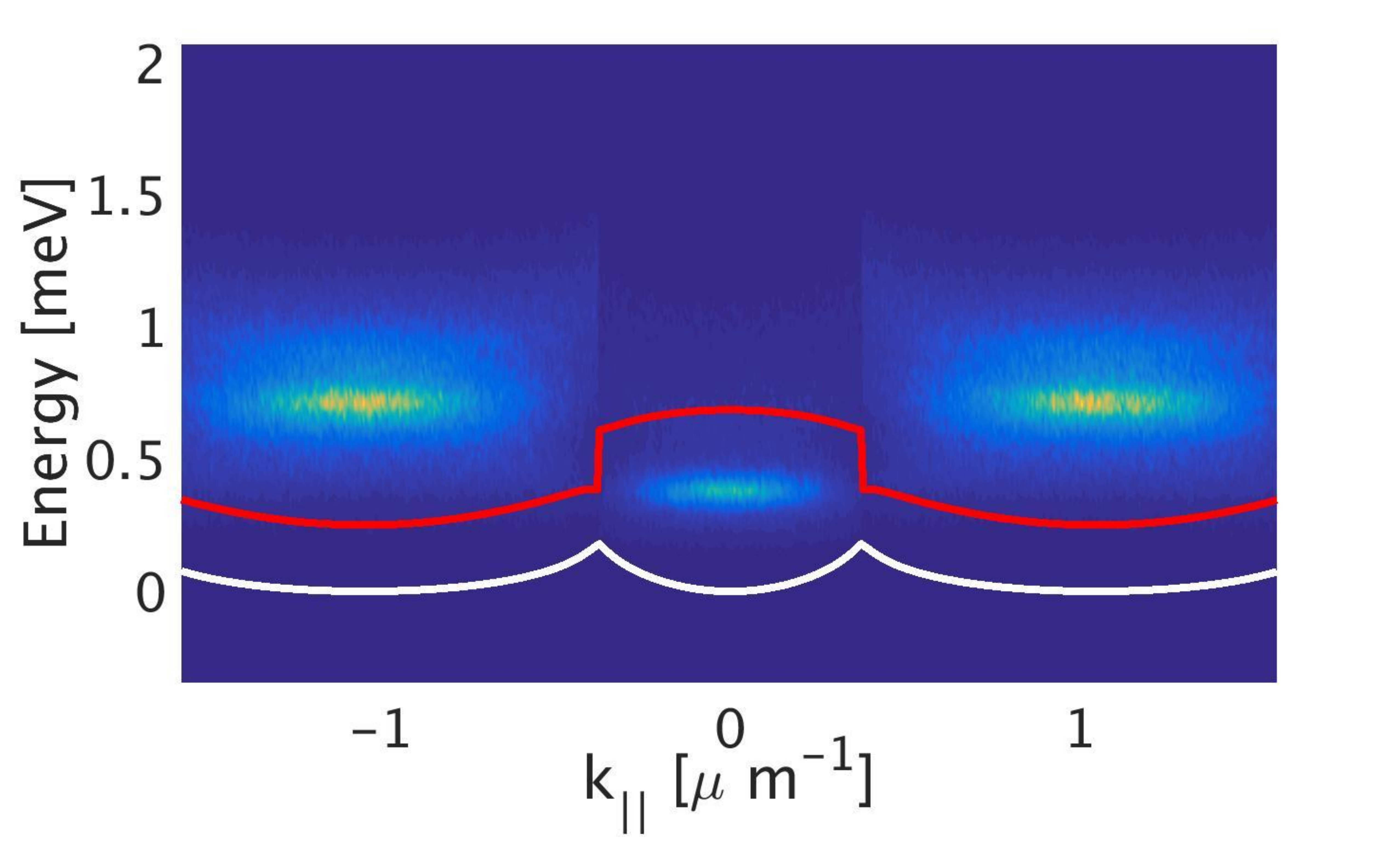}
	\caption[Condensation in the alternative configuration]{Condensate result from the energy dispersion in Fig.~\ref{fig:AP2_s-2}. The red and white lines represent the decay rate and the energy dispersion of the polariton, respectively. The figure is taken from~\cite{Sun:2017ab}.}
	\label{fig:AP2_s-3}
\end{figure}
By taking the following parameters, we yield the dispersion presented in Figs.~\ref{fig:AP2_s-2} and~\ref{fig:AP2_s-3}. Lattice period: $T=3.0$ $\mu$ m; potential profile for the cavity photons: square function $-0.25\sim0$ meV; potential profile for the excitons: sine function $-0.389 \sim 697.95$ meV; exciton-photon coupling constant: $\Omega=2.47$ meV; decay rate of the cavity photons: $\gamma =1$ meV; decay rate of the excitons: $0$ meV, effective mass of a cavity photon: $5*10^{-5} m_e$; effective mass of an exciton: $0.22 m_e$.

Figure~\ref{fig:AP2_s-3} is the result of the nonequilibrium model. Since the decay rate of the exciton-polaritons varies significantly with momentum $k$, it becomes easier for exciton-polaritons to condense at $k=\pm k_{BZ}$ rather than at $k=0$. As a result, the blueshifts differ at $k=0$ and $k=\pm k_{BZ}$. The minima in $k$-space become inequivalent in properties, making it difficult to consider entanglement between them. It should be noted, though, that the two minima at $k=\pm k_0$ in Fig.~\ref{fig:CH2_Fig4}(b) are equivalent in properties.

\begin{figure}[!tb]
\centering
\includegraphics[height=0.4\linewidth]{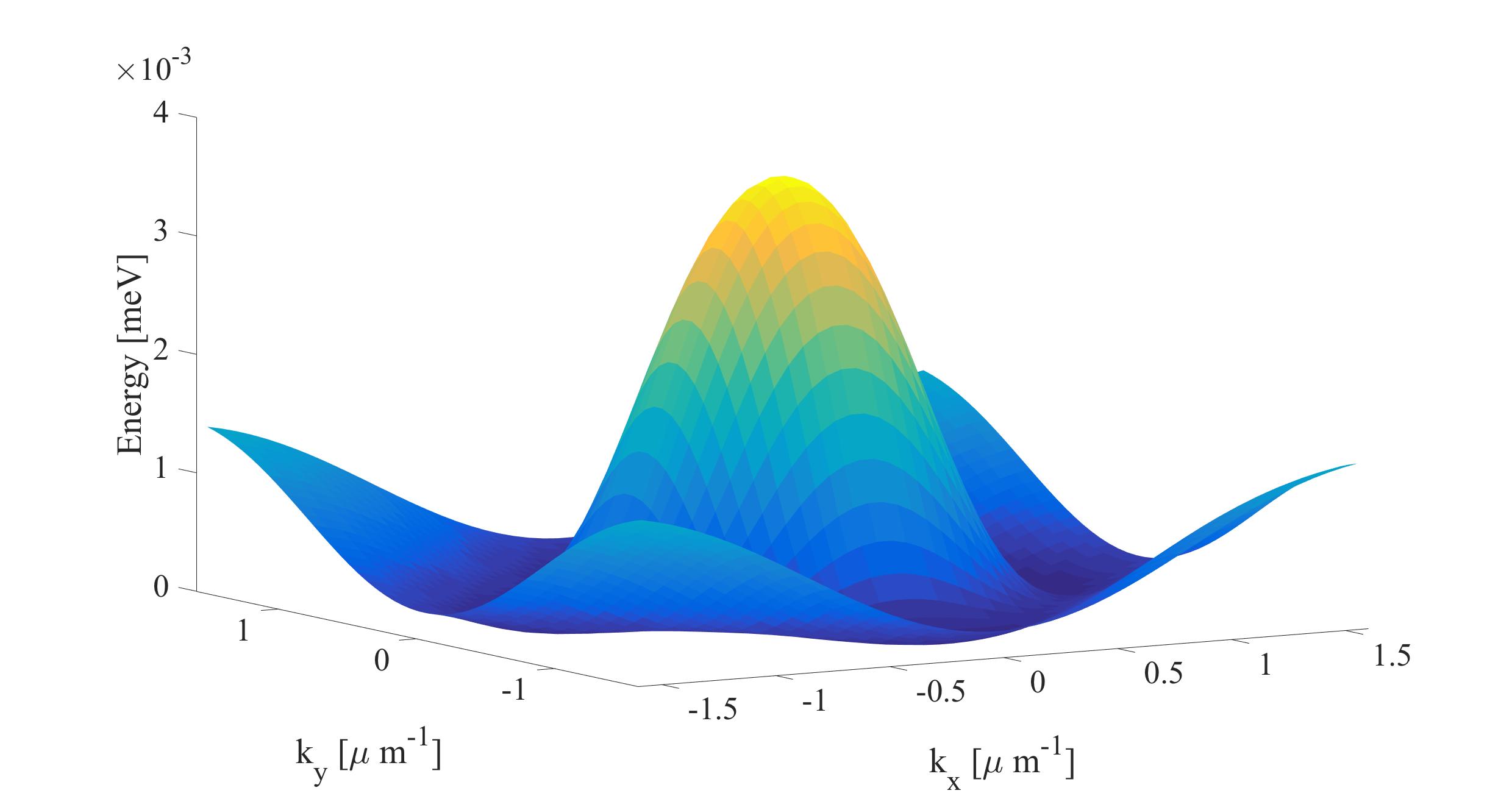}
\caption[Ground state dispersion in 3D]{Energy dispersion in the 2D lattice. The figure is taken from~\cite{Sun:2017ab}.}
\label{fig:AP2_s-4}
\end{figure}

Figure~\ref{fig:AP2_s-4} is a 3D plot of the energy dispersion in the 2D lattice corresponding to Fig.~\ref{fig:CH2_Fig2D}(a).

%% file: appendix6_revised.tex
\chapter{Appendix: Bogolon-mediated electron scattering in graphene}\label{AP:CH6}
\section{Derivation of resistivity via Bloch--Gr\"uneisen approach }
In this Appendix, we derive the low-temperature $T$-dependence of the conductivity of graphene in the hybrid Bose--Fermi system using the Bloch--Gr\"uneisen approach.
We start from the Boltzmann equation,
\begin{equation}\label{AP6_A1}
e_0\textbf{E}\cdot\frac{\partial f}{\hbar\partial \textbf{p}}=I\{f\},
\end{equation}
where $\mathbf{p}$ is the wave vector and we set $p\equiv\abs{\mathbf{p}}$, $\mathbf{E}$ is the perturbing electric field, and $f$ is the distribution function.
The scattering integral is given by
\begin{eqnarray}
\label{AP6_A2}
I\{f\}&=&-\frac{1}{\hbar}\int\frac{d\textbf{q}d\textbf{p}'}{(2\pi)^2}|M_q|^2\Bigl[N_qf_p(1-f_{p'})\delta(\varepsilon_p-\varepsilon_{p'}+\hbar\omega_q)\delta(\textbf{p}-\textbf{p}'+\textbf{q})\nonumber\\
&{}&{}+(N_q+1)f_p(1-f_{p'})\delta(\varepsilon_p-\varepsilon_{p'}-\hbar\omega_q)\delta(\textbf{p}-\textbf{p}'-\textbf{q})\nonumber\\
&{}&{}+N_qf_{p'}(1-f_{p})\delta(\varepsilon_{p'}-\varepsilon_{p}+\hbar\omega_q)\delta(\textbf{p}'-\textbf{p}+\textbf{q})\nonumber\\
&{}&{}+(N_q+1)f_{p'}(1-f_p)\delta(\varepsilon_{p'}-\varepsilon_{p}-\hbar\omega_q)\delta(\textbf{p}'-\textbf{p}-\textbf{q})\Bigr].
\end{eqnarray}
Note that in writing this integral we set the length of the sample $L$ equal to one. In performing a dimensionality analysis, one should include the length squared such that $I\{f\}$ has a dimension of inverse time, as it should.

For small enough electric fields, the electron distribution is not substantially different from the equilibrium Fermi distribution, and thus it can be presented in the form
\begin{eqnarray}
f=f^0(\varepsilon_p)-\left(-\frac{\partial f^0}{\partial\varepsilon_p}\right)f^{(1)}_\textbf{p},
\end{eqnarray}
where $f^0$ is the equilibrium Fermi--Dirac distribution and $f^{(1)}_\textbf{p}$ has a dimensionality of energy.
Following the steps of the derivation reported in~\cite{Zaitsev:2014aa}, we rewrite:
\begin{eqnarray}
\label{AP6_A3}
e_0\textbf{E}\cdot\frac{\partial f}{\hbar\partial \textbf{p}}&=&v_F\frac{e_0\textbf{E}\cdot\textbf{p}}{|\mathbf{p}|}\frac{\partial f^0}{\partial \varepsilon_p}=I\{f^{(1)}_\textbf{p}\};\\\nonumber
I\{f^{(1)}_\textbf{p}\}&=&-\frac{1}{\hbar}\int\frac{d\textbf{q}d\textbf{p}'}{(2\pi)^2}|M_q|^2\frac{1}{\hbar}\frac{\partial N_q}{\partial \omega_q}
\left(f^0(\varepsilon_p)-f^0(\varepsilon_{p'})\right)\\\nonumber
&{}&{}\times\left(f^{(1)}_{\textbf{p}}-f^{(1)}_{\textbf{p}'}\right)
\Bigl[\delta(\varepsilon_{p}-\varepsilon_{p'}-\hbar\omega_q)\delta(\textbf{p}-\textbf{p}'-\textbf{q})\nonumber\\
&{}&{}-\delta(\varepsilon_{p}-\varepsilon_{p'}+\hbar\omega_q)\delta(\textbf{p}-\textbf{p}'+\textbf{q})\Bigr],
\end{eqnarray}
where $$\frac{\partial N_q}{\partial \omega_q}=-\frac{\hbar}{k_BT}N_q(1+N_q).$$
By further integrating over the electron wave vector $\textbf{p}'$, we find
\begin{eqnarray}
\label{AP6_A4}
v_F\frac{e_0\textbf{E}\cdot\textbf{p}}{p}\frac{\partial f^0}{\partial \varepsilon_p}
=&-&\frac{1}{\hbar}\int\frac{d\textbf{q}}{(2\pi)^2}|M_q|^2\frac{1}{\hbar}\frac{\partial N_q}{\partial \omega_q}
\left(f^0(\varepsilon_p)-f^0(\varepsilon_{p}-\hbar\omega_q)\right)\\ \nonumber
&\times&\left(f^{(1)}_{\textbf{p}}-f^{(1)}_{\textbf{p}-\textbf{q}}\right)
\delta(\varepsilon_{\textbf{p}}-\varepsilon_{\textbf{p}-\textbf{q}}-\hbar\omega_\textbf{q})\\
\nonumber
&+&\frac{1}{\hbar}\int\frac{d\textbf{q}}{(2\pi)^2}|M_q|^2\frac{1}{\hbar}\frac{\partial N_q}{\partial \omega_q}
\left(f^0(\varepsilon_p)-f^0(\varepsilon_{p}+\hbar\omega_q)\right)\\ \nonumber
&\times&\left(f^{(1)}_{\textbf{p}}-f^{(1)}_{\textbf{p}+\textbf{q}}\right)
\delta(\varepsilon_{\textbf{p}}-\varepsilon_{\textbf{p}+\textbf{q}}+\hbar\omega_\textbf{q}).
\end{eqnarray}
Let the electric field be directed along the $x$-axis; then, we can use the correction function in the form
\begin{eqnarray}
f^{(1)}_\textbf{p}= v_F\frac{e_0E_xp_x}{k_F}\tau(\varepsilon_p),
\end{eqnarray}
where $p_F$ is the Fermi wave vector and $\tau(\varepsilon_p)$ is the relaxation time.  
We now have
\begin{eqnarray}\label{AP6_A5}
\frac{p_x}{p}\frac{\partial f^0}{\partial \varepsilon_p}
=&-&\frac{1}{\hbar}\int\frac{d\textbf{q}}{(2\pi)^2}|M_q|^2\frac{1}{\hbar}\frac{\partial N_q}{\partial \omega_q}
\left[f^0(\varepsilon_p)-f^0(\varepsilon_{p}-\hbar\omega_q)\right] \nonumber \\ 
&\times&\left[\frac{p_x}{k_F}\tau(\varepsilon_p)-\frac{p_x-q_x}{k_F}\tau(\varepsilon_p-\hbar\omega_q)\right]
\delta(\varepsilon_{\textbf{p}}-\varepsilon_{\textbf{p}-\textbf{q}}-\hbar\omega_\textbf{q})\nonumber\\
&+&\frac{1}{\hbar}\int\frac{d\textbf{q}}{(2\pi)^2}|M_q|^2\frac{1}{\hbar}\frac{\partial N_q}{\partial \omega_q}
\left[f^0(\varepsilon_p)-f^0(\varepsilon_{p}+\hbar\omega_q)\right] \nonumber \\ 
&\times&\left[\frac{p_x}{k_F}\tau(\varepsilon_p)-\frac{p_x+q_x}{k_F}\tau(\varepsilon_p+\hbar\omega_q)\right]
\delta(\varepsilon_{\textbf{p}}-\varepsilon_{\textbf{p}+\textbf{q}}+\hbar\omega_\textbf{q}).
\end{eqnarray}
Assuming that the relaxation time is constant~\cite{Ziman:2001aa} at $\tau=\tau_0$, we find
\begin{eqnarray}
\label{AP6_ABoltzmann1}
\frac{k_F}{p}p_x\frac{\partial f^0}{\partial \varepsilon_p}
=&-&\frac{\tau_0}{\hbar}\int \frac{d\textbf{q}}{(2\pi)^2}q_x|M_q|^2\frac{1}{\hbar}\frac{\partial N_q}{\partial \omega_q}\\ \nonumber
&\times&\left[f^0(\varepsilon_p)-f^0(\varepsilon_{p}-\hbar\omega_q)\right]
\delta(\varepsilon_{\textbf{p}}-\varepsilon_{\textbf{p}-\textbf{q}}-\hbar\omega_\textbf{q})\\
\nonumber
&+&\frac{\tau_0}{\hbar}\int \frac{d\textbf{q}}{(2\pi)^2}q_x|M_q|^2\frac{1}{\hbar}\frac{\partial N_q}{\partial \omega_q}\\ \nonumber
&\times&\left[f^0(\varepsilon_p)-f^0(\varepsilon_{p}+\hbar\omega_q)\right]
\delta(\varepsilon_{\textbf{p}}-\varepsilon_{\textbf{p}+\textbf{q}}+\hbar\omega_\textbf{q}).
\end{eqnarray}

Let us denote the angle between vectors $\textbf{p}$ and $\textbf{q}$ as $\varphi$, and the angle between vectors $\textbf{p}$ and $\textbf{E}$ as $\beta$. This gives $q_x=q\cos(\varphi+\beta)$ and $p_x=p\cos\beta$. Integrating over $\phi$, we obtain
\begin{eqnarray}
\label{AP6_AEq19}
\int_0^{2\pi} &d\phi&\cos (\phi+\beta)\delta(a-\sqrt{b^2\pm c^2\cos\phi})\\\nonumber
&=&\pm\frac{4|a|(b^2-a^2)\Theta[c^4-(b^2-a^2)^2])}{c^2\sqrt{c^4-(b^2-a^2)^2}}\cos\beta,
\end{eqnarray}
where $\Theta[x]$ is the Heaviside step function, $a=\hbar(v_Fp-sq)$, $b^2=\hbar^2v_F^2(p^2+q^2)$, and $c^2=2\hbar^2 v_F^2pq$. 
To derive Eq.~\eqref{AP6_AEq19}, we denote a new variable $x=\cos\phi$. This implies $d\phi=\mp dx[1-x^2]^{-1/2}$, where the $-$($+$) case is for $0\leq\phi <\pi$ ($\pi\leq\phi<2\pi$).
After integrating over the angle $\phi$, we can integrate Eq.~\eqref{AP6_ABoltzmann1} over $\xi_p=\varepsilon_p-\mu$ using
\begin{eqnarray}
\int\limits_{-\infty}^{\infty}d\xi_p\left(f^0(\varepsilon_p)-f^0(\varepsilon_{p}\pm\hbar\omega_q)\right)&=&\mp\hbar\omega_q,\nonumber \\
\int\limits_{-\infty}^{\infty}d\xi_p\frac{\partial f^0}{\partial \varepsilon_p}&=&-1,
\end{eqnarray}
and putting all electron wave vectors to be $p=k_F$.

Resistivity is inversely proportional to scattering time by
\begin{eqnarray}
\label{AP6_Arho1}
\rho\propto\frac{1}{\tau_0}&=&\frac{\hbar\xi_I^2}{8\pi^2k_FM}\frac{1}{kT}\int_0^\infty dq q^4e^{-2ql}q\nonumber\\
&\times&(\Gamma_--\Gamma_+)_{k_F}N_q(1+N_q),
\end{eqnarray}
where we introduce $\xi_I= e_0^2d\sqrt{n_c}/2\epsilon$ and
\begin{eqnarray}
\label{AP6_Agamma}
\Gamma_{\pm}=\frac{4|a_{\pm}|(a^2_{\pm}-b^2)\Theta[c^4-(a^2_{\pm}-b^2)^2]}{c^2\sqrt{c^4-(a^2_\pm-b^2)^2}}.
\end{eqnarray}
The subscript $k_F$ in the expression $(\Gamma_--\Gamma_+)_{k_F}$ in Eq.~\eqref{AP6_Arho1} means that all the electron wave vectors $p$ are to be substituted by the Fermi value $k_F$.

We now introduce a new dimensionless variable,
\begin{eqnarray}
\label{AP6_Dim}
u=\frac{\hbar sq}{k_BT},
\end{eqnarray}
in Eq.~\eqref{AP6_Arho1} and obtain
\begin{eqnarray}
\label{AP6_Arho2}
\frac{1}{\tau_0}&=&\frac{\xi_I^2}{8\pi^2k_FMs}\left(\frac{k_BT}{\hbar s}\right)^4\int_0^\infty du \frac{u^4e^{(1-2\tilde{l})u}}{(e^u-1)^2}\nonumber\\
&\times&(\Gamma_--\Gamma_+)_{k_F},
\end{eqnarray}
where
\begin{eqnarray}
\Tilde{l}=\frac{lk_BT}{\hbar s}\sim\frac{k_BT}{10\mbox{ meV}}.
\end{eqnarray}
We set $s=10^5$ m/s and $l= 5.0\times 10^{-8}$ m/s. 
Note that room temperature is $k_BT_R\sim 26$ meV, so that for far lower temperatures we have $\Tilde{l}\ll 1$. Hence, we can replace
\begin{eqnarray}
\label{AP6_approx1}
e^{(1-2\tilde{l})u}\rightarrow e^u.
\end{eqnarray}

Let us now look at the Heaviside theta function argument in Eq.~\eqref{AP6_Agamma}. It can be simplified to
\begin{eqnarray}
-(v_F^2-s^2)q^2\pm 4k_Fsv_Fq+4k_F^2v_F^2.
\end{eqnarray}
This expression is positive for
\begin{eqnarray}
0\leq q<\frac{2k_Fv_F}{v_F\mp s}\approx 2k_F,
\end{eqnarray}
or
\begin{eqnarray}
\label{AP6_Alambda}
0\leq u<\frac{T_{BG}}{T}\equiv\Lambda,
\end{eqnarray}
where $T_{BG}=2\hbar sk_F/k_B$ is the Bloch--Gr\"{u}neisen temperature for bogolons.

Let us consider the case in which temperature $T\ll T_{BG}$, which specifically means $\Lambda >10$. This inequality gives us the precise form of what we mean by \textit{low temperature}: $k_BT/E_F<10^{-2}$. For typical $E_F\sim 10^{-1}$ eV, this gives $T_{BG}=183$ K and  $T<18$ K (for a particular distance between the layers of $l>50$ nm). 
It should be mentioned that the condition $\tilde{l}\ll 1$ is not a requirement. However, the general case does not allow for the analytical extraction of temperature out of the integral, since we come up with the term $\sim\exp[lk_BTu/(\hbar s)]$ under the integration.

For large $u$ (or $q$), the factors $\Gamma_\pm$ in Eq.~\eqref{AP6_Arho2} approach constant values. In the meantime, the term $u^4\exp({-u})$ rapidly goes to zero for $u>10$.
Therefore, we can remove the theta function in Eq.~\eqref{AP6_Agamma} and only incur a small (imaginary) error. Doing this and also using $v_F^2-s^2\sim v_F^2$, Eq.~\eqref{AP6_Agamma} now becomes
\begin{eqnarray}
\label{AP6_Agamma2}
\Gamma_\pm=\frac{2|v_Fk_F\pm sq|(2v_Fsk_F-v_F^2q)}{\hbar v_F^3k_F q\sqrt{\pm 4k_Fsv_Fq+4k_F^2v_F^2-v_F^2q^2}},
\end{eqnarray}
where the expression in the numerator can be rewritten as
\begin{eqnarray}
2v_Fsk_F-v_F^2q&=&2v_Fsk_F-v_F^2\frac{k_BT}{\hbar s}u\nonumber\\
&=&\frac{v_Fk_BT}{\hbar s}(s\Lambda-v_Fu).
\end{eqnarray}
Here, $\Lambda$ does depend on $T$, as was defined in Eq.~\eqref{AP6_Alambda}. For $T\ll T_{BG}$ though, due to the factor $\exp({-u})$, we can simply replace $\Lambda\sim 10$ (or greater) without significantly affecting the result. 
We introduce the ratio of velocities $\alpha=s/v_F$ and for simplicity, we choose $\Lambda$ such that $\alpha\Lambda=1$ as long as $\Lambda>10$, to then get
\begin{eqnarray}
2v_Fsk_F-v_F^2q
&=&\frac{v_Fk_BT}{\alpha\hbar }(1-u).
\end{eqnarray}
We find
\begin{eqnarray}
\frac{1}{\tau_0}&=&\frac{10v_F\xi_I^2}{8\pi^2k_FMs}\left(\frac{k_BT}{\hbar s}\right)^5\nonumber\\
&\times&\int_0^\infty du \frac{u^4e^u(1-u)}{(e^u-1)^2}(\gamma_--\gamma_+)_{k_F},
\end{eqnarray}
where
\begin{eqnarray}
\label{AP6_AEqMath}
\gamma_\pm=\frac{2|v_Fk_F\pm sq|}{\hbar v_F^3k_F q\sqrt{\pm 4k_Fsv_Fq+4k_F^2v_F^2-v_F^2q^2}}.
\end{eqnarray}
The term in the absolute value can be rewritten as
\begin{eqnarray}
|v_Fk_F\pm sq|=\frac{k_BT}{\hbar}\left|\frac{\Lambda}{2\alpha}\pm u\right|.
\end{eqnarray}

Finally, the term under the square root in Eq.~\eqref{AP6_AEqMath} can be written as 
\begin{eqnarray}
\nonumber
\pm 4k_Fsv_Fq&+&4k_F^2v_F^2-v_F^2q^2\nonumber\\
&\sim&-v_F^2(q-2k_F)(q+2k_F)\nonumber\\
&=&-\frac{v_F^2k_B^2T^2}{\hbar^2s^2}(u-\Lambda)(u+\Lambda).
\end{eqnarray}

Summing up, we find
\begin{eqnarray}
\frac{1}{\tau_0}&=&\frac{I_0\xi_I^2k_F^2}{4\pi^2\hbar \alpha^4v_F^2M}\left(\frac{k_BT}{E_F}\right)^4\nonumber\\
&=&\frac{5I_0e_0^2}{8\pi^2\hbar^3 v_F}\frac{M}{n_cE_F^2}(k_BT)^4,
\end{eqnarray}
where $I_0$ is a dimensionless integral, which can be found numerically by
\begin{eqnarray}
I_0&=&\int_0^\infty du\frac{u^3(1-u)e^u}{(e^u-1)^2\sqrt{100-u^2}}\nonumber\\
&\times&\left(\left|\frac{\Lambda}{2\alpha}-u\right|-\left|\frac{\Lambda}{2\alpha}+u\right|\right)
\approx 26.2.
\end{eqnarray}
This gives
\begin{eqnarray}
\frac{1}{\tau_0}=(1.4\times 10^{16}\mbox{s}^{-1})\left(\frac{k_BT}{E_F}\right)^4.
\end{eqnarray}
In terms of the resistivity,
\begin{eqnarray}
\label{AP6_Arho3}
\rho=\frac{\pi\hbar^2}{e_0^2E_F}\frac{1}{\tau_0}=(1.0\times 10^6\;\Omega)\left(\frac{k_BT}{E_F}\right)^4.
\end{eqnarray}
Thus, we conclude that at low temperatures $T\ll T_{BG}$, the resistivity is $\propto T^4$.

%% file: appendix7_revised2.tex
\chapter{Appendix: Bogolon-mediated electron scattering in 2DEG}\label{AP:CH7}
In this Appendix, we derive the low- and high-temperature $T$-dependencies of the resistivity of a 2D electron gas (2DEG) with parabolic dispersion in a hybrid Bose--Fermi system by extending the Bloch--Gr\"uneisen approach.

%
%
\section[Bloch--Gr\"uneisen~formula~for single-bogolon processes]{Bloch-Gr\"uneisen~formula~for single-bogolon\\ processes}
Let $f(\mathbf{r},\mathbf{p},t)$ be the electron single-particle distribution function, which depends on coordinate $\mathbf{r}$, particle wave vector $\mathbf{p}$, and time $t$. Its full derivative over time reads
\begin{eqnarray}
\frac{df}{dt}=\frac{\partial f}{\partial t}+\frac{\partial f}{\partial \mathbf{r}}\frac{d\mathbf{r}}{dt}
+\frac{\partial f}{\hbar\partial \mathbf{p}}\frac{d\mathbf{p}}{dt}.
\end{eqnarray}
Assuming a homogeneous in space distribution of particles ($\partial f/\partial\mathbf{r}=0$), no explicit dependence of the distribution function on time ($\partial f/\partial t=0$), and using Newton's law, we can rewrite the above as
\begin{eqnarray}
\frac{df}{dt}=
\mathbf{F}\cdot
\frac{\partial f}{\hbar\partial \mathbf{p}},
\end{eqnarray}
where $\mathbf{F}$ is a force acting on the particle.
Assuming that the electron is under the influence of the Lorentz force $\mathbf{F}=e_0\mathbf{E}$, where $\mathbf{E}$ is an electric field with kinetics determined by interaction (collisions) with other particles in the media, we find
the Boltzmann equation
\begin{equation}\label{AP7_Eq1}
e_0\textbf{E}\cdot\frac{\partial f}{\hbar\partial \textbf{p}}=I\{f\},
\end{equation}
where the scattering integral in its most generic form is given by~\cite{Zaitsev:2014aa}:
\begin{eqnarray}
\label{AP7_Eq2}
I\{f\}&=&-\frac{1}{\hbar}\int\frac{d\textbf{q}d\textbf{p}'}{(2\pi)^2}|M^{(1)}_q|^2\Bigl[N_qf_p(1-f_{p'})\delta(\varepsilon_p-\varepsilon_{p'}+\hbar\omega_q)\delta(\textbf{p}-\textbf{p}'+\textbf{q})\nonumber\\
&{}&{}+(N_q+1)f_p(1-f_{p'})\delta(\varepsilon_p-\varepsilon_{p'}-\hbar\omega_q)\delta(\textbf{p}-\textbf{p}'-\textbf{q})\nonumber \\
&{}&{}+N_qf_{p'}(1-f_{p})\delta(\varepsilon_{p'}-\varepsilon_{p}+\hbar\omega_q)\delta(\textbf{p}'-\textbf{p}+\textbf{q})\nonumber\\
&{}&{}+(N_q+1)f_{p'}(1-f_p)\delta(\varepsilon_{p'}-\varepsilon_{p}-\hbar\omega_q)\delta(\textbf{p}'-\textbf{p}-\textbf{q})\Bigr],
\end{eqnarray}
where $M^{(1)}_q=\sqrt{n_c}g_q(u_q+v_q)/L$ is a matrix element of interaction (coming from Eq.~\eqref{CH6_eq.5-a} in the main text), and we consider the fact that the functions $u_q$ and $v_q$ depend on the absolute value of momentum $q$ only.
$N_q$ is a distribution of other particles (to be scattered on), and we use $p\equiv\abs{\mathbf{p}}$.
(Note that the sample length $L$ cancels out in the equation above.)

Further, we take the variations of the integrals in Eq.~\eqref{AP7_Eq2} over $f_p$ and $f_{p'}$ along the method discussed in~\cite{Zaitsev:2014aa}.
For example, let us consider the terms in the square brackets in the first and last lines of Eq.~\eqref{AP7_Eq2}: $N_{\textbf{q}}f_p(1-f_{p'})-(N_{\textbf{q}'}+1)f_{p'}(1-f_{p})$.
Taking a variation over $f_p$, we find:
\begin{eqnarray}
\label{EqBol1}
\delta f_p\left\{(1-f_{p^\prime})N_q
+f_{p^\prime}(N_{q}+1)\right\}
=
\delta f_pN_qn_F(\xi_{p'})
    \left(
e^{\frac{\xi_{p'}}{k_BT}}
+e^{\frac{\hbar sq}{k_BT}}
    \right)
,
\end{eqnarray}
where we denote the equilibrium Fermi distribution as $n_F(p)\equiv n_F(\xi_p)$ with $\xi_p=\varepsilon_p-\mu$ where $\mu$ is the chemical potential.
In the last equality we also consider energy conservation $\xi_{p}=\xi_{p^{\prime}}-\hbar sq$ (which is legitimate for this particular term).

We then assume that the distribution function of electrons $f_p$ is close to the equilibrium one, thus expanding $f_p=n_F(\xi_p)+\delta f_p=n_F(\xi_p)+(\partial n_F(p)/\partial \xi_p)\phi_p=n_F(\xi_p)+\frac{1}{k_BT}n_F(\xi_p)[1-n_F(\xi_p)]\phi_\mathbf{p}$, where we introduce the correction $\phi_\mathbf{p}$. Then Eq.~\eqref{AP7_Eq2} (after some straightforward algebra with the distribution functions) turns into
\begin{eqnarray}
\label{AP7_EqPhi}
\frac{\phi_\mathbf{p}}{k_BT}n_F(p^\prime)(1-n_F(p))
(N_q+1)
=\frac{\phi_\mathbf{p}}{k_BT}
(n_F(p)-n_F(p^\prime))
(N_q+1)N_q.
\end{eqnarray}
In a similar fashion, we treat the variation of the first and last lines in Eq.~\eqref{AP7_Eq2} over $f_{p^\prime}$ and find
\begin{eqnarray}
\label{AP7_EqPhiPrime1}
-\frac{\phi_{\mathbf{p}^\prime}}{k_BT}
(n_F(p)-n_F(p^\prime))
(N_q+1)N_q.
\end{eqnarray}
Taken together, we find the first term (out of two) to enter our target expression:
\begin{eqnarray}
-\frac{\phi_\mathbf{p}-\phi_{\mathbf{p}^\prime}}{k_BT}
(n_F(p)-n_F(p'))
(N_q+1)N_q
\delta(\varepsilon_{\textbf{p}'}-\varepsilon_{\textbf{p}}-\hbar\omega_{\textbf{q}}).
\end{eqnarray}
Repeating a similar variation procedure with all the other terms in Eq.~\eqref{AP7_Eq2}, we find the total formula:
\begin{eqnarray}
\label{AP7_3}
e_0\textbf{E}\cdot\frac{\partial f}{\hbar\partial \textbf{p}}&=&\frac{\hbar e_0}{m}\textbf{E}\cdot\textbf{p}\frac{\partial f^0}{\partial \varepsilon_p}=I\{\phi_\textbf{p}\};\\
I\{\phi_\textbf{p}\}&=&-\frac{1}{\hbar}\int\frac{d\textbf{q}d\textbf{p}'}{(2\pi)^2}|M^{(1)}_q|^2\frac{1}{\hbar}\frac{\partial N_q}{\partial \omega_q}
\left(f^0(\varepsilon_p)-f^0(\varepsilon_{p'})\right)
\left(\phi_{\textbf{p}}-\phi_{\textbf{p}'}\right) \\ \nonumber
&\times&\Bigl[\delta(\varepsilon_{p}-\varepsilon_{p'}-\hbar\omega_q)\delta(\textbf{p}-\textbf{p}'-\textbf{q})
-\delta(\varepsilon_{p}-\varepsilon_{p'}+\hbar\omega_q)\delta(\textbf{p}-\textbf{p}'+\textbf{q})\Bigr],
\end{eqnarray}
where
$$\frac{\partial N_q}{\partial \omega_q}=-\frac{\hbar}{k_BT}N_q(1+N_q)$$
(which can be checked by direct substitution) and $m$ is the effective electron mass in the 2DEG having the dispersion $\varepsilon_p=\hbar^2p^2/(2m)$.

Further using the wave vector-dependent delta-function, we integrate over the electron wave vector $\textbf{p}'$ and find
\begin{eqnarray}
\label{AP7_Eq4}
\frac{\hbar e_0}{m}\textbf{E}\cdot\textbf{p}\frac{\partial f^0}{\partial \varepsilon_p}
=&-&\frac{1}{\hbar}\int\frac{d\textbf{q}}{(2\pi)^2}|M^{(1)}_q|^2\frac{1}{\hbar}\frac{\partial N_q}{\partial \omega_q}
\left(f^0(\varepsilon_p)-f^0(\varepsilon_{p}-\hbar\omega_q)\right)\nonumber \\ \nonumber
&\times& \left(\phi_{\textbf{p}}-\phi_{\textbf{p}-\textbf{q}}\right)
\delta(\varepsilon_{\textbf{p}}-\varepsilon_{\textbf{p}-\textbf{q}}-\hbar\omega_\textbf{q})\\
\nonumber
&+&\frac{1}{\hbar}\int\frac{d\textbf{q}}{(2\pi)^2}|M^{(1)}_q|^2\frac{1}{\hbar}\frac{\partial N_q}{\partial \omega_q}
\left(f^0(\varepsilon_p)-f^0(\varepsilon_{p}+\hbar\omega_q)\right)\\ 
&\times& \left(\phi_{\textbf{p}}-\phi_{\textbf{p}+\textbf{q}}\right)
\delta(\varepsilon_{\textbf{p}}-\varepsilon_{\textbf{p}+\textbf{q}}+\hbar\omega_\textbf{q}).
\end{eqnarray}
Without a loss of generality, we set the electric field to be directed along the $x$-axis.
Then the correction function becomes
\begin{eqnarray}
\phi_\textbf{p}= \frac{\hbar e_0}{m}E_xp_x\tau(\varepsilon_p),
\end{eqnarray}
where $\tau(\varepsilon_p)$ is the relaxation time.
Cancelling out $e_0/m$ and several other terms on both the right- and left-hand sides of Eq.~\eqref{AP7_Eq4}, we have
\begin{eqnarray}\label{AP7_5}
\hbar p_x\frac{\partial f^0}{\partial \varepsilon_p}
=&-&\frac{1}{\hbar}\int\frac{d\textbf{q}}{(2\pi)^2}|M^{(1)}_q|^2\frac{1}{\hbar}\frac{\partial N_q}{\partial \omega_q}
\left[f^0(\varepsilon_p)-f^0(\varepsilon_{p}-\hbar\omega_q)\right] \\ 
&\times&\left[\frac{p_x}{k_F}\tau(\varepsilon_p)-\frac{p_x-q_x}{k_F}\tau(\varepsilon_p-\hbar\omega_q)\right]
\delta(\varepsilon_{\textbf{p}}-\varepsilon_{\textbf{p}-\textbf{q}}-\hbar\omega_\textbf{q})\nonumber\\
&+&\frac{1}{\hbar}\int\frac{d\textbf{q}}{(2\pi)^2}|M^{(1)}_q|^2\frac{1}{\hbar}\frac{\partial N_q}{\partial \omega_q}
\left[f^0(\varepsilon_p)-f^0(\varepsilon_{p}+\hbar\omega_q)\right] \nonumber\\ 
&\times&\left[\frac{p_x}{k_F}\tau(\varepsilon_p)-\frac{p_x+q_x}{k_F}\tau(\varepsilon_p+\hbar\omega_q)\right]
\delta(\varepsilon_{\textbf{p}}-\varepsilon_{\textbf{p}+\textbf{q}}+\hbar\omega_\textbf{q}).\nonumber
\end{eqnarray}
Replacing the relaxation time $\tau(\varepsilon)$ with its energy-averaged value $\tau_0$~\cite{Ziman:2001aa}, thus assuming $\tau(\varepsilon_p)\approx\tau(\varepsilon_p\pm\hbar\omega_q)=\tau_0$, one finds
\begin{eqnarray}
\label{AP7_Boltzmann1}
\hbar p_x\frac{\partial f^0}{\partial \varepsilon_p}
=&-&\tau_0\int \frac{d\textbf{q}}{(2\pi)^2}q_x|M^{(1)}_q|^2\frac{1}{\hbar}\frac{\partial N_q}{\partial \omega_q} \\ \nonumber
&\times&\left[f^0(\varepsilon_p)-f^0(\varepsilon_{p}-\hbar\omega_q)\right]
\delta(\varepsilon_{\textbf{p}}-\varepsilon_{\textbf{p}-\textbf{q}}-\hbar\omega_\textbf{q})\\
\nonumber
&-&\tau_0\int \frac{d\textbf{q}}{(2\pi)^2}q_x|M^{(1)}_q|^2\frac{1}{\hbar}\frac{\partial N_q}{\partial \omega_q} \\ \nonumber
&\times&\left[f^0(\varepsilon_p)-f^0(\varepsilon_{p}+\hbar\omega_q)\right]
\delta(\varepsilon_{\textbf{p}}-\varepsilon_{\textbf{p}+\textbf{q}}+\hbar\omega_\textbf{q}).
\end{eqnarray}

By denoting the angle between vectors $\textbf{p}$ and $\textbf{q}$ as $\varphi$ and the angle between vectors $\textbf{p}$ and $\textbf{E}$ as $\beta$, we have $q_x=q\cos(\varphi+\beta)$ and $p_x=p\cos\beta$.
Using the substitution technique, the $\varphi$-dependent part of Eq.~\eqref{AP7_Boltzmann1} gives
\begin{footnotesize}
\begin{eqnarray}
\label{AP7_Eq19}
    &~&\int_0^{2\pi} d\varphi\cos (\varphi+\beta)\delta(\varepsilon_p-\varepsilon_{|\mathbf{p}\pm\mathbf{q}|}\pm\hbar\omega_q)\\ \nonumber
    &=&\frac{2m}{\hbar^2p}\cos\beta\frac{\left(\mp\frac{q}{2p}+\frac{ms}{\hbar p}\right)\Theta\left[1-\left(\mp\frac{q}{2p}+\frac{ms}{\hbar p}\right)^2\right]}{\sqrt{1-\left(\mp\frac{q}{2p}+\frac{ms}{\hbar p}\right)^2}},
\end{eqnarray}
\end{footnotesize}
where $\Theta[x]$ is the Heaviside step function ($\Theta$-function in what follows).
\footnote{The way to derive Eq.~\eqref{AP7_Eq19} is the same as the way we derived Eq.~\eqref{AP6_AEq19}.}

After integrating over the angle $\varphi$, we can integrate Eq.~\eqref{AP7_Boltzmann1} over $\xi_p=\varepsilon_p-\mu$, using
\begin{eqnarray}
\int\limits_{-\infty}^{\infty}d\xi_p\frac{\partial f^0}{\partial \varepsilon_p}=-1,\nonumber\\
\int\limits_{-\infty}^{\infty}d\xi_p\left(f^0(\varepsilon_p)-f^0(\varepsilon_{p}\pm\hbar\omega_q)\right)=\pm\hbar\omega_q,
\end{eqnarray}
and let all electron wave vectors be on the Fermi surface.
%
We find
\begin{eqnarray}
\label{AP7_rho1}
\frac{1}{\tau_0}&=&\frac{m\xi_I^2}{\hbar k_F^2M}\frac{1}{k_BT}\int_0^\infty \frac{dq}{(2\pi)^2} \frac{q^3e^{-2ql}}{\epsilon(q)^2}(\Gamma_+-\Gamma_-)_{k_F}N_q(1+N_q),
\end{eqnarray}
where the subscript $k_F$ in the expression $(\Gamma_--\Gamma_+)_{k_F}$ means that all the electron wave vectors $p$ are to be substituted by the Fermi value $k_F$.
We also
introduce $\xi_I= e_0^2d\sqrt{n_c}/2\epsilon$ and
\begin{eqnarray}
\label{AP7_gamma}
\Gamma_{\pm}=\frac{\left(\mp\frac{q}{2p}+\frac{ms}{\hbar p}\right)\Theta\left[1-\left(\mp\frac{q}{2p}+\frac{ms}{\hbar p}\right)^2\right]}{\sqrt{1-\left(\mp\frac{q}{2p}+\frac{ms}{\hbar p}\right)^2}}.
\end{eqnarray}
Here, $\epsilon(q)$ is the static screening given by
\begin{eqnarray}
\epsilon(q)=\left(1+\frac{2}{a_Bq}\right)\left(1+\frac{1}{q^2\xi^2}\right),
\end{eqnarray}
where $a_B$ is the Bohr radius and, as we recall, $\xi$ is the healing length of the condensate.

We now apply the dimensionless variable introduced in Eq.~\eqref{AP6_Dim},
$u=\frac{\hbar sq}{k_BT}$, into Eq.~\eqref{AP7_rho1} and obtain
\begin{eqnarray}
\label{AP7_rho2}
\frac{1}{\tau_0}&=&\frac{m\xi_I^2}{\hbar k_F^2M}\frac{(k_BT)^3}{(\hbar s)^4}\int_0^\infty \frac{du}{(2\pi)^2} \frac{u^3e^{(1-2\tilde{l})u}}{(e^u-1)^2}(\Gamma_+-\Gamma_-)_{k_F},
\end{eqnarray}
where $s=10^5$ m/s and $l= 5.0\times 10^{-8}$ m/s, then
\begin{eqnarray}
\Tilde{l}=\frac{lk_BT}{\hbar s}\sim\frac{k_BT}{10\mbox{ meV}}.
\end{eqnarray}
Note that room temperature is $k_BT_R\sim 26$ meV, so that for far lower temperatures we have $\Tilde{l}\ll 1$.
Hence, we can replace
\begin{eqnarray}
\label{AP7_approx1}
e^{(1-2\tilde{l})u}\rightarrow e^u.
\end{eqnarray}
To keep things general, we instead expand
\begin{eqnarray}
\label{AP7_taylor}
e^{-2\tilde{l}u}=\sum_{n=0}^\infty\frac{(-1)^n(2\tilde{l}u)^n}{n!}.
\end{eqnarray}

Let us now look at the argument of the $\Theta$ function in Eq.~\eqref{AP7_gamma}. Its roots for both the cases are
\begin{eqnarray}
q_0=\pm 2k_F-\frac{2ms}{\hbar}\approx\pm2k_F,
\end{eqnarray}
which means that the Heaviside theta function is non-zero in the integration range
\begin{eqnarray}
0\leq q\lesssim 2k_F,
\end{eqnarray}
or in terms of $u$,
\begin{eqnarray}
\label{AP7_lambda}
0\leq u<\frac{T_\textrm{BG}}{T}\equiv\Lambda,
\end{eqnarray}
where $T_\textrm{BG}=2\hbar sk_F/k_B$ is the Bloch--Gr\"{u}neisen temperature for bogolons.

For large $u$ (or $q$), the factors $\Gamma_\pm$ in Eq.~\eqref{AP7_rho2} approach constant values.
In the meantime, the term $u^4\exp({-u})$ rapidly goes to zero for $u>10$.
Therefore, we can remove the $\Theta$ function in Eq.~\eqref{AP7_gamma} and only incur a small (imaginary) error.

The term inside the square root of Eq.~\eqref{AP7_gamma} can be rewritten as
\begin{eqnarray}
\label{AP7_sqrt}
1-\left(\mp\frac{q}{2p}+\frac{ms}{\hbar p}\right)^2&\approx &\frac{1}{4k_F^2}(2k_F-q)(2k_F+q)\nonumber\\
&=&\left(\frac{k_BT}{2k_F\hbar s}\right)^2(\Lambda-u)(\Lambda+u),
\end{eqnarray}
where $\Lambda$ depends on $T$, as was defined in Eq.~\eqref{AP7_lambda}. However, for $T\ll T_\textrm{BG}$, due to the factor $\exp({-u})$, we can simply replace $\Lambda\sim 10$ (or greater) without significantly affecting the result.

The resistivity is inversely proportional to the scattering time. Using Eqs.~\eqref{AP7_taylor} and~\eqref{AP7_sqrt}, and the arguments presented above allowing us to remove the $\Theta$ function, we find
\begin{eqnarray}
\label{AP7_28}
\rho^{(1)}=\frac{\pi\hbar^3\xi_I^2}{e_0^2ME_F}\sum_{n=0}^\infty\frac{(-2)^nl^n\gamma_n}{n!(\hbar s)^{n+4}}(k_BT)^{n+4},
\end{eqnarray}
where
\begin{eqnarray}
\label{AP7_EqMath}
\gamma_n=\int_0^\Lambda\frac{du}{(2\pi)^2}\frac{e^uu^{n+3}}{(e^u-1)^2\sqrt{(\Lambda-u)(\Lambda+u)}}.
\end{eqnarray}

This dimensionless integral can be evaluated in closed form when we note that (i) $\Lambda\gg 1$, and (ii) due to the exponential factors in the integrand, the relevant contribution to the integral comes from $0<u\lesssim 1$, so that
\begin{eqnarray}
\label{AP7_EqMath2}
\gamma_n\approx\frac{1}{\Lambda}\int_0^\infty\frac{du}{(2\pi)^2}\frac{e^uu^{n+3}}{(e^u-1)^2}=\frac{(n+3)!}{(2\pi)^2}\zeta(n+9)\frac{T}{T_\textrm{BG}}.
\end{eqnarray}
The leading term in Eq.~\eqref{AP7_28} if $T\ll T_\textrm{BG}$ reads
\begin{eqnarray}
\label{AP7_rho1b}
\rho^{(1)}\approx\frac{\pi\hbar^3\xi_I^2}{e_0^2ME_F}\frac{3!\zeta(3)}{(2\pi)^2k_BT_\textrm{BG}}\left(\frac{k_BT}{\hbar s}\right)^4,
\end{eqnarray}
and hence the single-bogolon resistivity behaves as $\rho^{(1)}\propto T^4$ at low temperatures.


\section{Bloch--Gr\"uneisen formula for two-bogolon processes}
The starting equation here is similar to Eq.~\eqref{AP7_Eq1}:
\begin{eqnarray}
\nonumber
e_0\mathbf{E}\cdot\frac{df_p}{\hbar d\mathbf{p}}=I\{f_p\}.
\end{eqnarray}
We consider the Hamiltonian
Eq.~\eqref{CH6_eq.5-b} from the main text:
\begin{equation}\label{1}
V_2=\frac{1}{L^2}\sum_{\mathbf{k},\mathbf{p}',\mathbf{q},\mathbf{q}'}
g_\textbf{k}
c^+_{\textbf{p}'}c_{\textbf{p}}\varphi^+_{\textbf{q}'}\varphi_{\textbf{q}},
\end{equation}
where
\begin{eqnarray}\label{AP7_2}
\varphi^+_{\textbf{q}'}\varphi_{\textbf{q}}&=&(u_{\textbf{q}'}b^+_{\textbf{q}'}+v_{\textbf{q}'}b_{-\textbf{q}'})
(u_{\textbf{q}}b_{\textbf{q}}+v_{\textbf{q}}b^+_{-\textbf{q}})\\ \nonumber
&=&u_{\textbf{q}'}b^+_{\textbf{q}'}u_{\textbf{q}}b_{\textbf{q}}+u_{\textbf{q}'}b^+_{\textbf{q}'}v_{\textbf{q}}b^+_{-\textbf{q}}+
v_{\textbf{q}'}b_{-\textbf{q}'}u_{\textbf{q}}b_{\textbf{q}}+v_{\textbf{q}'}b_{-\textbf{q}'}v_{\textbf{q}}b^+_{-\textbf{q}}.
\end{eqnarray}
Thus, the collision integral reads $I\{f_p\}=I_1-I_2$, where
\begin{footnotesize}
\begin{eqnarray}\label{AP7_I1I2}
I_1&=&-\sum_{\mathbf{k},\mathbf{p}',\mathbf{q},\mathbf{q}'}|g_\textbf{k}|^2f_p(1-f_{p'})\delta(\textbf{p}'-\textbf{p}-\textbf{k})\delta(\textbf{q}'-\textbf{q}+\textbf{k})\times\\\nonumber
&\times&\Bigl[u^2_{\textbf{q}'}u^2_{\textbf{q}}(N_{\textbf{q}'}+1)N_{\textbf{q}}
\delta(\varepsilon_{\textbf{p}'}-\varepsilon_{\textbf{p}}+\omega_{\textbf{q}'}-\omega_{\textbf{q}})
+u^2_{\textbf{q}'}v^2_{\textbf{q}}(N_{\textbf{q}'}+1)(N_{-\textbf{q}}+1)
\delta(\varepsilon_{\textbf{p}'}-\varepsilon_{\textbf{p}}+\omega_{\textbf{q}'}+\omega_{-\textbf{q}})\\\nonumber
&+&v^2_{\textbf{q}'}u^2_{\textbf{q}}N_{-\textbf{q}'}N_{\textbf{q}}
\delta(\varepsilon_{\textbf{p}'}-\varepsilon_{\textbf{p}}-\omega_{-\textbf{q}'}-\omega_{\textbf{q}})
+v^2_{\textbf{q}'}v^2_{\textbf{q}}N_{-\textbf{q}'}(N_{-\textbf{q}}+1)
\delta(\varepsilon_{\textbf{p}'}-\varepsilon_{\textbf{p}}-\omega_{-\textbf{q}'}+\omega_{-\textbf{q}})\Bigr],\\
\nonumber
I_2&=&-\sum_{\mathbf{k},\mathbf{p}',\mathbf{q},\mathbf{q}'}|g_\textbf{k}|^2f_{p'}(1-f_{p})\delta(\textbf{p}-\textbf{p}'-\textbf{k})\delta(\textbf{q}'-\textbf{q}+\textbf{k})\times\\\nonumber
&\times&\Bigl[u^2_{\textbf{q}'}u^2_{\textbf{q}}(N_{\textbf{q}'}+1)N_{\textbf{q}}
\delta(\varepsilon_{\textbf{p}}-\varepsilon_{\textbf{p}'}+\omega_{\textbf{q}'}-\omega_{\textbf{q}})
+u^2_{\textbf{q}'}v^2_{\textbf{q}}(N_{\textbf{q}'}+1)(N_{-\textbf{q}}+1)
\delta(\varepsilon_{\textbf{p}}-\varepsilon_{\textbf{p}'}+\omega_{\textbf{q}'}+\omega_{-\textbf{q}})\\\nonumber
&+&v^2_{\textbf{q}'}u^2_{\textbf{q}}N_{-\textbf{q}'}N_{\textbf{q}}
\delta(\varepsilon_{\textbf{p}}-\varepsilon_{\textbf{p}'}-\omega_{-\textbf{q}'}-\omega_{\textbf{q}})
+v^2_{\textbf{q}'}v^2_{\textbf{q}}N_{-\textbf{q}'}(N_{-\textbf{q}}+1)
\delta(\varepsilon_{\textbf{p}}-\varepsilon_{\textbf{p}'}-\omega_{-\textbf{q}'}+\omega_{-\textbf{q}})\Bigr],
\end{eqnarray}
\end{footnotesize}
where we assume that $N_\mathbf{x}$ are equilibrium Bose distribution functions. In $I_2$ we can change the signs of the vectors: $\textbf{k}\rightarrow-\textbf{k},\,\textbf{q}\rightarrow-\textbf{q},\,\textbf{q}'\rightarrow-\textbf{q}'$. Taking into account that the distribution functions and energies only depend on the absolute value of the wave vectors, we find
\begin{footnotesize}
\begin{eqnarray}
I_1&=&-\sum_{\mathbf{k},\mathbf{p}',\mathbf{q},\mathbf{q}'}|g_\textbf{k}|^2f_p(1-f_{p'})\delta(\textbf{p}'-\textbf{p}-\textbf{k})\delta(\textbf{q}'-\textbf{q}+\textbf{k})\times\\\nonumber
&\times&\Bigl[u^2_{\textbf{q}'}u^2_{\textbf{q}}(N_{\textbf{q}'}+1)N_{\textbf{q}}
\delta(\varepsilon_{\textbf{p}'}-\varepsilon_{\textbf{p}}+\omega_{\textbf{q}'}-\omega_{\textbf{q}})
+u^2_{\textbf{q}'}v^2_{\textbf{q}}(N_{\textbf{q}'}+1)(N_{\textbf{q}}+1)
\delta(\varepsilon_{\textbf{p}'}-\varepsilon_{\textbf{p}}+\omega_{\textbf{q}'}+\omega_{\textbf{q}})\\\nonumber
&+&v^2_{\textbf{q}'}u^2_{\textbf{q}}N_{\textbf{q}'}N_{\textbf{q}}
\delta(\varepsilon_{\textbf{p}'}-\varepsilon_{\textbf{p}}-\omega_{\textbf{q}'}-\omega_{\textbf{q}})
+v^2_{\textbf{q}'}v^2_{\textbf{q}}N_{\textbf{q}'}(N_{\textbf{q}}+1)
\delta(\varepsilon_{\textbf{p}'}-\varepsilon_{\textbf{p}}-\omega_{\textbf{q}'}+\omega_{\textbf{q}})\Bigr],\\
\nonumber
I_2&=&-\sum_{\mathbf{k},\mathbf{p}',\mathbf{q},\mathbf{q}'}|g_\textbf{k}|^2f_{p'}(1-f_{p})\delta(-\textbf{p}+\textbf{p}'+\textbf{k})\delta(-\textbf{q}'+\textbf{q}-\textbf{k})\times\\\nonumber
&\times&\Bigl[u^2_{\textbf{q}'}u^2_{\textbf{q}}(N_{\textbf{q}'}+1)N_{\textbf{q}}
\delta(\varepsilon_{\textbf{p}}-\varepsilon_{\textbf{p}'}+\omega_{\textbf{q}'}-\omega_{\textbf{q}})
+u^2_{\textbf{q}'}v^2_{\textbf{q}}(N_{\textbf{q}'}+1)(N_{\textbf{q}}+1)
\delta(\varepsilon_{\textbf{p}}-\varepsilon_{\textbf{p}'}+\omega_{\textbf{q}'}+\omega_{\textbf{q}})\\\nonumber
&+&v^2_{\textbf{q}'}u^2_{\textbf{q}}N_{\textbf{q}'}N_{\textbf{q}}
\delta(\varepsilon_{\textbf{p}}-\varepsilon_{\textbf{p}'}-\omega_{\textbf{q}'}-\omega_{\textbf{q}})
+v^2_{\textbf{q}'}v^2_{\textbf{q}}N_{\textbf{q}'}(N_{\textbf{q}}+1)
\delta(\varepsilon_{\textbf{p}}-\varepsilon_{\textbf{p}'}-\omega_{\textbf{q}'}+\omega_{\textbf{q}})\Bigr].
\end{eqnarray}
\end{footnotesize}
We see that the delta-functions describing the momentum conservation are the same. We also use that for the linear spectrum of bogolons,  $u^2_{\textbf{q}'}u^2_{\textbf{q}}=u^2_{\textbf{q}'}v^2_{\textbf{q}}=v^2_{\textbf{q}'}v^2_{\textbf{q}}=v^2_{\textbf{q}'}v^2_{\textbf{q}}$.
This yields:
\begin{footnotesize}
\begin{eqnarray}\label{APL_5}
I_1-I_2&=&-\sum_{\mathbf{k},\mathbf{p}',\mathbf{q},\mathbf{q}'}u^2_{\textbf{q}'}u^2_{\textbf{q}}|g_\textbf{k}|^2\delta(\textbf{p}'-\textbf{p}-\textbf{k})\delta(\textbf{q}'-\textbf{q}+\textbf{k})\\ \nonumber
&\times&  \Bigr[N_{\textbf{q}'}N_{\textbf{q}}f_p(1-f_{p'})-(N_{\textbf{q}'}+1)(N_{\textbf{q}}+1)f_{p'}(1-f_{p})\Bigl] \delta(\varepsilon_{\textbf{p}'}-\varepsilon_{\textbf{p}}-\omega_{\textbf{q}'}-\omega_{\textbf{q}})\\ \nonumber
&+&\Bigl[(N_{\textbf{q}'}+1)(N_{\textbf{q}}+1)f_p(1-f_{p'})-N_{\textbf{q}'}N_{\textbf{q}}f_{p'}(1-f_{p})\Bigr] \delta(\varepsilon_{\textbf{p}'}-\varepsilon_{\textbf{p}}+\omega_{\textbf{q}'}+\omega_{\textbf{q}})\\\nonumber
&+&\Bigr[N_{\textbf{q}'}(N_{\textbf{q}}+1)f_p(1-f_{p'})-(N_{\textbf{q}'}+1)N_{\textbf{q}}f_{p'}(1-f_{p})\Bigl] \delta(\varepsilon_{\textbf{p}'}-\varepsilon_{\textbf{p}}-\omega_{\textbf{q}'}+\omega_{\textbf{q}})\\\nonumber
&+&\Bigl[(N_{\textbf{q}'}+1)N_{\textbf{q}}f_p(1-f_{p'})-N_{\textbf{q}'}(N_{\textbf{q}}+1)f_{p'}(1-f_{p})\Bigr] \delta(\varepsilon_{\textbf{p}'}-\varepsilon_{\textbf{p}}+\omega_{\textbf{q}'}-\omega_{\textbf{q}}) .
\end{eqnarray}
\end{footnotesize}

The sums in Eq.~\eqref{APL_5} can be replaced by integrals in the continuous limit, as discussed in Eq.~\eqref{AP7_Eq2} for the single-bogolon case and~\cite{Zaitsev:2014aa}.
For example, let us consider the terms in the square brackets in the second line of Eq.~\eqref{APL_5}: $N_{\textbf{q}'}N_{\textbf{q}}f_p(1-f_{p'})-(N_{\textbf{q}'}+1)(N_{\textbf{q}}+1)f_{p'}(1-f_{p})$.
Taking a variation over $f_p$ we have:
\begin{eqnarray}
\label{EqBol}
&{}&{}\delta f_p\left\{(1-f_{p^\prime})N_qN_{q^\prime}+f_{p^\prime}(N_q+1)(N_{q^\prime}+1)\right\}\\ \nonumber
&=& \delta f_pN_qN_{q^\prime}n_F(p^\prime)\left\{\exp\left(\frac{\xi_{p^{\prime}}}{k_BT}\right)+\exp\left(\frac{\hbar s (q+q^\prime)}{k_BT}\right)\right\}\\ \nonumber
&=&\delta f_pN_qN_{q^\prime}n_F(p^\prime) \exp\left(\frac{\hbar s (q+q^\prime)}{k_BT}\right) \left\{\exp\left(\frac{\xi_{p^{\prime}}-\hbar s (q+q^\prime)}{k_BT}\right)+1\right\}\\ \nonumber
&=& \delta f_p(N_q+1)(N_{q^\prime}+1)n_F(p^\prime)\frac{1}{n_F(p)},
\end{eqnarray}
where w$n_F(p)\equiv n_F(\xi_p)$ is the equilibrium Fermi distribution. In the last equality, we also use the energy conservation $\xi_{p}=\xi_{p^{\prime}}-\hbar s (q+q^\prime)$.

We assume that the distribution function of electrons $f_p$ is close to the equilibrium distribution, thus expanding $f_p=n_F(\xi_p)+\delta f_p=n_F(\xi_p)+(\partial n_F(p)/\partial \xi_p)\phi_p=n_F(\xi_p)+\frac{1}{k_BT}n_F(\xi_p)[1-n_F(\xi_p)]\phi_p$, where we introduce the correction $\phi_p$. Then Eq.~\eqref{EqBol} turns into
\begin{eqnarray}
\label{AP7_EqPhi2}
&{}&{}\frac{\phi_p}{k_BT}n_F(p)(1-n_F(p))
(N_q+1)(N_{q^\prime}+1)n_F(p^\prime)\frac{1}{n_F(p)}\\\nonumber
&=&\frac{\phi_p}{k_BT}(n_F(p)-n_F(p^\prime))
(N_q+1)(N_{q^\prime}+1)N_{q+q^\prime}.
\end{eqnarray}

We similarly treat the variation of the first line in Eq.~\eqref{APL_5}  over $f_{p^\prime}$ and find
\begin{eqnarray}
\label{AP7_EqPhiPrime2}
-\frac{\phi_{p^\prime}}{k_BT}(n_F(p)-n_F(p^\prime))
N_qN_{q^\prime}(N_{q+q^\prime}+1).
\end{eqnarray}
Discovering that $N_qN_{q^\prime}(N_{q+q^\prime}+1)=(N_q+1)(N_{q^\prime}+1)N_{q+q^\prime}$, which means that $\phi_p$ and $\phi_{p^\prime}$ in Eqs.~\eqref{AP7_EqPhi2} and~\eqref{AP7_EqPhiPrime2} have equivalent prefactors, we find the first term (out of four) to enter our target expression:
\begin{eqnarray}
-\frac{\phi_p-\phi_{p^\prime}}{k_BT}(n_F(p)-n_F(p^\prime))
N_qN_{q^\prime}(N_{q+q^\prime}+1)\delta(\varepsilon_{\textbf{p}'}-\varepsilon_{\textbf{p}}-\omega_{\textbf{q}'}-\omega_{\textbf{q}}).
\end{eqnarray}

Repeating a similar variation procedure with all the other terms in Eq.~\eqref{APL_5}, we find the total formula:
\begin{gather}\label{EqZaiatz}
I_1-I_2=-\sum_{\mathbf{k},\mathbf{p}',\mathbf{q},\mathbf{q}'}u^2_{\textbf{q}'}u^2_{\textbf{q}}|g_\textbf{k}|^2
\frac{\phi_{\mathbf{p}}-\phi_{\mathbf{p^\prime}}}{k_BT}
[n_F(\mathbf{p})-n_F(\mathbf{p}^\prime)]
\delta(\textbf{p}'-\textbf{p}-\textbf{k})\delta(\textbf{q}'-\textbf{q}+\textbf{k})\\
\nonumber
\times
\{
N_{q}N_{q^\prime}(N_{q+q^\prime}+1)
\Bigr[\delta(\varepsilon_{\textbf{p}'}-\varepsilon_{\textbf{p}}-\omega_{\textbf{q}'}-\omega_{\textbf{q}})
-
\delta(\varepsilon_{\textbf{p}'}-\varepsilon_{\textbf{p}}+\omega_{\textbf{q}'}+\omega_{\textbf{q}})\Bigl]+\\
\nonumber
+
(N_{q}+1)N_{q^\prime}(N_{q^\prime-q}+1)
\Bigr[\delta(\varepsilon_{\textbf{p}'}-\varepsilon_{\textbf{p}}-\omega_{\textbf{q}'}+\omega_{\textbf{q}})
-
\delta(\varepsilon_{\textbf{p}'}-\varepsilon_{\textbf{p}}+\omega_{\textbf{q}'}-\omega_{\textbf{q}})\Bigl]
\}.
\end{gather}

This expression can be presented in the form
\begin{gather}\label{6}
I_1-I_2=-\sum_{\mathbf{k},\mathbf{p}'}|g_\textbf{k}|^2\frac{\phi_{\mathbf{p}}-\phi_{\mathbf{p^\prime}}}{k_BT}
[n_F(\mathbf{p})-n_F(\mathbf{p}^\prime)]
\delta(\textbf{p}'-\textbf{p}-\textbf{k})\int d \epsilon \delta(\varepsilon_{\textbf{p}'}-\varepsilon_{\textbf{p}}-\epsilon)F(\textbf{k},\epsilon),\\\nonumber
F(\textbf{k},\epsilon)=\sum_{\mathbf{q},\mathbf{q}'} u^2_{\textbf{q}'}u^2_{\textbf{q}}
\delta(\textbf{q}'-\textbf{q}+\textbf{k})
\Bigl(
N_{q}N_{q^\prime}(N_{q+q^\prime}+1)
\Bigl[\delta(\epsilon-\omega_{\textbf{q}'}-\omega_{\textbf{q}})
-
\delta(\epsilon+\omega_{\textbf{q}'}+\omega_{\textbf{q}})\Bigl]+\\
\nonumber
+
(N_{q}+1)N_{q^\prime}(N_{q^\prime-q}+1)
\Bigr[\delta(\epsilon-\omega_{\textbf{q}'}+\omega_{\textbf{q}})
-
\delta(\epsilon+\omega_{\textbf{q}'}-\omega_{\textbf{q}})\Bigr]
\Bigr).
\end{gather}
Now we can integrate over $\textbf{p}',\textbf{q}'$. Using the momentum-conserving delta functions, we find:
\begin{gather}\label{7}
I=-\sum_{\mathbf{k}}\int d \epsilon |g_\textbf{k}|^2\frac{\phi_{\mathbf{p}}-\phi_{\textbf{p}+\textbf{k}}}{T}
[n_F(\varepsilon_{\textbf{p}})-n_F(\varepsilon_{\textbf{p}}+\epsilon)]
 \delta(\varepsilon_{\textbf{p}+\textbf{k}}-\varepsilon_{\textbf{p}}-\epsilon)F(\textbf{k},\epsilon);\\\nonumber
F(\textbf{k},\epsilon)=\sum_{\mathbf{q}} u^2_{|\textbf{q}-\textbf{k}|}u^2_{\textbf{q}}
\Bigl(
N_{q}N_{|\textbf{q}-\textbf{k}|}(N_{q+|\textbf{q}-\textbf{k}|}+1)
\Bigl[\delta(\epsilon-\omega_{|\textbf{q}-\textbf{k}|}-\omega_{\textbf{q}})
-
\delta(\epsilon+\omega_{|\textbf{q}-\textbf{k}|}+\omega_{\textbf{q}})\Bigl]+\\
\nonumber
+
(N_{q}+1)N_{|\textbf{q}-\textbf{k}|}(N_{|\textbf{q}-\textbf{k}|-q}+1)
\Bigr[\delta(\epsilon-\omega_{|\textbf{q}-\textbf{k}|}+\omega_{\textbf{q}})
-
\delta(\epsilon+\omega_{|\textbf{q}-\textbf{k}|}-\omega_{\textbf{q}})\Bigr]
\Bigr).
\end{gather}
Let us consider the function $F(\mathbf{k},\epsilon)$. We make a replacement $\mathbf{q}\rightarrow \mathbf{q}+\mathbf{k}$ to find:
\begin{gather}
F(\textbf{k},\epsilon)=\sum_{\mathbf{q}} u^2_{q}u^2_{|\textbf{q}+\textbf{k}|}
\Bigl\{
N_{|\textbf{q}+\textbf{k}|}N_{q}(N_{|\textbf{q}+\textbf{k}|+q}+1)
\Bigl[\delta(\epsilon-\omega_{q}-\omega_{|\textbf{q}+\textbf{k}|})
-
\delta(\epsilon+\omega_{q}+\omega_{|\textbf{q}+\textbf{k}|})\Bigl]+\\
\nonumber
+
(N_{|\textbf{q}+\textbf{k}|}+1)N_{q}(N_{q-|\textbf{q}+\textbf{k}|}+1)
\Bigr[\delta(\epsilon-\omega_{q}+\omega_{|\textbf{q}+\textbf{k}|})
-
\delta(\epsilon+\omega_{q}-\omega_{|\textbf{q}+\textbf{k}|})\Bigr]
\Bigr\}.
\end{gather}
Furthermore, we switch from summation to integration $\sum_{\mathbf{q}}\rightarrow\int d\mathbf{q}$, and we introduce a new variable $q_1=|\textbf{q}+\textbf{k}|$.
Then in $\int d\mathbf{q}$ we will integrate over $q$ and $q_1$ instead of $q$ and the angle between the vectors, using
\begin{eqnarray}
\int\frac{d\mathbf{q}}{2\pi}=\frac{4}{(2\pi)^2}\int_0^\infty qdq\int_{|q-k|}^{q+k}q_1dq_1
\frac{1}{\sqrt{[(q+k)^2-q_1^2][q_1^2-(q-k)^2]}}.
\end{eqnarray}

This gives
\begin{gather}
F(\textbf{k},\epsilon)=\frac{4}{(2\pi)^2}\int_0^\infty
qdq 
u^2_{q}
\int_{|q-k|}^{q+k}
q_1dq_1
u^2_{q_1}
\frac{1}{\sqrt{[(q+k)^2-q_1^2][q_1^2-(q-k)^2]}}\\
\nonumber
\times\Bigl\{
N_{q_1}N_{q}(N_{q_1+q}+1)
\Bigl[\delta(\epsilon-\omega_{q}-\omega_{q_1})
-
\delta(\epsilon+\omega_{q}+\omega_{q_1})\Bigl]+\\
\nonumber
+
(N_{q_1}+1)N_{q}(N_{q-q_1}+1)
\Bigr[\delta(\epsilon-\omega_{q}+\omega_{q_1})
-
\delta(\epsilon+\omega_{q}-\omega_{q_1})\Bigr]
\Bigr\}.
\end{gather}
Now we can use the definitions of $u_q$~\cite{Giorgini:1998aa} and linear bogolon dispersion to find:
\begin{gather}
F(\textbf{k},\epsilon)=\frac{4}{(2\pi)^2}\frac{(ms)^2}{4}
\int_0^\infty dq 
\int_{|q-k|}^{q+k} dq_1
\frac{1}{\sqrt{[(q+k)^2-q_1^2][q_1^2-(q-k)^2]}}\\
\nonumber
\times
\Bigl\{
N_{q_1}N_{q}(N_{q_1+q}+1)
\Bigl[\delta(\epsilon-s(q+q_1))
-
\delta(\epsilon+s(q+q_1))\Bigl]+\\
\nonumber
+
(N_{q_1}+1)N_{q}(N_{q-q_1}+1)
\Bigr[\delta(\epsilon-s(q-q_1))
-
\delta(\epsilon+s(q-q_1))\Bigr]
\Bigr\}.
\end{gather}
For convenience we denote new variables $x=s(q+q_1)$ and $y=-s(q-q_1)$, and we introduce the cut-off $sL^{-1}$ in the integrals. This infrared cut-off is necessary for the convergence of the final integral, as will become clear later on. We emphasize that this cut-off has a physical grounding: it means that the momentum integration cannot include fluctuations with wavelengths larger than the sample size $L$. This yields
\begin{gather}
F(\textbf{k},\epsilon)=\frac{1}{2}\left(\frac{ms}{2\pi}\right)^2
\int_{sk+sL^{-1}}^\infty \frac{dx}{\sqrt{x^2-s^2k^2}}
\int_{-sk+sL^{-1}}^{sk-sL^{-1}} \frac{dy}{\sqrt{s^2k^2-y^2}}
\\
\nonumber
\times
\Bigl\{
N\left(\frac{x+y}{2s}\right)N\left(\frac{x-y}{2s}\right)(N\left(\frac{x}{s}\right)+1)
\Bigl[\delta(\epsilon-x)
-
\delta(\epsilon+x)\Bigl]+\\
\nonumber
+
(N\left(\frac{x+y}{2s}\right)+1)N\left(\frac{x-y}{2s}\right)(N\left(\frac{-y}{s}\right)+1)
\Bigr[\delta(\epsilon+y)
-
\delta(\epsilon-y)\Bigr]
\Bigr\}.
\end{gather}
We exchange $y\rightarrow-y$ to find:
\begin{gather}
F(\textbf{k},\epsilon)=\frac{1}{2}\left(\frac{ms}{2\pi}\right)^2
\int_{sk+sL^{-1}}^\infty \frac{dx}{\sqrt{x^2-s^2k^2}}
\int_{-sk+sL^{-1}}^{sk-sL^{-1}} \frac{dy}{\sqrt{s^2k^2-y^2}}
\\
\nonumber
\times
\Bigl\{
N\left(\frac{x+y}{2s}\right)N\left(\frac{x-y}{2s}\right)(N\left(\frac{x}{s}\right)+1)
\Bigl[\delta(\epsilon-x)
-
\delta(\epsilon+x)\Bigl]+\\
\nonumber
+
(N\left(\frac{x-y}{2s}\right)+1)N\left(\frac{x+y}{2s}\right)(N\left(\frac{y}{s}\right)+1)
\Bigr[\delta(\epsilon-y)
-
\delta(\epsilon+y)\Bigr]
\Bigr\}.
\end{gather}
Now we can split the function $F(\textbf{k},\epsilon)$ into two functions, $F_1(\textbf{k},\epsilon)$ and $F_2(\textbf{k},\epsilon)$, such that $F(\textbf{k},\epsilon)=F_1(\textbf{k},\epsilon)+F_2(\textbf{k},\epsilon)$, in the following way:
\begin{gather}
F_1(\textbf{k},\epsilon)=\frac{1}{2}\left(\frac{ms}{2\pi}\right)^2
\int_{sk+sL^{-1}}^\infty \frac{dx}{\sqrt{x^2-s^2k^2}}
\int_{-sk+sL^{-1}}^{sk-sL^{-1}} \frac{dy}{\sqrt{s^2k^2-y^2}}
\\
\nonumber
\times\Bigl(
N\left(\frac{x+y}{2s}\right)N\left(\frac{x-y}{2s}\right)(N\left(\frac{x}{s}\right)+1)
\Bigl[\delta(\epsilon-x)
-
\delta(\epsilon+x)\Bigl];
\\
\nonumber
F_2(\textbf{k},\epsilon)=\frac{1}{2}\left(\frac{ms}{2\pi}\right)^2
\int_{sk+sL^{-1}}^\infty \frac{dx}{\sqrt{x^2-s^2k^2}}
\int_{-sk+sL^{-1}}^{sk-sL^{-1}} \frac{dy}{\sqrt{s^2k^2-y^2}}
\\
\nonumber
\times\Bigl(N\left(\frac{x-y}{2s}\right)+1)N\left(\frac{x+y}{2s}\right)(N\left(\frac{y}{s}\right)+1)
\Bigr[\delta(\epsilon-y)
-
\delta(\epsilon+y)\Bigr]
\Bigr).
\end{gather}
As a result, we can separately perform $x$- and $y$-integrations using the delta functions.

Let us start with $F_1(\textbf{k},\epsilon)$. Since the integration is performed over $x>0$, we can use the relation $\delta(\epsilon-x)-\delta(\epsilon+x)=\textrm{sgn}(\epsilon)\delta(x-|\epsilon|)$ to find: 
\begin{gather}
F_1(\textbf{k},\epsilon)=\mathrm{sgn}(\epsilon)\frac{1}{2}\left(\frac{ms}{2\pi}\right)^2\frac{\Theta[|\epsilon|-sk]}{\sqrt{\epsilon^2-s^2k^2}}
\int_{-+sL^{-1}}^{sk-sL^{-1}} \frac{dy}{\sqrt{s^2k^2-y^2}}
\\
\nonumber
\times
N\left(\frac{|\epsilon|+y}{2s}\right)N\left(\frac{|\epsilon|-y}{2s}\right)(N\left(\frac{|\epsilon|}{s}\right)+1),
\end{gather}
and using another variable $y=skz$, we find:
\begin{eqnarray}
    F_1(\textbf{k},\epsilon)&=& \frac{\mathrm{sgn}(\epsilon)}{2}\left(\frac{ms}{2\pi}\right)^2\frac{e^{\frac{|\epsilon|}{2T}}}{e^{\frac{|\epsilon|}{T}}-1}\frac{\Theta[|\epsilon|-sk]}{\sqrt{\epsilon^2-s^2k^2}} \\ \nonumber
    &\times& \int_{0}^{1-L^{-1}/k} \frac{dz}{\sqrt{1-z^2}}
\frac{1}{\mathrm{cosh}
\left(\frac{|\epsilon|}{2T}\right) - \mathrm{cosh}\left(\frac{sk}{2T}z\right)}.
\end{eqnarray}

Now let us take care of the function $F_2(\textbf{k},\epsilon)$. The integral $\int_{-sk}^{sk}$ can be split in two: $\int_{-sk}^{0}$ and $\int_{0}^{sk}$. In the former we do a replacement $y\rightarrow-y$, after which we combine the two terms to get
\begin{small}
\begin{eqnarray}
&{}&{}F_2(\textbf{k},\epsilon)=\frac{1}{2}\left(\frac{ms}{2\pi}\right)^2
\int_{sk+sL^{-1}}^\infty \frac{dx}{\sqrt{x^2-s^2k^2}}
\int_{0}^{sk-sL^{-1}} \frac{dy}{\sqrt{s^2k^2-y^2}}
\Bigr[\delta(\epsilon-y)
-
\delta(\epsilon+y)\Bigr] \nonumber
\\
\nonumber
&\times&\left\{(N\left(\frac{x-y}{2s}\right)+1)N\left(\frac{x+y}{2s}\right)(N\left(\frac{y}{s}\right)+1)
-
(N\left(\frac{x+y}{2s}\right)+1)N\left(\frac{x-y}{2s}\right)(N\left(\frac{-y}{s}\right)+1)
\right\}\\
\nonumber
&=&
\frac{1}{2}\left(\frac{ms}{2\pi}\right)^2
\int_{1+L^{-1}/k}^\infty
\frac{dz}{\sqrt{1-z^2}}
\int_{0}^{sk-sL^{-1}}
\frac{dy}{\sqrt{s^2k^2-y^2}}
\Bigr[\delta(\epsilon-y)
-
\delta(\epsilon+y)\Bigr]\\ 
&\times&\frac{1/2}{\mathrm{cosh}\left(\frac{x}{2T}\right)-\mathrm{cosh}\left(\frac{y}{2T}\right)}\cdot\frac{2e^{\frac{y}{2T}}}{e^{\frac{y}{T}}-1}.
\end{eqnarray}
\end{small}
This integral is over positive $y$, hence (as before) we use $\delta(\epsilon-y)-\delta(\epsilon+y)=\mathrm{sgn}(\epsilon)\delta(y-|\epsilon|)$. This gives
\begin{footnotesize}
\begin{eqnarray}
F_2(\textbf{k},\epsilon)
    &=& - \frac{\mathrm{sgn}(\epsilon)}{2}\left(\frac{ms}{2\pi}\right)^2
\frac{e^{\frac{|\epsilon|}{2T}}}{e^{\frac{|\epsilon|}{T}}-1}
\frac{\Theta[sk-|\epsilon|]}{\sqrt{s^2k^2-\epsilon^2}}\nonumber \\
    &\times&\int_{1+L^{-1}/k}^\infty \frac{dz}{\sqrt{z^2-1}}
\frac{1}{\mathrm{cosh}\left(\frac{|\epsilon|}{2T}\right)-\mathrm{cosh}\left(\frac{sk}{2T}z\right)}.
\end{eqnarray}
\end{footnotesize}

Let us now return to Eq.~\eqref{7}.
We set $\phi_\mathbf{p}=\hbar e_0E_xp_x\tau_0/m$,
and then $\phi_\mathbf{p}-\phi_{\mathbf{p}+\mathbf{k}}=-\hbar e_0E_xk_x\tau_0/m$ where $k_x=k\cos(\beta+\varphi)$.
We find:
\begin{eqnarray}
\frac{\hbar e_0E_xp_0\cos(\beta)}{m}\frac{df_p}{d\epsilon_p} &=&-\sum_{\mathbf{k}}\int d\epsilon g_k^2\left(\frac{-\hbar e_0E_xk\cos(\beta+\varphi)\tau_0}{Tm}\right) \\ \nonumber
&\times&\left(-\epsilon\frac{df_p}{d\epsilon_p}\right)\delta(\epsilon_{\mathbf{p}+\mathbf{k}}-\epsilon_{\mathbf{p}}-\epsilon)F(\mathbf{k},\epsilon),
\end{eqnarray}
and by cancelling out the matching terms,
\begin{eqnarray}
\label{Eq40}
p_0\cos(\beta)
=
-\frac{\tau_0}{T}\sum_{\mathbf{k}}g_k^2\int \epsilon d\epsilon \cos(\beta+\varphi)
\delta(\epsilon_{\mathbf{p}+\mathbf{k}}-\epsilon_{\mathbf{p}}-\epsilon)F(\mathbf{k},\epsilon).
\end{eqnarray}
Since the function $F(\mathbf{k},\epsilon)$ depends on the absolute value $|\mathbf{k}|$, using
\begin{eqnarray}
\sum_{\mathbf{k}}=\int_0^\infty \frac{kdk}{2\pi}\int_0^{2\pi}\frac{d\varphi}{2\pi},
\end{eqnarray}
in the $\varphi$-dependent part of the integral we come to (denoting $v_0=p_0/m$):
\begin{eqnarray}
&{}&{}\int_0^{2\pi}\frac{d\varphi}{2\pi}\cos(\beta+\varphi)
\delta\left(\frac{p_0k}{m}\cos(\varphi)+\frac{k^2}{2m}-\epsilon\right) \\ \nonumber
&=&
\cos(\beta)\int_0^{2\pi}\frac{d\varphi}{2\pi}\cos(\varphi)
\delta\left(v_0k\cos(\varphi)+\frac{k^2}{2m}-\epsilon\right)
\\
\nonumber
&=&
\cos(\beta)
\left(\frac{\epsilon-k^2/(2m)}{v_0k}\right)
\frac{1}{\pi}
\frac{\Theta[v_0^2k^2-(\epsilon-k^2/(2m))^2]}{\sqrt{v_0^2k^2-(\epsilon-k^2/(2m))^2}} \\ \nonumber
&\approx&
\cos(\beta)
\left(\frac{\epsilon-k^2/(2m)}{v_0k}\right)
\frac{1}{\pi}
\frac{\Theta[v_0^2k^2-\epsilon^2]}{\sqrt{v_0^2k^2-\epsilon^2}}.
\end{eqnarray}
Substituting this result in Eq.~\eqref{Eq40} gives
\begin{footnotesize}
\begin{eqnarray}
p_0\cos(\beta)
=
-\frac{\tau_0}{T}
\frac{\cos(\beta)}{2\pi^2}
\int_0^\infty \frac{k^2dkg_k^2}{v_0k}\int_{-\infty}^\infty \epsilon d\epsilon \left(\epsilon-k^2/(2m)\right)
\frac{\Theta[v_0^2k^2-\epsilon^2]}{\sqrt{v_0^2k^2-\epsilon^2}}
F(k,\epsilon).
\end{eqnarray}
\end{footnotesize}
Since $F(k,\epsilon)$ is an odd function due to the term $\mathrm{sgn}(\epsilon)$, we find:
\begin{eqnarray}
p_0
=
-\frac{\tau_0}{\pi^2T2mv_0}
\int_0^\infty k^3dkg_k\int_0^{v_0k} d\epsilon \frac{\epsilon F(k,\epsilon)}{\sqrt{v_0^2k^2-\epsilon^2}}.
\end{eqnarray}
It is convenient here to introduce a new variable $t$: $\epsilon\rightarrow skt$, which yields:
\begin{footnotesize}
\begin{eqnarray}
2\pi^2p_0^2T
&=&
s^2\tau_0
\int_0^\infty k^4dkg_k
\int_0^{v_0/s}
\frac{tdt}{\sqrt{v_0^2-s^2t^2}}
F(k,skt)\\
\nonumber
&=&
s^2\tau_0
\int_0^\infty k^4dkg_k
\int_0^{v_0/s}
\frac{tdt}{\sqrt{v_0^2-s^2t^2}}
\left(\frac{ms}{4\pi}\right)^2
\frac{1}{\sinh(\frac{sk}{2T}t)}\\ \nonumber
&\times& \Bigg\{
\frac{\Theta(t-1)}{sk\sqrt{t^2-1}}
\int_0^1\frac{dz}{\sqrt{1-z^2}}
\frac{1}{\cosh(\frac{sk}{2T}t)-\cosh(\frac{sk}{2T}z)}\\ \nonumber
&-& \frac{\Theta(1-t)}{sk\sqrt{1-t^2}}
\int_1^\infty \frac{dz}{\sqrt{z^2-1}}   \frac{1}{\cosh(\frac{sk}{2T}t)-\cosh(\frac{sk}{2T}z)}
\Bigg\}.
\end{eqnarray}
\end{footnotesize}
Cancelling out $sk$, we get:
\begin{eqnarray}
2\pi^2p_0^2T
&=&
s\tau_0
\left(\frac{ms}{4\pi}\right)^2
\int_0^\infty k^3dkg^2_k
\int_0^{v_0/s}
\frac{tdt}{\sqrt{v_0^2-s^2t^2}}
\frac{1}{\sinh(\frac{sk}{2T}t)} \\ \nonumber
&\times& \Big\{
\frac{\Theta(t-1)}{\sqrt{t^2-1}}
\int_0^1\frac{dz}{\sqrt{1-z^2}}
\frac{1}{\cosh(\frac{sk}{2T}t)-\cosh(\frac{sk}{2T}z)}\\ \nonumber
&-& \frac{\Theta(1-t)}{\sqrt{1-t^2}}
\int_1^\infty \frac{dz}{\sqrt{z^2-1}}   \frac{1}{\cosh(\frac{sk}{2T}t)-\cosh(\frac{sk}{2T}z)} \Big\}.\nonumber
\end{eqnarray}

Now let us consider the integral
\begin{eqnarray}\label{AP7_int.1}
J&=&\int_0^\infty k^3dkg^2_k
\int_0^{v_0/s}
\frac{tdt}{\sqrt{v_0^2-s^2t^2}}
\frac{1}{\sinh(\frac{sk}{2T}t)}\times\\\nonumber
&\times&\Big\{
\frac{\Theta(t-1)}{\sqrt{t^2-1}}
\int_0^1\frac{dz}{\sqrt{1-z^2}}
\frac{1}{\cosh(\frac{sk}{2T}t)-\cosh(\frac{sk}{2T}z)} \\ \nonumber
&-& \frac{\Theta(1-t)}{\sqrt{1-t^2}}
\int_1^\infty \frac{dz}{\sqrt{z^2-1}}   \frac{1}{\cosh(\frac{sk}{2T}t)-\cosh(\frac{sk}{2T}z)}
\Big\}.
\end{eqnarray}
We assume $v_0>s$, as is typical for real structures. Thus, we have to deal with the expression
\begin{small}
\begin{eqnarray}
\label{int.2}
J&=&\int_0^\infty k^3dkg^2_k
\\
\nonumber
    &\times& \Biggl[
\int\limits_{1+L^{-1}/k}^{v_0/s}
\frac{tdt}{\sqrt{v_0^2-s^2t^2}}
\frac{1}{\sinh(\frac{sk}{2T}t)}\frac{1}{\sqrt{t^2-1}}
\int\limits_0^{1-L^{-1}/k}\frac{dz}{\sqrt{1-z^2}}
\frac{1}{\cosh(\frac{sk}{2T}t)-\cosh(\frac{sk}{2T}z)}\\
\nonumber
    &-&
\int\limits_0^{1-L^{-1}/k}
\frac{tdt}{\sqrt{v_0^2-s^2t^2}}
\frac{1}{\sinh(\frac{sk}{2T}t)}\frac{1}{\sqrt{1-t^2}}
\int\limits_{1+L^{-1}/k}^\infty \frac{dz}{\sqrt{z^2-1}}   \frac{1}{\cosh(\frac{sk}{2T}t)-\cosh(\frac{sk}{2T}z)}
\Biggr].
\end{eqnarray}
\end{small}
The two terms in the second and third lines diverge at $t\sim z\sim 1$. We therefore introduce new variables: $t-1=u$ and $1-z=v$ in the first term, and $1-t=u$ and $z-1=v$ in the second term. The two terms read:

\begin{eqnarray}\label{AP7_int.3}
\text{First }&=&\int\limits_{L^{-1}/k}^{v_0/s-1}
\frac{(1+u)du}{\sqrt{v_0^2-s^2(1+u)^2}}
\frac{1}{\sinh\left[\frac{sk}{2T}(1+u)\right]}\frac{1}{\sqrt{u(2+u)}}\\\nonumber
&\times&\int_{L^{-1}/k}^1\frac{dv}{\sqrt{v(2-v)}}
\frac{1}{\cosh\left[\frac{sk}{2T}(1+u)\right]-\cosh\left[\frac{sk}{2T}(1-v)\right]},
\end{eqnarray}
and
\begin{eqnarray}\label{AP7_int.4}
\text{Second }&=&\int\limits_{L^{-1}/k}^{1}
\frac{(1-u)du}{\sqrt{v_0^2-s^2(1-u)^2}}
\frac{1}{\sinh\left[\frac{sk}{2T}(1-u)\right]}\frac{1}{\sqrt{u(2-u)}}\\\nonumber
&\times&\int\limits_{L^{-1}/k}^\infty\frac{dv}{\sqrt{v(2+v)}}
\frac{1}{\cosh\left[\frac{sk}{2T}(1-u)\right]-\cosh\left[\frac{sk}{2T}(1+v)\right]}.
\end{eqnarray}
Now expanding these expressions for small $u$ and $v$, we find for the first term
\begin{gather}\label{AP7_int.5}
\frac{2T}{2sk\sqrt{v_0^2-s^2}}\frac{1}{\sinh^2\left[\frac{sk}{2T}\right]}\int\limits_{L^{-1}/k}^{v_0/s-1}
\frac{du}{\sqrt{u}}
\int_{L^{-1}/k}^1\frac{dv}{\sqrt{v}}
\frac{1}{u+v},
\end{gather}
and for the second term
\begin{gather}\label{AP7_int.6}
-\frac{2T}{2sk\sqrt{v_0^2-s^2}}\frac{1}{\sinh^2\left[\frac{sk}{2T}\right]}\int\limits_{L^{-1}/k}^{1}
\frac{du}{\sqrt{u}}
\int\limits_{L^{-1}/k}^\infty\frac{dv}{\sqrt{v}}
\frac{1}{u+v}.
\end{gather}
If $v_0\gg s$, we ultimately find
\begin{eqnarray}\label{AP7_int.7}
J&=&\frac{2T}{sv_0}\int\limits_{L^{-1}}^\infty\frac{k^2g_k^2dk}{\sinh^2\left[\frac{sk}{2T}\right]}\int\limits_{L^{-1}/k}^{1}
\frac{du}{\sqrt{u}}
\int\limits_{L^{-1}/k}^\infty\frac{dv}{\sqrt{v}}
\frac{1}{u+v}\\
&=&\frac{2\pi T}{sv_0}\int\limits_{L^{-1}}^\infty\frac{k^2g_k^2dk}{\sinh^2\left[\frac{sk}{2T}\right]}\ln(kL).
\end{eqnarray}
Here it becomes clear why we had to introduce the cut-offs $sL^{-1}$. Otherwise the integrals over $u$ and $v$ in Eq.~\eqref{AP7_int.7} would be diverging like $\ln(1/0)$.
It is important to note that $L$ does not considerably influence the resistivity, in contrast to the interlayer separation $l$, as can be seen in Fig.~\ref{AP7_Fig1S} below.

The resistivity, after restoring the constants, becomes
%
\begin{eqnarray}
\rho=\frac{M^2s^2}{8\pi^2e_0^2m^3v_F^5}\int\limits_{L^{-1}}^\infty
\frac{k^2g_k^2dk}{\sinh^2\left[\frac{\hbar sk}{2k_BT}\right]}\ln(kL).
\end{eqnarray}

We can evaluate the closed form of the integral for the two limiting cases of low and high temperatures.
First, we change the integration variable $x=2kl$ and write
\begin{eqnarray}
I\equiv\int_0^\infty\frac{k^2e^{-2kl}dk}{\sinh^2\left[\frac{\hbar sk}{2k_BT}\right]}\ln(kL)=\int_0^\infty\left(\frac{x}{2l}\right)^2\frac{e^{-x}\ln(\frac{Lx}{2l})}{\sinh^2\left(\frac{T_\textrm{BG}}{T}x\right)}dx.
\end{eqnarray}
For high temperatures $T\gg T_\textrm{BG}$,
\begin{eqnarray}
\sinh^2\left(\frac{T_\textrm{BG}}{T}x\right)\approx\left(\frac{T_\textrm{BG}}{T}x\right)^2,
\end{eqnarray}
and the fact that the main contribution of the integral comes from $0\leq x\lesssim 1$ gives
\begin{eqnarray}
I\approx\left(\frac{T}{T_{BG}}\right)^2\frac{1}{(2l)^3}\left[\ln\left(\frac{L}{2l}\right)-\gamma_C\right],
\end{eqnarray}
where $\gamma_C$ is the Euler gamma function.

For low temperatures $T\ll T_\textrm{BG}$, we obtain
\begin{eqnarray}
I\approx\left(\frac{T}{T_\textrm{BG}}\right)^3\frac{1}{(2l)^3}\ln\left(\frac{L}{2l}\right)\frac{\pi^2}{6}.
\end{eqnarray}
\begin{figure*}[ht]
\includegraphics[width=0.99\textwidth]{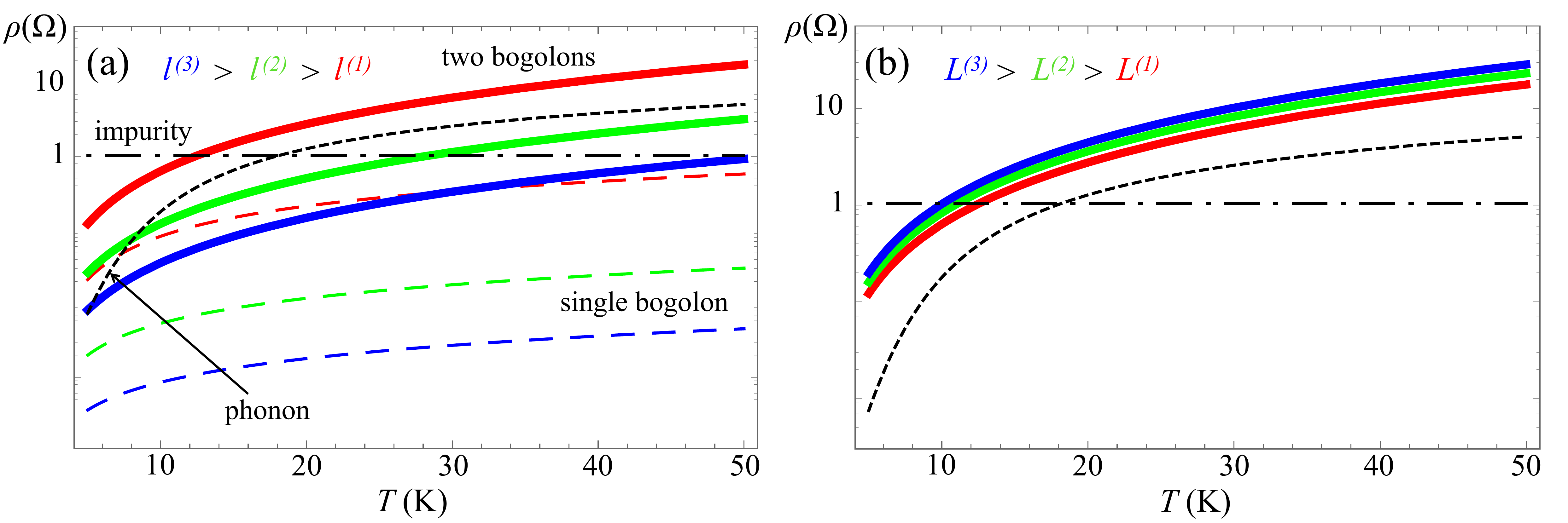}
\caption[Temperature dependence of single- and double-bogolon resistivity]{Temperature dependence of single- and double-bogolon resistivities at
(a) different interlayer separations: $l=5$ (red), $l=10$ (green), and $l=15$~nm (blue) for fixed $L=10$~$\mu$m,
and at
(b) different sample sizes: $L=10$ (red), $L=100$, and $L=1000$~$\mu$m (blue) for fixed $l=5$~nm.
The density of the MoS$_2$ condensate is taken as $n_c=10^9$~cm$^{-2}$, and the electron density as $n_e=10^{13}$~cm$^{-2}$.
The dashdotted and dashed curves show the corresponding impurity and phonon-mediated resistivities. The figure is taken from~\cite{Villegas:2019aa}.}
\label{AP7_Fig1S}
\end{figure*}
%
%
%

%
%

\section{Screening}
In this section, we calculate the screening factor $\epsilon_k$. In the presence of a condensate, it takes a usual form~\cite{Fetter:1971aa}
\begin{eqnarray}
\epsilon_k=(1-v_k\Pi_k)\left(1-\frac{e_0^2d}{\epsilon_0}P_k\right)-g_k^2\Pi_kP_k,
\end{eqnarray}
where $\Pi_k=-m/\pi$ and $P_k=-4Mn_c/k^2$ are the polarization operators for the electrons and exciton condensate, respectively, $v_k=2\pi e_0^2/k$ is the Coulomb interaction between electrons, and $g_k=e_0^2de^{-kl}/(2\epsilon_0)$ is the electron--exciton interaction.
After some algebra, we obtain
\begin{eqnarray}
\epsilon_k=1+\frac{2}{a_Bk}+\frac{1}{k^2\xi^2}+\frac{2}{a_Bk}\frac{1}{k^2\xi^2}\left(1-\frac{kd}{2}e^{-2kl}\right),
\end{eqnarray}
where $a_B$ is the Bohr radius.
For $l/d>1$ ,
\begin{eqnarray}
1-\frac{kd}{2}e^{-2kl}\approx 1,
\end{eqnarray}
and hence we find
\begin{eqnarray}
\epsilon_k=\left(1+\frac{2}{a_Bk}\right)\left(1+\frac{1}{k^2\xi^2}\right).
\end{eqnarray}
In order to account for the screening in our calculation of resistivity in Appendices A and B, we should simply replace
\begin{eqnarray}
|g_k|^2\rightarrow\left|\frac{g_k}{\epsilon_k} \right|^2.
\end{eqnarray}
%

%
%
\section{Validity of the linear spectrum for bogolons}
In this section, we provide a more quantitative analysis of the linear bogolon dispersion approximation.
Bogolon dispersion can be treated as linear if
\begin{eqnarray}
\xi < k^{-1}.
\end{eqnarray}
Due to the appearance of the factor
\begin{eqnarray}
g_k^2\sim e^{-2lk}
\end{eqnarray}
in the integral over $k$, the relevant contribution to the integral is only over the range
\begin{eqnarray}
2l<k^{-1}.
\end{eqnarray}
The linear spectrum approximation is guaranteed to be valid when:
\begin{eqnarray}
\xi &<& 2l\nonumber\\
\frac{\hbar}{2Ms}&<&2l\nonumber\\
\frac{\hbar}{4Ms}&<&s=\sqrt{\frac{\kappa n_c}{M}}.
\end{eqnarray}
This gives us the lower bound of the condensate density as
\begin{eqnarray}
n_c>\frac{\hbar^2}{16Ml^2\kappa}\approx 10^8~ \mbox{cm}^{-2},
\end{eqnarray}
for which the linear spectrum approximation is valid.